%% file: umetsu14.tex
\documentclass[a4paper,apj,numberedappendix]{emulateapj}
\usepackage[colorlinks=true,linkcolor=blue,anchorcolor=blue,citecolor=blue,urlcolor=blue]{hyperref}
\usepackage{color}
\usepackage{graphicx}
\usepackage{amssymb,amsmath}
\usepackage{natbib}
 
\newcommand{\simgt}{\lower.5ex\hbox{$\; \buildrel > \over \sim \;$}}
\newcommand{\simlt}{\lower.5ex\hbox{$\; \buildrel < \over \sim \;$}}
\newcommand{\llangle}{\langle\!\langle}
\newcommand{\rrangle}{\rangle\!\rangle}

\def\bc{\mbox{\boldmath $c$}}
\def\btheta{\mbox{\boldmath $\theta$}}

\def\bSigma{\mbox{\boldmath $\Sigma$}} 
\def\bDSigma{\mbox{\boldmath $\Delta\Sigma$}} 
\def\bg{\mbox{\boldmath $g$}}

\def\bd{\mbox{\boldmath $d$}}
\def\bs{\mbox{\boldmath $s$}}
\def\bp{\mbox{\boldmath $p$}}

\lefthead{Umetsu et al.}
\righthead{CLASH Weak Lensing Analysis of 20 Galaxy Clusters}

\begin{document}

\title{CLASH: Weak-Lensing Shear-and-Magnification Analysis of 20 Galaxy
Clusters\altaffilmark{*}}

\author{Keiichi Umetsu\altaffilmark{1}}  
\author{Elinor Medezinski\altaffilmark{2}}   
\author{Mario Nonino\altaffilmark{3}}        
\author{Julian Merten\altaffilmark{4,5}}     
\author{Marc Postman\altaffilmark{6}}        
\author{Massimo Meneghetti\altaffilmark{7,8,9}}  
\author{Megan Donahue\altaffilmark{10}}      
\author{Nicole Czakon\altaffilmark{1}}       
\author{Alberto Molino\altaffilmark{11}}     
\author{Stella Seitz\altaffilmark{12,13}}    
\author{Daniel Gruen\altaffilmark{12,13}}    
\author{Doron Lemze\altaffilmark{2}}         
\author{Italo Balestra\altaffilmark{14,3}}   
\author{Narciso Ben\'itez\altaffilmark{11}}  
\author{Andrea Biviano\altaffilmark{3}}      
\author{Tom Broadhurst\altaffilmark{15}}     
\author{Holland Ford\altaffilmark{2}}        
\author{Claudio Grillo\altaffilmark{16}}     
\author{Anton Koekemoer\altaffilmark{6}}     
\author{Peter Melchior\altaffilmark{17}}     
\author{Amata Mercurio\altaffilmark{14}}     
\author{John Moustakas\altaffilmark{18}}     
\author{Piero Rosati\altaffilmark{19}}       
\author{Adi Zitrin\altaffilmark{5,20}}       

\altaffiltext{*}
 {Based in part on data collected at the Subaru Telescope,
  which is operated by the National Astronomical Society of Japan.}
\email{keiichi@asiaa.sinica.edu.tw}
\altaffiltext{1}
 {Institute of Astronomy and Astrophysics, Academia Sinica,
  P.~O. Box 23-141, Taipei 10617, Taiwan}
\altaffiltext{2}{Department of Physics and Astronomy, The Johns Hopkins
University, 3400 North Charles Street, Baltimore, MD 21218, USA} 
\altaffiltext{3}{INAF-Osservatorio Astronomico di Trieste, via
G.B. Tiepolo 11, 34143 Trieste, Italy}
\altaffiltext{4}{Jet Propulsion Laboratory, California Institute of
Technology,  Pasadena, CA 91109, USA} 
\altaffiltext{5}{Cahill Center for Astronomy and Astrophysics,
California Institute of Technology, MS 249-17, Pasadena, CA 91125, USA}
\altaffiltext{6}{Space Telescope Science Institute, 3700 San Martin
Drive, Baltimore, MD 21208, USA} 
\altaffiltext{7}{Jet Propulsion Laboratory, California Institute of Technology, 4800 Oak Grove Drive, Pasadena, CA 91109, USA}
\altaffiltext{8}{INAF-Osservatorio Astronomico di Bologna, via Ranzani 1, 40127 Bologna, Italy}
\altaffiltext{9}{INFN, Sezione di Bologna, Viale Berti Pichat 6/2, 40127 Bologna, Italy}
\altaffiltext{10}{Department of Physics and Astronomy, Michigan State
University, East Lansing, MI 48824, USA} 
\altaffiltext{11}{Instituto de Astrof\'isica de Andaluc\'ia (CSIC),
Granada, Spain} 
\altaffiltext{12}{Universit\"ats-Sternwarte, M\"unchen, Scheinerstr. 1,
D-81679 M\"unchen. Germany}
\altaffiltext{13}{Max-Planck-Institut f\"ur extraterrestrische Physik,
Giessenbachstr. 1, 85748 Garching} 
\altaffiltext{14}{INAF-Osservatorio Astronomico di Capodimonte, Via
Moiariello 16 I-80131 Napoli, Italy} 
\altaffiltext{15}{Ikerbasque, Basque Foundation for Science, Alameda
Urquijo, 36-5 Plaza Bizkaia 48011, Bilbao, Spain}
\altaffiltext{16}{Dark Cosmology Centre, Niels Bohr Institute,
University of Copenhagen, Juliane Maries Vej 30, DK-2100 Copenhagen,
Denmark} 
\altaffiltext{17}{Center for Cosmology and Astro-Particle Physics \&
Department of Physics, The Ohio State University, Columbus, OH, USA.} 
\altaffiltext{18}{Department of Physics and Astronomy, Siena College, 515
Loudon Road, Loudonville, NY 12211, USA} 
\altaffiltext{19}{Universit\`a di Ferrara, Via Saragat 1, I-44122
Ferrara, Italy} 
\altaffiltext{20}{Hubble Fellow}


\begin{abstract}
We present a joint shear-and-magnification weak-lensing analysis of a sample
 of 16 X-ray-regular and 4 high-magnification galaxy clusters
at $0.19\simlt z\simlt 0.69$
 selected from the Cluster Lensing And Supernova survey with Hubble (CLASH).
Our analysis uses wide-field multi-color imaging,
taken primarily with Suprime-Cam on the Subaru  Telescope. 
From a stacked shear-only analysis of the X-ray-selected subsample,
we detect the ensemble-averaged lensing signal
with a total signal-to-noise ratio of $\simeq 25$
in the radial range of $200$ to $3500$\,kpc\,$h^{-1}$,
providing integrated constraints on the halo profile
 shape and concentration--mass relation.
The stacked tangential-shear signal is well described by a family of
standard density profiles predicted for dark-matter-dominated halos in
 gravitational equilibrium,
namely the
Navarro-Frenk-White (NFW),
truncated variants of NFW,
and Einasto models.
For the NFW model, we measure a mean concentration of
$c_{200{\rm c}}=4.01^{+0.35}_{-0.32}$ 
at an effective halo mass of $M_{200{\rm c}}=1.34^{+0.10}_{-0.09}\times 10^{15}M_\odot$.
We show this is in excellent agreement with $\Lambda$ cold-dark-matter
 ($\Lambda$CDM) predictions when the CLASH X-ray selection function and
 projection effects are taken into account.
The best-fit Einasto shape parameter is $\alpha_{\rm E}=0.191^{+0.071}_{-0.068}$, 
which is consistent with
the NFW-equivalent Einasto parameter of $\sim 0.18$. 
We reconstruct projected mass density profiles of all CLASH clusters
from a joint likelihood analysis of shear-and-magnification data,
and measure cluster masses at several characteristic radii
assuming an NFW density profile.
We also derive  an ensemble-averaged total projected mass profile of the
 X-ray-selected subsample by stacking their individual mass profiles.
The stacked total mass profile, constrained by the shear+magnification data,
is shown to be consistent with our shear-based halo-model predictions 
including the effects of surrounding large-scale structure as a two-halo
 term, establishing further consistency in the context of the
 $\Lambda$CDM model.
\end{abstract} 
 


\section{Introduction}
\label{sec:intro}

Clusters of galaxies represent the largest cosmic
structures that have reached a state in the vicinity of gravitational equilibrium.
The abundance of massive clusters as a function of redshift is
highly sensitive to the amplitude and growth rate of primordial
density fluctuations  
as well as the cosmic volume-redshift relation \citep{2001ApJ...553..545H}.
Clusters therefore play a fundamental role in examining cosmological models,
allowing several independent tests of any viable cosmology,
including the current concordance
$\Lambda$ cold dark matter ($\Lambda$CDM) model defined in the framework
of general relativity.

Clusters, by virtue of their enormous mass, serve as giant physics
laboratories for astronomers to explore the role and nature of dark
matter,
the physics governing the final state of
self-gravitating collisionless systems in an expanding universe
\citep{1972ApJ...176....1G,1996ApJ...462..563N,Taylor+Navarro2001,Lapi+Cavaliere2009a,Hjorth+2010DARKexp},
and screening mechanisms in
long-range modified models of gravity whereby general relativity is restored 
\citep{Narikawa+Yamamoto2012}.

A key ingredient of such cosmological tests is the mass distribution of
clusters.
In the standard picture of hierarchical
structure formation, cluster halos are located at dense
nodes where the filaments intersect and are still forming through
successive mergers of smaller halos as well 
as through smooth accretion of matter along their surrounding
large-scale structure (LSS).

The standard $\Lambda$CDM model and its variants
provide observationally testable predictions for the structure of
DM-dominated halos.
Cosmological $N$-body simulations of collisionless DM have established
a nearly self-similar form for the spherically-averaged density
profile $\rho(r)$ of equilibrium halos \citep[][hereafter,
NFW]{1996ApJ...462..563N} over a wide range of halo masses, with  
some intrinsic variance associated with mass accretion histories of individual halos
\citep{Jing+Suto2000,Merritt+2006,Graham+2006,Navarro+2010,Gao+2012Phoenix,Ludlow+2013}.
The degree of mass concentration, 
$c_{200{\rm c}}=r_{200{\rm c}}/r_{\rm s}$,\footnote{The quantity
$r_{200{\rm c}}$ is defined as the radius within which the mean density
is $200\times$ the critical density $\rho_c(z)$ 
of the universe at the cluster redshift $z$, and $r_{\rm s}$ is a scale
radius at which $d\ln{\rho}/d\ln{r}=-2$.}, 
is predicted to correlate with halo mass, since DM
halos that are more massive collapse later on average,
when the mean background density of the universe
is correspondingly lower
\citep{2001MNRAS.321..559B,2007MNRAS.381.1450N}.  
Accordingly, cluster-sized halos are predicted to 
be less concentrated than less massive systems,
and to have typical concentrations of 
$c_{200{\rm c}}\simeq 3$--$4$, 
compared to $c_{200\rm c}\simeq 5$ for group-szied halos
\citep{Duffy+2008,Bhatt+2013}.

Unlike individual galaxies, massive clusters
are not expected to be significantly affected by baryonic gas cooling  
\citep{Blumenthal+1986,Mead+2010AGN,Duffy+2010,Lau+2011,Blanchard+2013}
because the majority ($\sim 80\%$) of baryons in clusters comprise a
hot, X-ray-emitting phase of the intracluster medium, 
in which the high temperature
and low density prevent efficient cooling and gas contraction.
Consequently, for clusters in a state of quasi equilibrium, the form of their total mass profiles
reflects closely the underlying DM distribution.

Clusters act as gravitational lenses, producing various detectable
lensing effects, including deflection, shearing and magnifying of the
images of distant background sources. 
There is a weak-lensing regime where lensing effects can be
linearly related to the gravitational potential so that it is possible to
determine mass distributions in a model-free way.  
Weak lensing shear offers a direct means of 
mapping the mass distribution of clusters 
\citep{1999PThPS.133...53U,2001PhR...340..291B,Hoekstra+2013}
irrespective of the physical nature, 
composition, and state of lensing matter,
providing a direct probe for testing well-defined predictions of halo 
structure.

Lensing magnification provides
complementary, independent
observational alternatives to gravitational shear
\citep{1995ApJ...438...49B,UB2008,vanWaerbeke+2010,Umetsu+2011,Hildebrandt+2011,Ford+2012,Zemcov+2013,Coupon+2013}. 
Gravitational magnification influences the surface
density of background sources, expanding the area of sky, and
enhancing the observed flux of background sources 
\citep{1995ApJ...438...49B}. 
The former effect reduces the effective observing area in the source
plane, decreasing the source counts per solid angle. 
The latter effect increases the number of sources above the limiting
flux because the limiting luminosity $L_{\rm lim}(z)$
at any background redshift $z$
lies effectively at a fainter limit, $L_{\rm lim}(z)/\mu(z)$, 
with $\mu(z)$ the magnification factor.
The net effect is known as {\it magnification bias} and depends on the  
steepness of the luminosity function.

In practice, 
magnification bias can be used
in combination with weak lensing shear
to obtain a model-free determination of the projected mass profiles of
clusters  \citep{Schneider+2000,UB2008,Umetsu+2011,Umetsu2013}, 
effectively breaking degeneracies inherent in a standard weak-lensing
analysis based on shape information alone 
\cite[Section \ref{subsec:wl}; see also][]{Schneider+Seitz1995}.
Our earlier work has established that deep multi-colour imaging allows
us to simultaneously detect the observationally independent shear and
magnifications signals efficiently from the same data. 
The combination of shear and magnification allows us
not only to perform consistency checks of observational systematics but
also to enhance the precision and accuracy of cluster mass estimates 
\citep{Rozo+Schmidt2010,Umetsu+2012,Umetsu2013}.

The Cluster Lensing And Supernova survey with Hubble
\citep[CLASH,][]{Postman+2012CLASH}
has been designed to map the DM distribution in a
representative sample of 25 clusters,
by using high-quality strong- and weak-lensing
data, in combination with wide-field imaging from Suprime-Cam on the Subaru Telescope
\citep[e.g.,][]{Umetsu+2011,Umetsu+2011stack,Umetsu+2012}.
CLASH is a 524-orbit multi-cycle treasury {\it Hubble Space Telescope}
({\it HST}) program to observe 25 clusters at
$0.18 < z < 0.89$, each in 16 filters with the Wide Field Camera
3 \citep[WFC3,][]{Kimble+2008}
and the Advanced Camera for
Surveys 
\citep[ACS,][]{Holland+2003ACS}.

The CLASH sample is drawn largely from the Abell and MACS cluster
catalogs
\citep{Abell1958,Abell1989,Ebeling+2001MACS,Ebeling+2007,Ebeling+2010}.
Twenty CLASH clusters were X-ray selected to be massive and to
have a regular X-ray morphology.
This selection is suggested to minimize the strong bias toward high concentrations in previously
well-studied clusters selected for their strong-lensing
strength, allowing us to meaningfully examine the $c$--$M$ relation 
for a cluster sample that is largely free of lensing bias
\citep{Postman+2012CLASH}. 
A further sample of five clusters
were selected by their high lens
magnification properties, with the primary goal of detecting and studying
high-redshift background galaxies magnified by the cluster potential.

In this paper we present a joint shear-and-magnification weak-lensing 
analysis of a sample of 
16 X-ray-regular and 4 high-magnification clusters targeted in the CLASH survey.
Our analysis uses wide-field multi-band imaging
obtained primarily with Subaru/Suprime-Cam.
In particular, we aim at using the combination of shear and
magnification information to study ensemble-averaged mass density
profiles of CLASH clusters and compare with theoretical expectations 
in the context of the $\Lambda$CDM cosmology.
This work has two companion papers: The strong-lensing and weak-shear study of
CLASH clusters by \citet{Merten2014clash} and the detailed characterization
of numerical simulations of CLASH clusters by \citet[][hereafter, M14]{Meneghetti2014clash}.


The paper is organized as follows. 
In Section \ref{sec:basics}, we summarize the basic theory of
cluster weak gravitational lensing. 
In Section \ref{sec:method}, we present the formalism we use for our
weak-lensing analysis which combines shear
and magnification information.
In Section \ref{sec:data}, we describe the observational dataset, its
reduction, weak-lensing shape measurements, and the selection of background
galaxies. 
In Section \ref{sec:clash} we describe our joint shear-and-magnification
analysis of 20 CLASH clusters.
In Section \ref{sec:clash_stack} we carry out stacked weak-lensing
analyses of our X-ray-selected subsample to study their
ensemble-averaged mass distribution.  
Section \ref{sec:discussion} is devoted to the discussion of the results.
Finally, a summary is  given in Section \ref{sec:summary}.
 
Throughout this paper, we use the AB magnitude system, and
adopt a concordance $\Lambda$CDM cosmology
with 
$\Omega_{\rm m}=0.27$,  
$\Omega_{\Lambda}=0.73$, 
and
$h\equiv 0.7h_{70}=0.7$ \citep{Komatsu+2011WMAP7},
where $H_0 = 100h\, {\rm km\, s^{-1}\,Mpc^{-1}}$.
We use the standard notation $M_{\Delta_{\rm c}}$ ($M_{\Delta_{\rm m}}$)
to denote the total mass enclosed within a
sphere of radius $r_{\Delta_{\rm c}}$ ($r_{\Delta_{\rm m}}$), within
which the mean density is $\Delta_{\rm c}$ ($\Delta_{\rm m}$)
times the critical (mean background) density of the universe at the cluster
redshift. 
All quoted errors are 68.3\% ($1\sigma$)
confidence limits (CL) unless otherwise stated.

\section{Weak Lensing Basics}
\label{sec:basics}

\subsection{Convergence, Shear, and Magnification}
\label{subsec:wl}

The image deformation due to weak lensing
is characterized by the convergence $\kappa$, 
which describes the isotropic focusing of light rays,
and
the gravitational shear $\gamma(\btheta)=|\gamma|e^{2i\phi}$  
with spin-2 rotational symmetry
\citep[e.g.,][]{2001PhR...340..291B}.
The lensing convergence is
$\kappa(\btheta)=\Sigma(\btheta)/\Sigma_{\rm c}$,
the projected mass density $\Sigma(\btheta)$
in units of the critical surface mass density for lensing,
\begin{equation} 
\label{eq:sigmacrit}
\Sigma_{\rm c} = \frac{c^2}{4\pi G}\frac{D_{\rm s}}{D_{\rm l} D_{\rm ls}}
\equiv  \frac{c^2}{4\pi G D_{\rm l}}\beta^{-1}
\end{equation}
with $D_{\rm l}$, $D_{\rm s}$, and $D_{\rm ls}$ the proper angular diameter
distances from the observer to the lens, 
the observer to the source, and the lens to the source, respectively.
The distance ratio $\beta=D_{\rm ls}/D_{\rm s}$ represents the geometric
strength of cluster lensing for a source at redshift $z$; $\beta(z)=0$
for unlensed objects, $z\le z_{\rm l}$.

The shear $\gamma(\btheta)$ induces a quadrupole anisotropy of the  
background images, which can be observed from ellipticities 
of background galaxies.
Given an arbitrary circular loop of radius $\theta$,
the tangential shear $\gamma_{+}$
and the $45^\circ$-rotated component $\gamma_{\times}$ 
averaged around the loop satisfy
the following {\it identity}
\citep{Kaiser1995}:
\begin{equation} 
\label{eq:loop}
\gamma_{+}(\theta) =
\kappa(<\theta)-\kappa(\theta) \equiv 
\Delta\Sigma_+(\theta)/\Sigma_{\rm c}, \ \ \
\gamma_{\times}(\theta)=0 
\end{equation}
with $\kappa(\theta)=\Sigma(\theta)/\Sigma_{\rm c}$ 
the azimuthally-averaged convergence at radius $\theta$,
$\kappa(<\theta)=\Sigma(<\theta)/\Sigma_{\rm c}$ 
the average convergence interior to $\theta$, and
$\Delta\Sigma_+(\theta)=\Sigma_{\rm c}\gamma_+(\theta)$
the {\it differential} surface mass density.
In general, the observable quantity for quadrupole weak lensing
is not $\gamma$ but the {\it reduced} gravitational shear,
\begin{equation}
\label{eq:redshear}
g(\btheta)=\frac{\gamma(\btheta)}{1-\kappa(\btheta)},
\end{equation}
which is invariant under 
$\kappa(\btheta) \to \lambda \kappa(\btheta) + 1-\lambda$ 
and
$\gamma(\btheta) \to \lambda \gamma(\btheta)$
with an arbitrary constant $\lambda\ne 0$,
known as the mass-sheet degeneracy
\citep[see][]{Schneider+Seitz1995}.
This degeneracy can be broken,
for example,\footnote{Alternatively, the constant $\lambda$ can be
determined such that the $\kappa$ averaged over the outermost cluster region
vanishes, if a sufficiently wide sky coverage is available.
Or, one may constrain $\lambda$ such that the enclosed mass
within a certain aperture is consistent with cluster mass
estimates from some other observations
\citep[e.g.,][]{Umetsu+Futamase1999}. 
}
by measuring the magnification 
\begin{equation}
\label{eq:mu2d}
\mu(\btheta) = \frac{1}{[1-\kappa(\btheta)]^2-|\gamma(\btheta)|^2},
\end{equation}
which transforms as $\mu(\btheta)\to \lambda^2\mu(\btheta)$.

\subsection{Source Redshift Distribution}
\label{subsec:pop}

For statistical weak-lensing measurements, 
we consider populations of source galaxies with respective redshift distribution
functions $\overline{N}(z)$.
The mean lensing depth for a given population is given by
\begin{equation}
\label{eq:depth}
\langle\beta\rangle =\left[
\int_0^\infty\!dz\, w(z)\overline{N}(z) \beta(z)\right]
\left[
\int_0^\infty\!dz\,w(z)\overline{N}(z)
\right]^{-1},
\end{equation}
where $w(z)$ is a weight factor.
In general,  we apply different size, magnitude, and color cuts in source selection
for measuring the shear and magnification effects, resulting
in different $\overline{N}(z)$. In contrast to the former effect, the
latter does not require source galaxies to be spatially  resolved, but
does require a stringent flux limit against incompleteness effects.

We introduce the relative lensing strength of a source population 
with respect to a fiducial source in the far background as
\citep{2001PhR...340..291B}
\begin{equation}
\langle W\rangle  
= \langle\beta\rangle  / \beta_\infty,
\end{equation}
with $\beta_\infty\equiv \beta(z\to \infty; z_{\rm l})$.
The associated critical surface mass density is
$\Sigma_{{\rm c},\infty}(z_{\rm l})=c^2/(4\pi G D_{\rm l}\beta_{\infty})$.
The source-averaged convergence and shear fields are then expressed as
$\langle\kappa(\btheta) \rangle =\langle W\rangle \kappa_\infty(\btheta)$ and 
$\langle\gamma(\btheta) \rangle =\langle W\rangle \gamma_\infty(\btheta)$, 
using those in the far-background limit.
Hereafter, we use the far-background lensing fields,
$\kappa_\infty$ and $\gamma_\infty$, 
to describe the projected mass distribution of clusters.

\section{Cluster Analysis Methodology}
\label{sec:method}

In this section, we present the formalism that we use for our
weak-lensing analysis, which combines complementary shear and
magnification information.
In Sections \ref{subsec:gt} and \ref{subsec:magbias} we first describe
our methods for measuring cluster lensing profiles as a function of
cluster radius.
In Section \ref{subsec:bayesian}, we outline our Bayesian approach 
for reconstructing the projected mass profile 
from a joint-likelihood analysis of shear+magnification measurements.
In Section \ref{subsec:stack}, 
we describe our stacked analysis
formalism and procedures for determining the ensemble-averaged lensing profiles.

\subsection{Tangential-distortion Profile}
\label{subsec:gt}

We construct azimuthally-averaged radial profiles of
the tangential distortion $g_+$ and  
the $45^\circ$-rotated component $g_\times$ as functions of
cluster radius $\theta$ \citep{UB2008,Umetsu2013}.
In the absence of higher order effects, weak lensing induces only
curl-free tangential distortions (Section \ref{subsec:wl}). 
In practice, the presence of $\times$ modes can be
used to check for systematic errors.

The tangential distortion averaged over the source redshift distribution 
$\overline{N}_{g}(z)$ is expressed as
$\langle g_+\rangle =
 \left[\int_0^\infty\!dz\,g_+(W[z])\overline{N}_{g}(z) \right] 
 \left[\int_0^\infty\!dz\,\overline{N}_{g}(z) \right]^{-1}$.
The averaging operator with respect to $\overline{N}(z)$ acts
nonlinearly on the redshift-dependent components 
in the cluster lensing observables. 
In the mildly-nonlinear regime, it is often sufficient 
to apply a low-order approximation using low-order moments of the
source-averaged lensing depth.

Specifically, we use the following approximation for the nonlinear
corrections to the source-averaged reduced shear profile 
\citep{Seitz1997,Umetsu2013}:
\begin{eqnarray}
\label{eq:nlcor}
\langle g_+\rangle
\approx
\frac{\langle W\rangle_{g} 
\left[
\kappa_\infty(<\theta) -\kappa_\infty(\theta)\right]}
{1- \kappa_\infty(\theta) 
\langle W^2\rangle_{g}/\langle W\rangle_{g}}
=
\frac{\langle\gamma_+\rangle}{1-f_{W,g}\langle \kappa\rangle}
\end{eqnarray}
where $\langle W\rangle_{g}$ is the relative lensing strength (see
Section \ref{subsec:pop}) averaged over the population $N_{g}(z)$ of
source galaxies, 
$f_{W,g}\equiv \langle W^2\rangle_{g}/\langle W\rangle_{g}^2$ 
is a dimensionless quantity of the order unity,  
$\langle \kappa\rangle = \langle W\rangle_{g} \kappa_\infty$,
and
$\langle \gamma_+\rangle = \langle W\rangle_{g} \gamma_{+,\infty}$.

\subsection{Magnification-bias Profile}
\label{subsec:magbias}

\subsubsection{Magnification Bias}


Since a given flux limit corresponds
to different intrinsic luminosities at different source redshifts,
count measurements of distinctly different background
populations probe different regimes of magnification-bias
effects.
Deep multi-band photometry allows us to explore the faint end of the
intrinsic luminosity function of red galaxies at $z\sim 1$ 
\citep[e.g., Figures 11--13 of][]{Ilbert2010}.
For a flux-limited sample of the faint red background
population, the effect of magnification bias is dominated by the geometric area
distortion because relatively few fainter galaxies can be magnified into the sample,
thus resulting in a net count depletion 
\citep{1998ApJ...501..539T,BTU+05,UB2008,Umetsu+2011,Umetsu+2012,Coe+2012A2261,Medezinski+2013}. 

In the present work, we perform count-depletion measurements 
using flux-limited samples of red background galaxies.
If the magnitude shift $\delta m=2.5\log_{10}\mu$ induced
by magnification is small compared to that on which 
the logarithmic slope of the source luminosity function
varies, 
the source number counts can be locally approximated by a power law at
the limiting magnitude $m$.  The magnification bias 
at redshift $z$ is then given by
\citep{1995ApJ...438...49B}
\begin{eqnarray} 
\label{eq:magbias}
N_\mu(\btheta,z; <m) &=& 
\overline{N}_\mu\,\mu^{2.5s-1}(\btheta,z)
\equiv\overline{N}_\mu b_\mu(\btheta,z; s), 
\end{eqnarray}   
where $\overline{N}_\mu=\overline{N}_\mu(z;<m)$ is the
unlensed mean source counts and
$s$ is the logarithmic count slope evaluated at $m$,
$s=d\log_{10} \overline{N}_\mu(z; <m)/dm$.
In the regime where $2.5s\ll 1$, the
effect of magnification bias is dominated by the geometric expansion of
the sky area, and hence is not sensitive to the exact form of the
intrinsic source luminosity function.


Taking into account the spread of $\overline{N}_\mu(z)$, we express the
population-averaged magnification bias as 
$\langle b_\mu\rangle =\left[\int_0^\infty\!dz\,b_\mu(W[z])\overline{N}_{\mu}(z) \right] 
\left[\int_0^\infty\!dz\,\overline{N}_{\mu}(z) \right]^{-1}$.
In this work, we use the following equation to interpret the observed
count-depletion measurements \citep{Umetsu2013}: 
\begin{equation}
\label{eq:bmu_approx}
\langle b_\mu(\btheta)\rangle =N_\mu(\btheta; <m)/\overline{N}_\mu(<m)\approx \langle \mu^{-1}(\btheta)\rangle^{1-2.5s_{\rm eff}}
\end{equation}
with  $s_{\rm eff}=d\log_{10}\overline{N}_{\mu}(<m)/dm$ the effective
count slope defined in analogy to Equation (\ref{eq:magbias}).
Equation (\ref{eq:bmu_approx}) is exact for $s_{\rm eff}=0$, and gives a good approximation for depleted background populations with $s_{\rm eff}\ll 0.4$
\citep{Umetsu2013}.
Furthermore, the source-averaged inverse magnification is approximated
as \citep{Umetsu2013}
\begin{eqnarray}
\label{eq:mu}
\langle \mu^{-1}\rangle
&=&
(1-\langle \kappa\rangle)^2 - |\langle\gamma\rangle|^2
+(f_{W,\mu}-1)
(\langle \kappa\rangle^2 - |\langle\gamma\rangle|^2)\nonumber\\
&\approx&
(1-\langle \kappa\rangle)^2 - |\langle\gamma\rangle|^2,
\end{eqnarray}
where $f_{W,\mu}\equiv \langle W^2\rangle_\mu/\langle W\rangle_\mu^2$ is of the order unity,
$\langle\kappa\rangle=\langle W\rangle_\mu\kappa_\infty$,
and
$\langle\gamma\rangle=\langle W\rangle_\mu\gamma_{\infty}$.
The error associated with the approximation above is $\langle\Delta
\mu^{-1}\rangle=(f_{W,\mu}-1)(\langle\kappa\rangle^2-|\langle\gamma\rangle|^2)
\equiv \Delta
f_{W,\mu}(\langle\kappa\rangle^2-|\langle\gamma\rangle|^2)$, which is
much smaller than unity for source populations of our concern 
($\Delta f_{W,\mu}\sim O(10^{-2})$) in the regime where
$\langle\kappa\rangle\sim |\langle\gamma\rangle|\sim O(10^{-1}$).

\subsubsection{Number-count Depletion}
\label{subsubsec:depletion}

In practical observations, 
the nonvanishing and unresolved angular correlation on small angular
scales can lead to a significant increase in the variance of counts
in cells \citep{2000MNRAS.313..524V}, 
which can be much larger than the lensing signal in a given cell. 
To obtain a clean measure of the
lensing signal, such intrinsic variance needs to be downweighted
\citep{UB2008}
and averaged over a sufficiently large sky area.

This local clustering noise can be largely overcome by performing
an azimuthal average around the cluster
\citep{Umetsu+2011,Umetsu2013}.
Here we calculate the surface number
density of background sources $n_\mu(\theta)=dN_\mu(\theta)/d\Omega$
as a function of cluster radius, by azimuthally
averaging the source counts in cells, $N_\mu(\btheta; <m)$.
The source-averaged magnification bias is then expressed as 
 $\langle n_\mu(\theta)\rangle = \overline{n}_\mu \langle \mu^{-1}(\theta)\rangle ^{1-2.5s_{\rm eff}}$ with $\overline{n}_\mu=d\overline{N}_\mu(<m)/d\Omega$
the unlensed mean surface number density of background sources.

In this work,
we adopt the following prescription: 
\begin{enumerate}
\item A positive tail of $>\nu\sigma$ cells is excluded in each annulus
      by iterative $\sigma$ clipping with $\nu=2.5$
      to reduce the effect of intrinsic angular clustering of
      source galaxies. 
      We take the systematic change between the mean counts estimated
      with and without $\nu\sigma$ clipping as a systematic error,
      $\sigma_\mu^{\rm sys}(\theta)=|n_\mu^{(\nu)}(\theta)-n_\mu^{(\infty)}(\theta)|/\nu$,
      where $n_\mu^{(\nu)}(\theta)$ and $n_\mu^{(\infty)}(\theta)$ are
      the clipped and unclipped mean counts in the annulus $\theta$,
      respectively.
      The statistical Poisson uncertainty $\sigma_{\mu}^{\rm
      stat}(\theta)$ is estimated from the clipped mean counts. 
\item An additional contribution to the uncertainty from the
      intrinsic clustering of source galaxies, $\sigma_\mu^{\rm int}(\theta)$, is estimated empirically
      from the variance in each annulus due to variations of the counts $N_\mu(\btheta)$
      along the azimuthal direction.
\item Each grid cell is weighted by the fraction of its area lying
      within the respective annular bins.
      We use Monte Carlo integration to calculate the area
      fractions for individual cells. 
\item Masking of background galaxies due to cluster galaxies,
      foreground objects, and saturated pixels is properly taken into
      account and corrected for,
      by calculating at each annulus
      $f_{\rm mask}(\theta)=\Delta\Omega_{\rm mask}(\theta)/\Delta\Omega_{\rm tot}(\theta)$
      with $\Delta\Omega_{\rm mask}(\theta)$ the area of masked
      regions in the annulus and
       $\Delta\Omega_{\rm tot}(\theta)$ the total area of the annulus.
      In our analysis, we use Method B of Appendix A developed by
      \citet{Umetsu+2011}. 
\end{enumerate}

The errors 
are combined in quadrature as $\sigma_\mu^2 = (\sigma_\mu^{\rm stat})^2
+ (\sigma_\mu^{\rm int})^2 + (\sigma_\mu^{\rm sys})^2$.
We note
the $\sigma_\mu^{\rm sys}$ contribution may account for 
(i) strong contamination by background clusters projected near the line
of sight,
and
(ii) spurious excess counts of red galaxies due perhaps to spatial variation
of the photometric zeropoint and/or to residual flatfield errors.

Finally, we apply the correction to the number counts for the masking
effects by
$n_\mu(\theta)\to n_\mu(\theta)/[1-f_{\rm mask}(\theta)]$
and
$\sigma_\mu(\theta)\to \sigma_\mu(\theta)/[1-f_{\rm mask}(\theta)]$.
Similarly, this correction is applied to the mean background counts
$\overline{n}_\mu$ and its total uncertainty.
The typical level of this correction is about $8\%$ in our weak-lensing
observations (see Section \ref{subsec:wlmass}).  
 
\subsection{Bayesian Mass-profile Reconstruction}
\label{subsec:bayesian}

\subsubsection{Joint Likelihood Function}
\label{subsubsec:likelihood}

In the Bayesian framework of \cite{Umetsu+2011}, 
the signal is described by a vector $\bs$ of parameters 
containing the binned convergence profile
$\{\kappa_{\infty,i}\}_{i=1}^N$ given by 
$N$ binned $\kappa$ values
and the average convergence enclosed by the innermost aperture radius
$\theta_{\rm min}$,
$\kappa_{\infty,{\rm min}}\equiv \kappa_{\infty}(<\theta_{\rm min})$,
so that 
\begin{equation}
\bs=\{\kappa_{\infty, {\rm min}}, \kappa_{\infty,i}\}_{i=1}^N
\equiv \Sigma_{{\rm c},\infty}^{-1}\bSigma
\end{equation}
specified by $(N+1)$ parameters. 

The joint likelihood function ${\cal L}(\bs)$ for
multi-probe lensing observations is given as a product of 
their separate likelihoods, 
${\cal L}={\cal  L}_g{\cal  L}_\mu$,
with ${\cal L}_{g}$ and ${\cal L}_{\mu}$
the likelihood functions for 
$\{\langle g_{+,i}\rangle\}_{i=1}^N$
and
$\{\langle n_{\mu,i}\rangle\}_{i=1}^N$,  
defined as 
\begin{eqnarray}
\ln{\cal L}_{g}(\bs)&=& -\frac{1}{2}\sum_{i=1}^{N}\frac{[\langle g_{+,i}\rangle-\hat{g}_{+,i}(\bs,\bc)]^2}{\sigma^2_{+,i}},\\
\ln{\cal L}_\mu(\bs)&=& -\frac{1}{2}\sum_{i=1}^{N}\frac{[\langle n_{\mu,i}\rangle-\hat{n}_{\mu,i}(\bs,\bc)]^2}{\sigma^2_{\mu,i}},
\end{eqnarray}
with ($\hat{g}_+,\hat{n}_\mu$) the
theoretical predictions for the corresponding observations
and $\bc$ the calibration nuisance parameters to marginalize over,
\begin{equation} 
\label{eq:cparam}
\bc=\{\langle W\rangle_{g}, f_{W,g}, \langle W\rangle_\mu, \overline{n}_\mu, s_{\rm eff}\}.
\end{equation}

For each parameter of the model $\bs$, we consider a flat uninformative
prior with a lower bound of $\bs=0$.
Additionally, we account for the calibration uncertainty in the
observational parameters $\bc$. 

\subsubsection{Estimators and Covariance Matrix}
\label{subsubsec:cmat}

We implement our method using a Markov Chain Monte Carlo (MCMC) 
approach with Metropolis-Hastings sampling, 
by following the prescription outlined in \citet{Umetsu+2011}. 
The shear+magnification method has been tested \citep{Umetsu2013}
against synthetic weak-lensing catalogs from simulations of analytical
NFW lenses performed using the public package {\sc glafic}
\citep{Oguri2010glafic}.  
The results suggest that, when the mass-sheet degeneracy is
broken, both maximum-likelihood (ML) and Bayesian marginal maximum a
posteriori probability (MMAP) solutions produce reliable
reconstructions with unbiased profile  measurements,
so that this method is not sensitive to the choice and form of priors.
In the presence of a systematic bias in the background-density constraint
($\overline{n}_\mu$), the global ML estimator is less sensitive
to systematic effects than MMAP, and provides robust reconstructions 
(Section \ref{subsubsec:backdens}). 

On the basis of our simulations, we thus use the global ML estimator for
determination of the mass profile.  In our error analysis we take into
account statistical, systematic, and cosmic-noise contributions to the
total covariance matrix $C_{ij}\equiv {\rm Cov}(s_i,s_j)$ as
\begin{equation}
\label{eq:Ckappa}
C= C^{\rm stat} + C^{\rm sys} + C^{\rm lss},
\end{equation}
where
$C^{\rm stat}$ is estimated from the posterior MCMC samples, 
$C^{\rm sys}$ accounts for systematic errors due  primary  to the mass-sheet uncertainty,
\begin{equation}
\label{eq:Csys}
C^{\rm sys}_{ij}=(\bs_{\rm ML}-\bs_{\rm MMAP})_i^2\delta_{ij},
\end{equation}
with $\bs_{\rm ML}$ and $\bs_{\rm MMAP}$ the ML and MMAP solutions,
respectively,
and $C^{\rm lss}$ is the cosmic noise covariance due to
uncorrelated LSS projected along the line of sight
\citep{2003MNRAS.339.1155H,Hoekstra+2011,Umetsu+2011stack}. 
For a given depth of weak-lensing observations, the impact of cosmic
noise is most important when the cluster signal itself is small
\citep{2003MNRAS.339.1155H}, 
that is, when nearby clusters are studied, or when data at large cluster
radii are examined.

The $C^{\rm lss}$ matrix is computed for a given source population
as outlined in
\citet{Umetsu+2011stack},\footnote{Note $C^{\rm lss}$ is calculated for our weak-lensing source
populations, and then scaled to the reference far-background source plane.}
using the nonlinear matter power spectrum of 
\citet{Smith+2003halofit} 
for the {\it Wilkinson Microwave Anisotropy Probe} (WMAP)
seven-year cosmology \citep{Komatsu+2011WMAP7}.



\subsection{Stacked Lensing Formalism}
\label{subsec:stack}


\subsubsection{Stacked Tangential-shear Profile}
\label{subsubsec:stackgt}

First we derive an expression for the averaged $\Delta\Sigma_+$ profile 
from stacking of tangential distortion signals around the cluster
centers, 
following the general procedure of \citet{Umetsu+2011stack}. 
For a given cluster sample, we center the shear catalogs on the
respective cluster centers, and construct their individual distortion
profiles  $\langle\bg_+\rangle$
in {\it physical proper length} units 
across the range $R=D_{\rm l}\theta =[R_{\rm min},R_{\rm max}]$.
As we shall see below,  
our choice of stacking in physical length units is to reduce
systematic errors 
in determining the ensemble-averaged cluster mass
profile from stacked lensing measurements \citep{Okabe+2013},
although this choice is not optimized for the signal-to-noise ratio
(S/N) of the stacked signal.

For each cluster, we express the covariance matrix $(C_+)_{ij}$
of $\langle\bg_+\rangle$ as a sum of the contributions from the shape
noise ($C_+^{\rm stat}$) 
and the cosmic shear ($C_+^{\rm lss}$) due to uncorrelated LSS projected
along the line of sight \citep{2003MNRAS.339.1155H},
$C_+ = C_+^{\rm stat}+C_+^{\rm lss}$,
where $(C_+^{\rm stat})_{ij}=\sigma_{+,i}^2\delta_{ij}$ is estimated
from bootstrap resampling of the background source catalog, and
$C_+^{\rm lss}$ is computed for a given source population
(see Section \ref{subsubsec:cmat}).
This cosmic noise is correlated between radial bins, but
can be overcome by stacking an ensemble of clusters along
independent lines of sight \citep{Umetsu+2011stack}.


Since the noise in different clusters is uncorrelated, the tangential
distortion profiles of individual clusters
can be co-added according to \citep{Umetsu+2011stack}
\begin{equation}
\label{eq:stack_g}
\llangle \widehat{\bDSigma_+}\rrangle = 
\left(\displaystyle\sum_n {{\cal W}_{+,n}} \right)^{-1}
 \,
\left(
\displaystyle\sum_n{ {{\cal W}_{+,n}} \widehat{\bDSigma}_{+,n} }
\right), 
\end{equation}
where $\llangle ...\rrangle$ denotes the sensitivity-weighted average over all
clusters in the sample,
$\widehat{\bDSigma_{+,n}} = \Sigma_{{\rm c},n} \, \langle\bg_{+,n}\rangle$,
and
${{\cal W}_{+,n}}$ is the {\it shear-sensitivity} matrix of the
$n$th cluster
\begin{equation}
\label{eq:wmat}
({{\cal W}_{+,n}})_{ij} \equiv \Sigma_{{\rm c},n}^{-2}\,\left({C_{+,n}}^{-1}\right)_{ij} 
\end{equation}
with $C_{+,n}$ the covariance matrix of the $n$th 
$\langle \bg_+\rangle$ profile.

The statistical covariance matrix ${\cal C}_+^{\rm stat}$ for 
$\llangle \widehat{\bDSigma}_+\rrangle$ is estimated from bootstrap
resampling of the cluster sample in Equation (\ref{eq:stack_g}), which
accounts for the statistical total variation of the mean mass profile
averaged over the sample. 
Additionally, we include in our error analysis the photo-$z$
uncertainties on the mean depth calibration,
${\cal C}^{\rm sys}$, and the {\it net} cosmic noise covariance,  
\begin{equation}
 {\cal C}_+^{\rm lss} =\left[
			\displaystyle \sum_n \Sigma_{{\rm c},n}^{-2} ({C^{\rm lss}_{+,n}})^{-1}
		       \right]^{-1}.
\end{equation}
Finally, the full covariance matrix for  
$\llangle\widehat{\bDSigma}_+\rrangle$ is  obtained as
\begin{equation}
\label{eq:covtot_gstack}
{\cal C}_+ = {\cal C}_+^{\rm stat} + {\cal C}_+^{\rm sys} + {\cal C}_+^{\rm lss}.
\end{equation}

The relation between the observable lens distortion and the lensing
fields is nonlinear (Equation (\ref{eq:redshear})),
 so that the stacked $\llangle \widehat{\Delta\Sigma_+}\rrangle$ profile
 is nonlinearly related to the averaged lensing fields.
Expanding the right hand side of Equation (\ref{eq:stack_g}) 
and taking the ensemble average,
we obtain the next-to-leading order correction as
\begin{eqnarray}
\label{eq:stack_nlincor}
\llangle \widehat{\bDSigma_{+}}\rrangle  
&=&
{\rm E}[\bDSigma_{+}]
+
\llangle \Sigma_{\rm c}^{-1} \rrangle
{\rm E}[\bSigma\bDSigma_+]\nonumber\\
&\approx&
{\rm E}[\bDSigma_{+}] \big/ \left(
1-\llangle \Sigma_{\rm c}^{-1} \rrangle {\rm E}[\bSigma]
\right),
\end{eqnarray}
where ${\rm E}[...]$ denotes the ensemble average, and
we have used the trace approximation 
$({{\cal W}_{+,n}})_{ij} \propto \delta_{ij}{\rm tr}({{\cal W}_{+,n}})$ 
in the first line of Equation (\ref{eq:stack_nlincor});
$\llangle \Sigma_{\rm c}^{-1}\rrangle$ is defined as
\begin{equation}
\label{eq:w_eff}
\llangle \Sigma_{\rm c}^{-1}\rrangle =
\frac{\sum_n {\rm tr}({{\cal W}_{+,n}}) 
\Sigma_{{\rm c},n}^{-1}}{\sum_n {\rm tr}({{\cal W}_{+,n}})},
\end{equation}
where the statistical weight ${\rm tr}({{\cal W}_{+,n}})$ 
is proportional to  $1/\Sigma_{{\rm c},n}^2$ (Equation (\ref{eq:wmat})).
We note that ${\rm tr}({{\cal W}_{+,n}})$ is independent of the
cluster mass when stacking in {\it physical length units} 
\citep{Okabe+2010WL,Okabe+2013,Umetsu+2011stack,Oguri+2012SGAS}.
On the other hand, as discussed by \citet{Okabe+2013},
stacking in length units scaled to $r_{\Delta}$
weights the contribution of each cluster to each radial bin in a
nonlinear and model-dependent manner, such that 
${\rm tr}{\cal W}_+ \propto r_\Delta^2 \propto M_\Delta^{2/3}$
when $C_+$is dominated by the shape noise contribution ($C_+^{\rm stat}$).

In this work, we shall use Equations (\ref{eq:stack_nlincor}) and
(\ref{eq:w_eff}) to obtain a best-fit model for a given 
$\llangle \widehat{\bDSigma}_+\rrangle$. 
The $\llangle {\bDSigma}_+\rrangle$ is then obtained as 
\begin{equation}
\label{eq:DSigma_approx}
\llangle \bDSigma_+\rrangle  \approx \llangle
 \widehat{\bDSigma_+}\rrangle
-
\llangle \Sigma_{\rm c}^{-1}\rrangle 
 (\bSigma\bDSigma_+),
\end{equation}
where the nonlinear correction $(\bSigma\bDSigma_+)$ is calculated using
the best-fit solution to $\llangle \widehat{\bDSigma}_+\rrangle$ in Equation (\ref{eq:stack_nlincor}).


\subsubsection{Stacked Mass Profile}
\label{subsubsec:stackwl}

Having obtained the mass density profiles $\bSigma$ of
individual clusters from combined weak-lensing shear and magnification
measurements (Section \ref{sec:method}), we can 
stack the clusters to produce an averaged radial mass profile.

Following \citet{Umetsu+2011stack},
we re-evaluate the mass profiles of individual clusters 
in proper length units across the range 
$R=[R_{\rm min},R_{\rm max}]$,
and construct 
$\bSigma=\{\Sigma(<R_{\rm min}),\Sigma(R_i)\}_{i=1}^N$
on the same radial grid for all clusters.
Stacking an ensemble of clusters  is expressed as
\begin{equation}
\label{eq:stack_mass}
\llangle \bSigma \rrangle = 
\left(\displaystyle\sum_n {\cal W}_n \right)^{-1}
 \,
\left(
\displaystyle\sum_n{ {\cal W}_n \bSigma_n}
\right), 
\end{equation}
where ${\cal W}_n$ is the $n$th sensitivity matrix defined as
$({\cal W}_n)_{ij} \equiv \Sigma_{({\rm c},\infty)n}^{-2} \, \left({C}^{-1}_n\right)_{ij}$
with $C_n$ the total covariance matrix (Equation
(\ref{eq:Ckappa})) of the $n$th cluster.\footnote{Since the covariance matrix $C$ is defined
for the far-background convergence $\kappa_\infty$, the associated
critical surface mass density too is a far-background quantity,
$\Sigma_{{\rm c},\infty} = \Sigma_{\rm c}(z\to\infty)$.}

Finally, 
the full covariance matrix ${\cal C}$ for the stacked $\llangle \bSigma\rrangle$ profile
can be obtained in a similar manner as for $\llangle \bDSigma_+\rrangle$ 
(Equation (\ref{eq:covtot_gstack})), accounting for the
profile variations in individual clusters,
observational uncertainties, 
photo-$z$ uncertainties on the mean-depth calibration, 
and the net cosmic noise contribution.

\section{Cluster Sample and Observations}
\label{sec:data}

\subsection{Cluster Sample}
\label{subsec:sample}

\input{table1.tex}
\input{table2.tex}

Our cluster sample comprises two subsamples, one with 16 X-ray-selected
clusters and another with 4 high-magnification clusters, both taken from
the CLASH sample \citep{Postman+2012CLASH}.
Table \ref{tab:sample} provides a summary of the cluster properties in
our sample.
In this work, 
the optical cluster center is taken to be the location of the brightest
cluster galaxy (BCG) when a single dominant central galaxy is
found. Otherwise it is defined as the center of the brightest
red-sequence selected cluster galaxies  
(MACS\,J0717.5+3745, \citet{Medezinski+2013}; MACS\,J0416.1-2403).

All clusters in the X-ray-selected subsample have
X-ray temperatures greater than 5\,keV and show
a smooth, regular morphology in their X-ray brightness
distribution \citep{Postman+2012CLASH}.
Importantly, the X-ray selection allows us to 
reduce a bias toward higher concentrations as found in  
lensing-selected clusters,
where selecting clusters according to their lensing
properties can introduce an orientation bias
in favor of prolate structure pointed to the observer, as this
orientation boosts the projected mass density,
and a selection bias toward intrinsically over-concentrated clusters  
\citep{BUM+08,Oguri+2009Subaru,Meneghetti+2010MARENOSTRUM}.
Our X-ray criteria also ensure well-defined cluster centers,
reducing the effects of cluster miscentering, where
smoothing from the miscentering effects flattens the recovered lensing
profiles below the offset scale and thus leads to an underestimation of
the derived concentration and inner-slope parameters.

\subsection{Wide-field Imaging Observations}
\label{subsec:reduction}

\input{table3.tex}
\input{table4.tex}

Our CLASH wide-field imaging data rely primarily on observations taken with
the Suprime-Cam imager
\citep[$34^\prime\times 27^\prime$;][]{2002PASJ...54..833M}
at the prime focus of the 8.3-m Subaru Telescope.
We combine both existing archival data taken from
SMOKA\footnote{\href{http://smoka.nao.ac.jp}{http://smoka.nao.ac.jp}}
with observations acquired by the team on the nights of 
March 17-18, 2010 (S10A-019),  
November 4-6, 2010 (S10B-059),
and July 22, 2012 (S12A-063).
A good fraction of the Subaru data 
were taken as part of the {\it Weighing the Giants} (WtG) project,
and were independently analyzed in their series of papers \citep{WtG1,WtG2,WtG3}. 
The complete multi-band filter information for all the clusters is
summarized in Table \ref{tab:sample}. 
The filter naming conventions and description are given in Table \ref{tab:filters}.

To improve the accuracy of our photometric redshift (photo-$z$) measurements, 
we also retrieved, reduced, and used optical data taken 
with the Megaprime/MegaCam
(RXJ1347.5-1145, MACS\,J0717.5+3745)
and near-IR data with the
WIRCam (MACS\,J1149.5+2223, MACS\,J0717.5+3745)
on the Canada-France-Hawaii Telescope (CFHT),
where available from the CFHT archive.\footnote{This research used the
facilities of the Canadian Astronomy Data Centre operated  by the
National Research Council of Canada with the support of the Canadian
Space Agency.} 
For Abell\,2261,
additional bands were available from the KPNO telescope 
archive\footnote{\href{http://portal-nvo.noao.edu/home/main}{http://portal-nvo.noao.edu/home/main}} 
to augment the existing Subaru data \citep[see][]{Coe+2012A2261}.

For our southernmost
cluster, RXC\,J2248-4431, which is not observable
from Subaru, we rely on data obtained by \citet{Gruen+2013RXJ2248}
with the Wide-Field Imager (WFI)
at the ESO 2.2-m MPG/ESO telescope at La Silla. 
We use the same co-added
mosaic images 
built by \citet{Gruen+2013RXJ2248}, but conduct an independent analysis
adopting substantially different approaches in performing shape
measurements, shear calibration, 
photometry,
photo-$z$ measurements,
background selection, and 
lensing reconstruction,
which is based
on the combination of shear and magnification effects.
We also obtained data
with the IMACS camera \citep{IMACS} on the Magellan 6-m telescope.

In general, each cluster was observed deeply in at least 3-6 bands in
the optical, with exposure times ranging 1000---10,000\,s
per passbands. Typical seeing in the $R_{\rm C}$-band, mostly used for
weak-lensing shape measurements, is around 0.6--0.8\,arcsec. 
Typical limiting magnitudes are
$\sim 26$--$26.5$ in the  $R_{\rm C}$-band for a $3\sigma$ detection within
a $2\arcsec$ diameter aperture \citep{Umetsu+2012,Medezinski+2013}. 

Basic information regarding the weak-lensing band is given in Table
\ref{tab:sample}.   
In the present analysis we use the $R_{\rm C}$-band ($R_{844}$ for 
RXJ2248.7-4431) to measure the shapes of background galaxies for all
clusters except Abell\,383 \citep[][]{Zitrin+2011A383}
and MACS\,J1206.2-0847 \citep[][]{Umetsu+2012},
both of which are based on our
published CLASH lensing work. 
\citet{Zitrin+2011A383}
and
\citet{Umetsu+2012} used the 
$i'$- and $I_{\rm C}$-band images, respectively, for their weak-lensing
shape analyses because these are of higher quality than the respective
$R_{\rm C}$-band in terms of the stability and coherence of the PSF
anisotropy pattern, taken in fairly good seeing conditions.

Details regarding our reduction and analysis pipelines can be found in
\citet{Umetsu+2012} and \citet{Medezinski+2013}. 
We thus refer the reader to those papers, and give a basic summary here. 
Our reduction pipeline derives from \cite{Nonino+2009},
and has been optimized separately for accurate photometry and
weak-lensing shape measurements.

For photometric measurements, standard reduction
steps include bias subtraction, super-flat-field correction  and
point-spread function (PSF) matching between exposures in the same band. 
An accurate astrometric solution is derived with the SCAMP software 
\citep{SCAMP},
using 2MASS\footnote{This
publication makes use of data products from the Two Micron All Sky
Survey, which is a joint project of the University of Massachusetts and
the Infrared Processing and Analysis Center/California Institute of
Technology, funded by the National Aeronautics and Space Administration
and the National Science Foundation.} 
as an external reference catalog, or 
the Sloan Digital Sky Survey
 \citep[SDSS,][]{SDSSDR6photometry} where available.\footnote{This research has made use of the VizieR catalogue access tool, CDS, Strasbourg, France.}
Finally, the {\sc Swarp} software \citep{Bertin+2002Swarp} is utilized to
stack the single exposures on a common World Coordinate System (WCS)
grid with pixel-scale of $0.2\arcsec$.  


The photometric zero-points for the co-added images were derived using
{\it HST}/ACS magnitudes of cluster elliptical-type galaxies.
These zero points were further refined by fitting SED (spectral energy
distribution) templates with the BPZ code
\citep[Bayesian photometric redshift estimation,][]{Benitez2000,Benitez+2004} to  galaxies
having spectroscopic redshifts where available. 
This leads to a final photometric accuracy of $\sim 0.01$\,mag in all
passbands. 
The magnitudes were corrected for Galactic extinction according to \citet{1998ApJ...500..525S}.
The multi-band photometry was measured
using SExtractor \citep{SExtractor} in dual-image mode on
PSF-matched images created by ColorPro
\citep{colorpro}.

For weak-lensing shape measurements, we separately stack data collected
at different epochs and different camera rotation angles. We do not
smear the single exposures before stacking, so as not to degrade the
weak-lensing information derived from the shapes of galaxies. 
A shape catalog is created for each epoch and camera rotation separately.
Then, these subcatalogs are combined by properly
weighting and stacking the calibrated distortion measurements  
for galaxies in the overlapping region (see Section \ref{subsec:shape}).


\subsection{Shape Measurement}
\label{subsec:shape}

For shape measurements of background galaxies,
we use our well-tested weak-lensing pipeline based on the {\sc imcat}
package \citep[][KSB]{1995ApJ...449..460K}   
incorporating improvements developed in \citet{Umetsu+2010CL0024}.
Our KSB+ implementation has been applied extensively
to Subaru observations of a large number of clusters
\citep[e.g.,][]{UB2008,Okabe+Umetsu08,Umetsu+2009,Umetsu+2010CL0024,Umetsu+2011,Umetsu+2012,Medezinski+2010,Medezinski+2011,Medezinski+2013,Coe+2012A2261}. 

Full details of our CLASH weak-lensing shape analysis pipeline are given
in \citet{Umetsu+2012}. 
Here, we only highlight several aspects of our weak-lensing analysis pipeline:
\begin{itemize}
\item {\it Object detection}: Objects are detected using the IMCAT
      peak-finding algorithm {\it hfindpeaks} which for each
      object yields object parameters such as the peak position, 
      Gaussian scale length, $r_g$, and 
      significance of the peak detection, $\nu$.

\item {\it Rejection of close pairs}: Objects having any detectable
      neighbor within $3r_g$ are identified. All such close pairs of
      objects are rejected to avoid possible shape measurement errors
      due to crowding. The detection threshold is set to $\nu=7$ for
      close-pair identification. After this close-pair rejection, objects
      with low detection significance $\nu<10$ are excluded from our analysis.   


\item {\it Shear calibration}: Following \citet{Umetsu+2010CL0024}, we calibrate
      KSB's isotropic correction factor $P_g$ as a function of $r_g$ and magnitude, 
      using galaxies detected with high significance $\nu>30$.
      This is to minimize the inherent shear calibration bias in the
      presence of measurement noise \citep{Okura+Futamase2012}.
\item {\it Combining subcatalogs}:
      For each galaxy we combine shape measurements and associated
      errors from different epochs and camera orientations by 
      $g =(\sum_{k=1}^{N_{\rm sub}} w_{k} g_{k}) / (\sum_{k=1}^{N_{\rm sub}} w_{k})$
      and
      $\sigma_{g}^2=(\sum_{k=1}^{N{\rm sub}} w_{k}\sigma_{g,k}^2)/(\sum_{k=1}^{N_{\rm sub}} w_{k})$,
      where $N_{\rm sub}$ is the number of subcatalogs,
      $g_{k}$ is the complex reduced-shear estimate for the galaxy in
      the $k$th subcatalog, and $w_{k}$ is its statistical weight, 
      defined such that $w_{k}=(\sigma_{g,k}^2+\alpha_g^2)^{-1}$ if the
      galaxy exists in the $k$th subcatalog, and $w_{k}=0$ otherwise. 
      Here $\alpha_g^2$ is the softening constant variance, taken to be
      $\alpha_g=0.4$ \citep[e.g.,][]{UB2008,Umetsu+2009,Umetsu+2012,Okabe+2010WL}.
\end{itemize}

On the basis of simulated Subaru/Suprime-Cam images 
\citep{Oguri+2012SGAS,2007MNRAS.376...13M},
we found in \citet{Umetsu+2010CL0024} that
the lensing signal can be recovered with $|m|\sim 5\%$ of the
multiplicative shear calibration bias 
\citep[as defined by][]{2006MNRAS.368.1323H,2007MNRAS.376...13M},
and $c\sim 10^{-3}$ of
the residual shear offset, which is about one order of magnitude
smaller than the typical distortion signal in cluster outskirts.
Accordingly, we include for each galaxy a shear
calibration factor of $1/0.95$ as
to account for residual calibration.\footnote{Our earlier CLASH
weak-lensing work of
\citet[][Abell 383]{Zitrin+2011A383}, 
\citet[][MACS\,J1206.2-0847]{Umetsu+2012}, and 
\citet[][Abell 2261]{Coe+2012A2261}
did not include the 5\% residual correction.}

\subsection{Background Galaxy Selection}
\label{subsec:back}

In general, the attainable number density $\overline{n}_g$ of background
galaxies for use in weak-lensing shape measurements is 
sensitive to the image quality (seeing FWHM) and depth of observations.  
In the shot-noise limited regime, the statistical precision of the weak-lensing
measurements scales as $1/\sqrt{\overline{n}_g}$. 
However, a careful background selection is even more critical
for a cluster weak-lensing analysis, so that unlensed cluster members and
foreground galaxies do not dilute the true lensing signal of the
background. In particular, this lensing {\it dilution} effect due to
contamination by cluster members can lead to 
a substantial underestimation of the true signal for 
$R \simlt 400$\,kpc\,$h^{-1}$ by a factor of 2--5, as demonstrated in
\citet{BTU+05} and \citet{Medezinski+2010}.
The relative
importance of the dilution effect indicates that, the impact of 
background purity and depth is more important than that of
shot noise ($\propto \overline{n}_g^{-1/2}$).      
 
We use the color-color (CC) selection method of \citet{Medezinski+2010} to
define undiluted samples of background galaxies from which to measure
the weak-lensing shear and magnification effects.
Here we refer the reader to \citet{Medezinski+2010} for further details.
This method is designed to  avoid the inclusion of unlensed cluster
members and to minimize foreground contamination on the basis of 
empirical correlations in CC-magnitude space,
which have been established by reference to evolutionary color tracks of
galaxies \citep{Medezinski+2010,Medezinski+2011,Hanami+2012}
as well as to the 30-band photo-$z$ distribution in the COSMOS field
\citep{Ilbert+2009COSMOS}.
Using CC-selected samples of differing depths, 
we showed in \citet{Medezinski+2011} that the redshift scaling of the
observed shear signal behind massive clusters follows the
expected form of the lensing distance-redshift relation $\beta(z)$, 
providing independent consistency checks.
Our color-cut approach and its variants have been successfully
applied to a large 
number of clusters 
\citep{Medezinski+2010,Medezinski+2011,Medezinski+2013,Umetsu+2010CL0024,Umetsu+2011,Umetsu+2012,Coe+2012A2261,Oguri+2012SGAS,Covone+2014}.

%
In the present analysis, 
we typically use the Subaru $B_{\rm J}R_{\rm C}z'$ photometry where available,
which spans the full optical wavelength range to
perform CC selection of background samples.
The specific CC-selection bands used for each cluster are indicated in
Table \ref{tab:sample}. 
For shape measurements, we select and combine two distinct populations
that encompass the red and blue branches of background galaxies in CC 
space
\citep{Medezinski+2010,Medezinski+2013,Umetsu+2010CL0024,Umetsu+2012},
each with typical redshift distributions peaked around $z \sim 1$ and
$\sim 2$, respectively \citep[see Figures 5 and 6
of][]{Medezinski+2011,Lilly+2007}.

Our conservative selection criteria yield a typical (median) surface
number density of $\overline{n}_{g}\simeq 12$ galaxies arcmin$^{-2}$ for
the weak-lensing-matched background catalogs (Table \ref{tab:wlsample}),
consistent with the values found by \citet{Oguri+2012SGAS}.  
For RXJ2248.7-4431 based on the 2.2-m/WFI data, we have
$\overline{n}_g=4.6$ galaxies arcmin$^{-2}$, which is about a factor 2.5
smaller than the median value of our sample.
That is, the shot-noise level for the cluster is about $40\%$ higher
than that obtained with the typical depth of our Subaru
observations. Accordingly, our weak-lensing measurements of
RXJ2248.7-4431  are highly shot-noise limited. On the other hand, the
low number density of background galaxies in the MACS\,J1931.8-2635
field, $\overline{n}_g=4.9$\,galaxies arcmin$^{-2}$, is due to its low
Galactic latitude, $b=-20.09^\circ$, which implies a high stellar
density and a correspondingly large area masked by bright saturated stars. 
Our magnification-bias measurements are based on the flux-limited
samples of red background galaxies at $\langle z \rangle\sim 1$,
with a typical count slope of $\langle s_{\rm eff}\rangle\sim 0.15$ 
(see Table \ref{tab:magsample}).

We estimate  the respective depths 
$\langle\beta \rangle$ and $\langle \beta^2\rangle$
of the different galaxy samples (Tables \ref{tab:wlsample} and
\ref{tab:magsample}), 
when converting the
observed lensing signal into physical mass units.
For this we utilize BPZ to measure photo-$z$s ($z_{\rm phot}$) using our   
deep PSF-corrected multi-band photometry.  
Following \citet{Umetsu+2012},
we employ BPZ's ODDS parameter 
as the weight factor $w(z)$ in Equation (\ref{eq:depth}).

We discard galaxies having photo-$z$'s in the
range $z_{\rm phot}>2.5$ and having ${\rm ODDS}<0.8$, as we find those to be
spurious and unreliable. We derive this scheme by comparing our
photo-$z$'s with measured spectroscopic redshifts ($z_{\rm spec}$),
compiled from both the NASA/IPAC Extragalactic Database
(NED)\footnote{\href{http://ned.ipac.caltech.edu/}{http://ned.ipac.caltech.edu/}}
and our VLT-CLASH large spectroscopic program (ID: 186.A-0798; PI: P. Rosati).
We find that when using the full photo-$z$ catalog, although we obtain a
reasonable accuracy 
with a normalized median absolute deviation (NMAD) of
$1.48{\rm MAD}((z_{\rm phot}-z_{\rm spec})/(1+z_{\rm spec}))=3.1\%$,
the outlier fraction is high,
$f_{\rm outliers}=14\%$,
where the outliers are defined as galaxies with 
$|z_{\rm phot}-z_{\rm spec}|/(1+z_{\rm spec})>0.15$ \citep{Jouvel+2013CLASH}.
These outliers mostly stem from galaxies identified to have
$z_{\rm phot}>2.5$, whereas their true redshifts are low, $z_{\rm spec}<0.6$.  
Excluding galaxies with $z_{\rm phot}>2.5$, we find an NMAD of $2.7\%$
and an outlier fraction of $f_{\rm outlier}=9.5\%$.

For a consistency check, we also make use of the COSMOS
catalog \citep{Ilbert+2009COSMOS} with robust photometry and photo-$z$
measurements for the majority of galaxies with $i'<25$\,mag.
For each cluster, we apply the same CC selection to the COSMOS
photometry and obtain the redshift distribution $\overline{N}(z)$ of field
galaxies.  
Since COSMOS is only complete to $i'<25$\,mag,
we derive the mean depth $\langle\beta\rangle$ as a
function of magnitude up to that limit, and extrapolate
to our limiting magnitudes \citep[][see their Section 3.3]{Medezinski+2013}.
For our sample of 20 CLASH fields, we find an excellent
statistical agreement between the BPZ- and COSMOS-based depth estimates
$\langle\beta\rangle$, 
with a median relative offset of 
$0.27\%$
and
an rms cluster-to-cluster scatter of $5.0\%$.


\section{CLASH Shear-and-Magnification Analysis}
\label{sec:clash}


\begin{figure*}[!htb] 
 \begin{center}
 $
 \begin{array}
  {c@{\hspace{.001in}}c@{\hspace{.001in}}c@{\hspace{.001in}}c@{\hspace{.001in}}c}
  \includegraphics[width=0.23\textwidth,angle=0,clip]{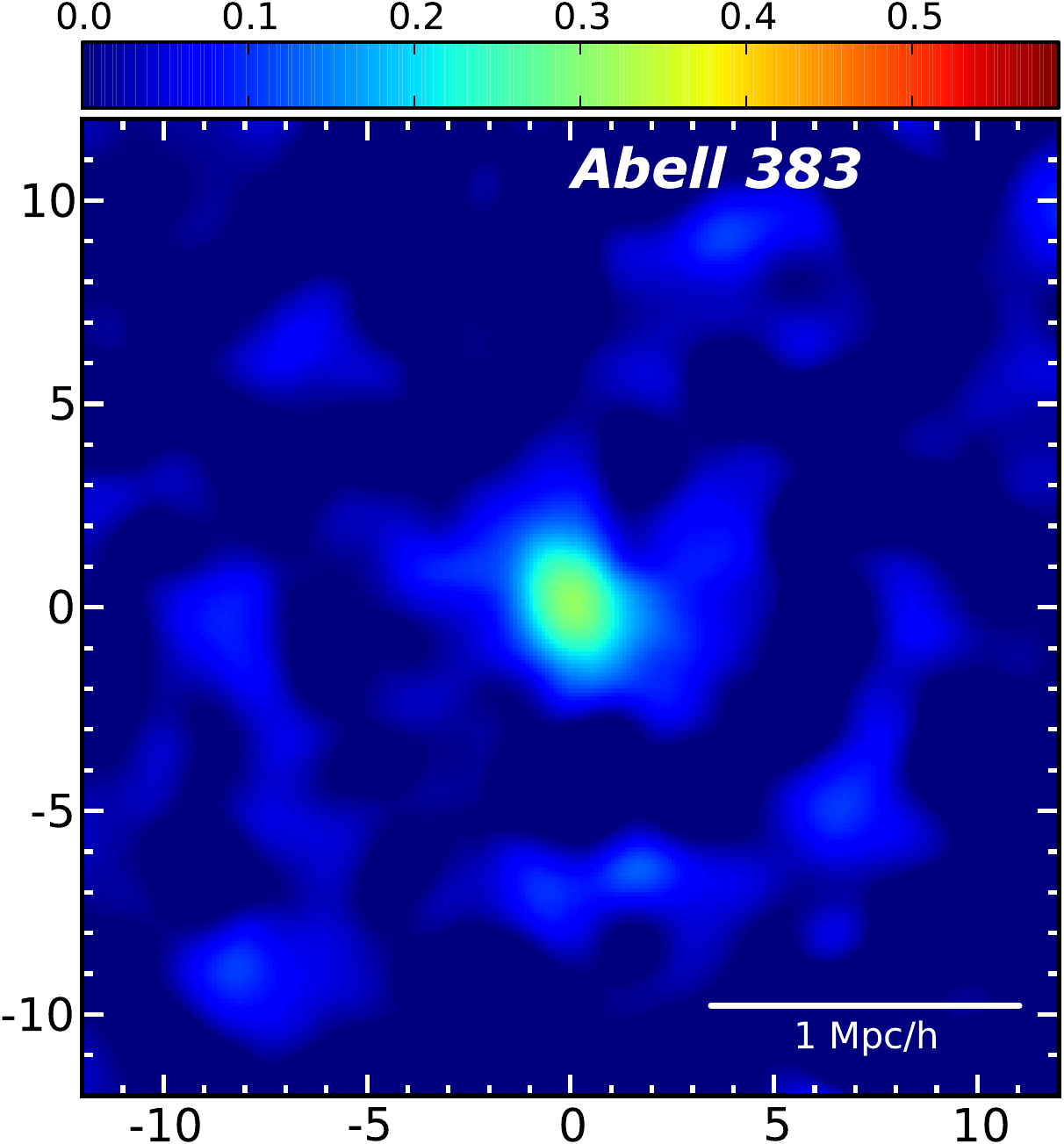} &
  \includegraphics[width=0.23\textwidth,angle=0,clip]{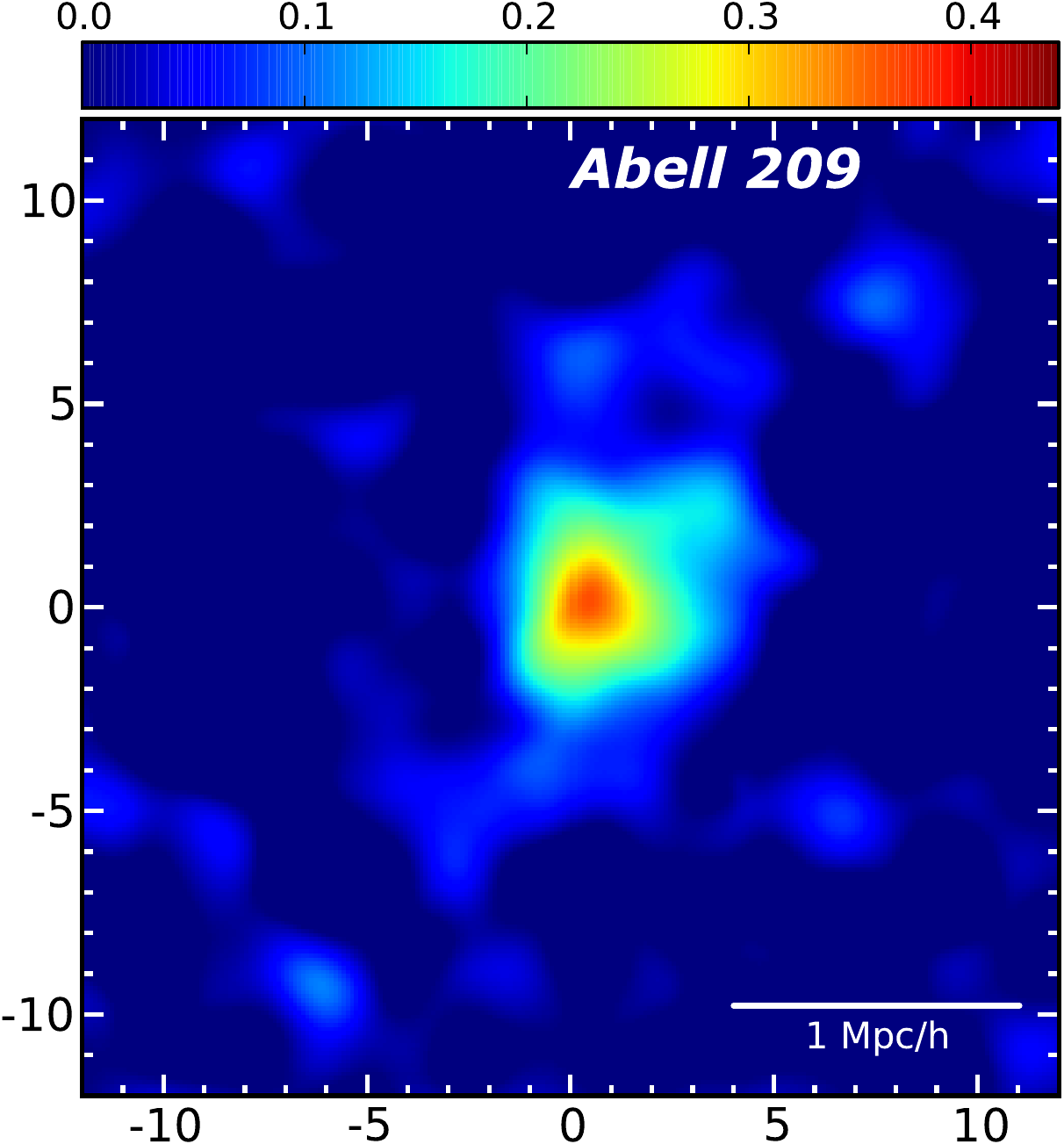} &
  \includegraphics[width=0.23\textwidth,angle=0,clip]{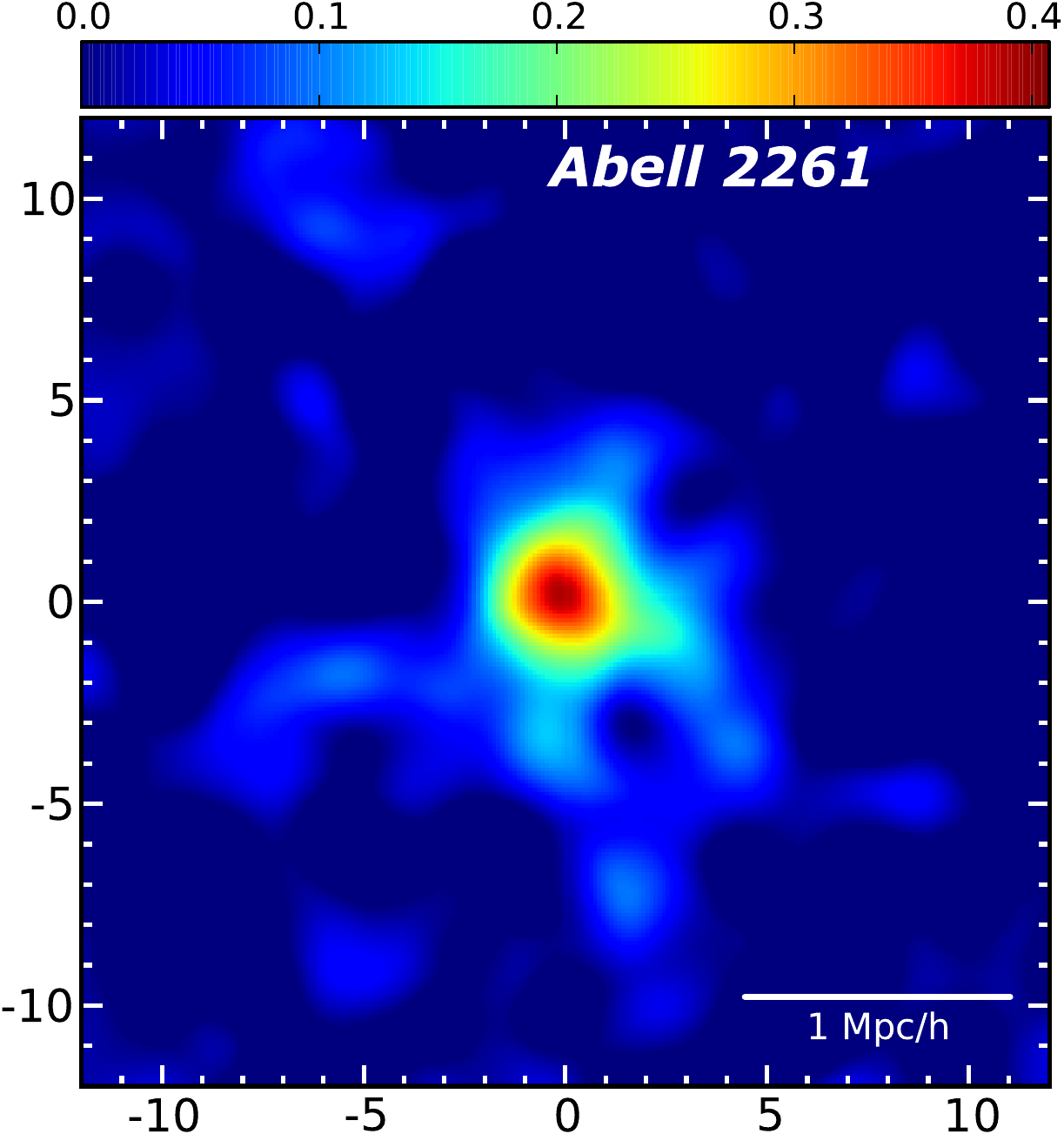}&
  \includegraphics[width=0.23\textwidth,angle=0,clip]{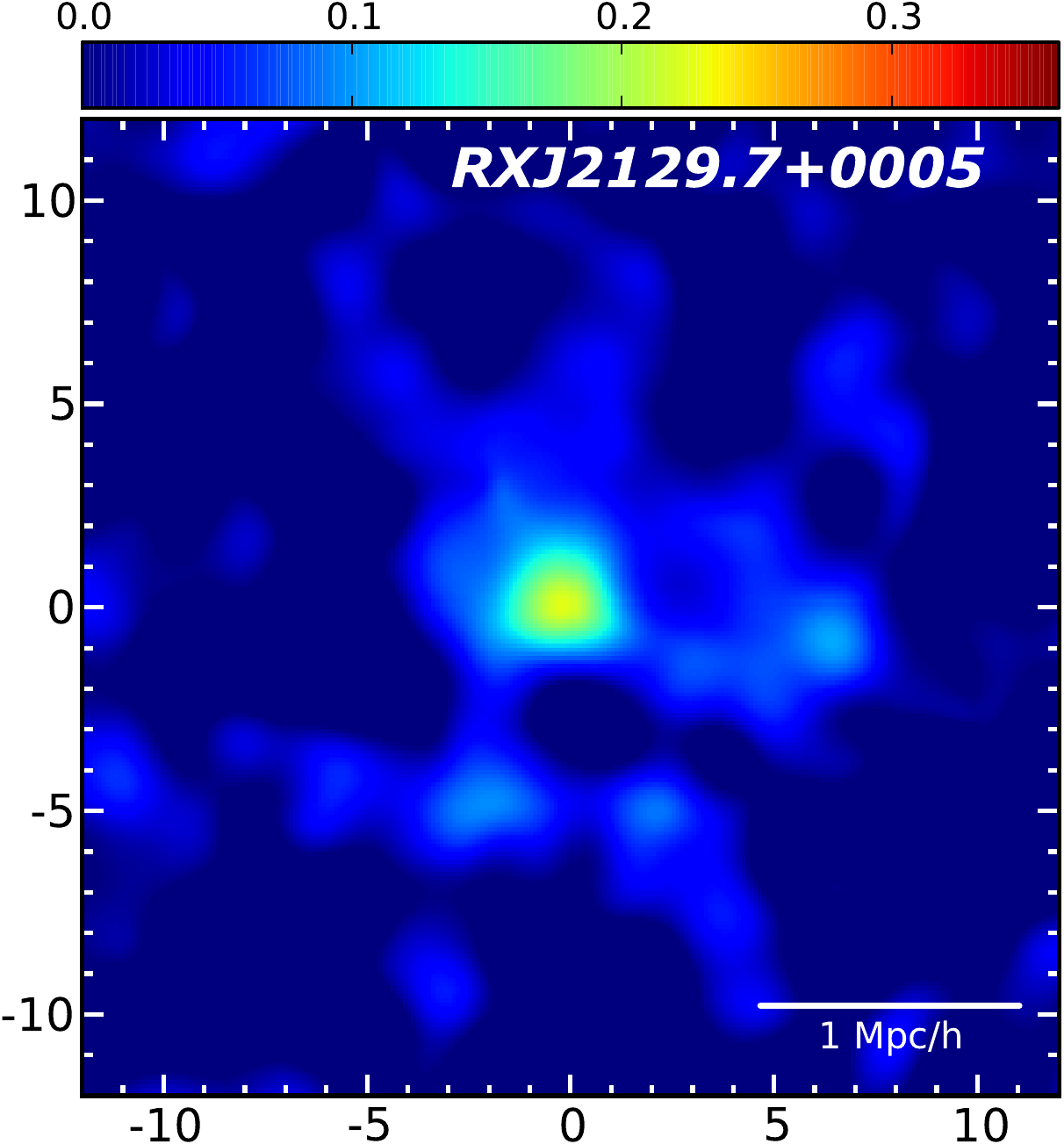}
 \end{array}
 $
 $
 \begin{array}
  {c@{\hspace{.001in}}c@{\hspace{.001in}}c@{\hspace{.001in}}c@{\hspace{.001in}}c}
  \includegraphics[width=0.23\textwidth,angle=0,clip]{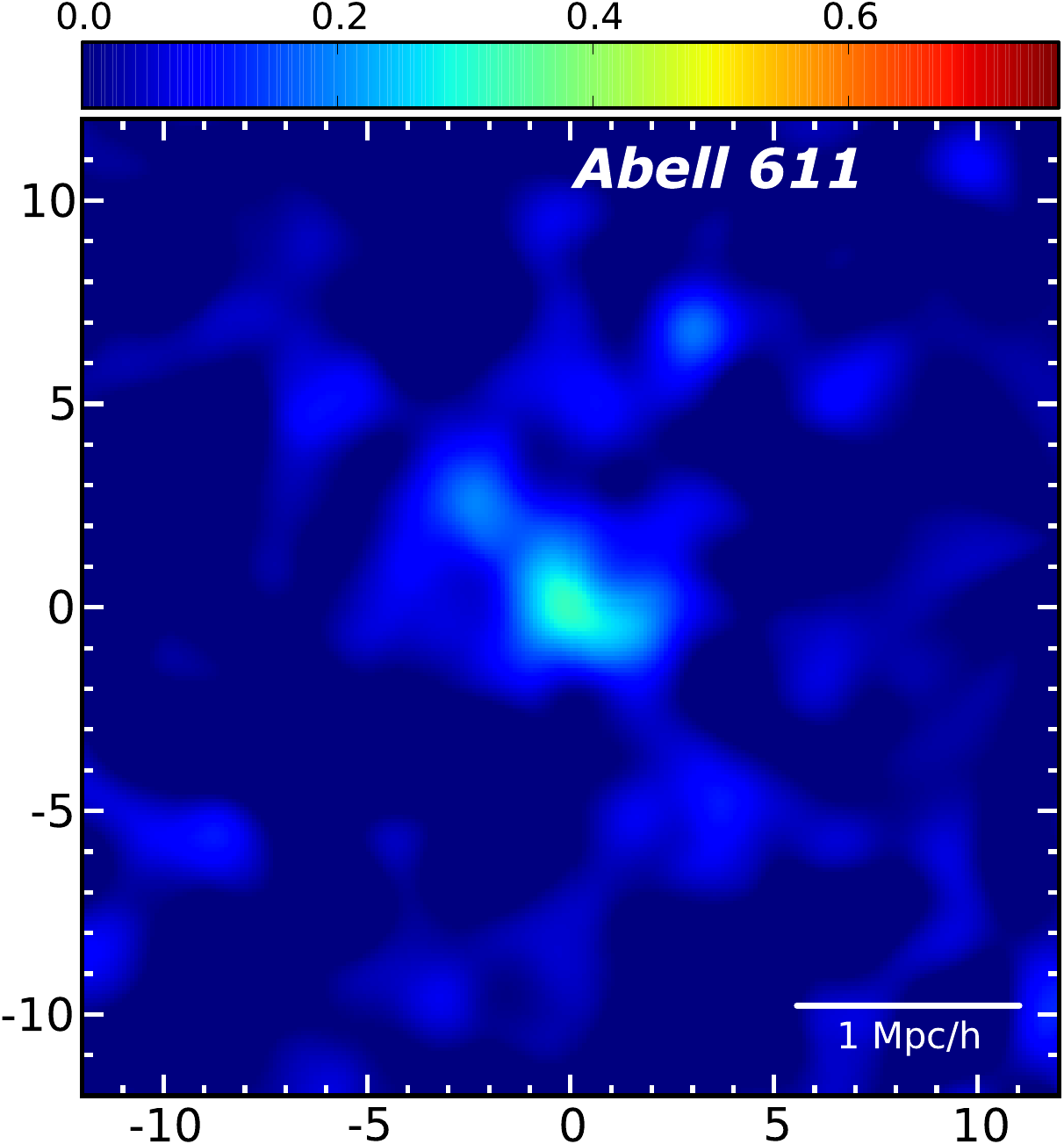} &
  \includegraphics[width=0.23\textwidth,angle=0,clip]{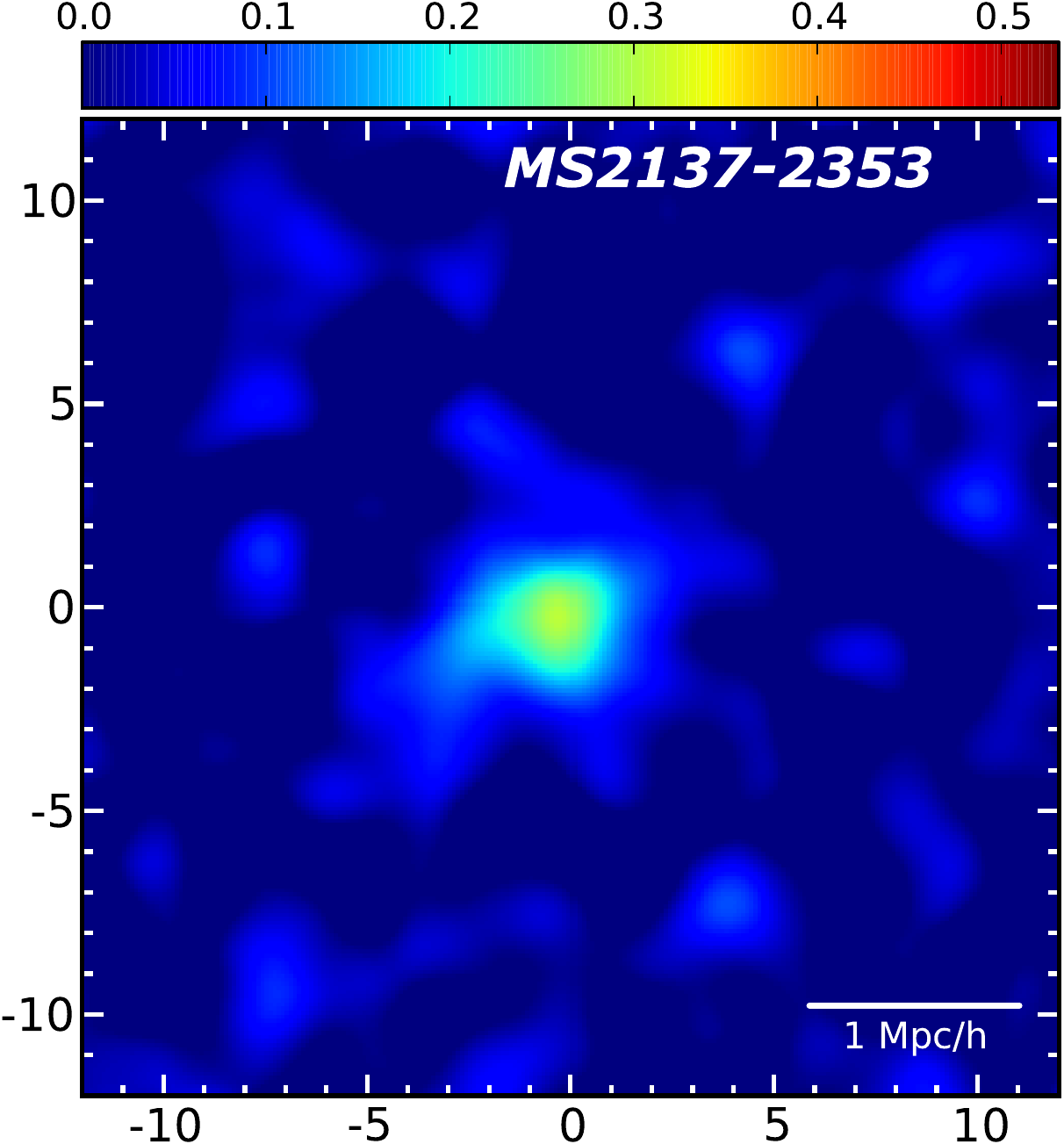}& 
  \includegraphics[width=0.23\textwidth,angle=0,clip]{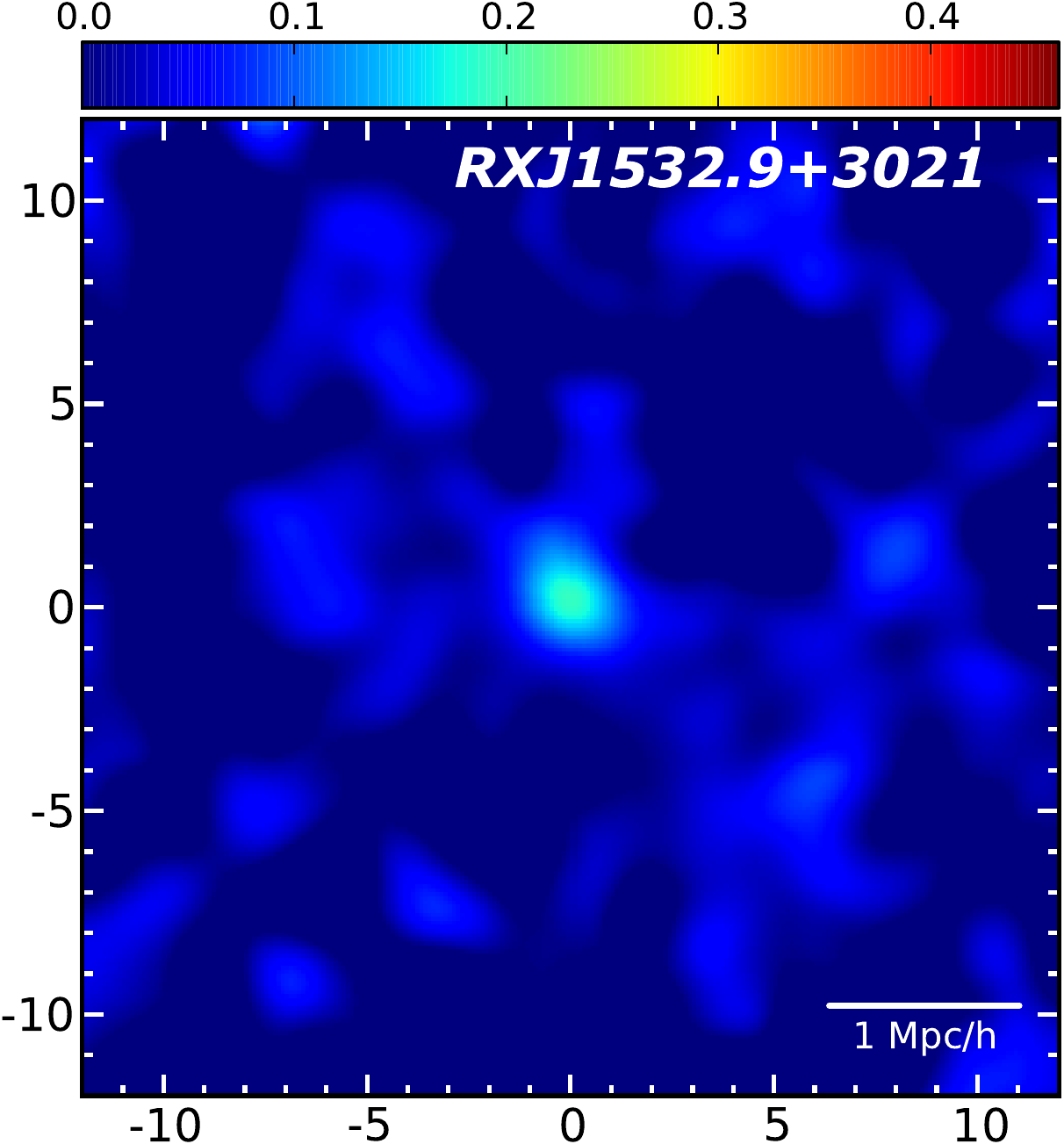}&
  \includegraphics[width=0.23\textwidth,angle=0,clip]{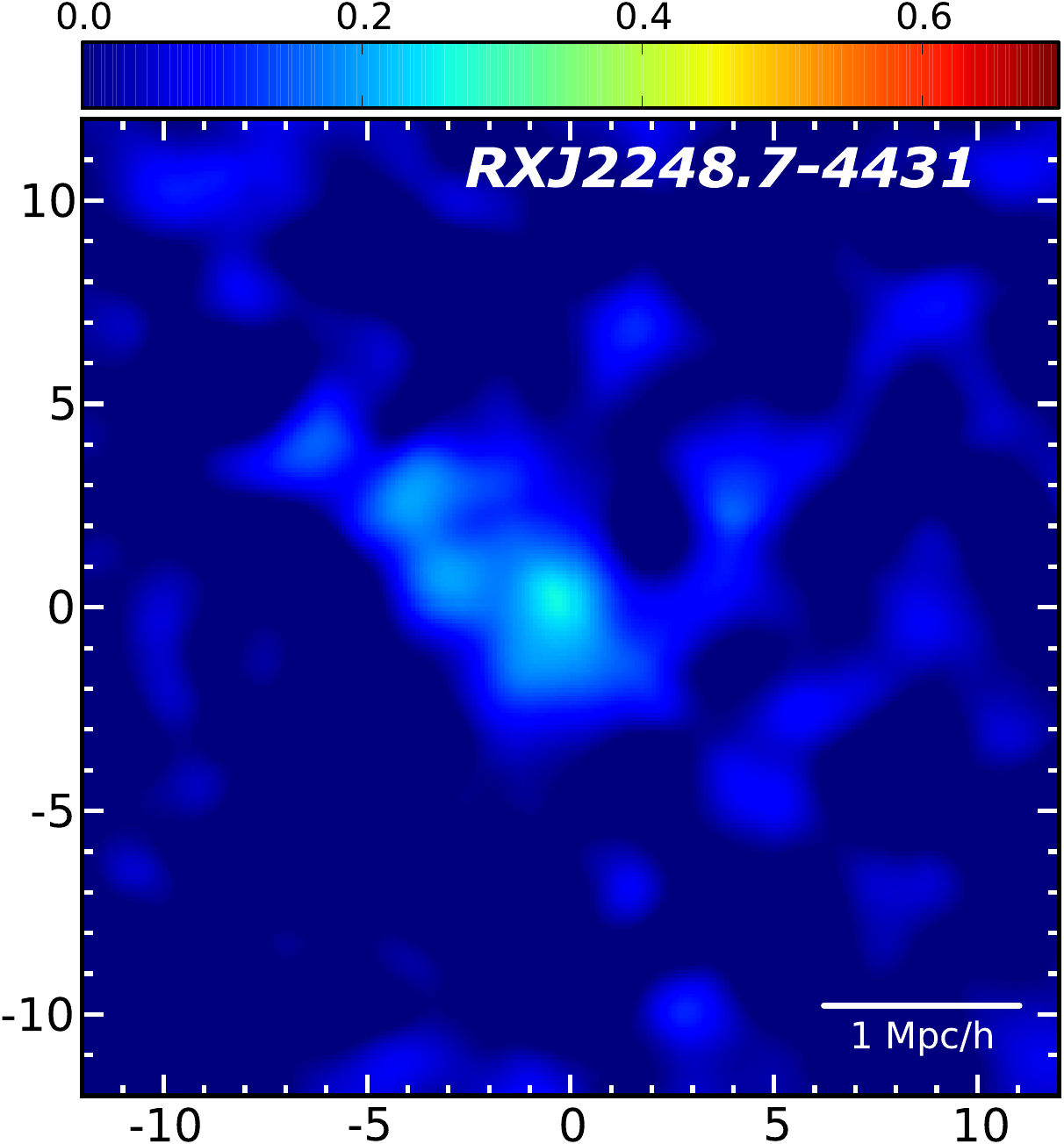}
 \end{array}
  $
  $
 \begin{array}
  {c@{\hspace{.001in}}c@{\hspace{.001in}}c@{\hspace{.001in}}c@{\hspace{.001in}}c}
   \includegraphics[width=0.23\textwidth,angle=0,clip]{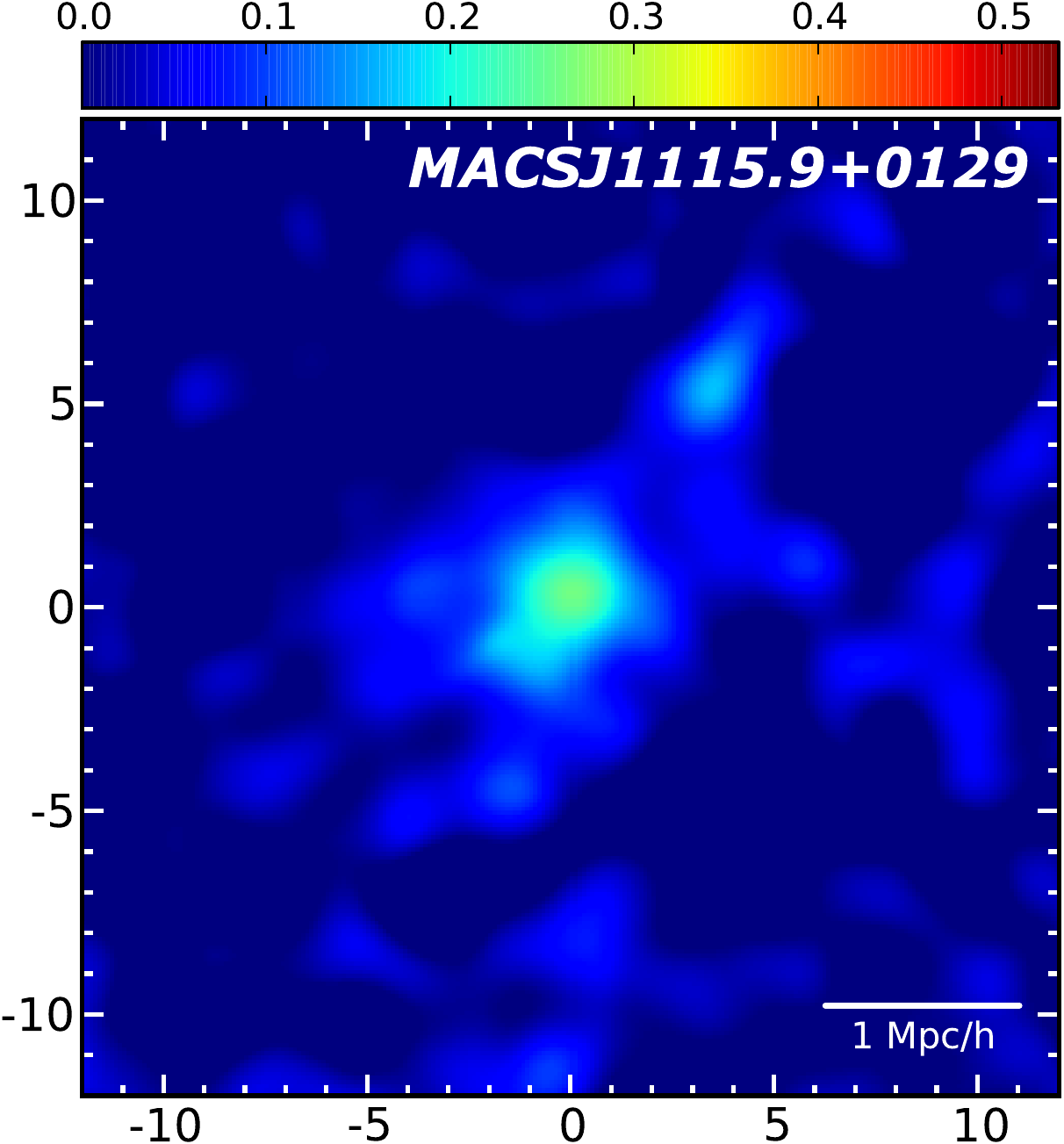}& 
   \includegraphics[width=0.23\textwidth,angle=0,clip]{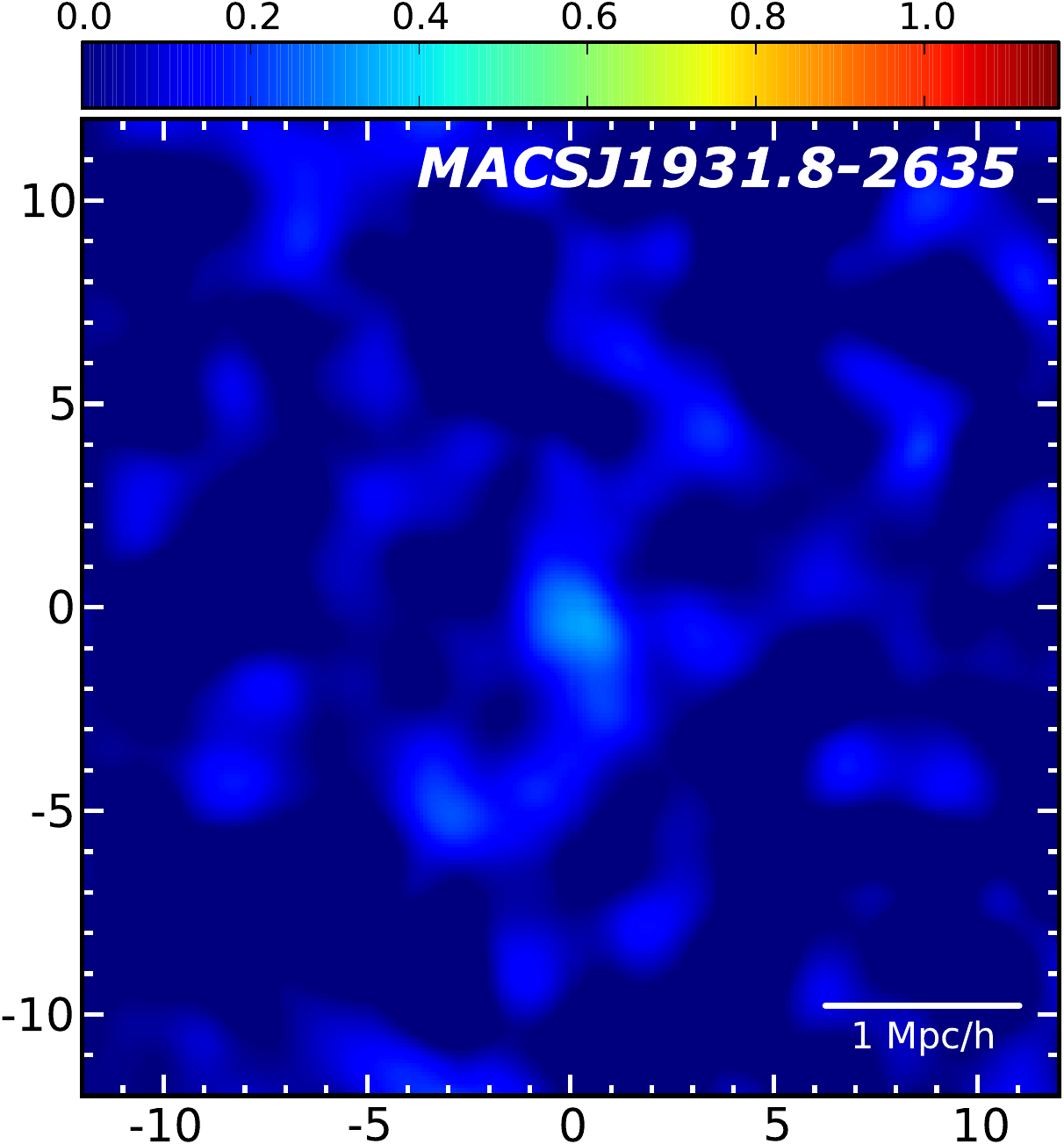}&
   \includegraphics[width=0.23\textwidth,angle=0,clip]{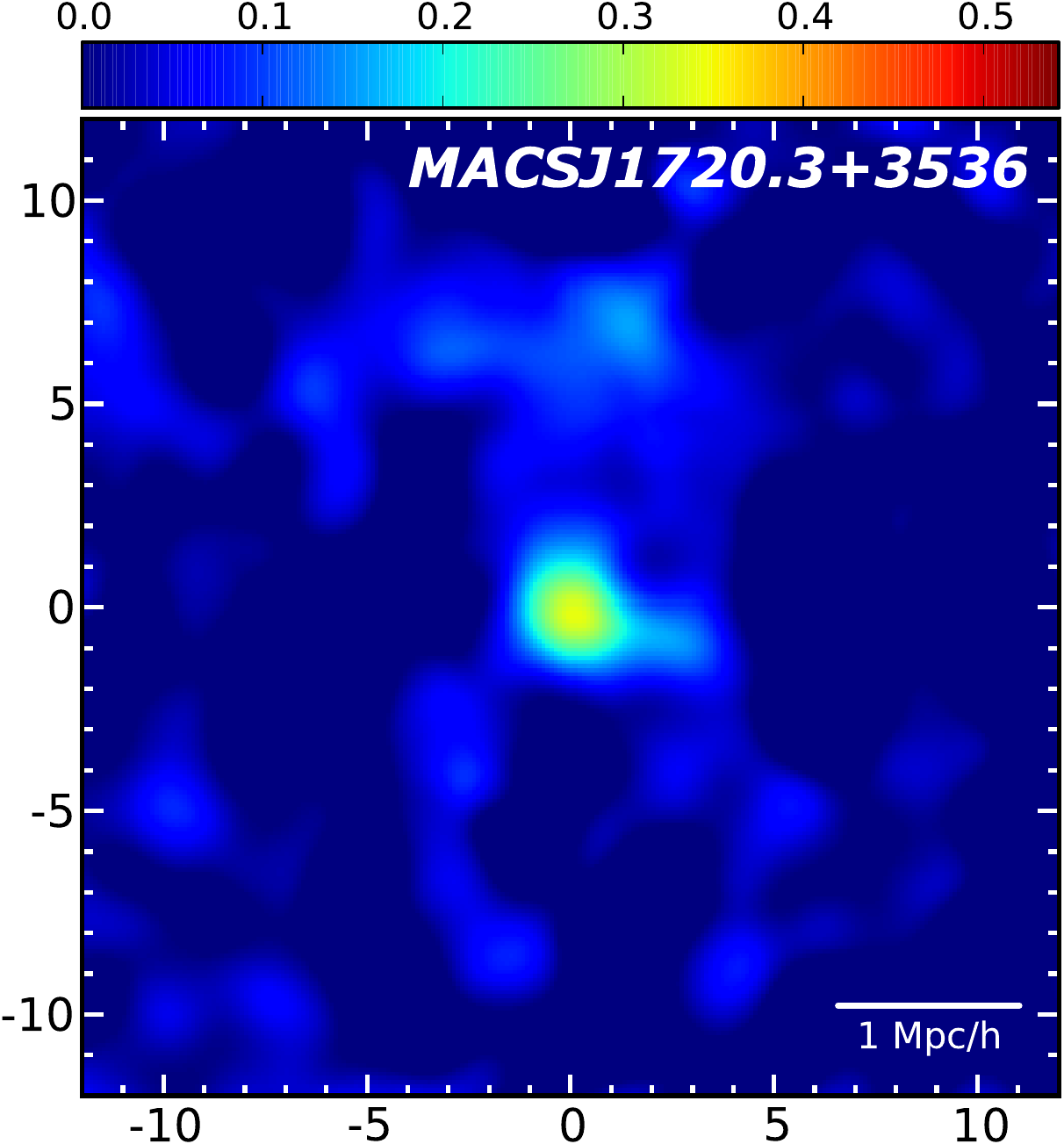}&  
   \includegraphics[width=0.23\textwidth,angle=0,clip]{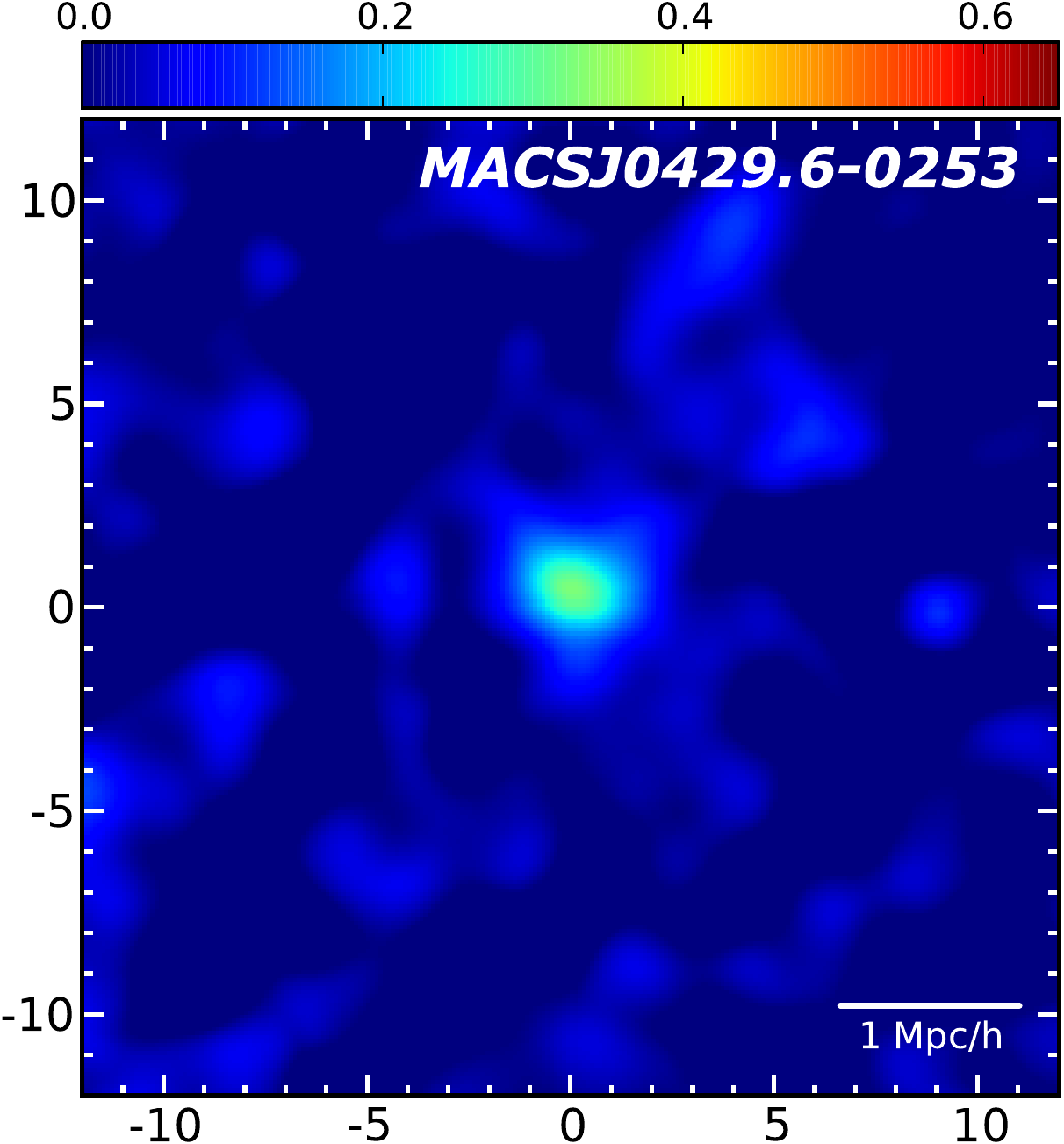}
  \end{array}
 $
 $
 \begin{array}
  {c@{\hspace{.001in}}c@{\hspace{.001in}}c@{\hspace{.001in}}c@{\hspace{.001in}}c}
   \includegraphics[width=0.23\textwidth,angle=0,clip]{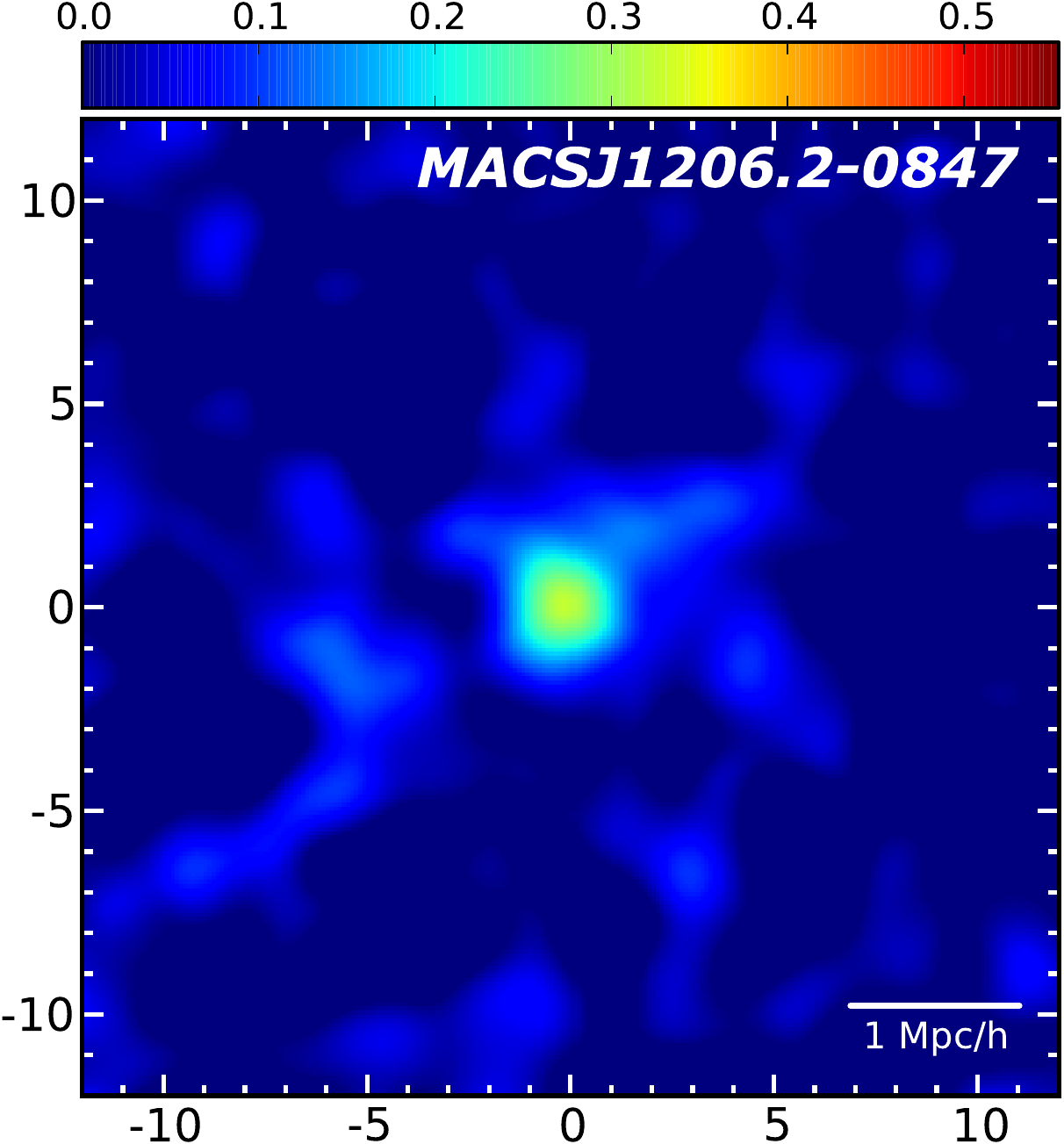}& 
   \includegraphics[width=0.23\textwidth,angle=0,clip]{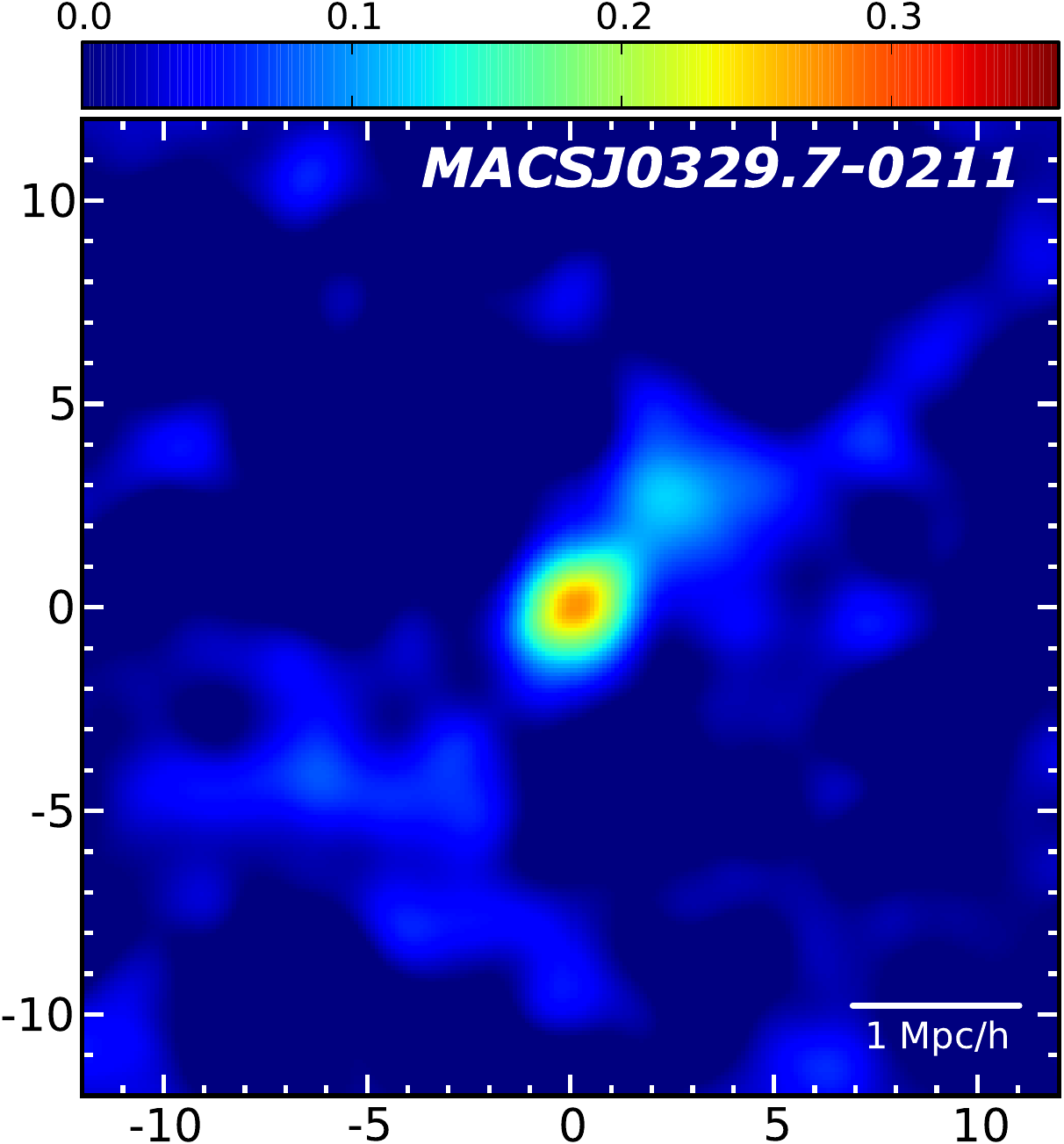}&
   \includegraphics[width=0.23\textwidth,angle=0,clip]{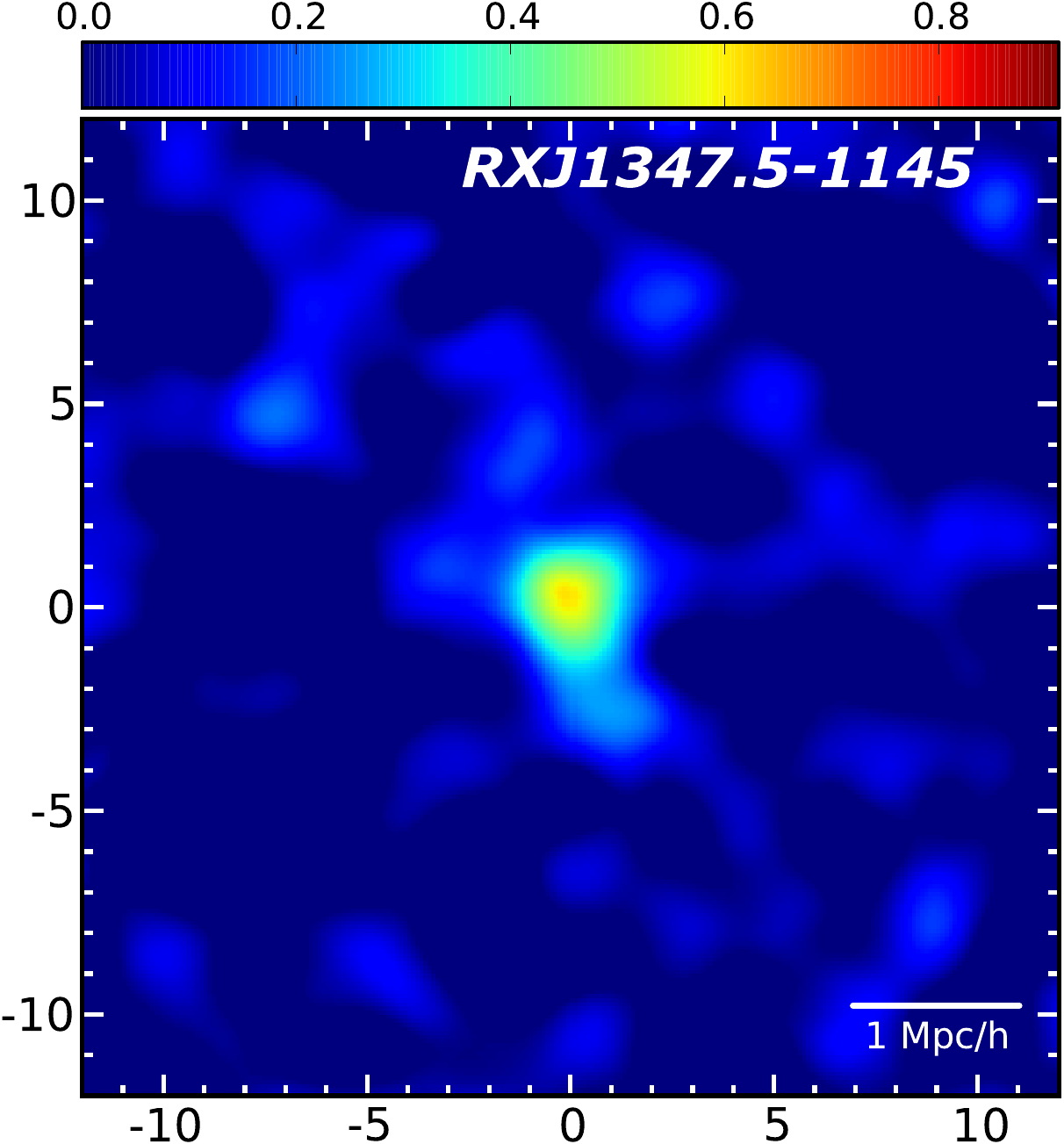} &
   \includegraphics[width=0.23\textwidth,angle=0,clip]{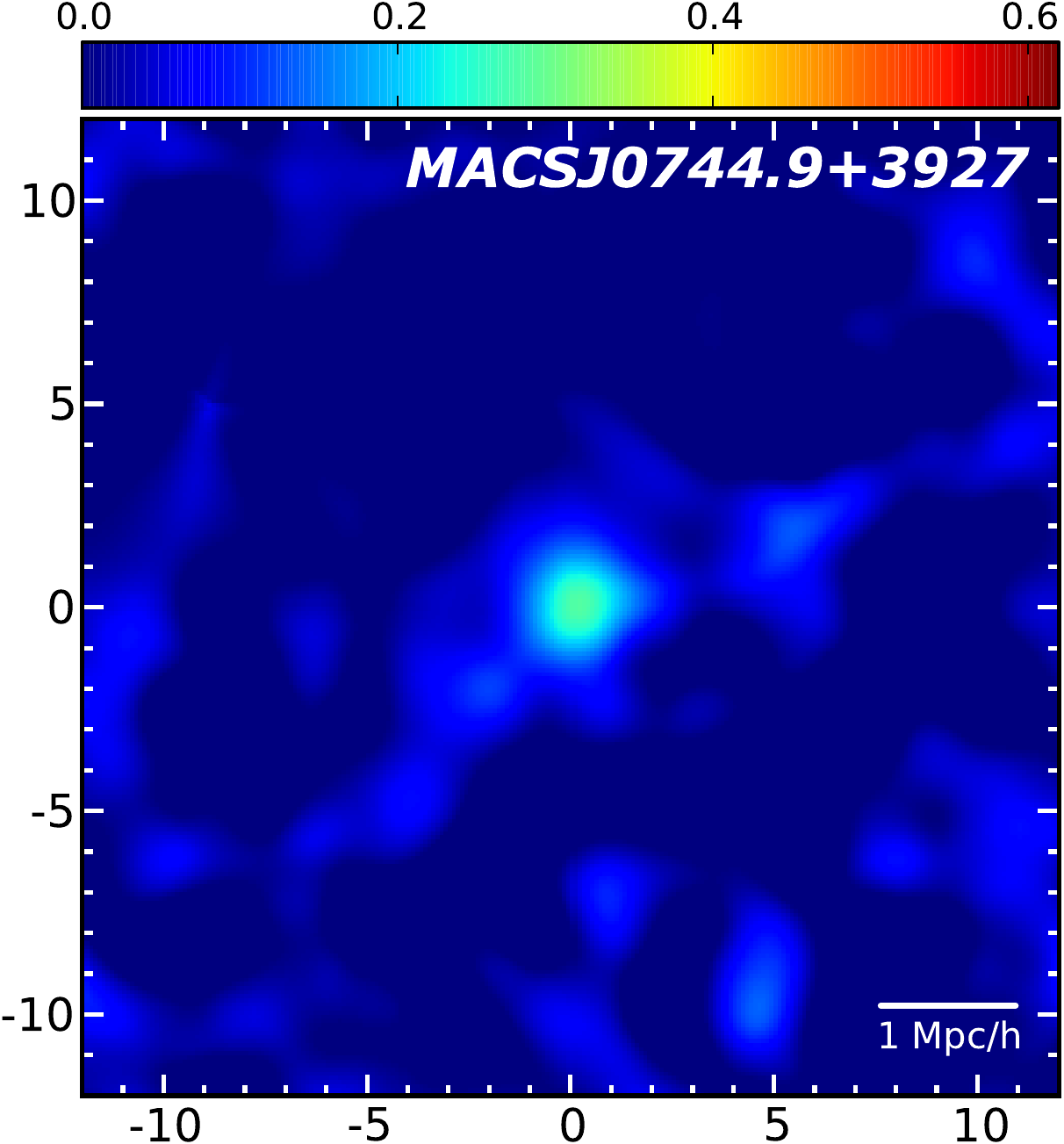} 
 \end{array}
 $
 $
 \begin{array}
  {c@{\hspace{.001in}}c@{\hspace{.001in}}c@{\hspace{.001in}}c@{\hspace{.001in}}c}
   \includegraphics[width=0.23\textwidth,angle=0,clip]{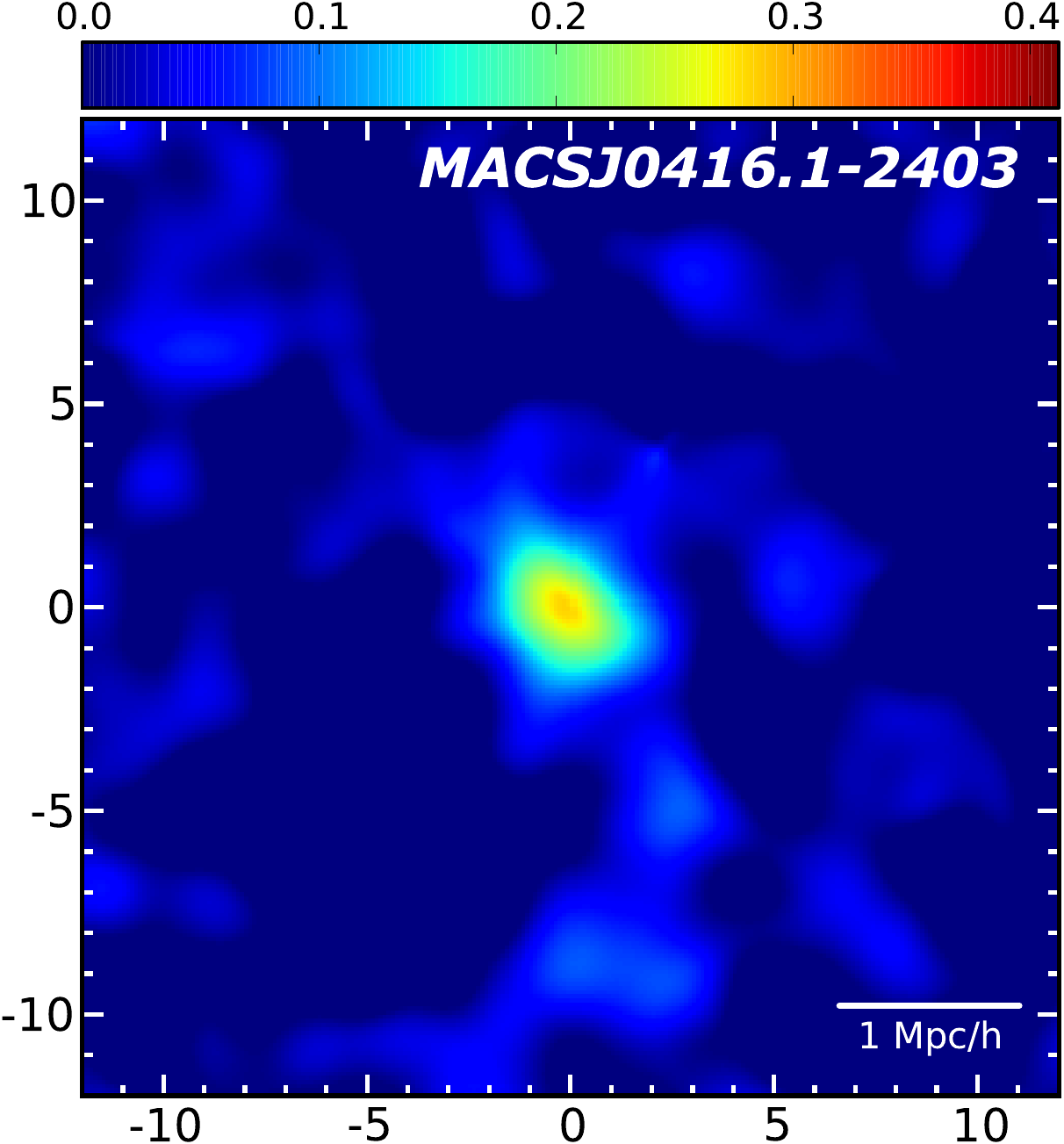}& 
   \includegraphics[width=0.23\textwidth,angle=0,clip]{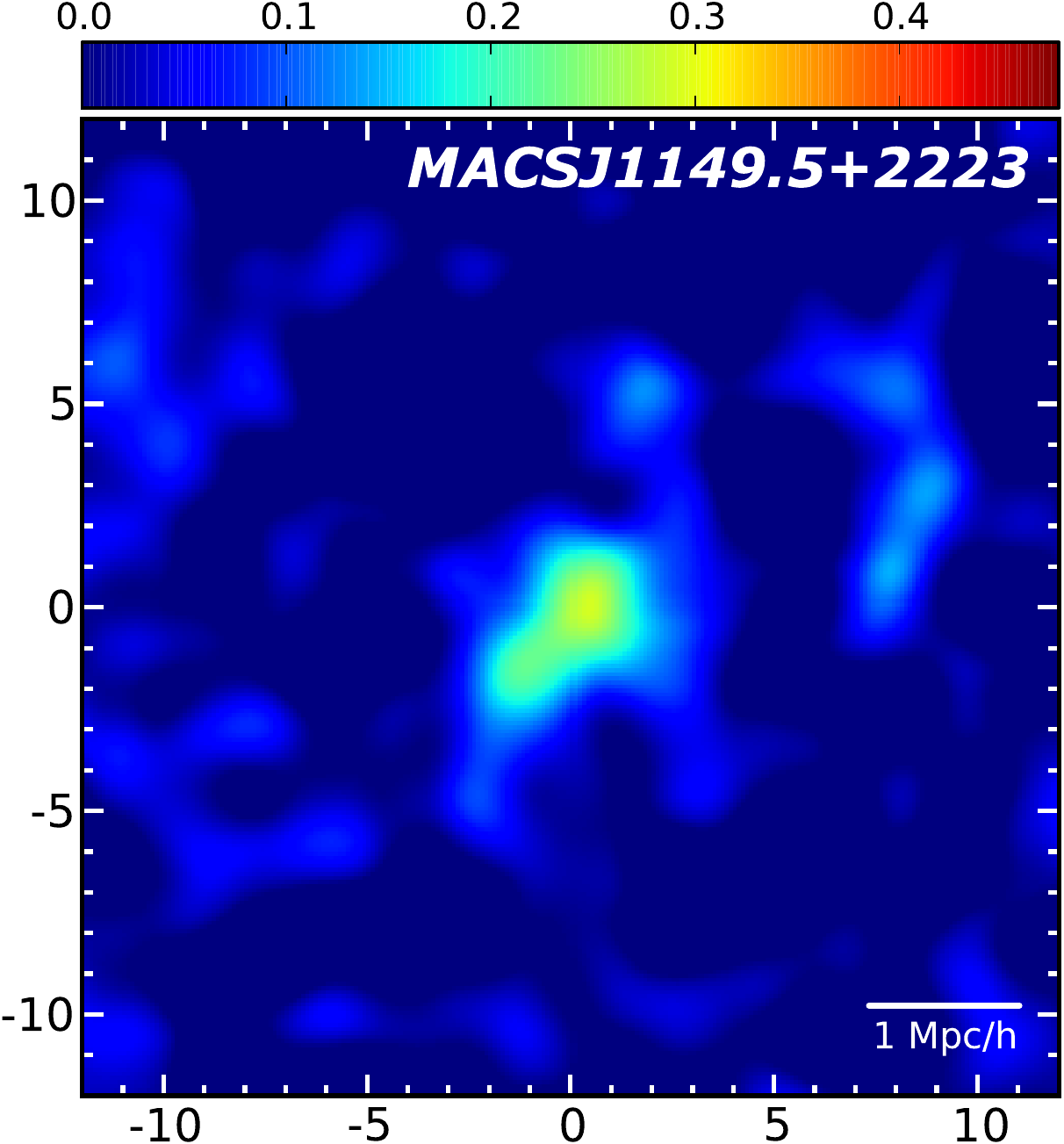}&
   \includegraphics[width=0.23\textwidth,angle=0,clip]{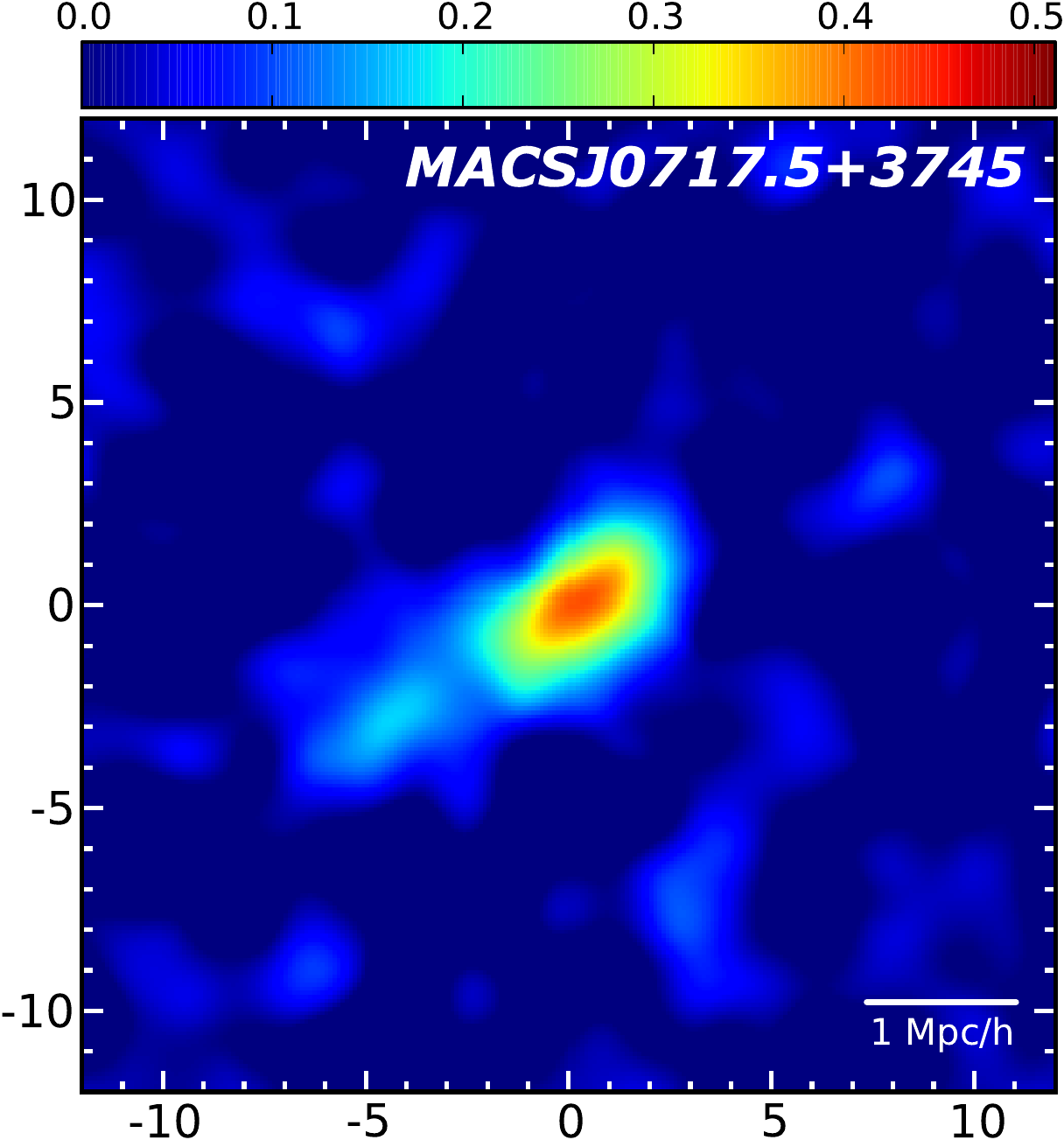} &
   \includegraphics[width=0.23\textwidth,angle=0,clip]{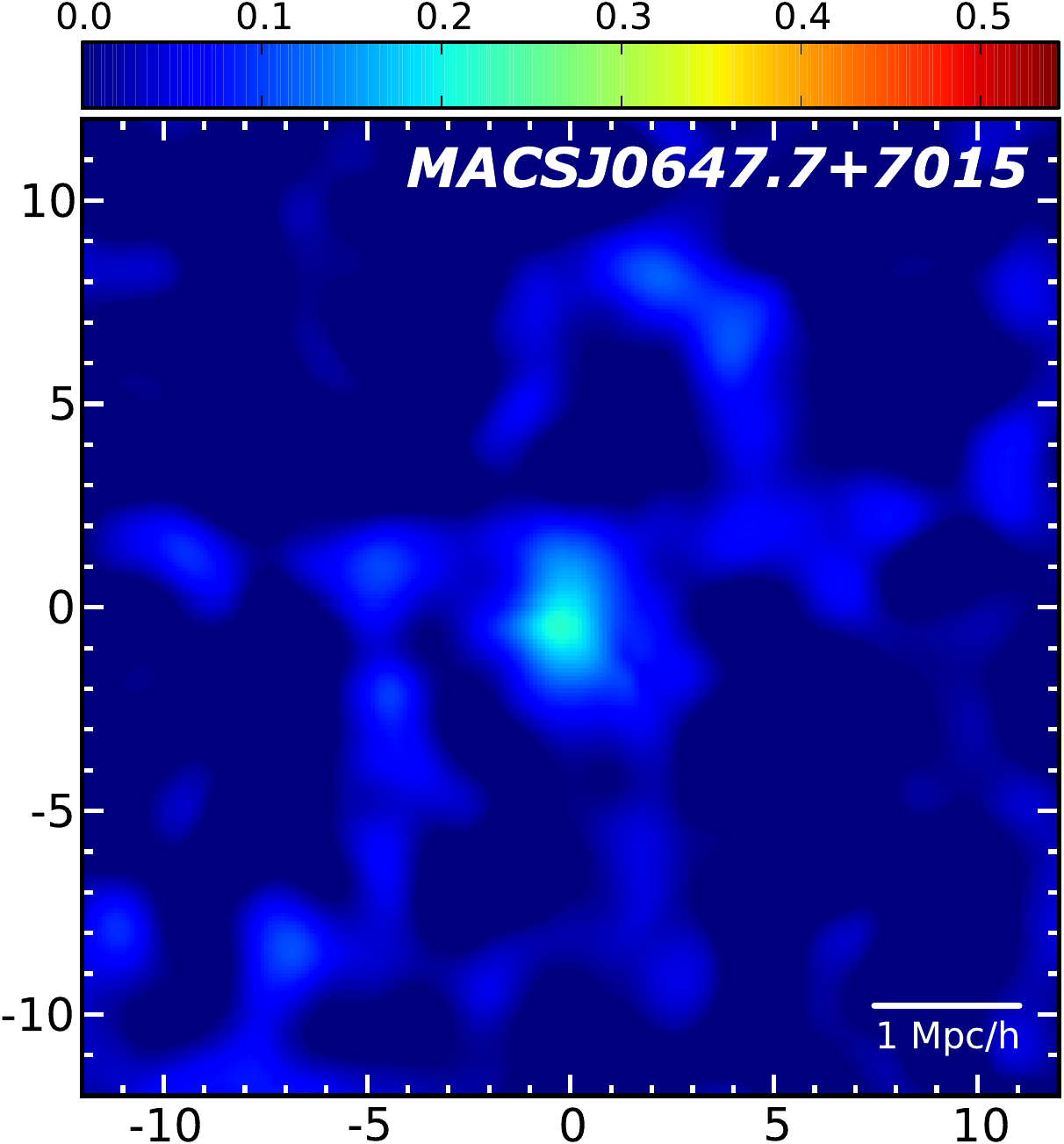} 
 \end{array}
 $
 \end{center}
  \caption{
Two-dimensional weak-lensing mass maps $\Sigma(\btheta)$
for a sample of 20 CLASH clusters reconstructed using wide-field
 multi-color imaging observations.  
For visualization purposes,
all images are smoothed with a circular Gaussian of FWHM $1.8\arcmin$.
 The images are $24\arcmin\times 24\arcmin$ in size, and centered on the
 respective optical cluster centers.
The color bar indicates the lensing convergence
 $\kappa=\Sigma/\Sigma_{\rm c}$, the surface
 mass density $\Sigma$ in units of the critical surface mass density
 $\Sigma_{\rm c}$.
The color scale is linear from ${\rm S/N}\equiv \kappa/\sigma_\kappa$ of
 $0$ (dark blue) to $15$ (dark red) for all clusters. North is to the
 top, east to the left. 
The horizontal bar in each panel represents $1$\,Mpc\,$h^{-1}$ at the
 cluster redshift.
\label{fig:map}} 
\end{figure*}

In this section we carry out a joint 
shear-and-magnification analysis of a sample of  
20 CLASH clusters. 
In Figure \ref{fig:map} we present our weak-lensing distortion data
for our sample
in the form of two-dimensional mass maps, 
where we have used the linear map-making method
outlined in \citet{Umetsu+2009}.
The mass maps are
smoothed with a Gaussian with $1.8\arcmin$ FWHM, and presented primarily
for visualization purposes.

\subsection{Cluster Center}
\label{subsec:center}

As summarized in Table \ref{tab:sample},
our cluster sample exhibits on average a small
offset $d_{\rm off}= |\bd_{\rm off}|$ between the BCG and X-ray peak,
characterized by a median offset of 
$\overline{d}_{\rm off}\simeq 10\,{\rm kpc}\,h^{-1}$
and an rms offset of 
$\sigma_{\rm off}\simeq 30\,{\rm kpc}\,h^{-1}$.
For the X-ray-selected subsample, we find a much smaller rms of
$\sigma_{\rm off}\simeq 11$\,kpc\,$h^{-1}$. 
\citet{Johnston+2007b} demonstrated that
$\kappa$ is much less affected by cluster miscentering than $\gamma_+$, 
and that the miscentering effects on $\kappa$ nearly vanish
at twice the typical positional offset from the cluster mass centroid.
This indicates that our mass-profile reconstructions 
would not be affected, on average, by
the miscentering effects beyond a radius of $R\sim 60$\,kpc\,$h^{-1}$
($R\sim 20$\,kpc\,$h^{-1}$ for the X-ray-selected subsample).
This level of centering offset is much smaller than the range of
overdensity radii of interest for cluster mass measurements,
and hence will not significantly affect our
cluster mass profile measurements.\footnote{This level of offset could
potentially lead to an underestimation of the central cusp slope, which
however is beyond the scope of this weak-lensing analysis.}
In what follows, we will adopt the
BCG position as the cluster center (Table \ref{tab:sample}).


\subsection{Cluster Mass Profiles}
\label{subsec:wlmass}


\begin{figure*}[!htb] 
 \begin{center}
 $
 \begin{array}
  {c@{\hspace{.1in}}c@{\hspace{.1in}}c@{\hspace{.1in}}c@{\hspace{.1in}}c}
  \includegraphics[width=0.22\textwidth,angle=0,clip]{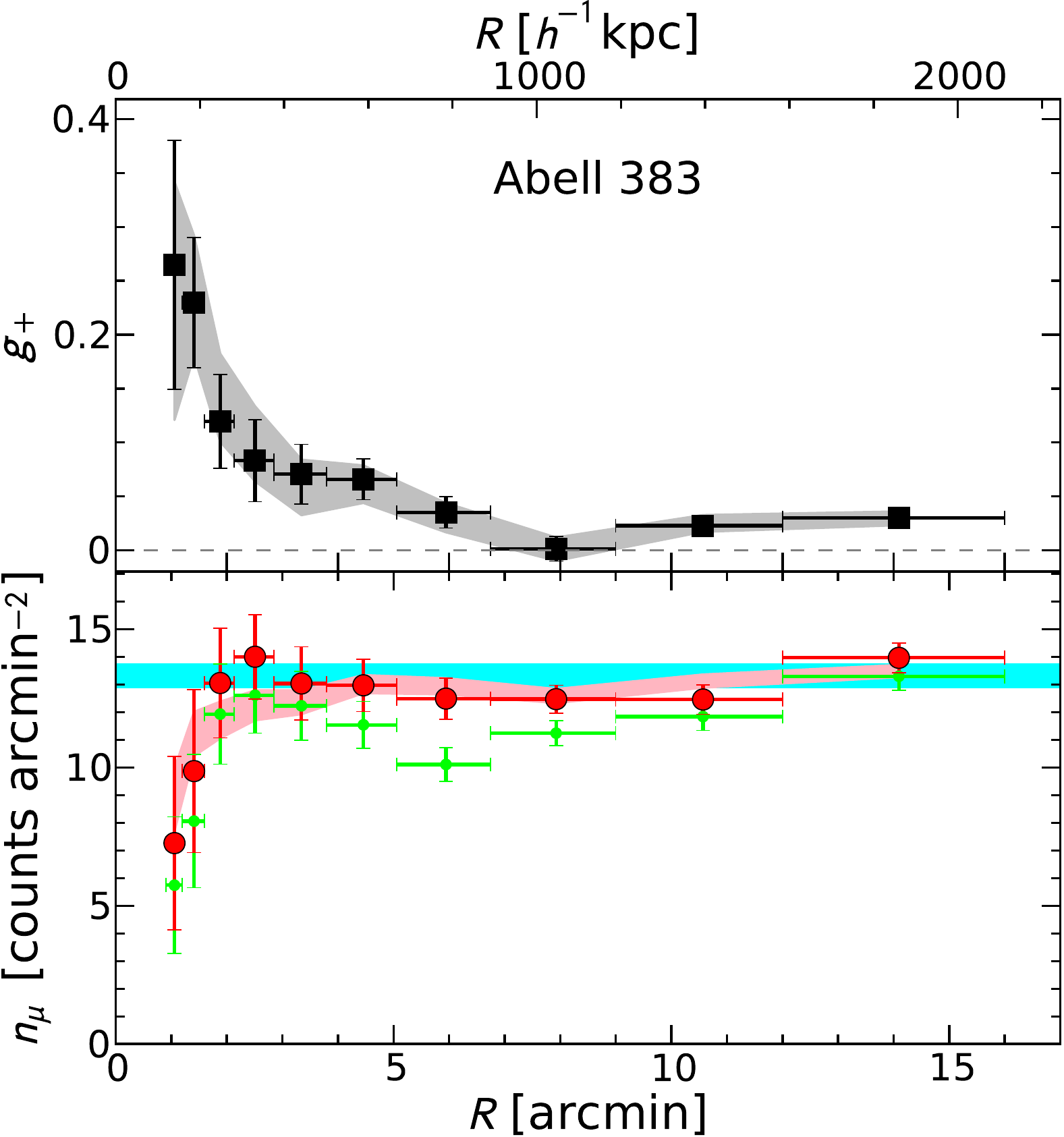} &
  \includegraphics[width=0.22\textwidth,angle=0,clip]{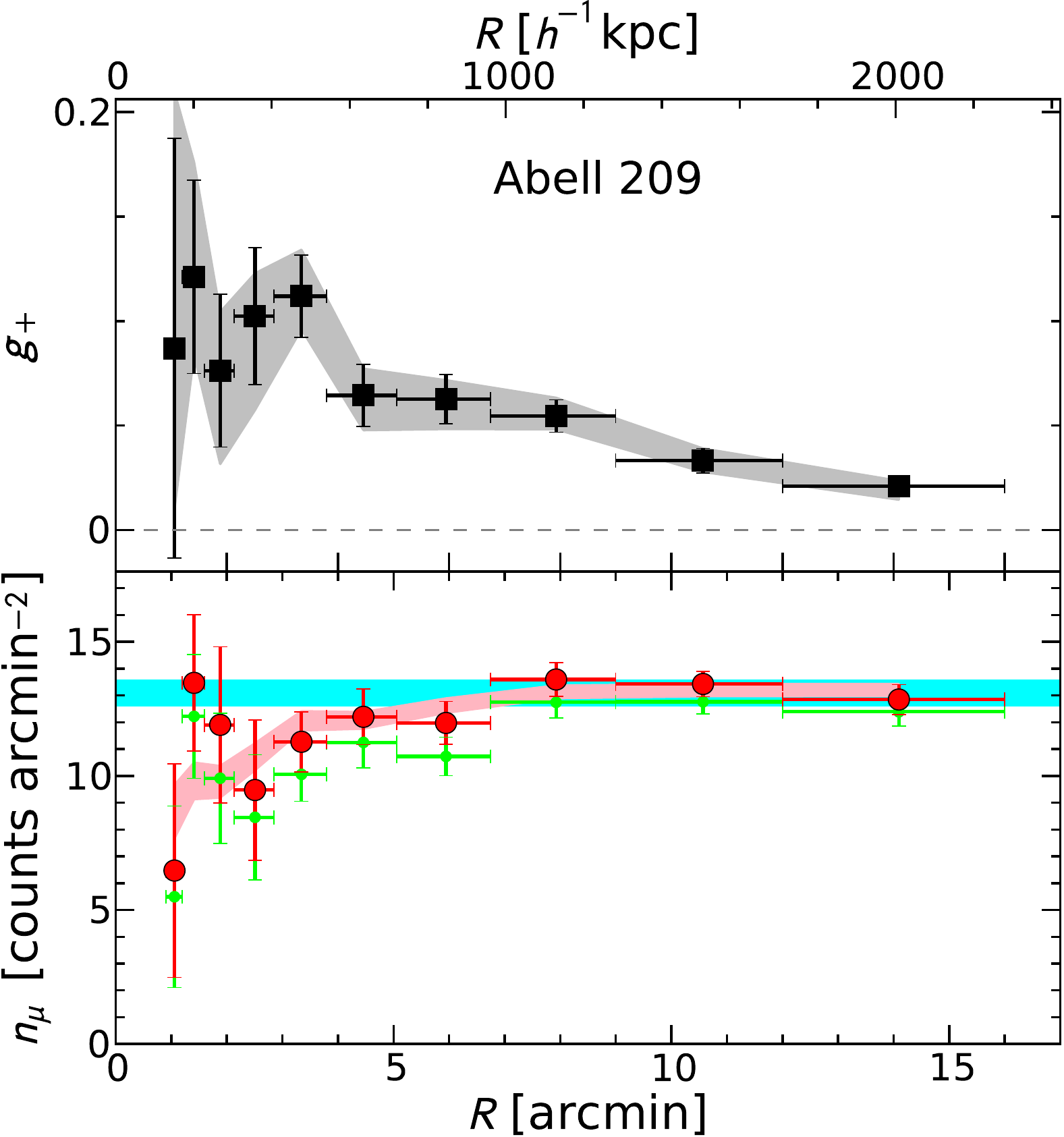} &
  \includegraphics[width=0.22\textwidth,angle=0,clip]{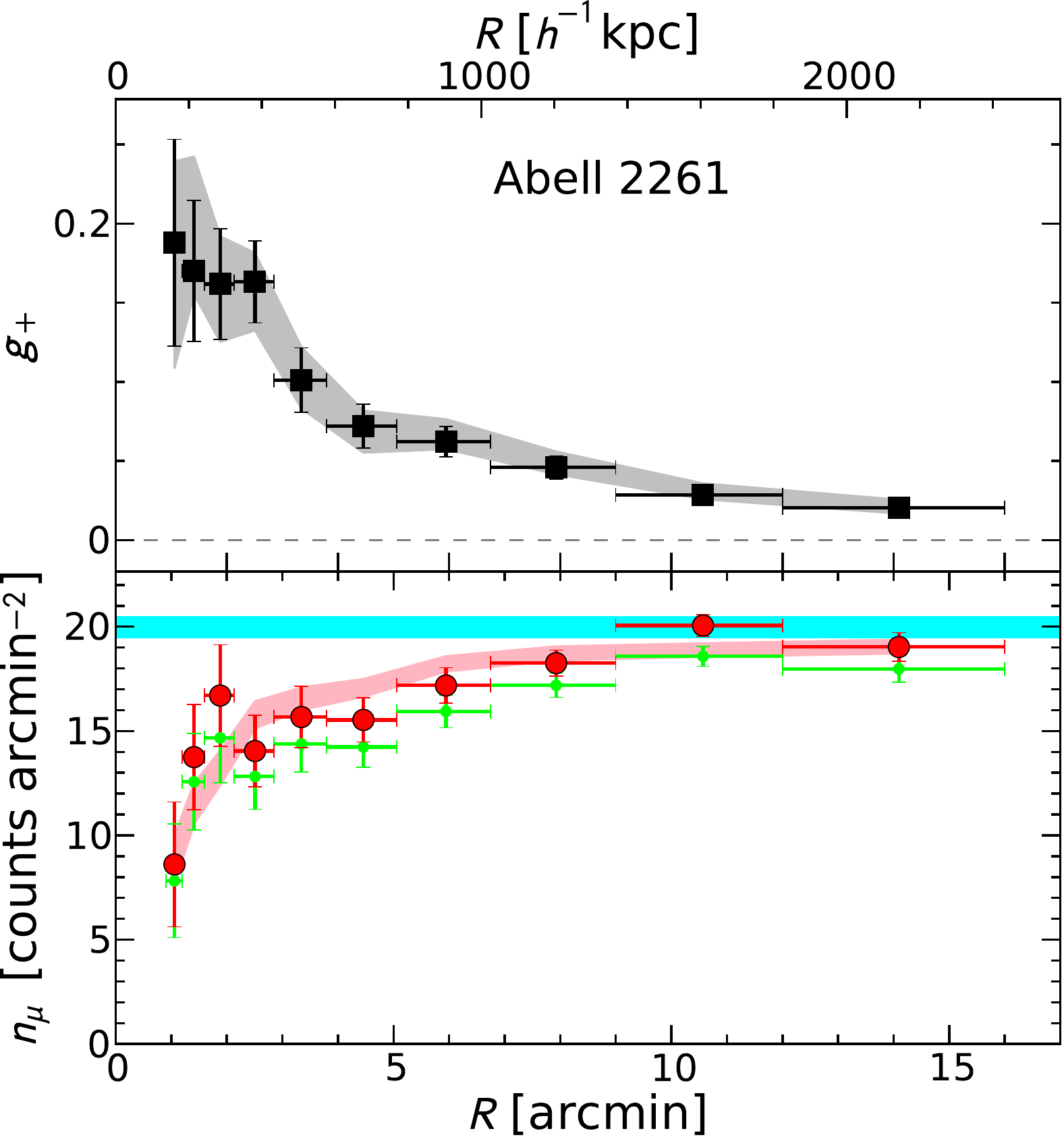}&
  \includegraphics[width=0.22\textwidth,angle=0,clip]{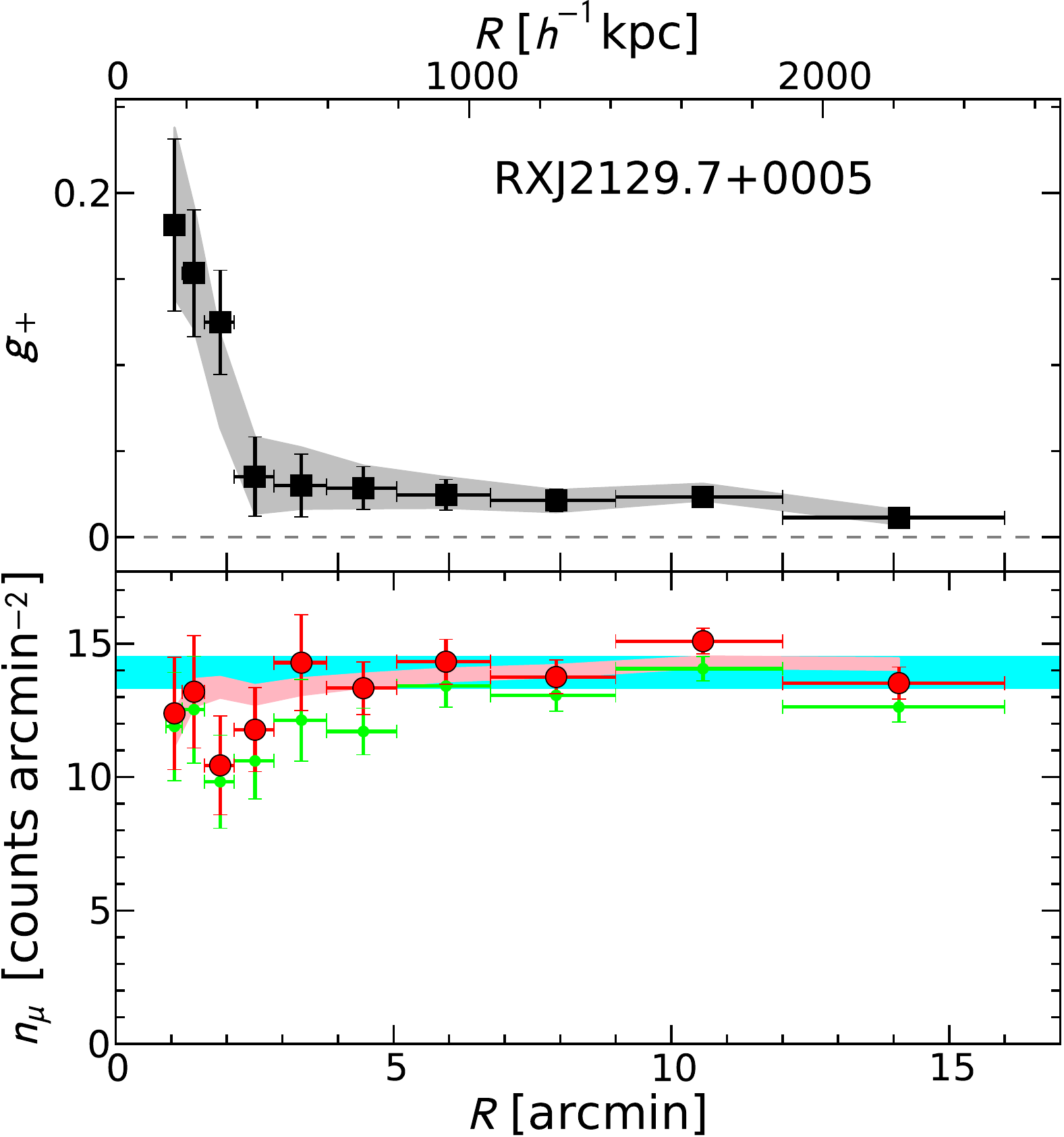}
 \end{array}
 $
 $
 \begin{array}
  {c@{\hspace{.1in}}c@{\hspace{.1in}}c@{\hspace{.1in}}c@{\hspace{.1in}}c}
  \includegraphics[width=0.22\textwidth,angle=0,clip]{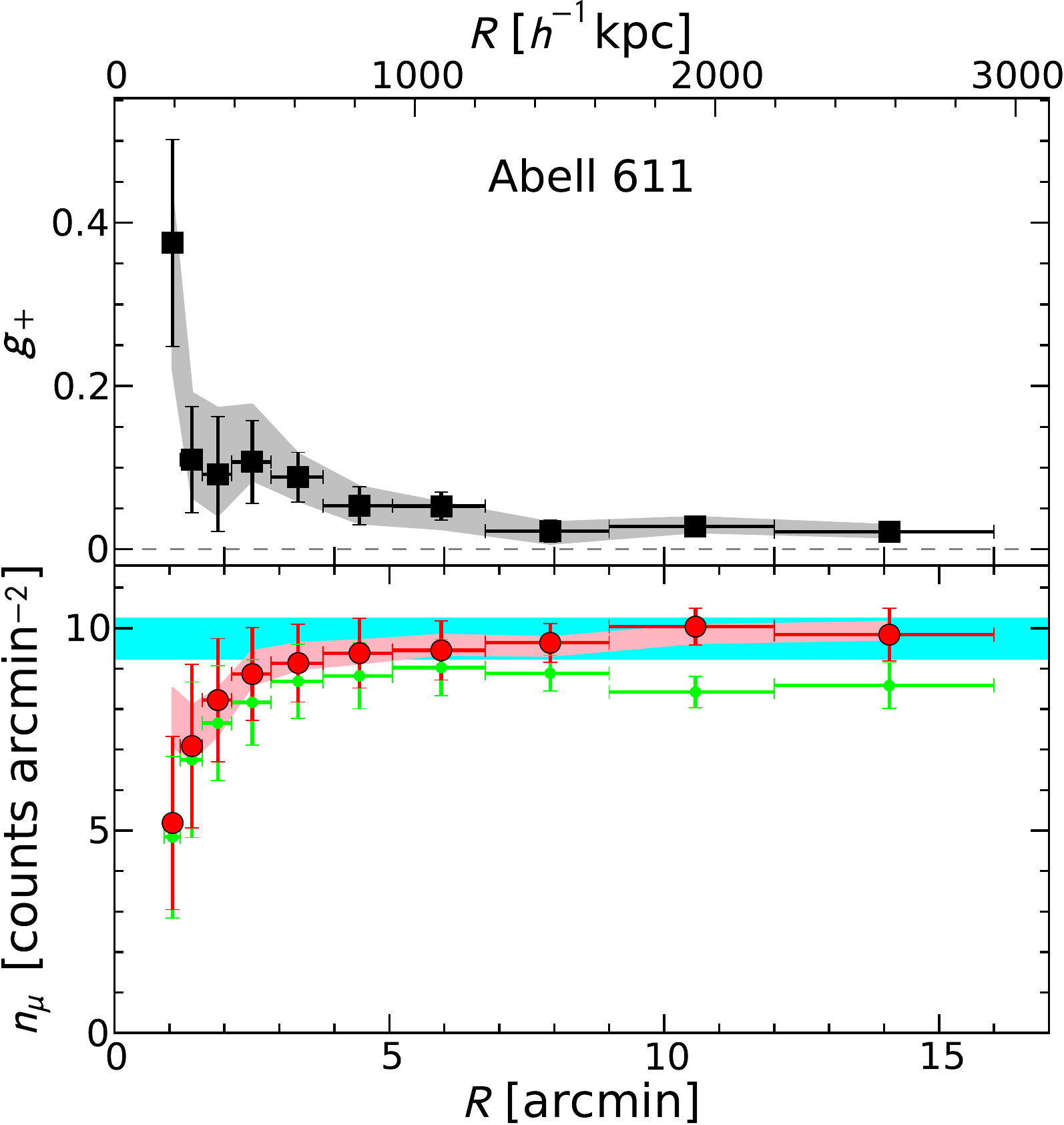} &
  \includegraphics[width=0.22\textwidth,angle=0,clip]{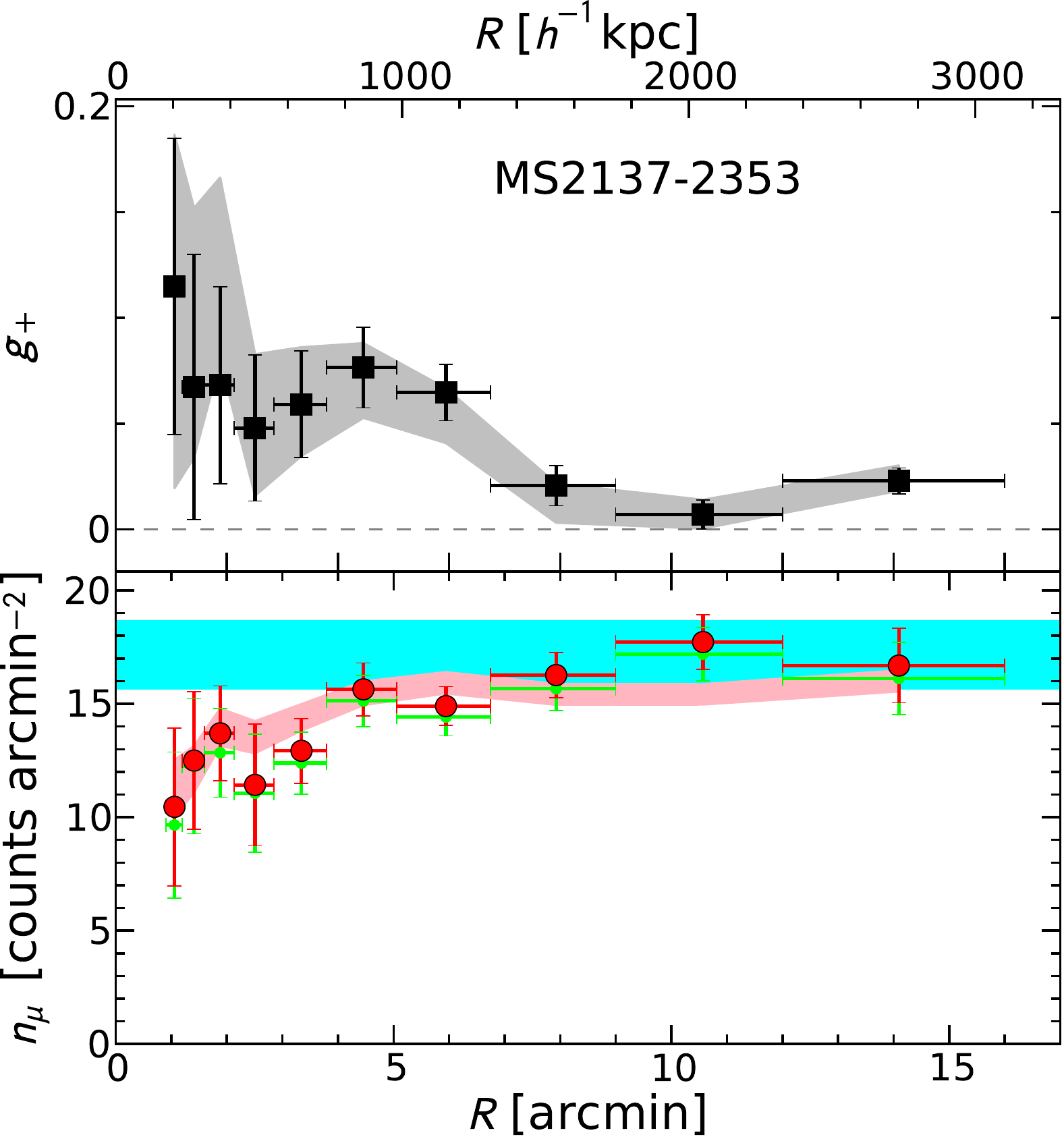}& 
  \includegraphics[width=0.22\textwidth,angle=0,clip]{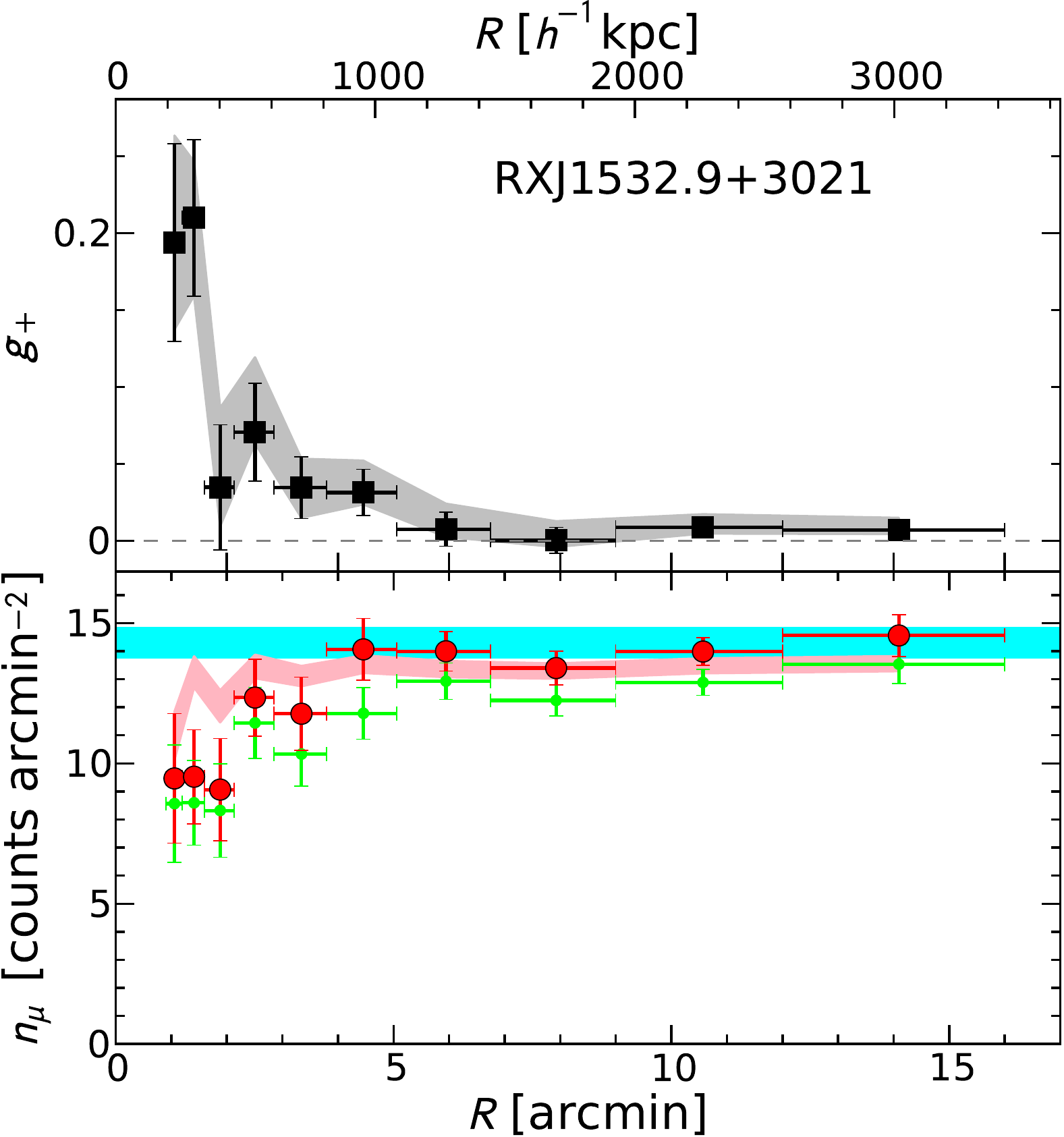}&
  \includegraphics[width=0.22\textwidth,angle=0,clip]{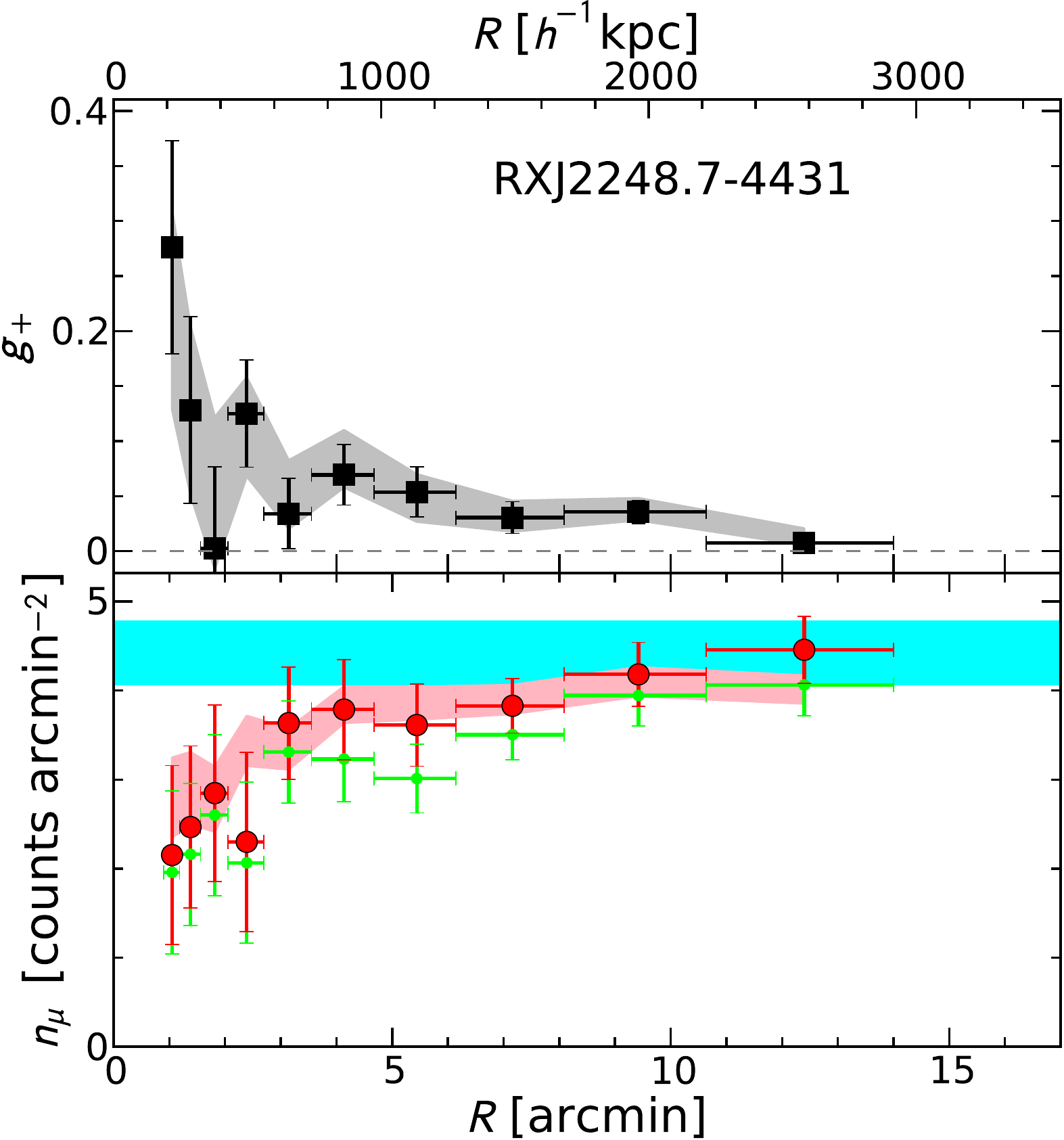} 
 \end{array}
  $
  $
 \begin{array}
  {c@{\hspace{.1in}}c@{\hspace{.1in}}c@{\hspace{.1in}}c@{\hspace{.1in}}c}
   \includegraphics[width=0.22\textwidth,angle=0,clip]{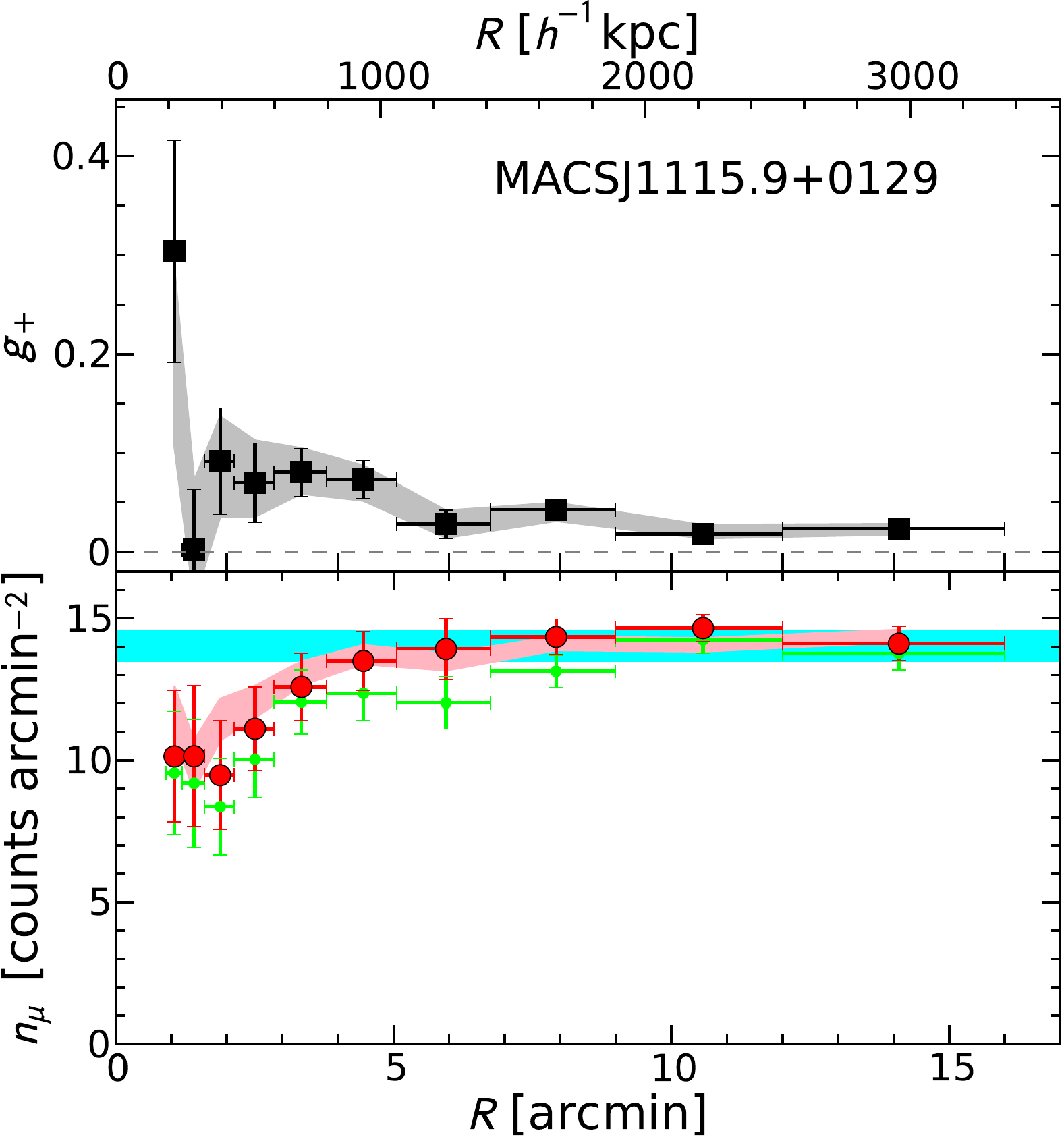}& 
   \includegraphics[width=0.22\textwidth,angle=0,clip]{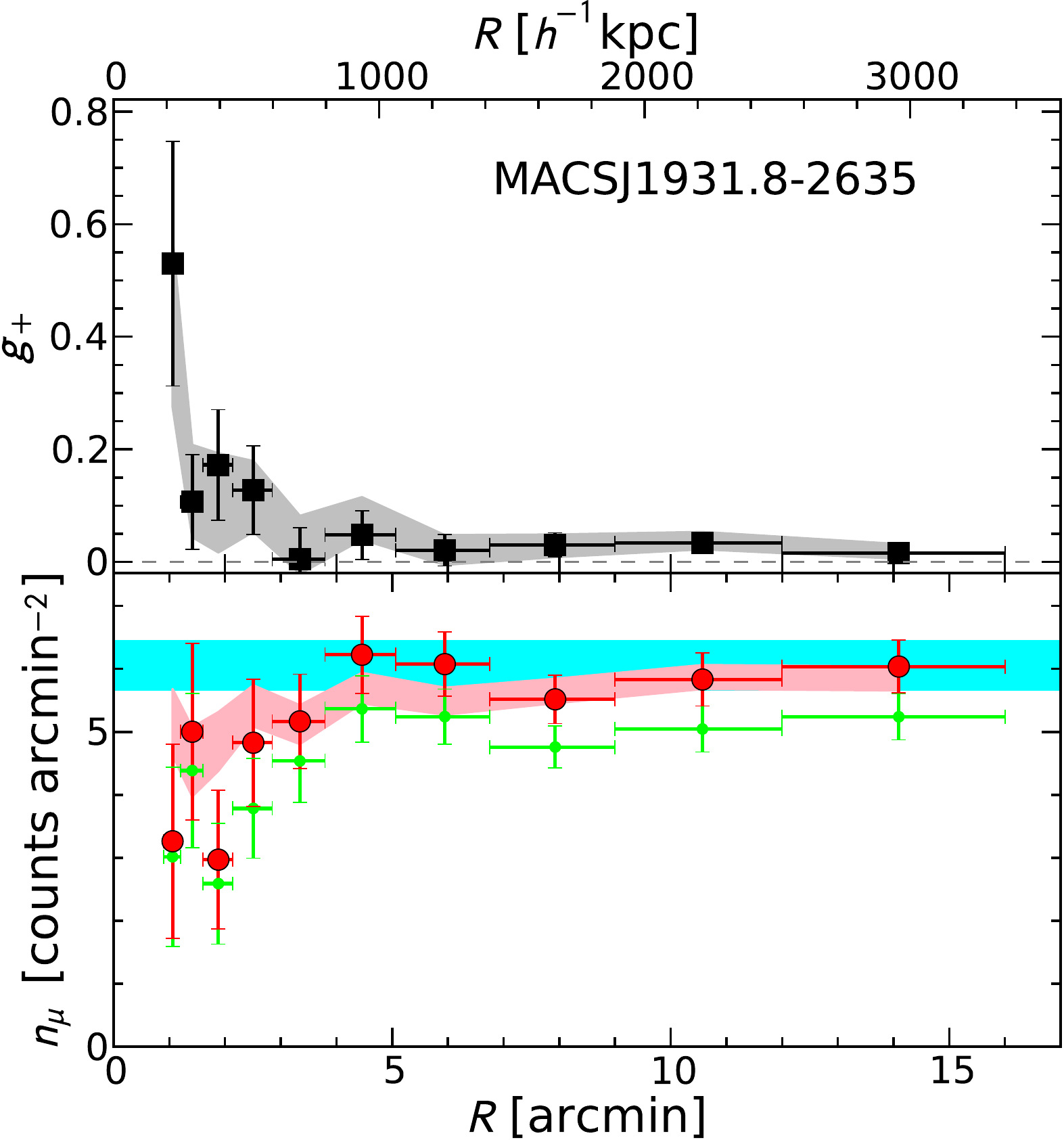}&
   \includegraphics[width=0.22\textwidth,angle=0,clip]{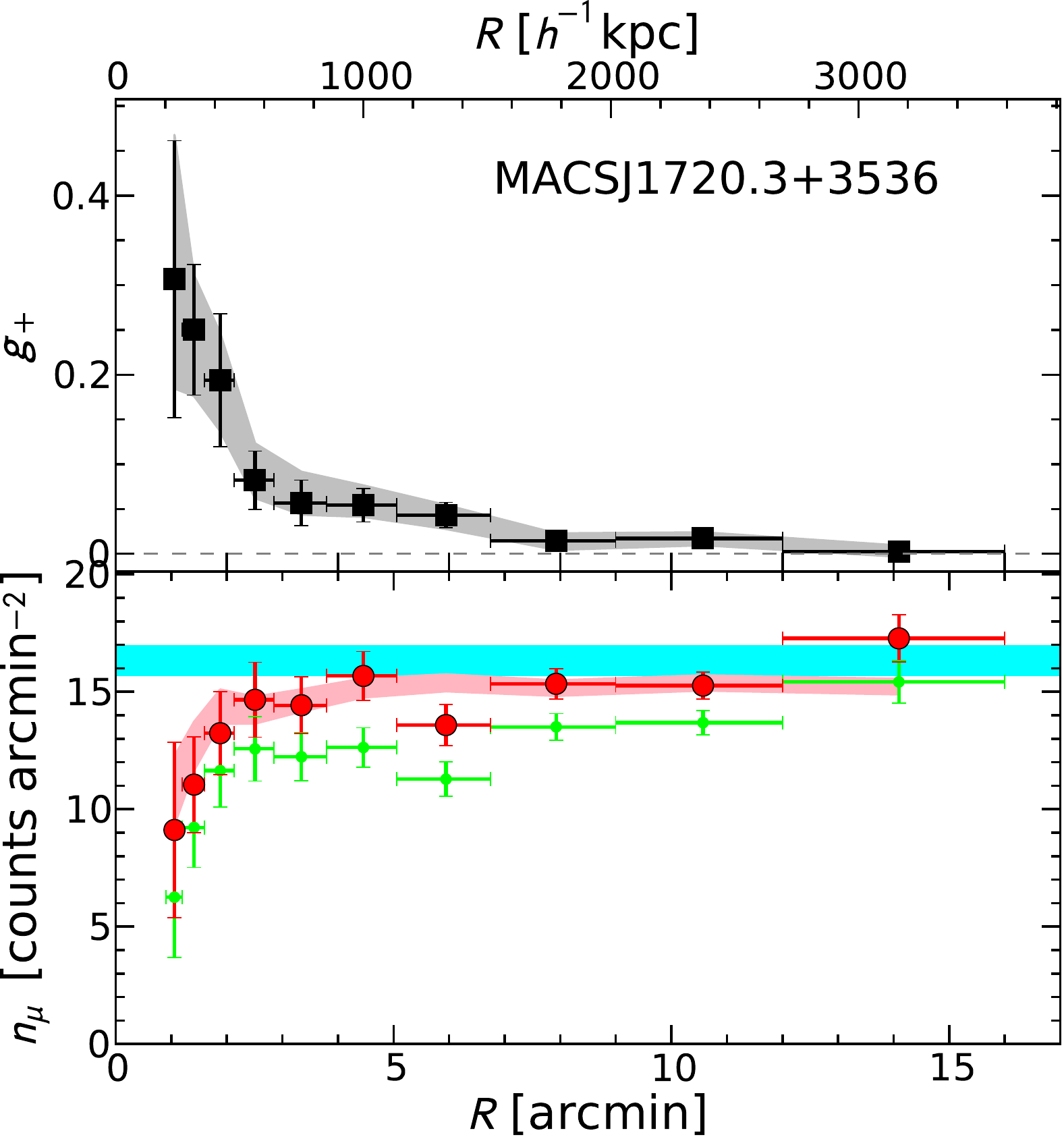}&
   \includegraphics[width=0.22\textwidth,angle=0,clip]{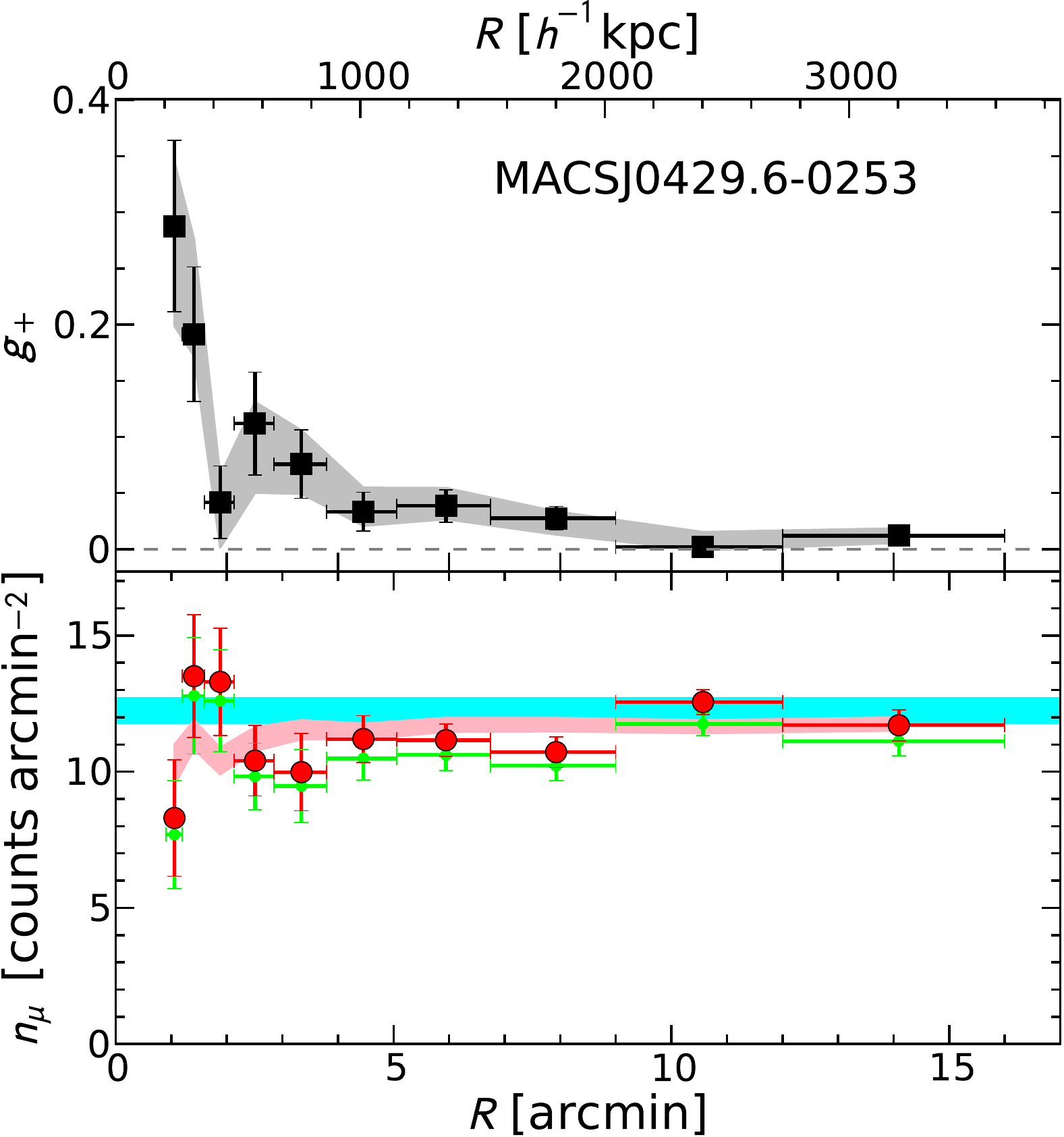}
  \end{array}
 $
 $
 \begin{array}
  {c@{\hspace{.1in}}c@{\hspace{.1in}}c@{\hspace{.1in}}c@{\hspace{.1in}}c}
   \includegraphics[width=0.22\textwidth,angle=0,clip]{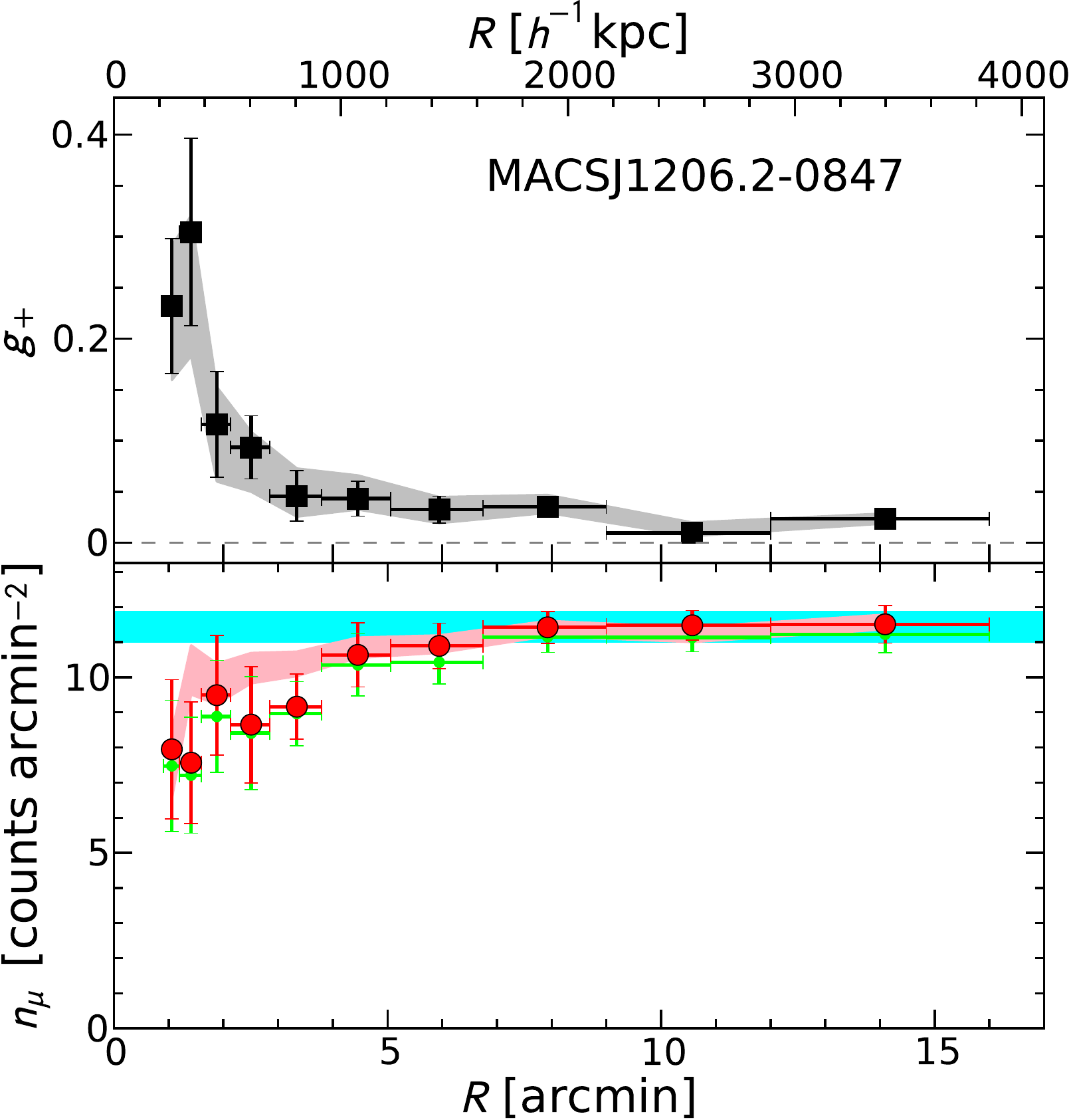}& 
   \includegraphics[width=0.22\textwidth,angle=0,clip]{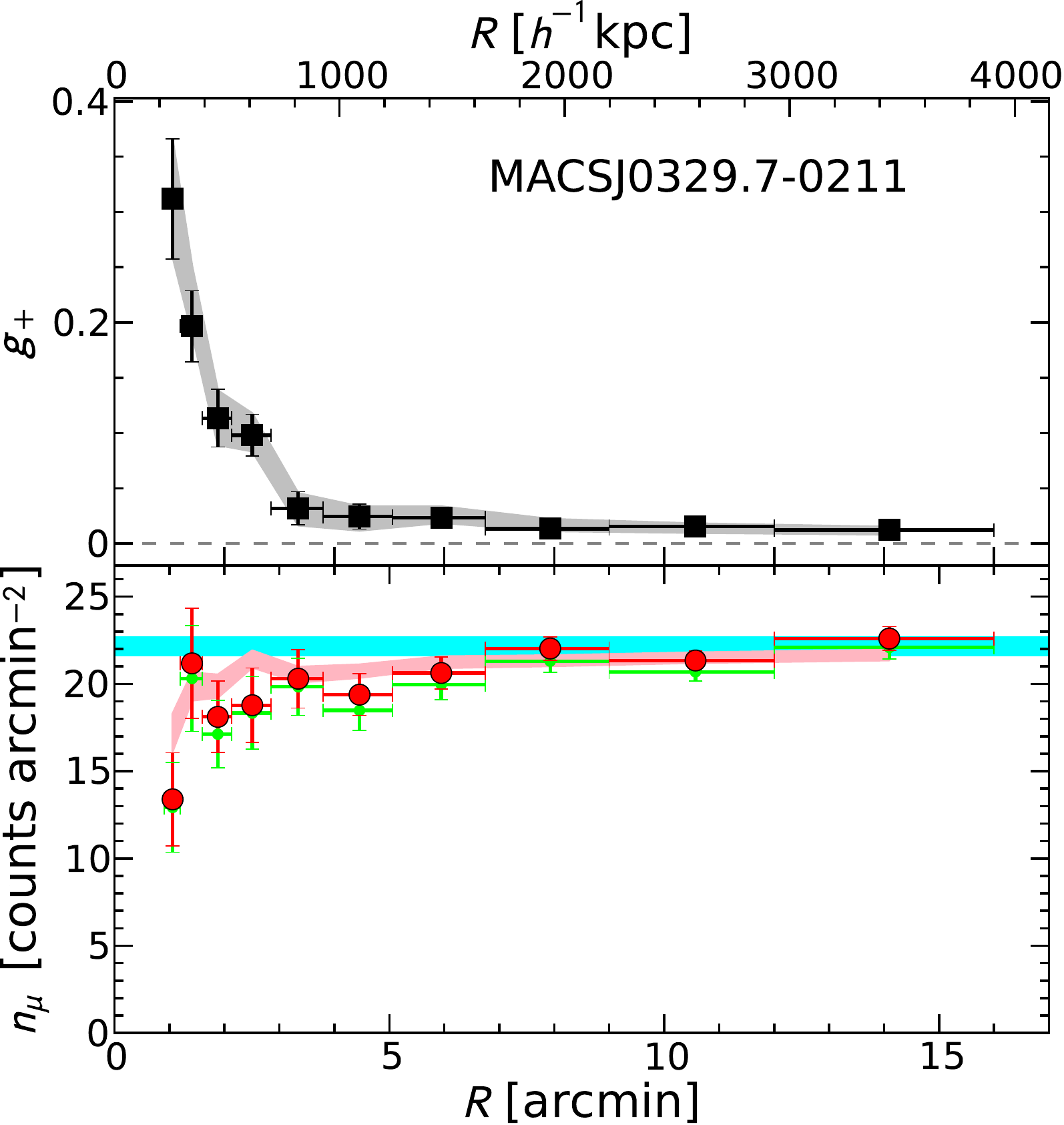}&
   \includegraphics[width=0.22\textwidth,angle=0,clip]{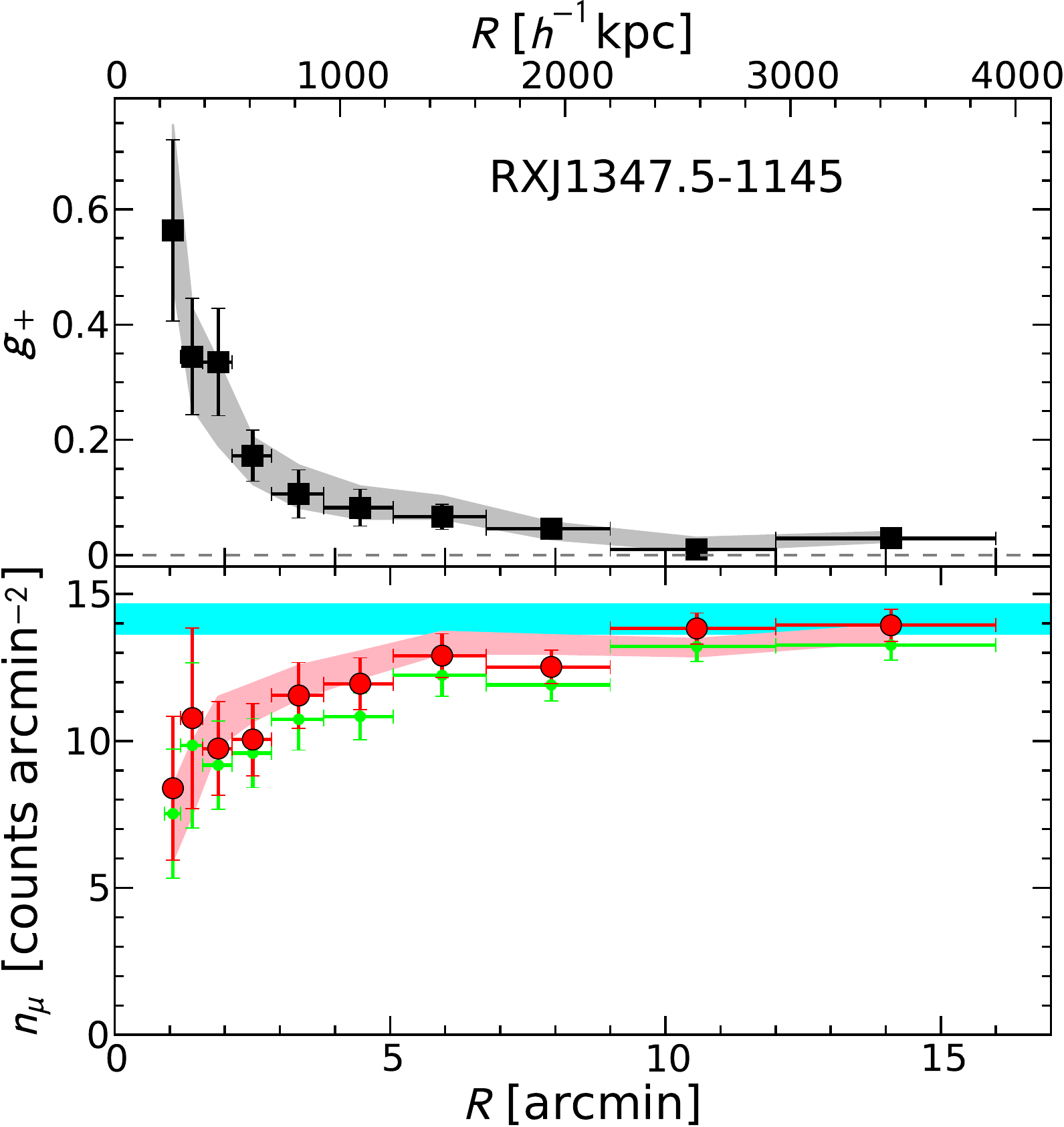} &
   \includegraphics[width=0.22\textwidth,angle=0,clip]{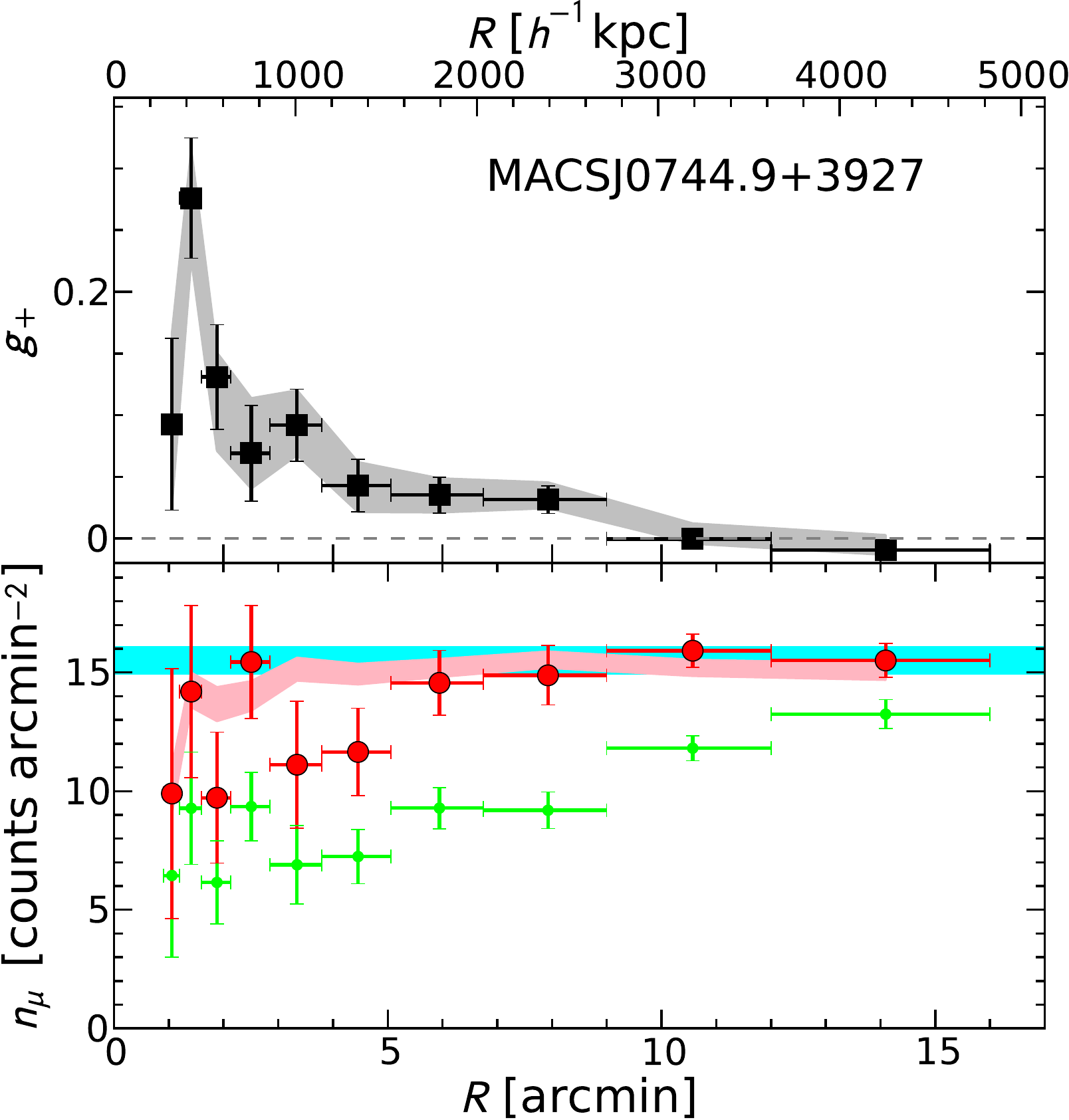} 
 \end{array}
 $
 $
 \begin{array}
  {c@{\hspace{.1in}}c@{\hspace{.1in}}c@{\hspace{.1in}}c@{\hspace{.1in}}c}
   \includegraphics[width=0.22\textwidth,angle=0,clip]{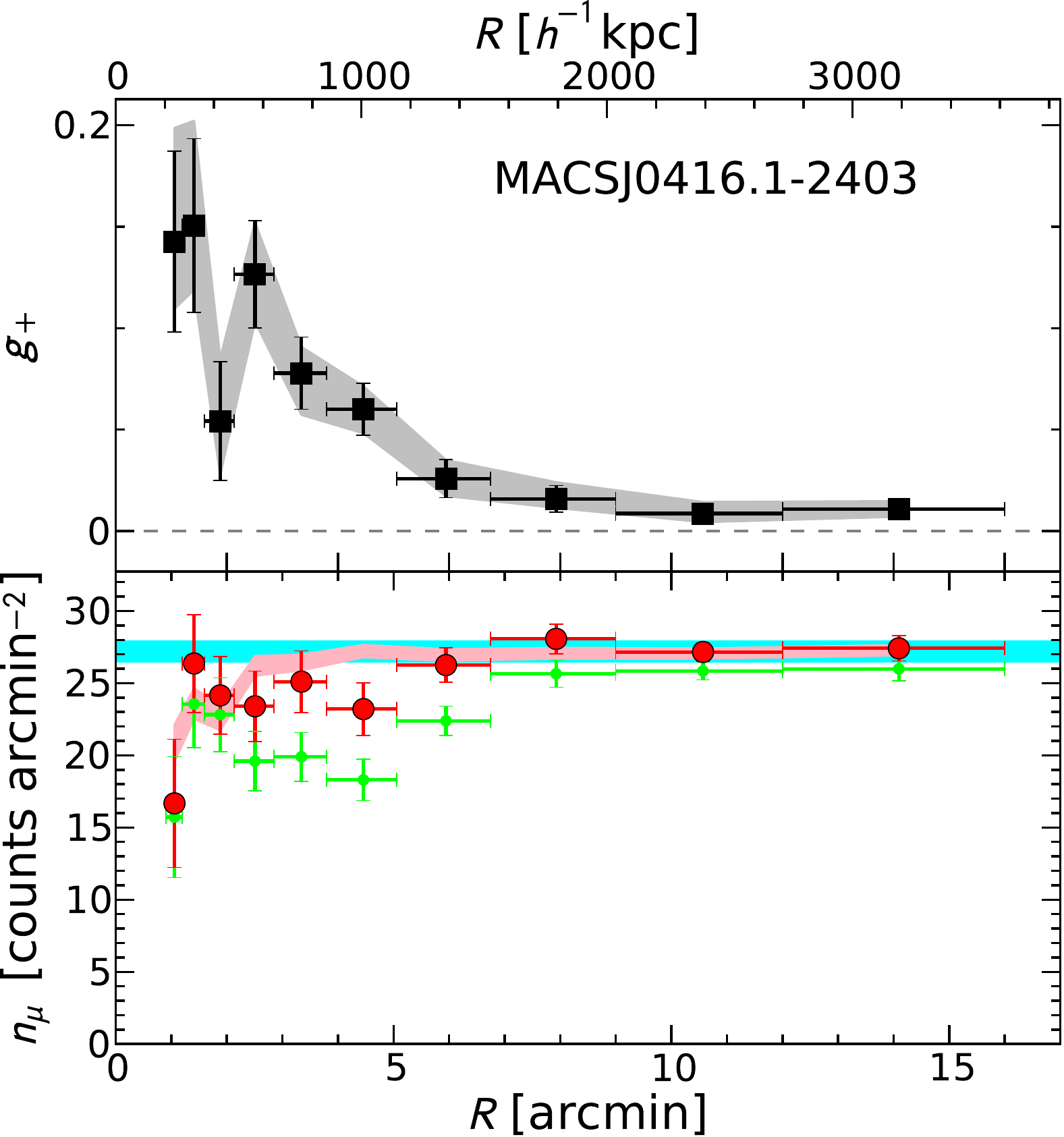}& 
   \includegraphics[width=0.22\textwidth,angle=0,clip]{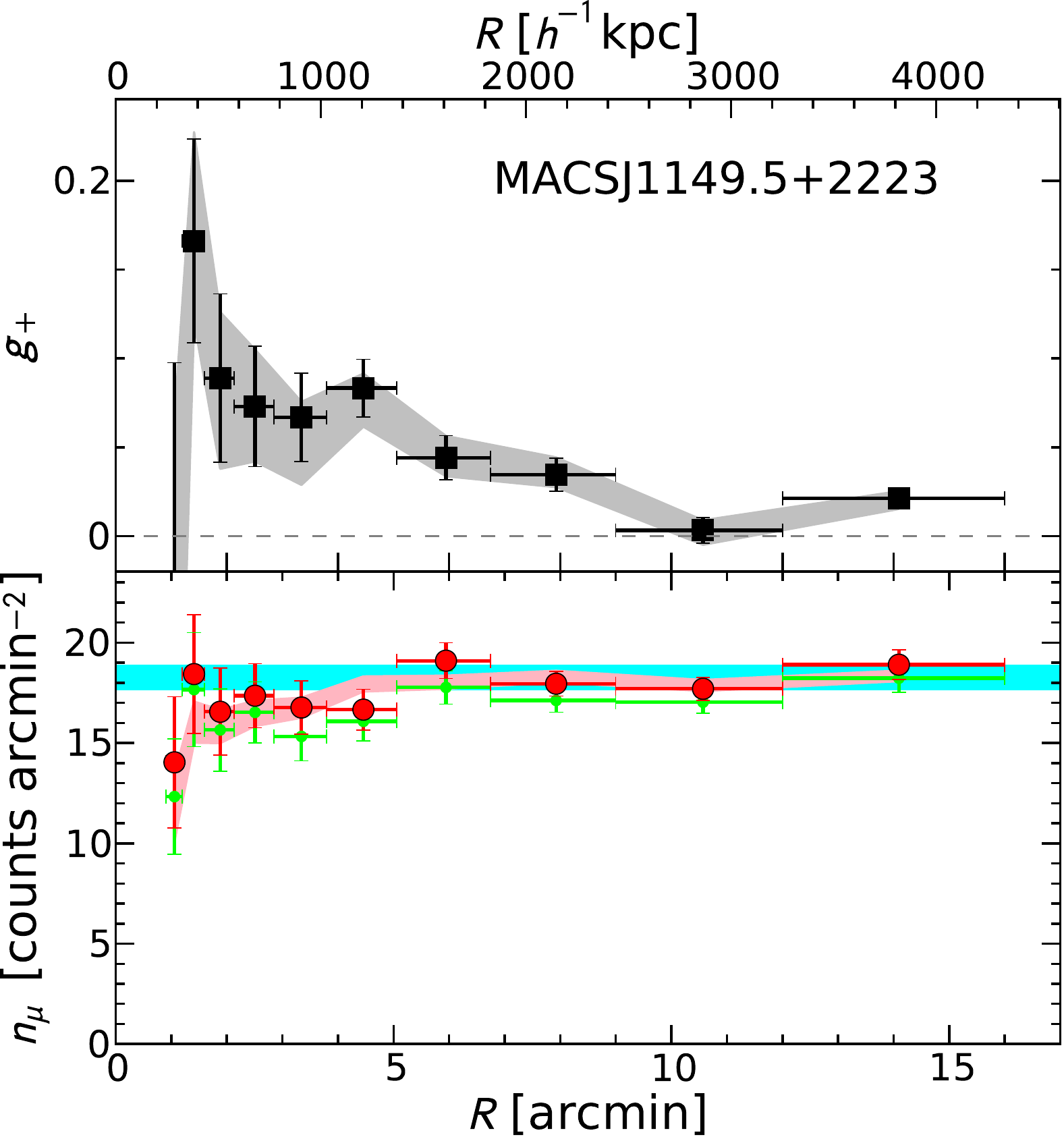}&
   \includegraphics[width=0.22\textwidth,angle=0,clip]{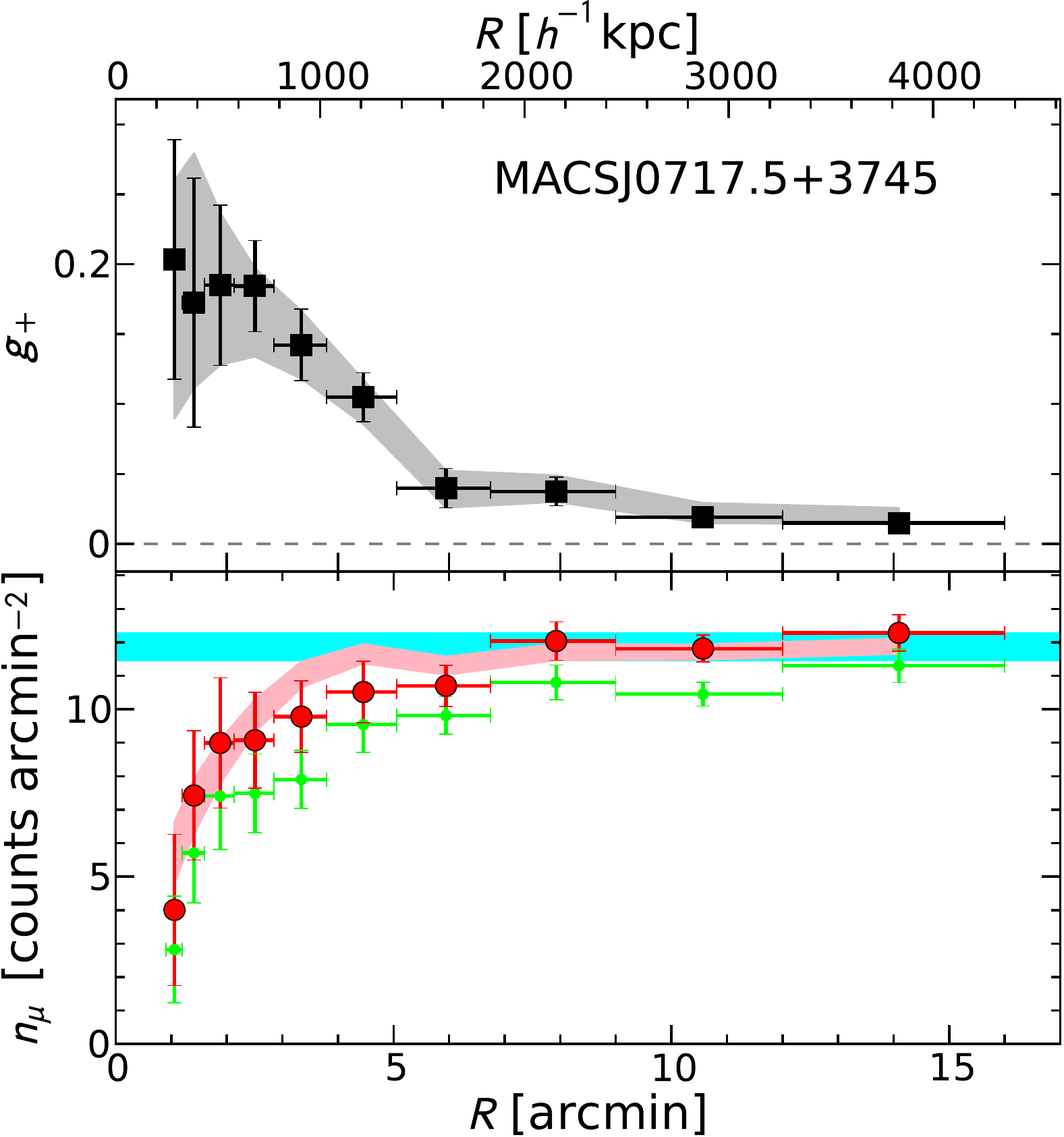} &
   \includegraphics[width=0.22\textwidth,angle=0,clip]{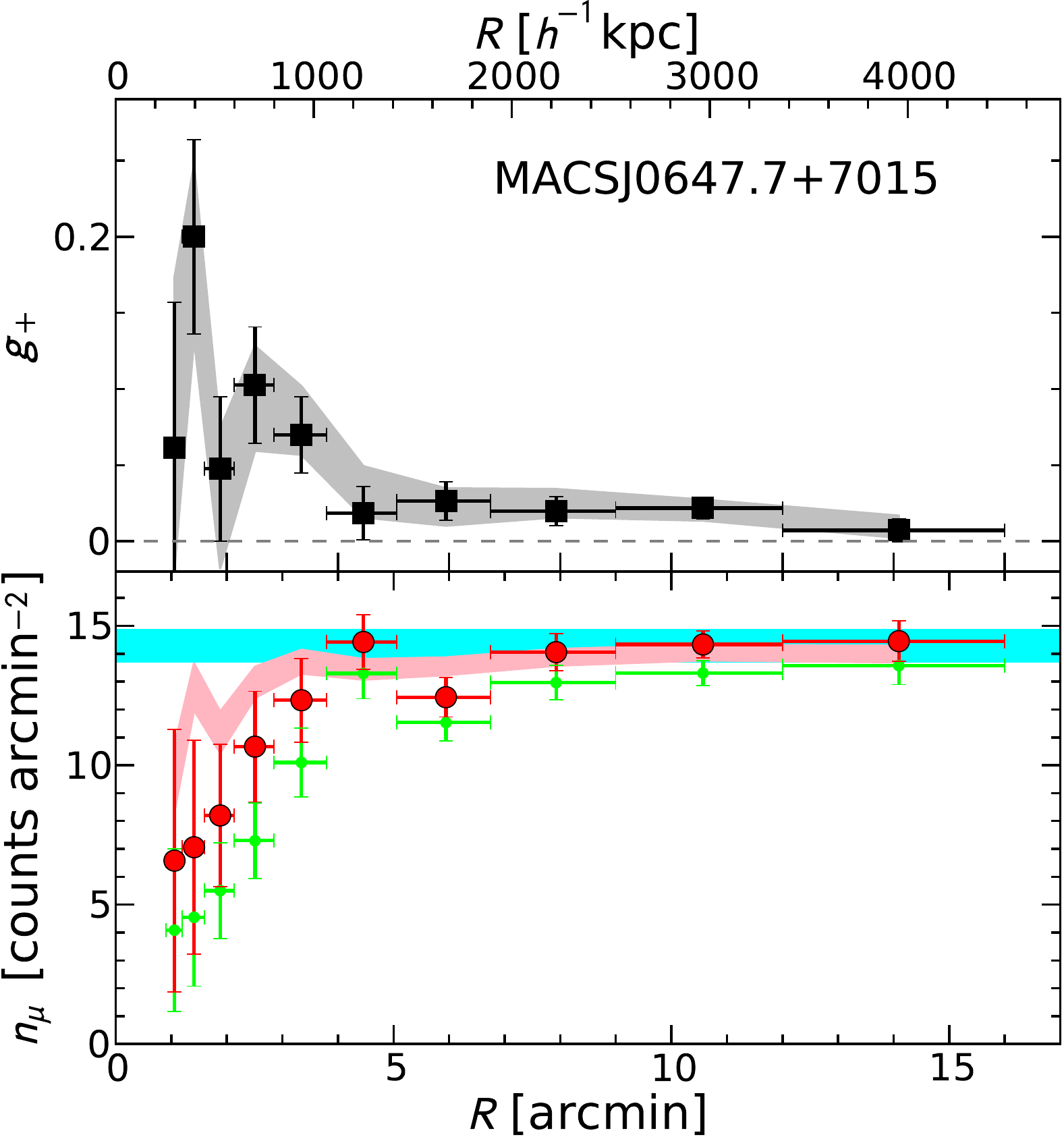} 
 \end{array}
 $
 \end{center}
\caption{
Azimuthally-averaged radial profiles of the tangential-distortion
 ($g_+$) and magnification-bias ($n_\mu$) measurements obtained from
 wide-field multi-color imaging observations, shown individually for our
 sample of 20 CLASH clusters. 
For each cluster, the upper panel shows the tangential reduced shear
 $g_+$ (black squares) of background galaxies, 
and the lower panel shows the count-depletion
 profile $n_\mu$ of red background  galaxies, with (red circles) and
 without (green dots) the mask  correction  
 due to bright foreground objects and cluster members. 
A systematic radial depletion of the source counts is seen toward the cluster center
due to magnification of the sky area.
The error bars include
contributions from Poisson counting uncertainties and contamination due to
intrinsic angular clustering of red galaxies.
For each observed profile, the shaded area represents the 
joint Bayesian  reconstruction ($68\%$ CL) from the combined 
tangential-distortion and  magnification-bias measurements.
The horizontal bar (cyan shaded region) shows the constraints on the unlensed count
normalization $\overline{n}_\mu$
estimated from the source counts in cluster outskirts.
A large correction for the incomplete area coverage, accounting for
 masked regions due to bright saturated stars,
was applied to the number count profiles of low Galactic latitude
 clusters ($|b| < 30^\circ$), such as
 MACS\,J1931.8-2635 and MACS\,J0744.9+3927. 
\label{fig:wldata}}
\end{figure*}


\begin{figure*}[!htb] 
 \begin{center}
 $
 \begin{array}
  {c@{\hspace{.1in}}c@{\hspace{.1in}}c@{\hspace{.1in}}c@{\hspace{.1in}}c}
  \includegraphics[width=0.22\textwidth,angle=0,clip]{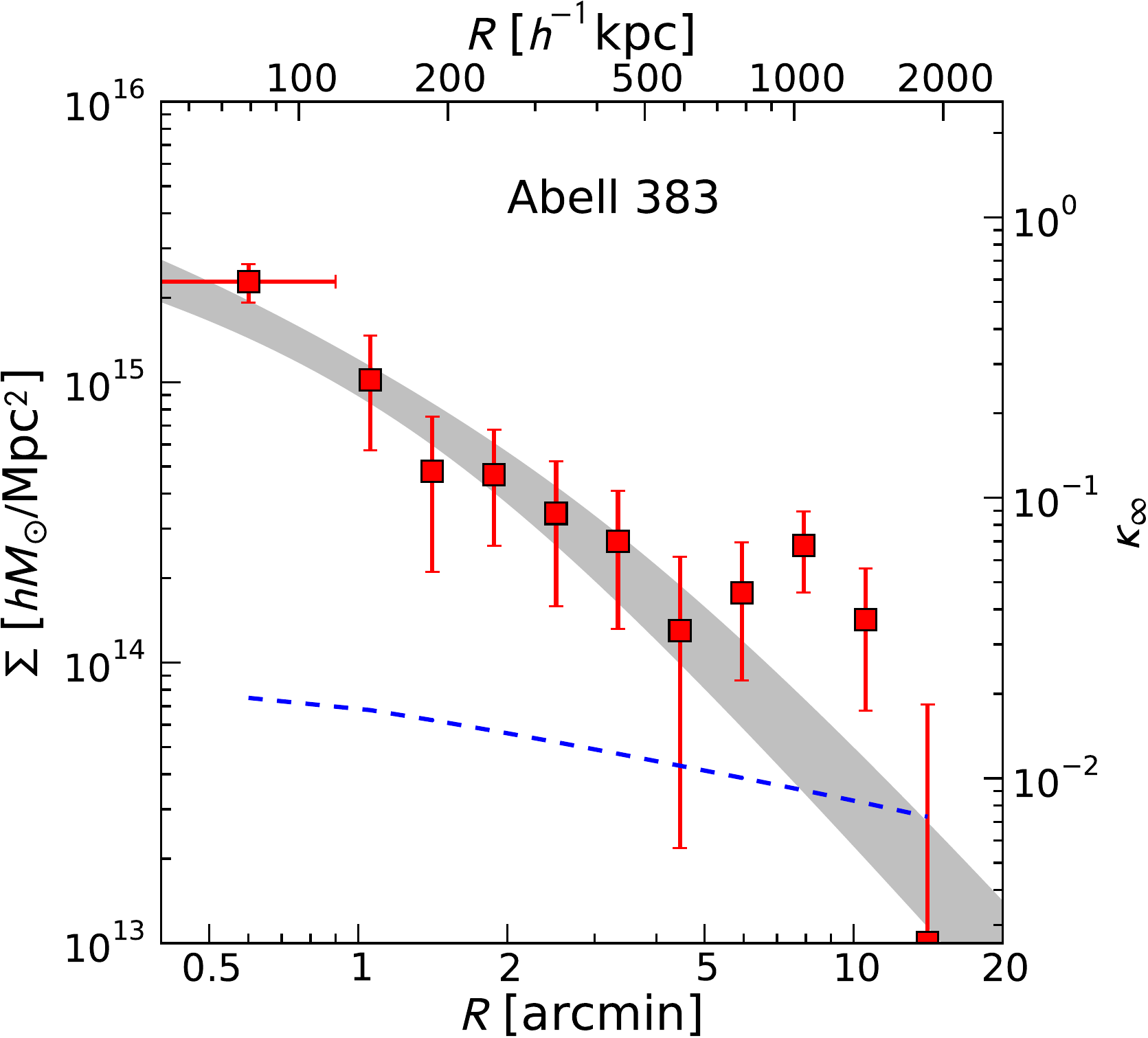} &
  \includegraphics[width=0.22\textwidth,angle=0,clip]{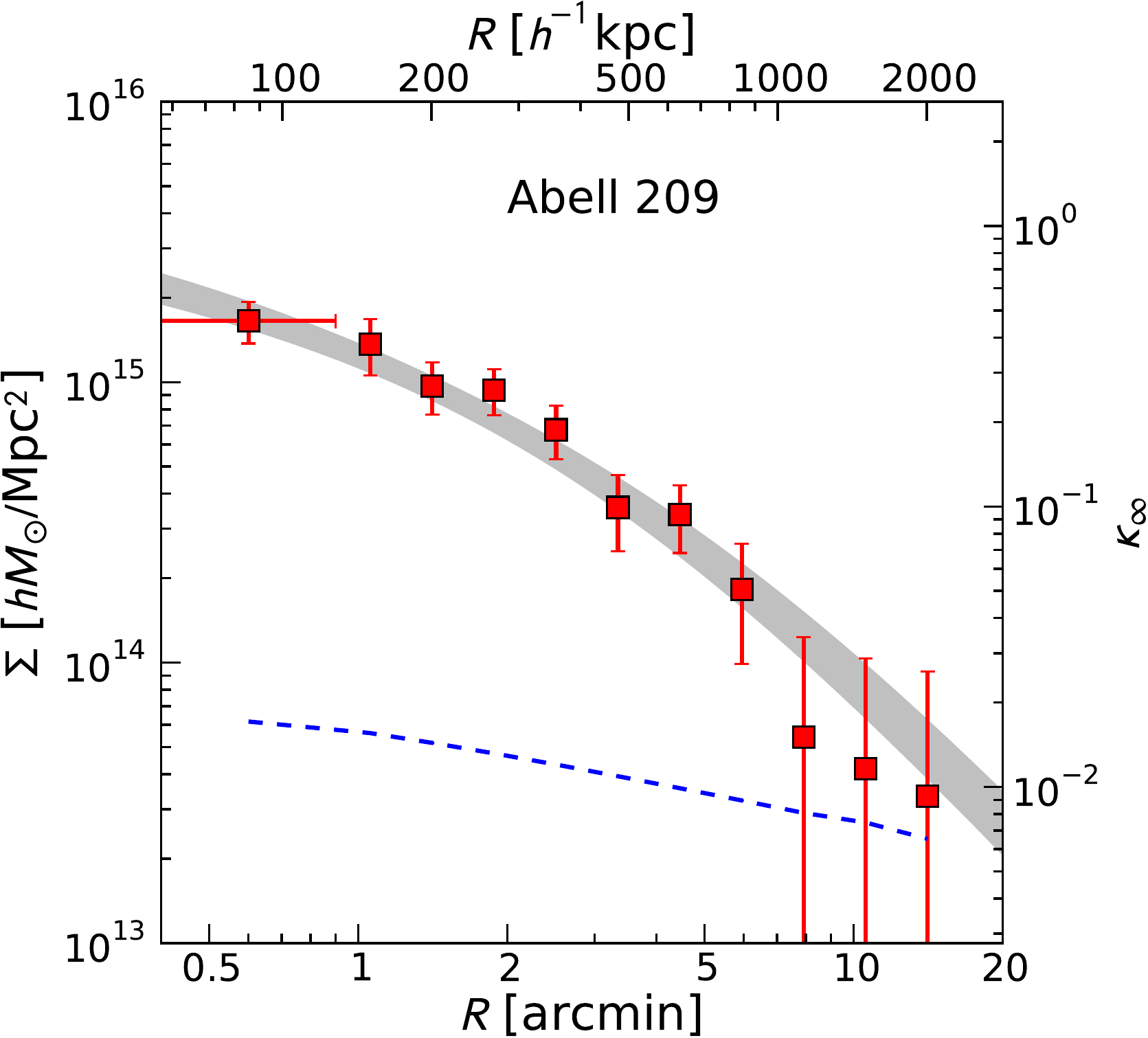} &
  \includegraphics[width=0.22\textwidth,angle=0,clip]{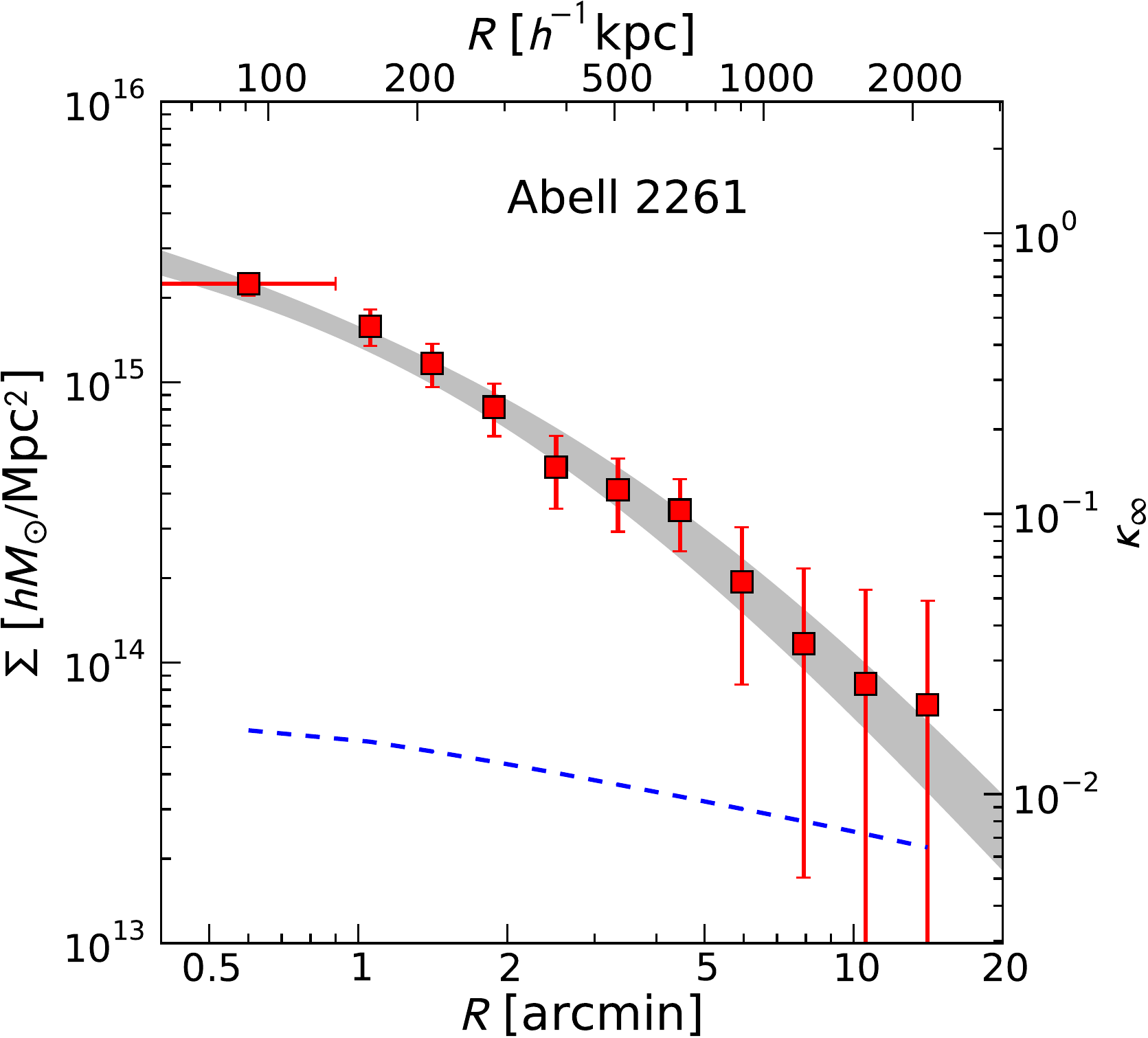}&
  \includegraphics[width=0.22\textwidth,angle=0,clip]{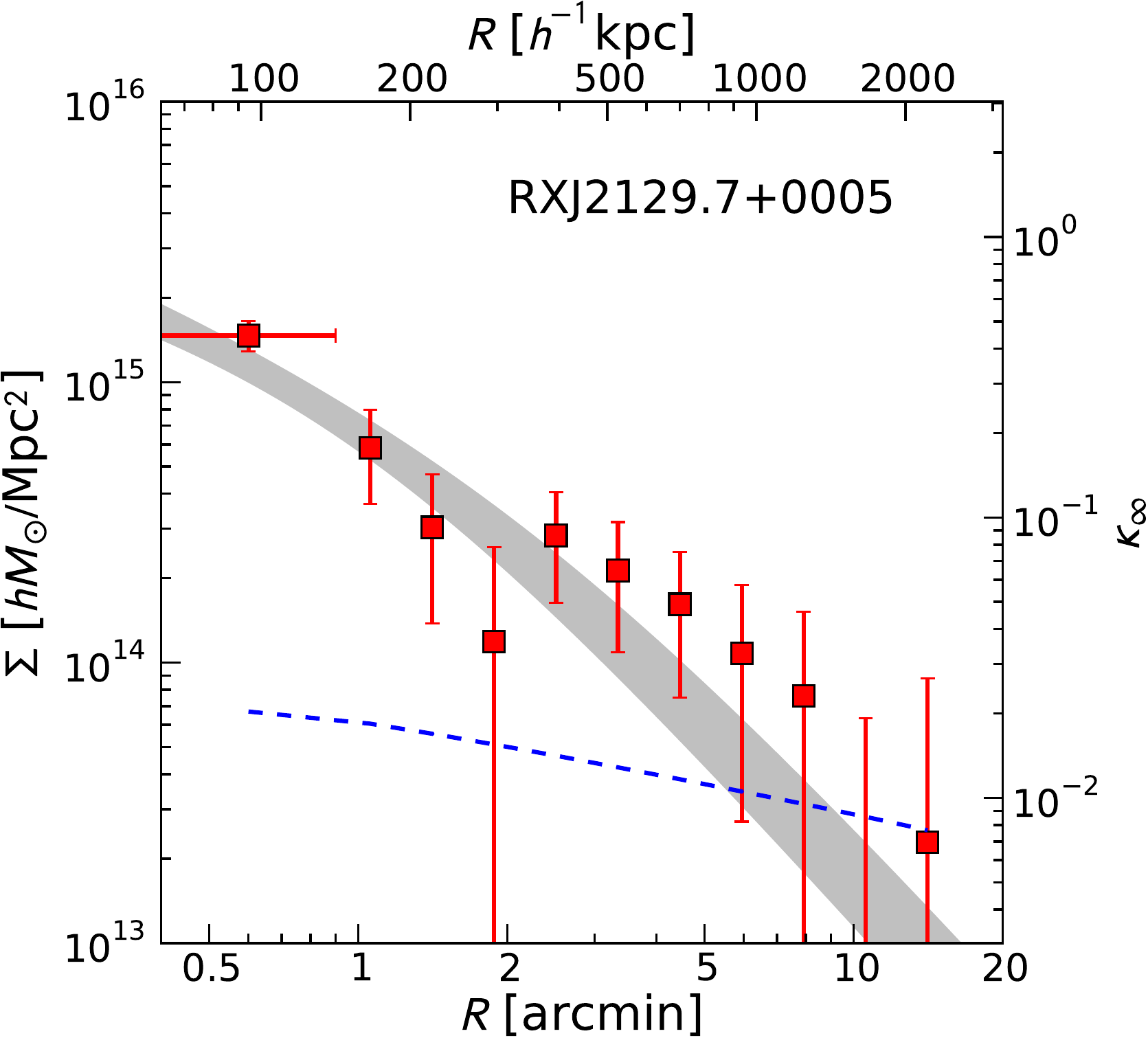}
 \end{array}
 $
 $
 \begin{array}
  {c@{\hspace{.1in}}c@{\hspace{.1in}}c@{\hspace{.1in}}c@{\hspace{.1in}}c}
  \includegraphics[width=0.22\textwidth,angle=0,clip]{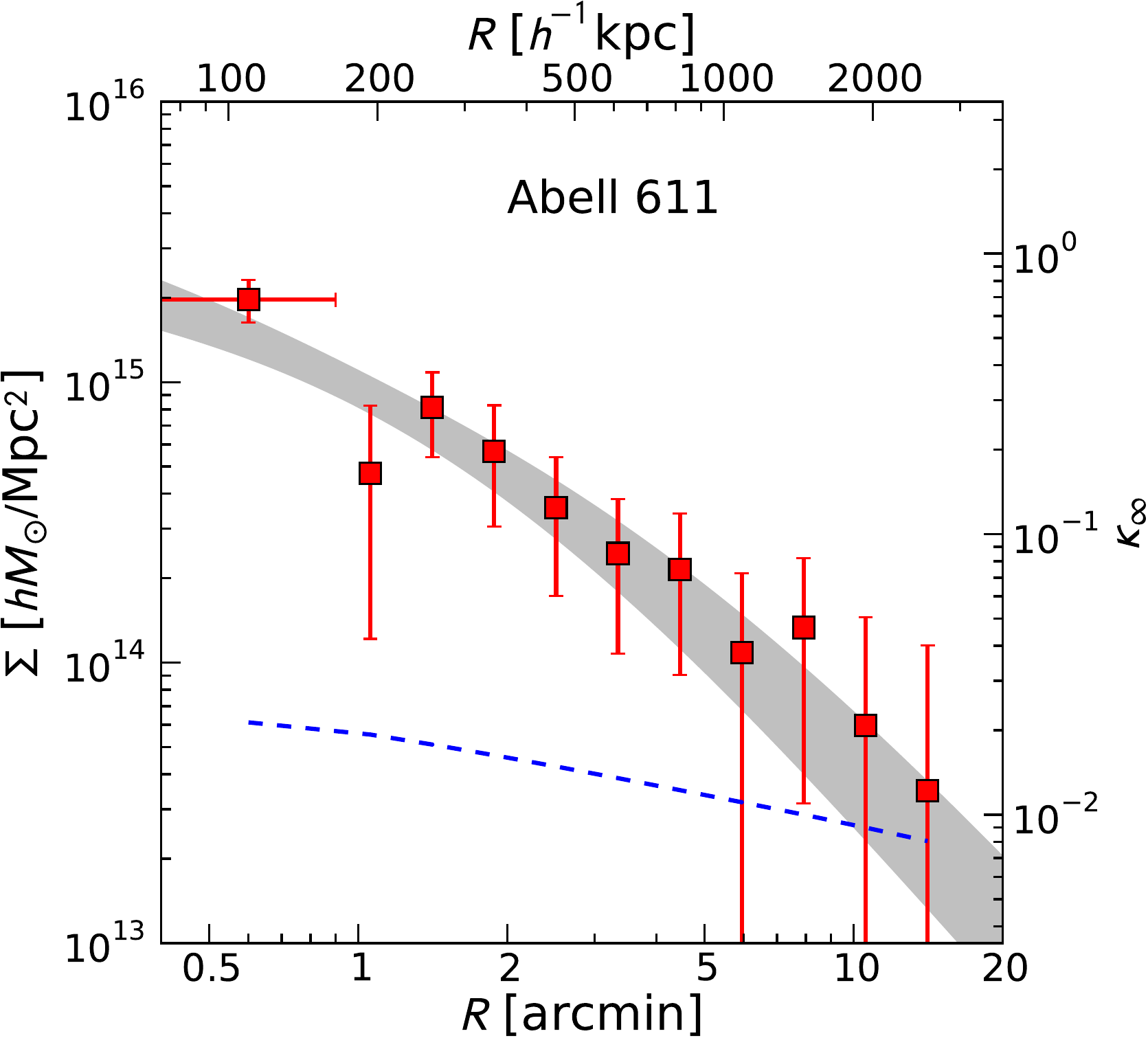} &
  \includegraphics[width=0.22\textwidth,angle=0,clip]{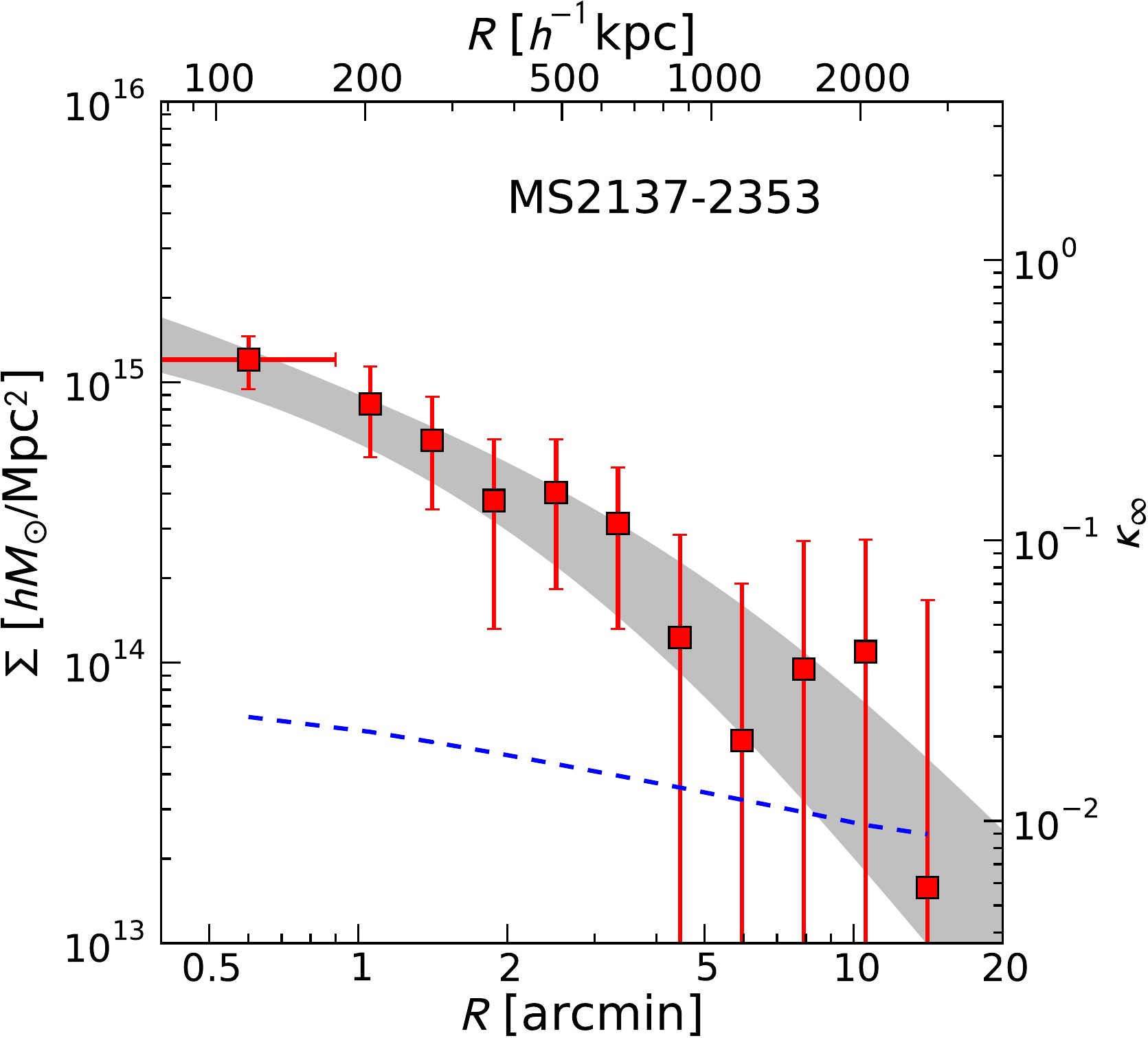}& 
  \includegraphics[width=0.22\textwidth,angle=0,clip]{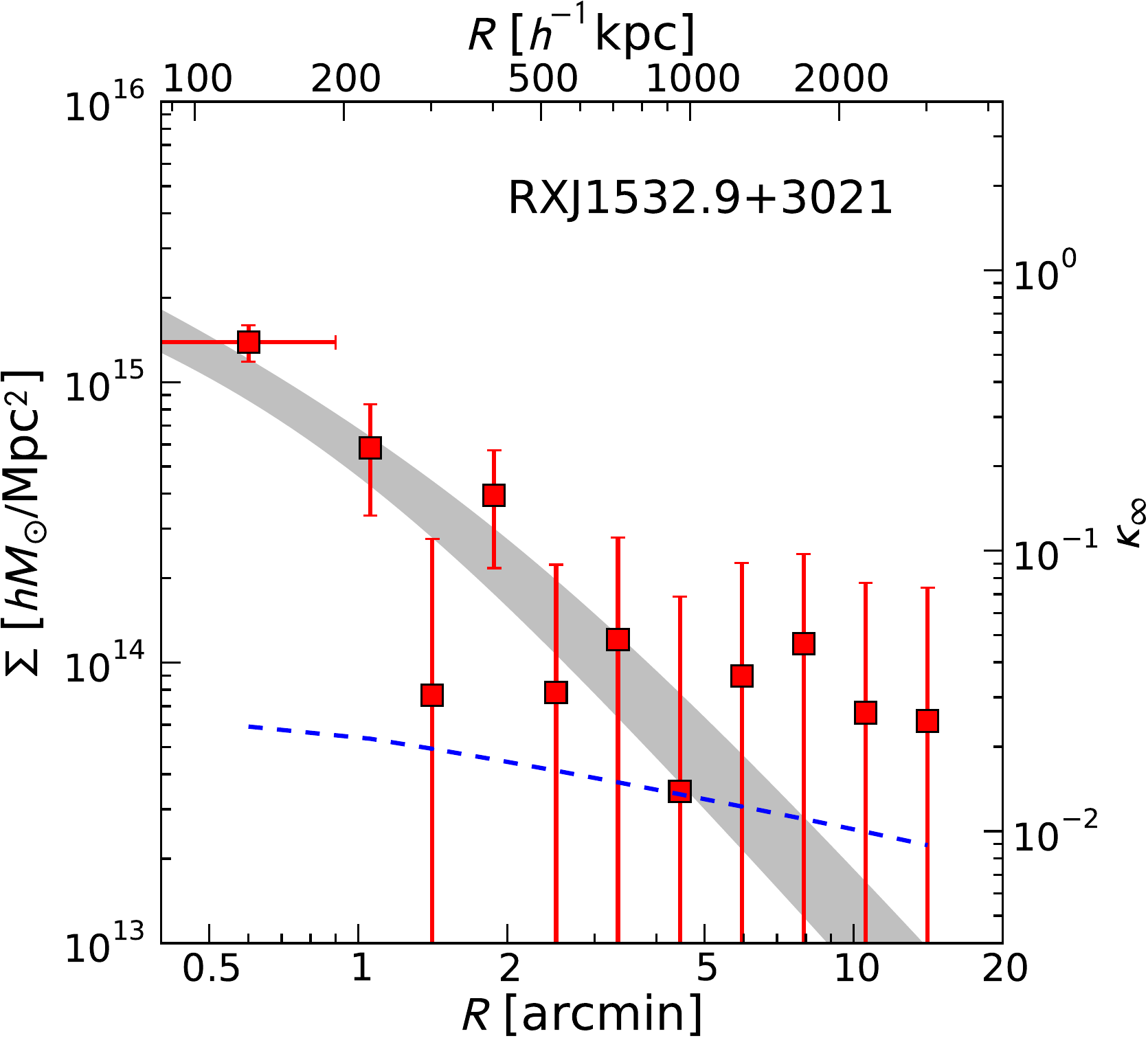}&
  \includegraphics[width=0.22\textwidth,angle=0,clip]{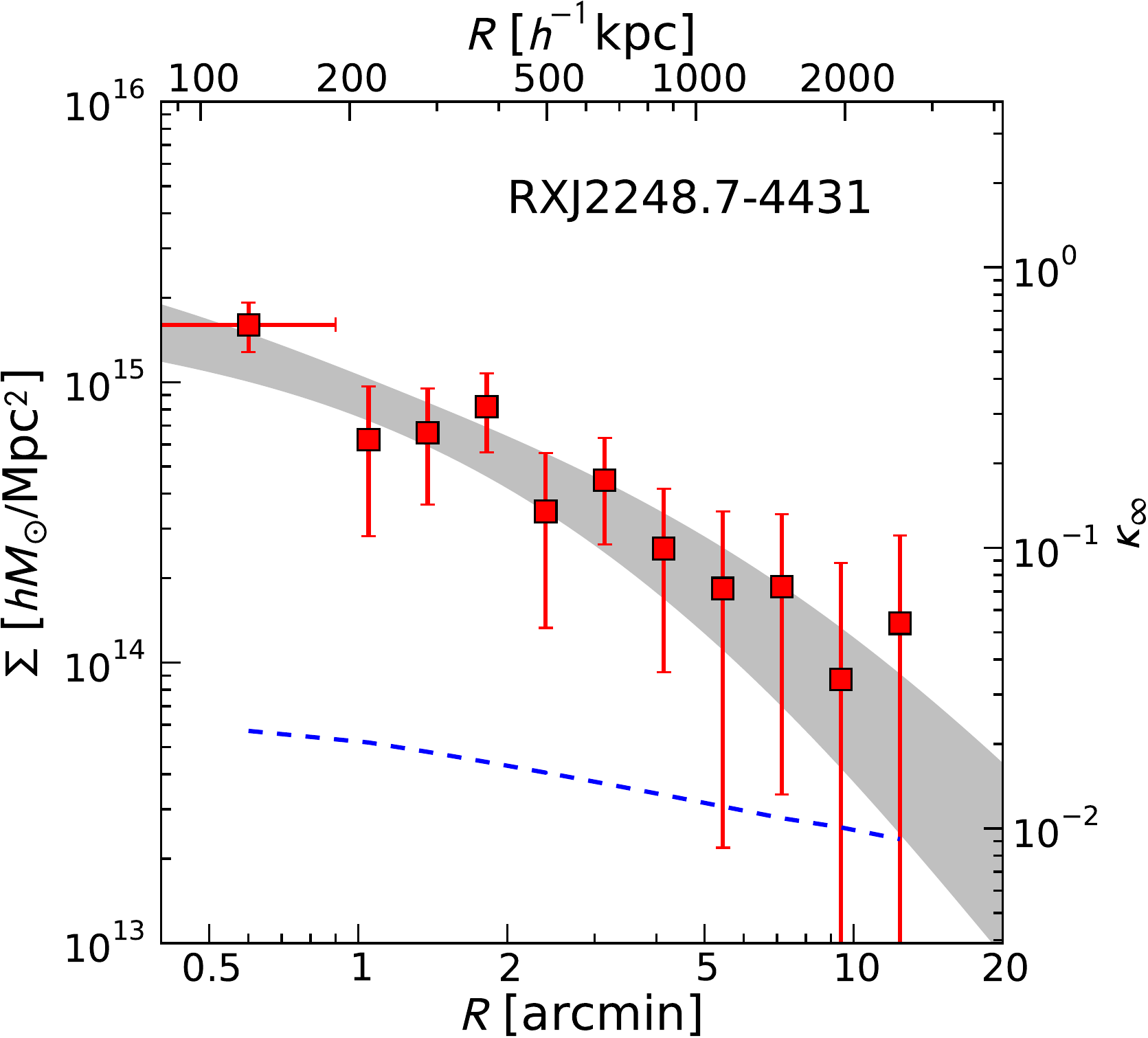} 
 \end{array}
  $
  $
 \begin{array}
  {c@{\hspace{.1in}}c@{\hspace{.1in}}c@{\hspace{.1in}}c@{\hspace{.1in}}c}
   \includegraphics[width=0.22\textwidth,angle=0,clip]{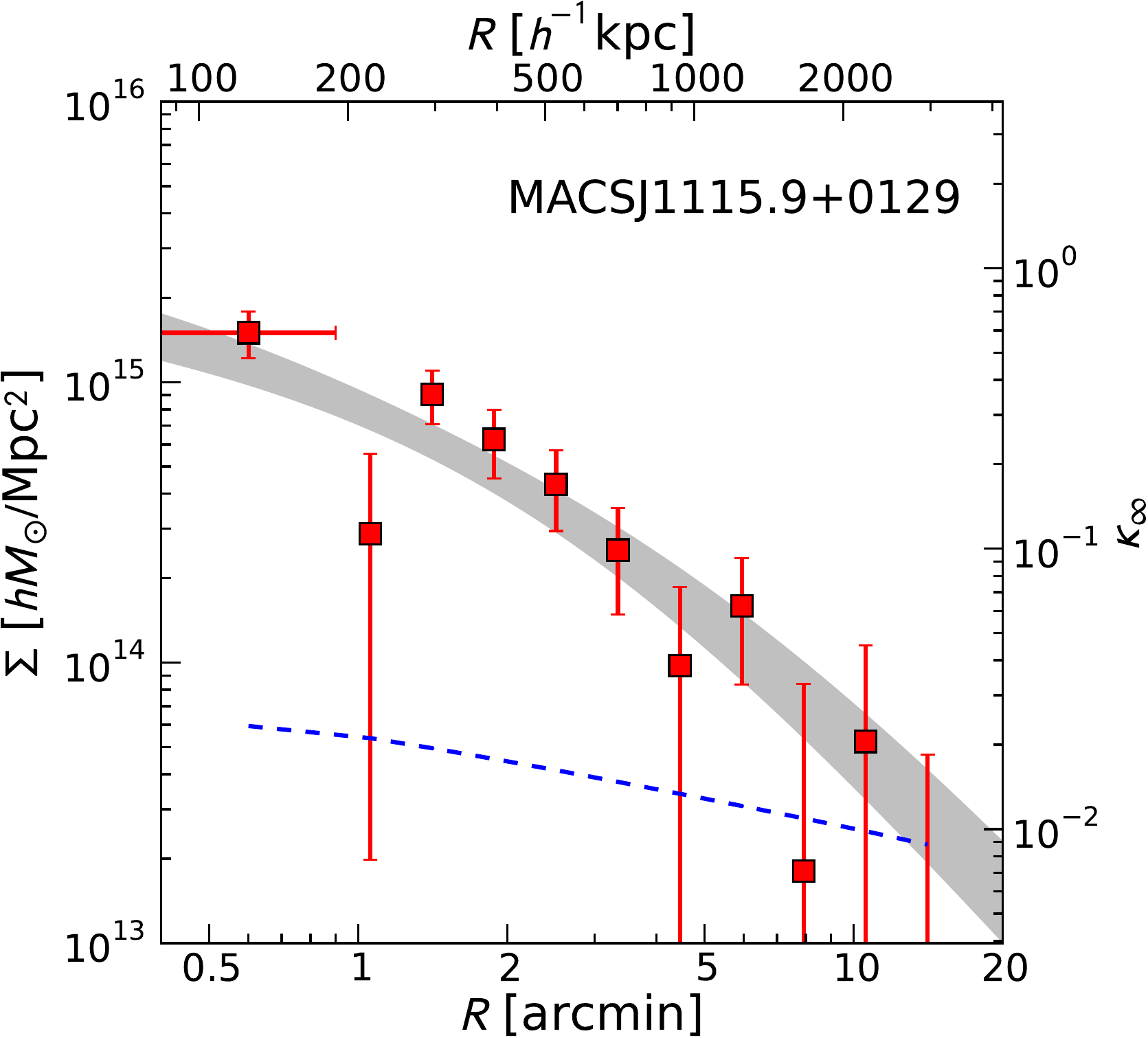}& 
   \includegraphics[width=0.22\textwidth,angle=0,clip]{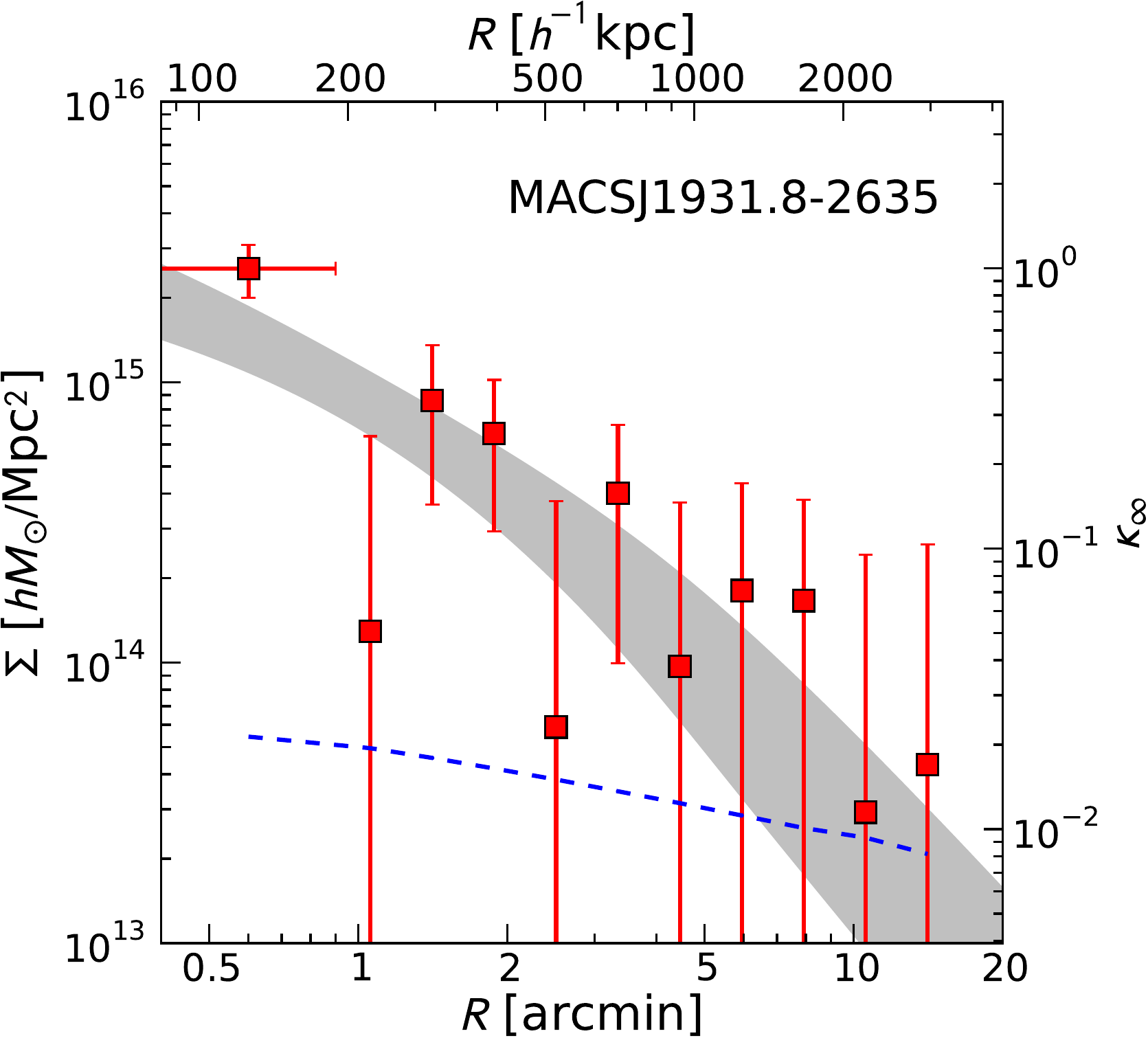}&
   \includegraphics[width=0.22\textwidth,angle=0,clip]{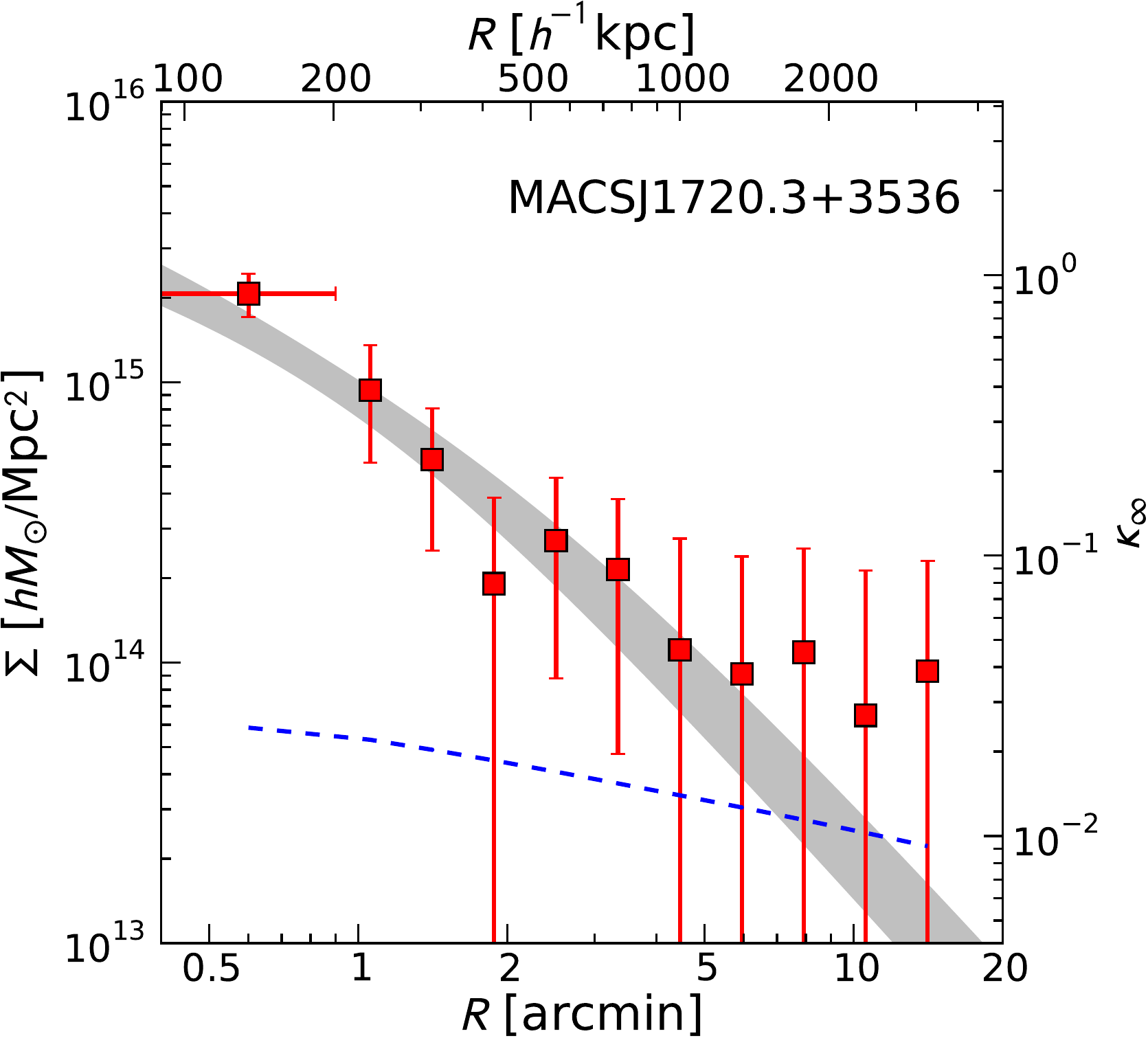}&
   \includegraphics[width=0.22\textwidth,angle=0,clip]{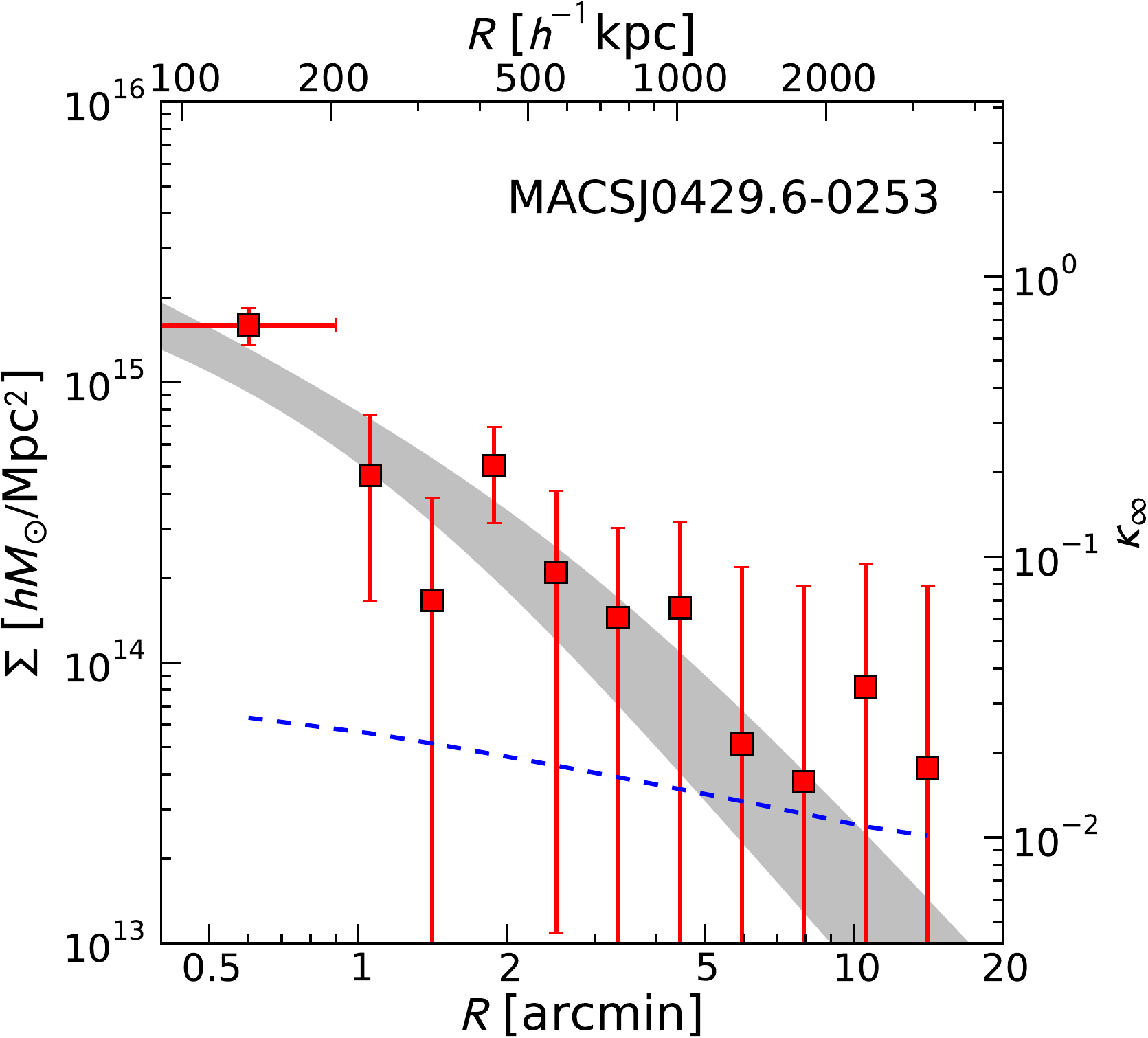}
  \end{array}
 $
 $
 \begin{array}
  {c@{\hspace{.1in}}c@{\hspace{.1in}}c@{\hspace{.1in}}c@{\hspace{.1in}}c}
   \includegraphics[width=0.22\textwidth,angle=0,clip]{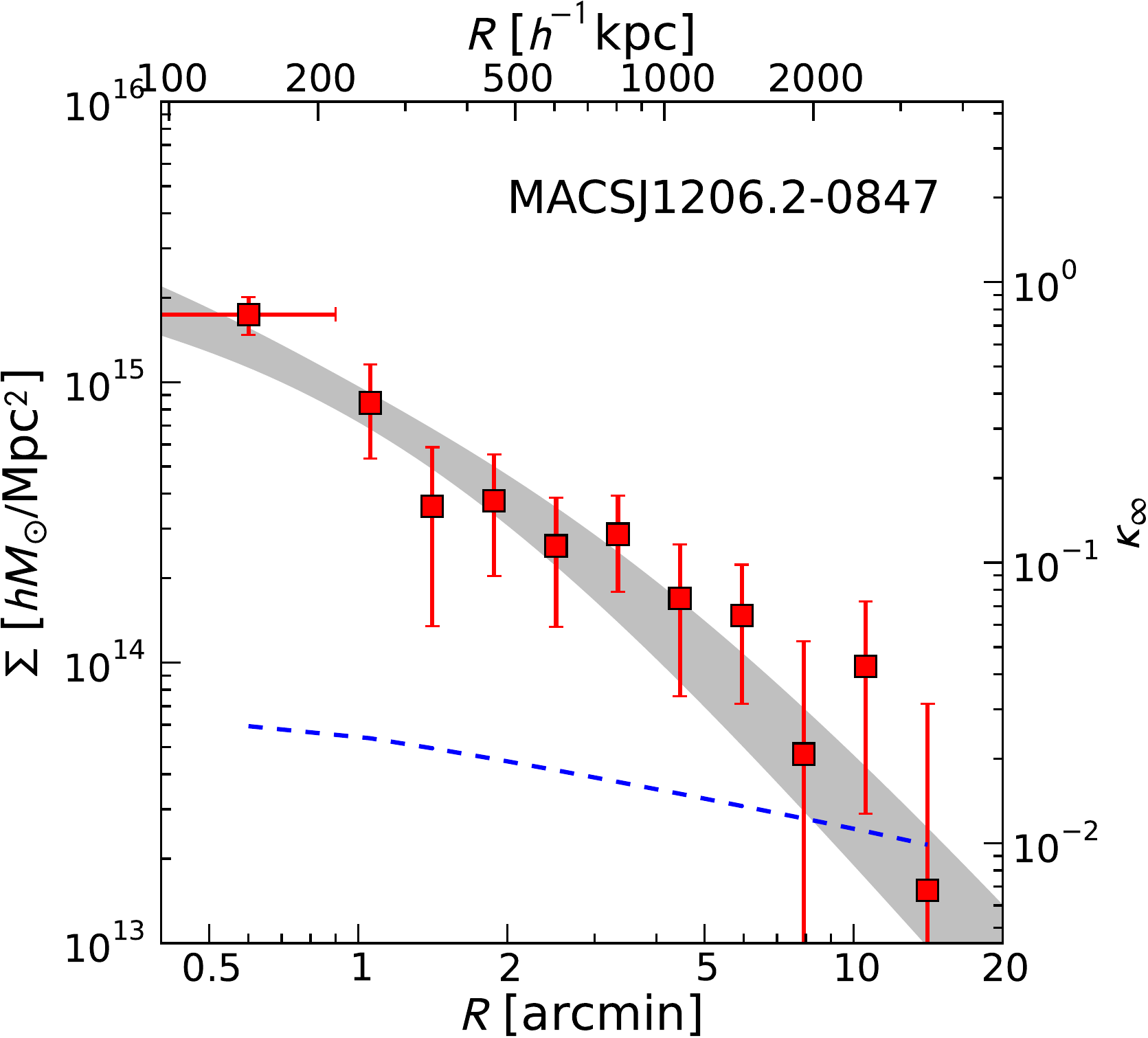}& 
   \includegraphics[width=0.22\textwidth,angle=0,clip]{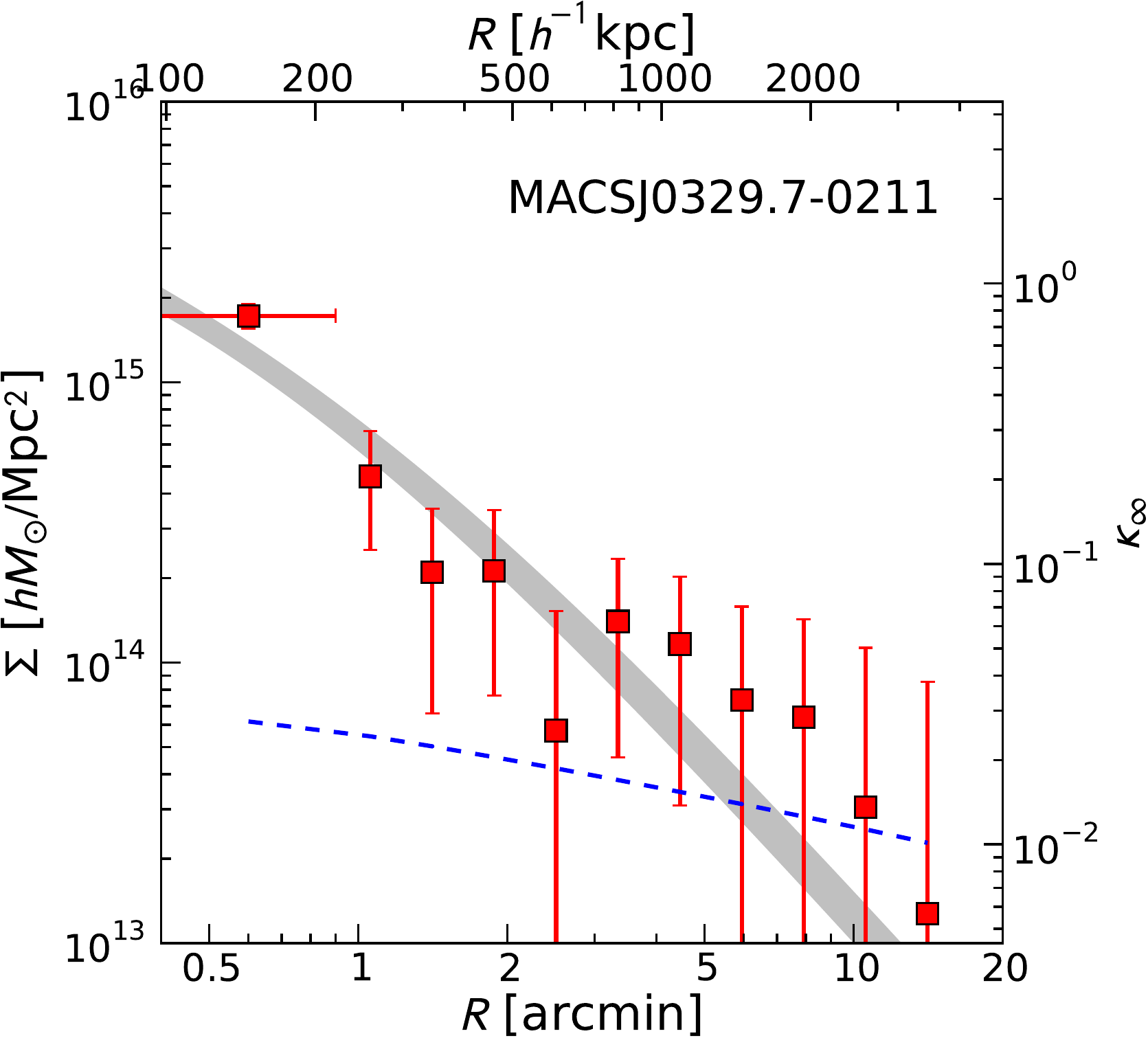}&
   \includegraphics[width=0.22\textwidth,angle=0,clip]{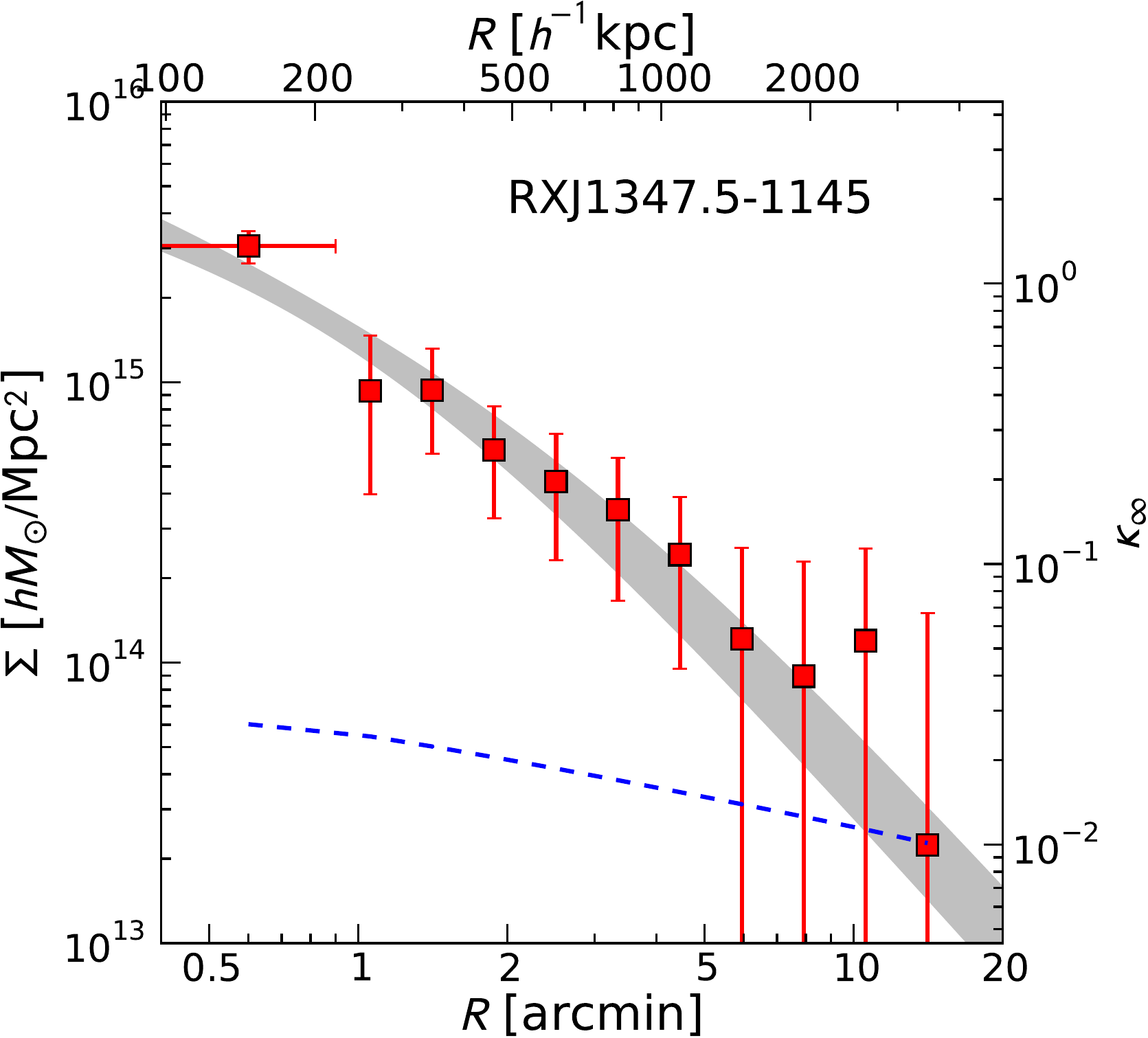} &
   \includegraphics[width=0.22\textwidth,angle=0,clip]{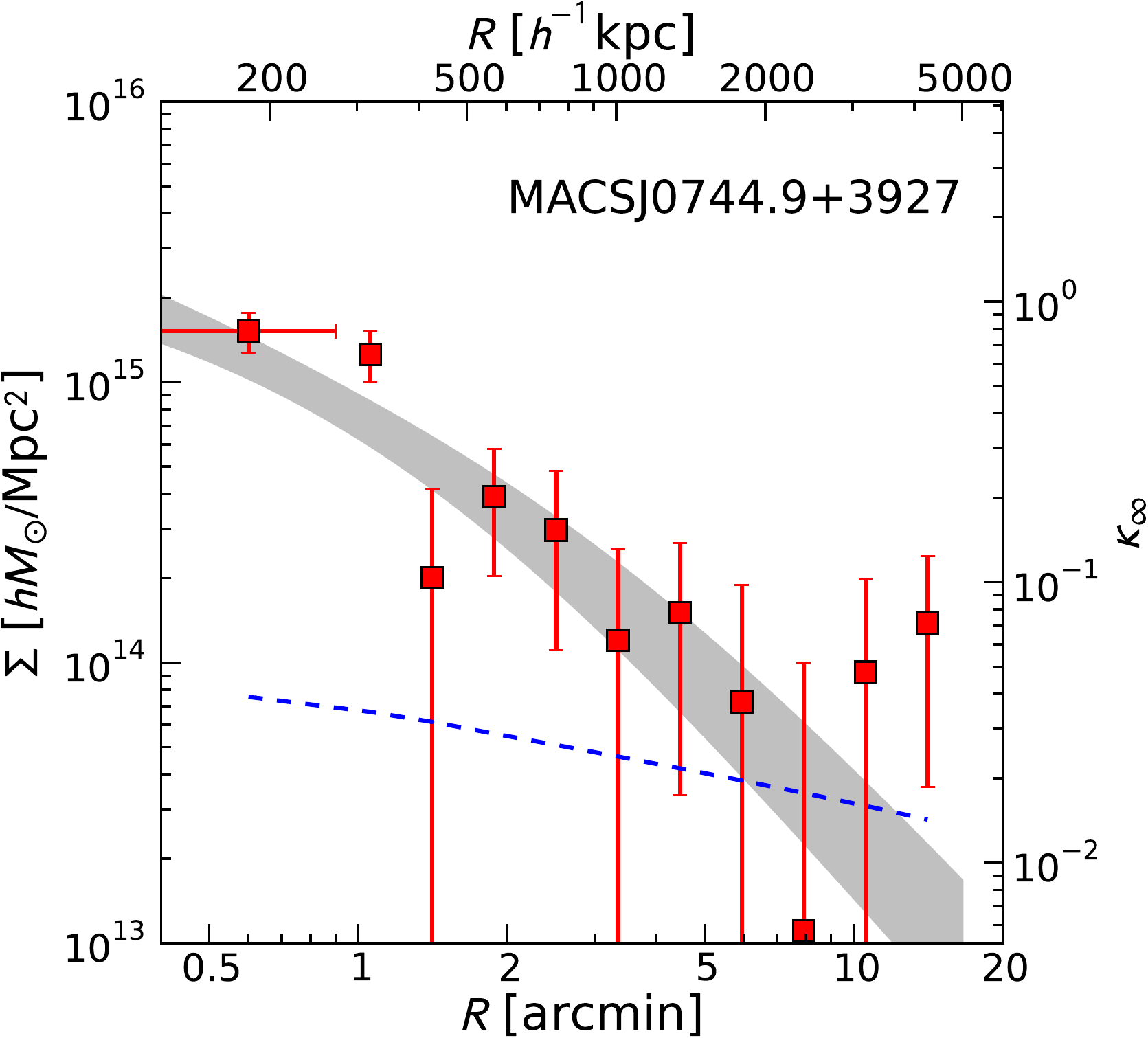} 
 \end{array}
 $
 $
 \begin{array}
  {c@{\hspace{.1in}}c@{\hspace{.1in}}c@{\hspace{.1in}}c@{\hspace{.1in}}c}
   \includegraphics[width=0.22\textwidth,angle=0,clip]{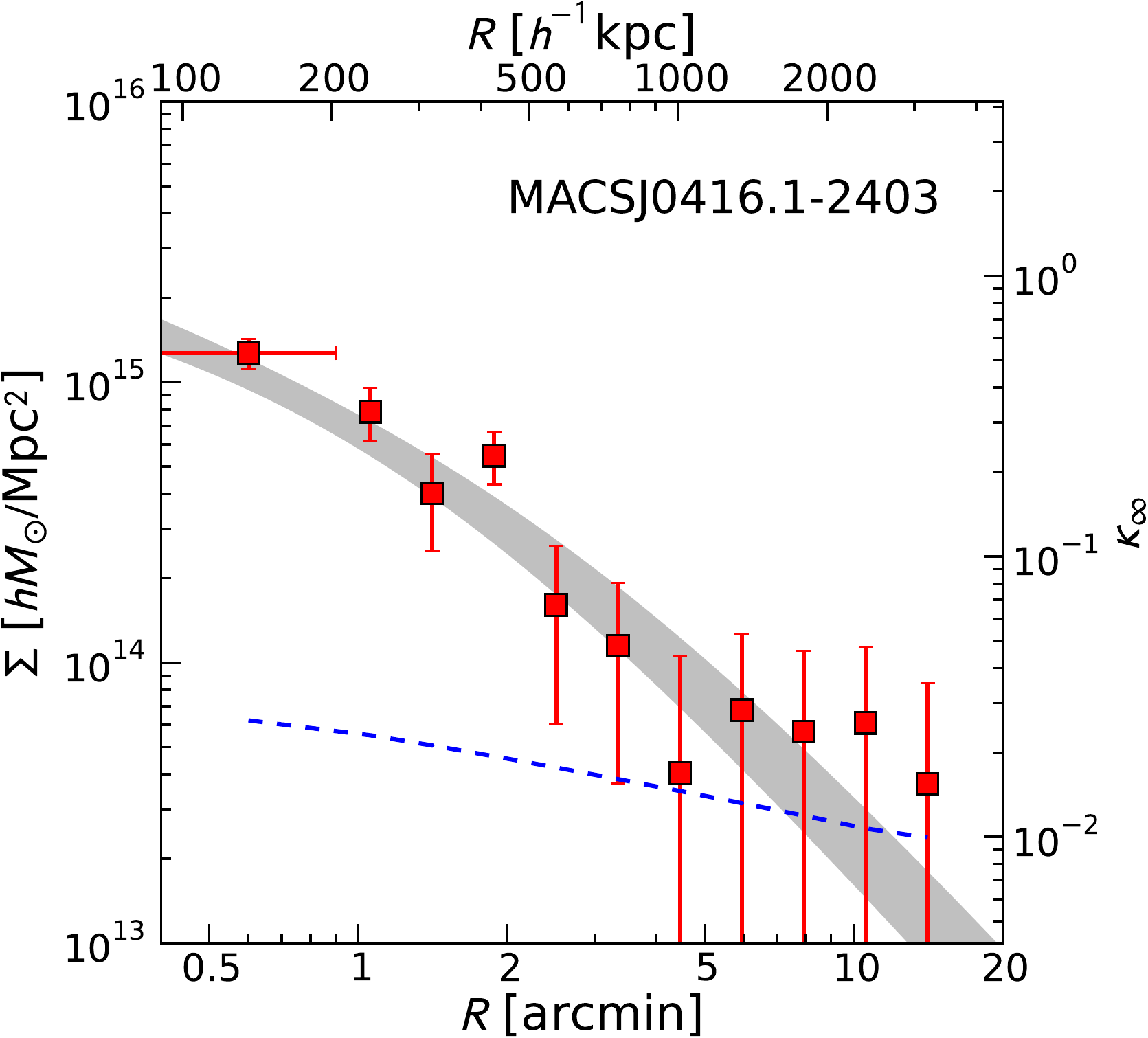}& 
   \includegraphics[width=0.22\textwidth,angle=0,clip]{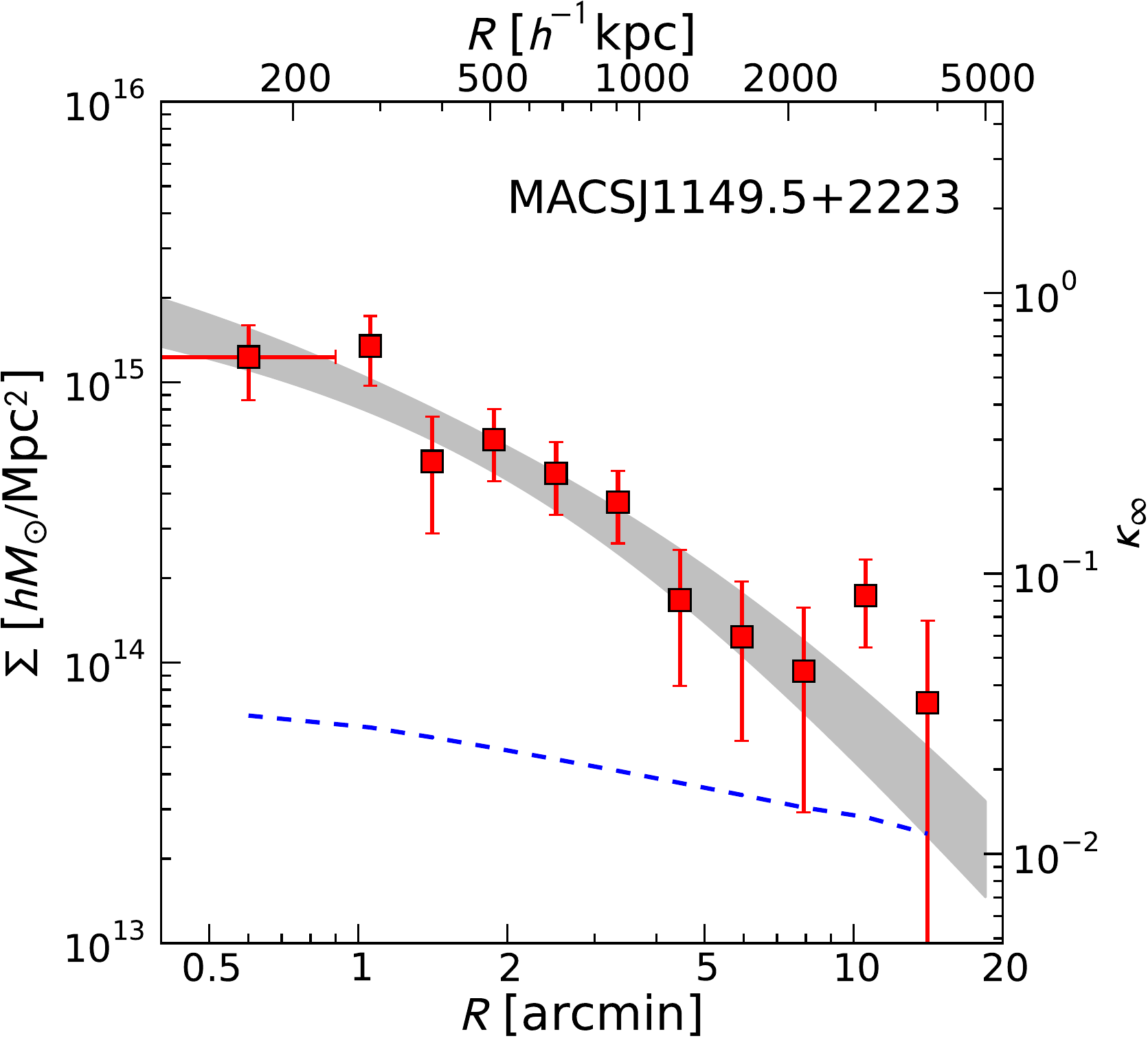}&
   \includegraphics[width=0.22\textwidth,angle=0,clip]{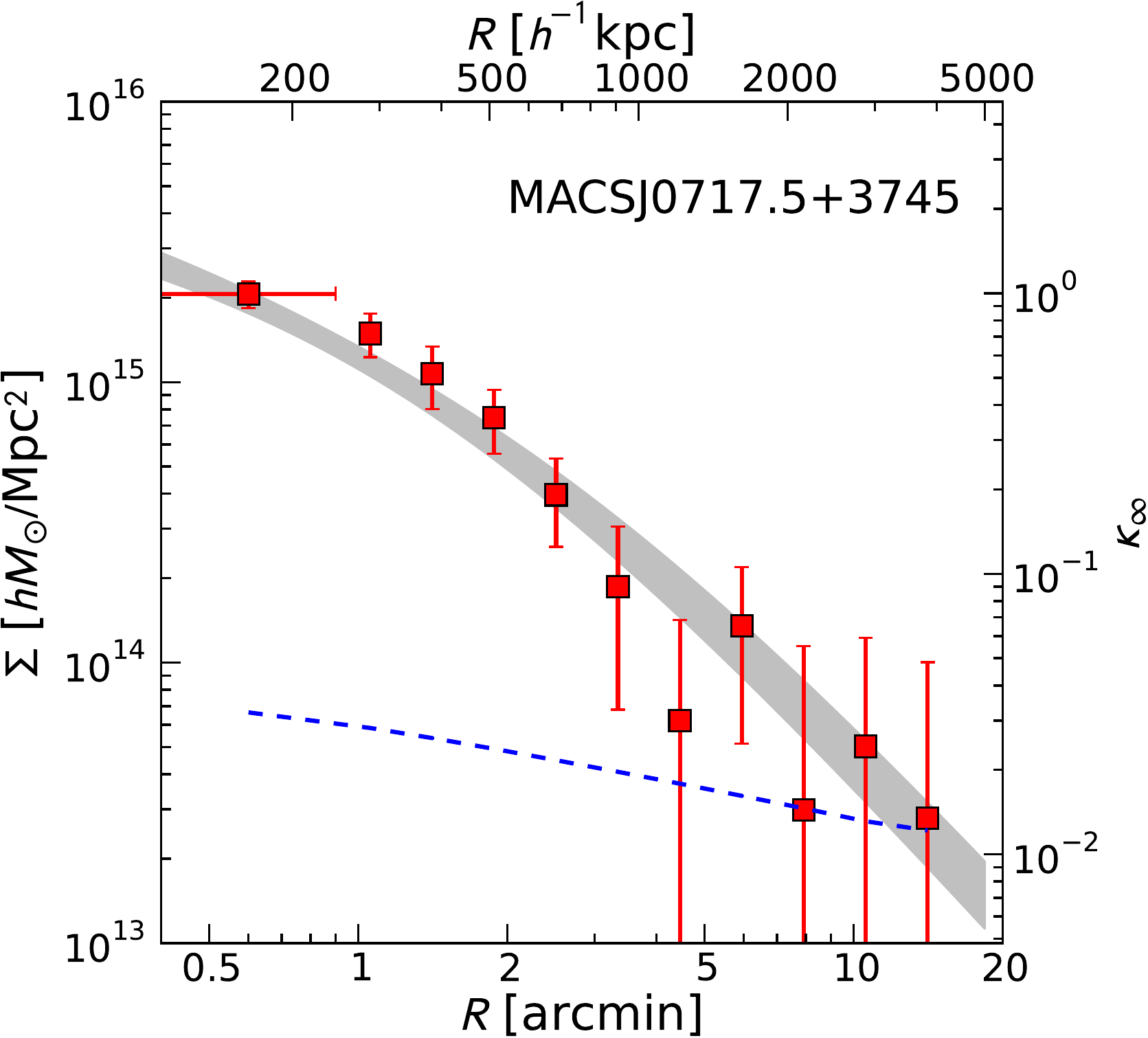} &
   \includegraphics[width=0.22\textwidth,angle=0,clip]{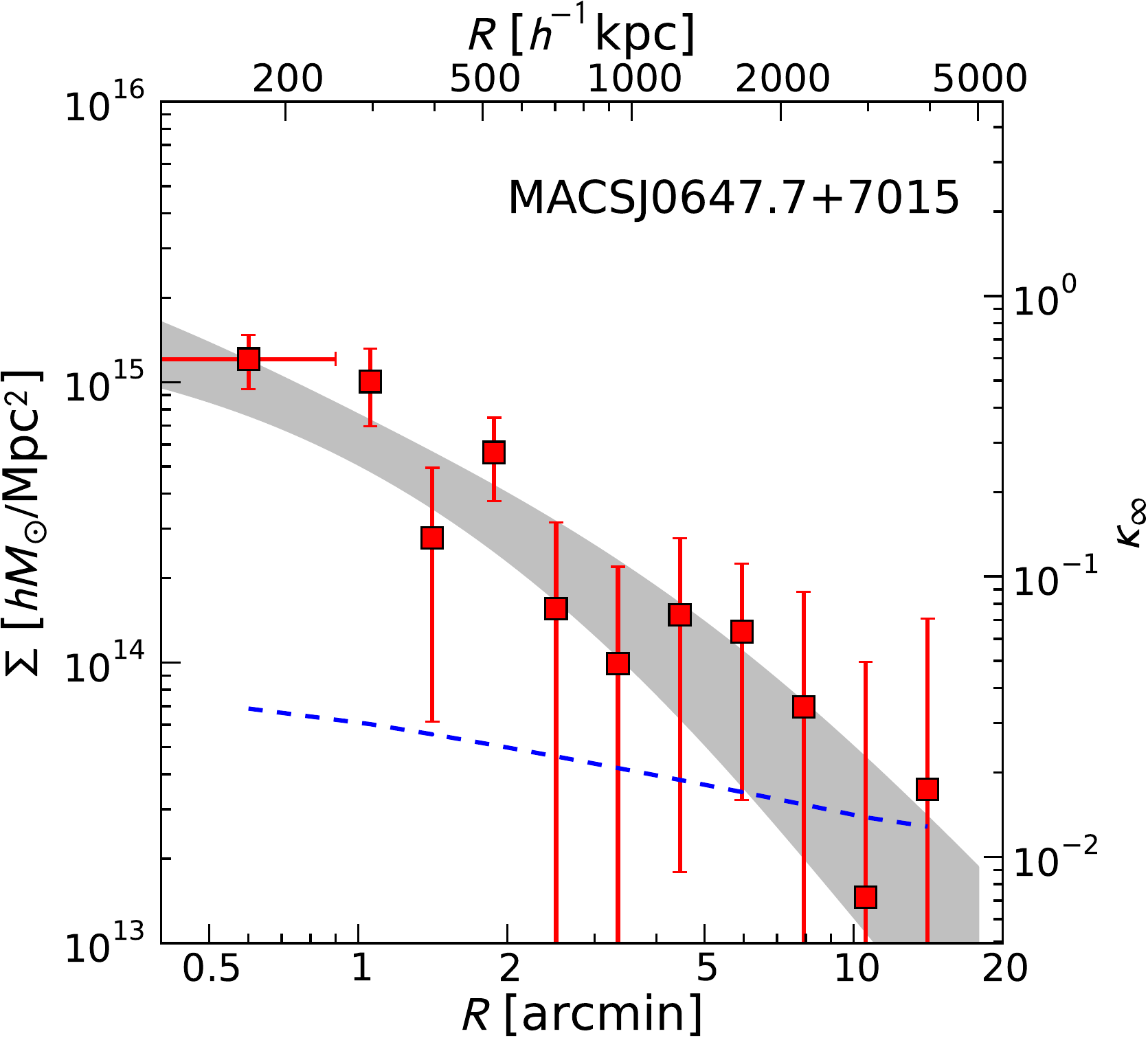} 
 \end{array}
 $
 \end{center}
  \caption{
Cluster mass profile dataset derived from our joint likelihood analysis
 of weak-lensing shear and magnification measurements shown in Figure 
 \ref{fig:wldata}. 
For each cluster, the central bin 
$\Sigma(<R_{\rm  min})$ is marked with a horizontal bar.  
The location of each binned $\Sigma$ point (squares) represents the
 area-weighted center of the radial band.
The error bars represent the 
 $1\sigma$ uncertainty from the diagonal part of the total covariance matrix
including statistical, systematic, and projected uncorrelated LSS contributions, 
$C=C^{\rm stat}+C^{\rm sys}+C^{\rm lss}$.
The gray area in each plot shows the best-fit Navarro-Frenk-White model
 (68\% CL) from the reconstructed $\Sigma$ profile.
The dashed line shows the estimated contribution to the variance
from uncorrelated LSS $C^{\rm lss}$ projected along the line of sight.
The scale on the right vertical axis shows the corresponding lensing
 convergence $\kappa_\infty$ scaled to the reference far-background
 source plane.
\label{fig:mdata}}
\end{figure*}

\input{table5.tex}

First we derive azimuthally-averaged lens distortion and count depletion
profiles (Section \ref{sec:method}) for our cluster sample
from our wide-field imaging data (Section \ref{sec:data}). 
The radial binning scheme is summarized in Table \ref{tab:wlsn}.

For each cluster we calculate the lensing profiles in $N$ discrete
radial bins from the cluster center (Table \ref{tab:sample}), spanning the range 
$[\theta_{\rm min},\theta_{\rm max}]$ with a constant
logarithmic radial spacing $\Delta\ln\theta = \ln(\theta_{\rm
max}/\theta_{\rm min})/N$, 
where the inner radial boundary $\theta_{\rm min}$ is taken for all
clusters to be 
$\theta_{\rm min}=0.9\arcmin$, 
which is larger than the
range of Einstein radii for our sample
 (A. Zitrin et al. 2014, in preparation). 
For all clusters in our sample, the inner radial limits satisfy 
$D_{\rm l}\theta_{\rm min}> 2 d_{\rm off}$,  
so that the miscentering effects on our mass profile measurements are
negligible \citep{Johnston+2007b,Umetsu+2011stack}.
The choice of $\theta_{\rm min}$ also ensures high-purity samples of
background galaxies with a low level of contamination by cluster members
defined by our CC-selection method (Section \ref{subsec:back}).
The outer boundary is set to
$\theta_{\rm max}=16\arcmin$ \citep{Umetsu+2012,Coe+2012A2261,Medezinski+2013}
for all clusters,
except RXJ2248.7-4431 observed with ESO/WFI,
for which we take $\theta_{\rm max}=14\arcmin$. 
These are sufficiently larger than the range of virial radii for our
sample ($r_{\rm vir}\simlt 2\,$Mpc\,$h^{-1}$), but sufficiently small
with respect to the sizes of the camera 
field of view so as to ensure accurate PSF anisotropy correction. 
The number of radial bins is set to $N=10$ for all
clusters,  such that the per-pixel detection S/N is of the order of
unity, which is optimal for an inversion problem.  
Here we quantify the significance of a 
detection for a given lensing profile in analogy to Equation (38) of
\citet{UB2008}.

Unlike the distortion effect,
the magnification signal falls off sharply with increasing distance from
the cluster center. 
For the magnification-bias analysis, 
the count normalization and slope 
($\overline{n}_\mu,s_{\rm eff}$) are estimated from the source counts in
cluster outskirts  
\citep{Umetsu+2011,Umetsu+2012,Coe+2012A2261,Medezinski+2013},
specifically at
$\theta=[10\arcmin,\theta_{\rm max}]$ (see Section \ref{subsubsec:backdens}).

In Figure \ref{fig:wldata} we show the distortion and 
magnification profiles for our sample of 20 CLASH clusters.
For most of the clusters, a systematic depletion of the red galaxy counts is seen
in the  high-density region of the cluster, 
and detected out to several arcminutes from the cluster center. 
For our sample of 20 cluster fields, we find a median masked-area
fraction $\overline{f}_{\rm mask}$ (Section \ref{subsubsec:depletion})
of 0.076 with a standard deviation of 0.055. 
The statistical significance of the detection of the tangential distortion 
ranges from $4.8\sigma$ to $16.4\sigma$.
On the other hand, the detection significance of the depletion signal is
in the range  $2.5\sigma$--$8.1\sigma$ 
(Table \ref{tab:wlsn}),
which is on average $45\%$ of the S/N of the tangential distortion.
This corresponds to an overall improvement of $\sim 10\%$ by combining the
shear and magnification measurements, compared to the
distortion-only case \citep{Umetsu+2011,Umetsu+2012,Umetsu2013,Coe+2012A2261}.

Following the methodology outlined in Section \ref{subsec:bayesian}
we derive for each cluster a mass-profile solution 
$\bSigma=\{\Sigma_{\rm min},\Sigma_i\}_{i=1}^{N}$
from a joint likelihood analysis of our shear+magnification data (Figure \ref{fig:wldata}).
In Table \ref{tab:wlsn} we present the minimum $\chi^2 (\equiv -2\ln{\cal L})$
values for the best-fit $\bSigma$ solutions, ranging from 
$\chi^2=2.2$ to 14.4 for $2N-(N+1)=9$ degrees of freedom
(a mean reduced $\chi^2$ of 0.92),
indicating good consistency between the shear and magnification
measurements having different systematics.
This is also demonstrated in Figure \ref{fig:wldata}, which compares the
observed lensing profiles with the respective joint Bayesian
reconstructions (68\% CL). 

The resulting mass-profile solutions $\bSigma$
are shown in Figure \ref{fig:mdata} for our 20 clusters
along with their NFW fitting results (see Section \ref{subsec:mass}). 
The error bars represent the $1\sigma$ uncertainties from the diagonal part
of the total covariance matrix $C=C^{\rm stat}+C^{\rm sys}+C^{\rm lss}$
(Equation (\ref{eq:Ckappa}))
including statistical, systematic, and projected uncorrelated LSS noise
contributions. 


\subsection{Cluster Mass Estimates}
\label{subsec:mass}

\input{table6.tex}

The standard weak-lensing approach to cluster mass estimates is based on
tangential-shear fitting with NFW functionals \citep[][]{Oguri+2009Subaru,Okabe+2010WL}.
This approach, however, has a disadvantage 
that the cluster mass estimates can be biased low by $5-10\%$ 
\citep{Meneghetti+2010a,Becker+Kravtsov2011,Rasia+2012}. 
This well-known bias arises from the fact that the tangential shear
responds negatively to local mass perturbations (Equation (\ref{eq:loop}))
which are abundant in rich cluster environments. 
This bias can be minimized if the fitting range is restricted to within
$\sim 2r_{\rm 500c}$ \citep{WtG3}.
An alternative approach is to use the combination of
 shear and magnification  
to reconstruct the projected mass density profile,
breaking the mass-sheet degeneracy. 

Here we use our mass-profile dataset (Figure \ref{fig:mdata}) to obtain
total mass estimates for individual clusters in our sample. 
To do this, we employ the spherical NFW model to 
facilitate comparison with results from complementary
X-ray,
Sunyaev-Zel'dovich effect (SZE), 
and 
dynamical observations.
The two-parameter NFW profile is given by
\begin{equation}
\label{eq:nfw}
\rho_{\rm NFW}(r)=\frac{\rho_{\rm s}}{(r/r_{\rm s})(1+r/r_{\rm s})^2},
\end{equation} 
where $\rho_{\rm s}$ and $r_{\rm s}$ are the characteristic density and radius,
respectively. 
For the NFW model, $r_{\rm s}$ marks the radius at which 
the logarithmic density slope equals the isothermal value, namely
$d\ln\rho(r)/d\ln{r}=-2$ at $r=r_{\rm s}$.
We specify the NFW model with
the halo mass $M_{\rm 200c}$
and the corresponding halo concentration $c_{200{\rm c}}=r_{200{\rm c}}/r_s$.

We use a Bayesian MCMC method  to obtain an accurate inference of the NFW
density profile (Equation (\ref{eq:nfw}))
from our data in the form of
the discrete cluster mass profile $\bs=\bSigma/\Sigma_{{\rm c},\infty}$
and its full covariance matrix $C=C^{\rm stat}+C^{\rm sys}+C^{\rm lss}$
 (Section \ref{subsec:bayesian}).
Here we employ the radial dependence of the projected NFW lensing profiles given
by \citet{2000ApJ...534...34W}, which provides a sufficiently good
description of the projected total matter distribution in cluster-sized halos
out to the virial radius $r_{\rm vir}$
\citep[][see also Section \ref{subsec:magbias}]{Oguri+Hamana2011,Umetsu+2011stack,Oguri+2012SGAS,Newman+2013a,Okabe+2013}.
For all clusters, we thus restrict the fitting range to $R\le
2\,$Mpc\,$h^{-1}$, which is approximately the virial radius of high-mass clusters. 
We assume uninformative log-uniform priors for the mass and concentration parameters
\citep[i.e., uniform priors for $\log_{10} M_{200{\rm c}}$ and
$\log_{10} c_{200{\rm c}}$, see][]{Sereno+Covone2013,Covone+2014}
in the respective intervals,
$0.1 \le M_{200{\rm c}}/(10^{15}M_\odot\,h^{-1}) \le 10$ 
and 
$0.1 \le c_{200{\rm c}}\le 10$.
The $\chi^2$ function for our weak-lensing observations is
\begin{equation}
\chi^2(\bp)
=\sum_{i,j} 
\big[
  s_i-\hat{s}_i(\bp)
\big]
 C^{-1}_{ij}
\big[
  s_j-\hat{s}_j(\bp)
\big],
\end{equation}
where 
$\bp=(M_{200{\rm c}},c_{200{\rm c}})$, and
$\hat{\bs}(\bp)=\hat{\bSigma}(\bp)/\Sigma_{{\rm c},\infty}$ is the model prediction
for the binned mass profile, given as
\begin{equation}
\hat{\Sigma}_i = \Sigma_{\rm NFW}(\theta_{i,1}<\theta<\theta_{i,2}) 
\end{equation}
with $\Sigma_{\rm NFW}(\theta_{i,1}<\theta<\theta_{i,2})$ 
the predicted mean surface mass density averaged over the $i$th annulus between 
$\theta_{i,1}$ and $\theta_{i,2}$, accounting for the
effect of bin averaging in observations.

From the posterior samples we derive marginalized constraints on the
total enclosed mass $M_\Delta = M_{\rm 3D}(<r_\Delta)$
at several characteristic interior overdensities $\Delta$ (see Section \ref{sec:intro}).
In Table \ref{tab:mass} we summarize the results of our weak-lensing cluster mass estimates,
where we employ the robust biweight estimators of 
\citet{1990AJ....100...32B}
for the central location (average) and scale (dispersion) of the
marginalized posterior probability distributions 
\citep[e.g.,][]{Sereno+Umetsu2011,Biviano+2013}.

\subsection{Systematic Mass Uncertainty}
\label{subsec:sysmass}

\subsubsection{Concentration--Mass Degeneracy}
\label{subsubsec:cMbias}

A robust determination of the concentration parameter requires sensitive
lensing measurements over a wide range of cluster radii because
the concentration is sensitive to the {\it radial curvature} in the mass
profile.  
In practice such a measurement can be achieved
by combining strong- and weak-lensing data
\citep{Merten+2009},
performing wide-field weak-lensing observations of nearby clusters 
\citep[e.g., A2142 in][]{Umetsu+2009},
or by stacking the lensing signal of a statistical sample of
clusters \citep{Johnston+2007b,Umetsu+2011stack,Oguri+2012SGAS}.

In this work we have not attempted to determine the concentration
for each cluster because the weak-lensing profiles
for individual clusters are highly degenerate in $M$ and $c$: the
observed lensing profile can be explained by halo density profiles with
larger $M$ and smaller $c$ than the best-fit values and 
vice versa.
This $c$--$M$ degeneracy is more significant for low-mass, high-redshift
systems, which are {\it unresolved}  
by weak-lensing observations and for which the scale radius
$r_{\rm s}$ is unconstrained by the data.
In such a case, the constraints on $c$ are essentially
imposed by prior information.
This inherent covariance can 
potentially bias the slope of the $c(M)$ relation determined
from weak lensing \citep{Hoekstra+2011}.

\subsubsection{Impact of the Choice of the Fitting Range}
\label{subsubsec:fitrange}

Unlike the shear which is sensitive to the mean interior 
density, the majority of the $\kappa$ signal
(with respect to the noise) comes from relatively inner regions and the
outer profile exhibits a high degree of positive correlation
($\sim 50\%$ in the last few bins).
The relative contribution of projected uncorrelated LSS noise
increases with increasing radius (Figure \ref{fig:mdata}),
 so that the effect of including 
$C^{\rm lss}$ is to further downweight the lensing signal in the outer regions
especially beyond $\theta\sim 10\arcmin$ \citep{2003MNRAS.339.1155H}.
On average, we find cosmic noise contributes $\sim 25\%$ to the total
error budget $\sqrt{C_{ii}}$ for the reconstructed $\kappa$ profile.

Without restricting the radial range for NFW fitting, 
we find $\sim 2\%$ lower virial masses ($M_{\rm vir}$)
relative to our baseline results
obtained with a maximum fitting radius of $R=2$\,Mpc\,$h^{-1}$.
This effect is less significant at higher overdensities,
$\Delta_{\rm c}\ge 200$.
We thus conclude that our cluster mass estimates $M_\Delta$ are
statistically insensitive to the choice of the outer fitting radius. 

%

\subsubsection{Halo Triaxiality}
\label{subsubsec:triaxiality}

Lensing mass measurements are sensitive to the halo triaxiality 
\citep{2005ApJ...632..841O}.
In the context of $\Lambda$CDM, prolate halo shapes are
expected to develop 
along filaments at early
stages of halo assembly, so that dynamically-young cluster
halos tend to have a prolate morphology \citep{Shaw+2006}.
Accordingly, a large fraction of clusters are expected to be elongated
in the plane of the sky. 
On average, this will lead to an underestimation of the cluster mass
when spherical symmetry is assumed \citep{Rasia+2012}.
Numerical simulations show that, for a sample of 
randomly-oriented prolate clusters,
their mass estimates are biased low by
$\simlt 5\%$ on average when the masses are recovered from the projected
mass distributions $\Sigma$ under the assumption of spherical symmetry
(M14).

\subsubsection{Shear--Magnification Consistency}
\label{subsubsec:msys}


\begin{figure}[!htb] 
 \begin{center}
 \includegraphics[width=0.45\textwidth,angle=0,clip]{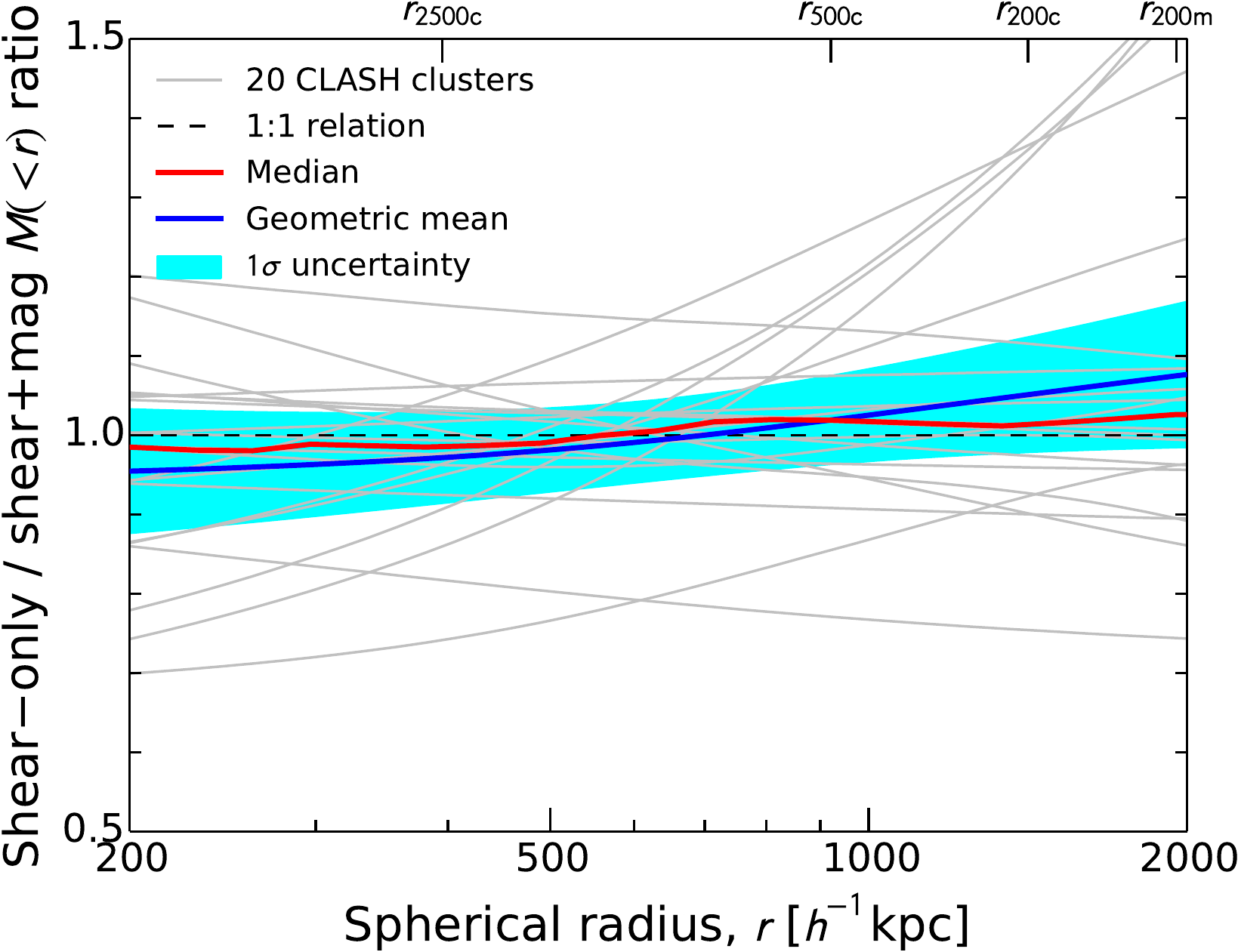}
 \end{center}
\caption{
\label{fig:mcomp}
The ratio of cluster masses $M_{\rm 3D}(<r)$ obtained from NFW fits to
 the tangential reduced-shear profile (shear-only)
and to the surface mass density
 profile reconstructed from the joint shear+magnification analysis (shear+mag).
The results are shown for our sample of  20 CLASH clusters (gray lines).
The red line represents the median mass ratio as a function of spherical
 radius. 
The blue line and cyan-shaded area show the geometric-mean mass ratio
 and its $1\sigma$ uncertainty, respectively.
The dashed horizontal line marks the 1:1 relation.
}
\end{figure}


Measuring the shear and magnification
effects provides a consistency check of weak-lensing measurements, 
thereby allowing us to assess the robustness of our cluster mass
estimates.
Here we compare our mass estimates 
based on the $\bSigma$ profiles recovered
from the joint shear+magnification analysis 
with those 
obtained using the standard shear-only approach.
Since background samples defined in different color regions (Section
\ref{subsec:back}) suffer different degrees of 
(if any) contamination by cluster members, this comparison is sensitive
to the presence of residual contamination of the background as
well as the shear calibration uncertainty. 
To do this,
we adopt the Bayesian inference approach described
in Section \ref{subsec:mass}, and
 fit the NFW model to the tangential reduced-shear profiles 
$\langle \bg_+\rangle$ 
in the range $R\le 2$\,Mpc\,$h^{-1}$.

In Figure \ref{fig:mcomp} we show for our sample the ratio of cluster
masses obtained using these two weak-lensing methods as a function of
spherical radius.
At each cluster radius, we compute the 
unweighted geometric mean and median of the shear-only to
shear+magnification mass ratios.
Note that we use geometric averaging, which satisfies 
$\langle Y/X\rangle = 1/\langle X/Y\rangle$ 
\citep[see also][]{Donahue2014clash}.
We see that the averaged mass ratio  is consistent with unity within the
errors,  
but increases monotonically with cluster radius from 
$0.95\pm 0.08$ at $r\simeq 200$\,kpc\,$h^{-1}$ to
$1.08\pm 0.09$ at $r\simeq 2$\,Mpc\,$h^{-1}$.
This systematic radial trend can be explained by the $c$--$M$ 
degeneracy discussed in Section \ref{subsubsec:cMbias}.
On the basis of this comparison, we estimate the
systematic uncertainty in the overall mass calibration to be of the order
of $\pm 8\%$.


\section{CLASH Stacked Lensing Analysis}
\label{sec:clash_stack}

Stacking an ensemble of clusters helps average out the projection
effects of cluster asphericity and substructure, as well as the cosmic
noise from projected uncorrelated LSS, inherent in lensing measurements.   
The statistical precision can be greatly improved by stacking together a
large number of clusters, especially on small angular scales
\citep{Okabe+2010WL}, allowing a tighter comparison of the averaged
lensing profile with theoretical models.

Here our stacked lensing analysis will focus on the CLASH X-ray-selected
subsample of \citet{Postman+2012CLASH}, which comprises a population of
high-mass X-ray regular clusters.  
The four high-magnification clusters are thus excluded from this part
of the analysis. 

In Section \ref{subsec:clash_stackgt} we present a stacked
tangential-distortion (shear-only) analysis of the 16 X-ray regular clusters,
and examine the form of their underlying halo  
mass profile using the ensemble-averaged $\llangle \bDSigma_+\rrangle$ profile.
In Section \ref{subsec:clash_stackwl} we derive the ensemble-averaged
total mass profile $\llangle\bSigma\rrangle$ from our
cluster mass-profile dataset (Figure \ref{fig:mdata}), 
for comparison with theoretical predictions taking into account both
one- and two-halo term contributions.

\subsection{CLASH Stacked Shear-only Analysis}
\label{subsec:clash_stackgt}

\subsubsection{Stacking the Weak Shear Signal}
\label{subsubsec:clash_stackgt}


\begin{figure}[!htb] 
 \begin{center}
 \includegraphics[width=0.5\textwidth,angle=0,clip]{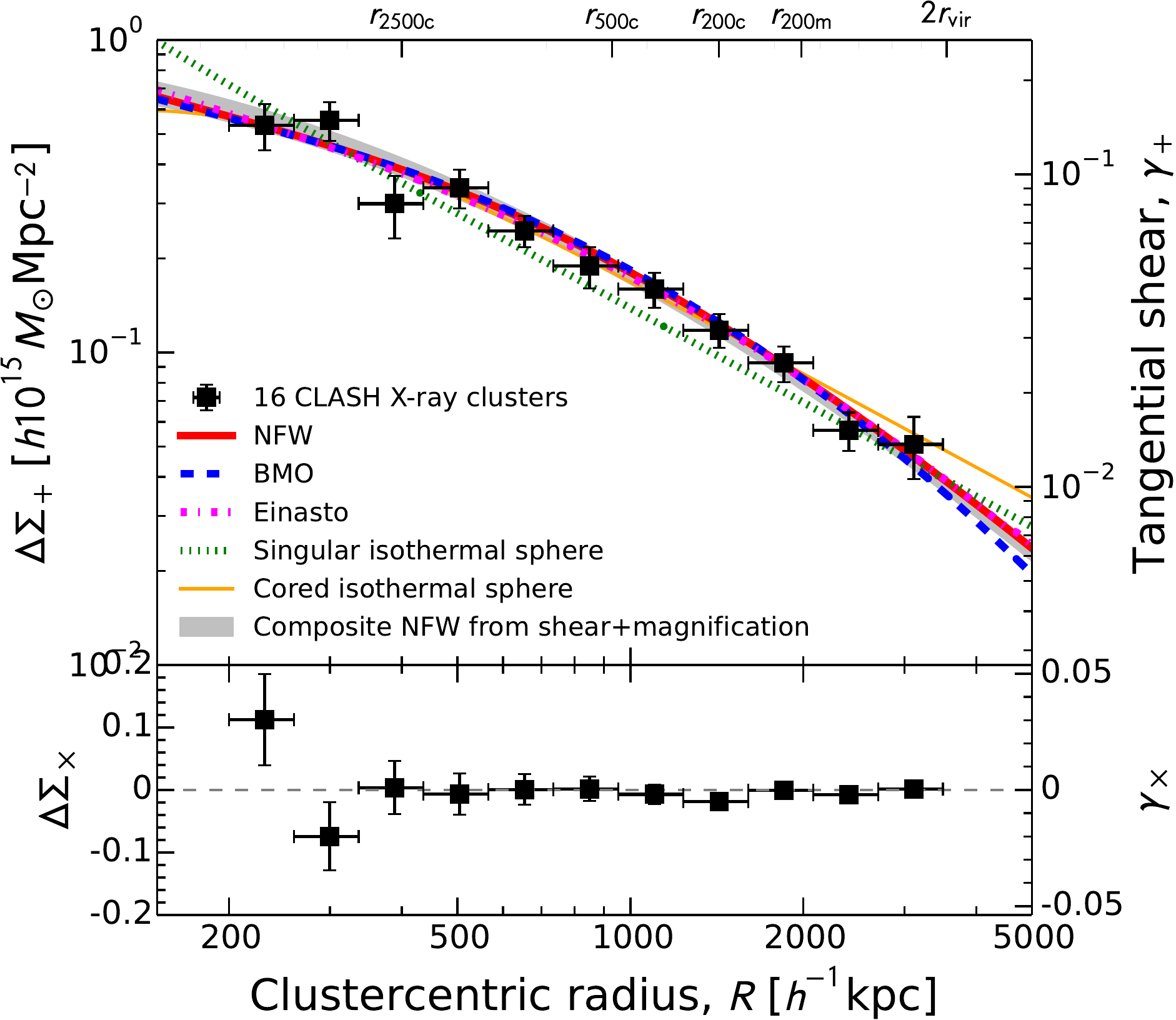} 
 \end{center}
\caption{
The average tangential-shear profile $\llangle\Delta\Sigma_+\rrangle$
 (upper panel, black squares)
 obtained from stacking the X-ray-selected subsample of 16 clusters,
shown in units of projected mass density. 
The thick-solid (red), dashed (blue), dashed-dotted (magenta), dotted
 (green), and thin-solid (orange) lines
 correspond to the best-fit NFW, truncated-NFW \citep[][BMO]{BMO},
 Einasto, SIS, and cored
 isothermal sphere profiles,  respectively. 
The gray-shaded area shows the composite halo mass-profile prediction
 ($1\sigma$)  derived from a  
 weighted average of individual NFW profiles based on the joint
 shear+magnification analysis (Figure \ref{fig:mdata}), in
 good agreement with the stacked shear-only constraints (Figure
 \ref{fig:mdelta}).
The lower panel shows the $45^\circ$-rotated $\times$ component
 $\llangle\Delta\Sigma_\times\rrangle$, which is 
consistent with a null signal 
well within $2\sigma$ 
at all radii, indicating the reliability of our distortion analysis.
The right vertical axes represent the corresponding shear components,
 $\gamma_+=\llangle \Sigma_{\rm c}^{-1}\rrangle\llangle\Delta\Sigma_+\rrangle$ 
and
 $\gamma_\times=\llangle \Sigma_{\rm c}^{-1}\rrangle\llangle\Delta\Sigma_\times\rrangle$,
scaled to the mean depth of weak-lensing observations. 
\label{fig:stackgt}
}
\end{figure}


The azimuthally-averaged tangential distortion 
is a measure of the {\it radially-modulated} 
surface mass density 
and is insensitive to sheet-like mass overdensities,
which resemble the projected two-halo term
within a couple of virial radii \citep{Oguri+Hamana2011}.
Hence, the stacked tangential-distortion signal around a large sample of
clusters is a sensitive probe of the cluster-only (one-halo
term) mass distribution.  

In Figure \ref{fig:stackgt} we show the stacked tangential-shear profile
$\llangle \bDSigma_+\rrangle$ 
derived for our sample where individual clusters and background
galaxies are weighted by the shear-sensitivity kernel 
(${\rm  tr}{\cal W}_+$ in Section  \ref{subsubsec:stackgt}).
The individual profiles are co-added 
in physical length units across the range
$R=[R_{\rm min},R_{\rm max}]=[200,3500]$\,kpc\,$h^{-1}$,
in $11$ log-spaced bins.
Here, the radial limits $[R_{\rm min}, R_{\rm max}]$ of our stacking analysis
represent approximately the respective median values of 
the radial boundaries $[\theta_{\rm min},\theta_{\rm max}]$
covered by the data for our clusters at $0.19\simlt z_{\rm l}\simlt 0.69$.
For individual clusters,
we impose their respective radial cuts  
$[\theta_{\rm min},\theta_{\rm max}]$ on the background samples, to be consistent with our individual
cluster analysis.
For our sample, we find a sensitivity-weighted average redshift of
$\llangle z_{\rm l}\rrangle=0.345$, in close agreement with the 
median redshift of $\overline{z}_{\rm l}=0.352$.  
The effective lensing sensitivity 
$\llangle \Sigma_{\rm c}^{-1}\rrangle$ (Equation (\ref{eq:w_eff})) is 
$1/\llangle \Sigma_{\rm c}^{-1}\rrangle \simeq 3.88\times 10^{15} hM_\sun$\,Mpc$^{-2}$.

We detect the stacked lensing signal at a total S/N of
$\simeq 25$ 
using the full covariance matrix ${\cal C}_+$ 
(Equation (\ref{eq:covtot_gstack}))
to take into account
intrinsic ellipticity and projected uncorrelated LSS noise,
photo-$z$ uncertainties in the mean-depth calibration,
and profile variations in individual clusters.
The $45^\circ$-rotated $\times$ component
$\llangle\bDSigma_\times\rrangle$ is consistent with a null
signal within $2\sigma$ at all radii, with a total S/N of $\simeq 2.8$, 
indicating that residual systematic errors are at least an order of magnitude
smaller than the measured lensing signal.

\subsubsection{Modeling the Stacked Weak Shear Signal}
\label{subsubsec:clash_prof}

\input{table7.tex}

Here we quantify and characterize the ensemble-averaged mass
distribution of our cluster sample using the stacked
tangential-distortion signal. 
We examine the following five models for the halo mass density profile,
$\rho(r)=dM_{\rm 3D}(<r)/dr/(4\pi r^2)$, each described by $N_{\rm p}$ parameters:
\begin{enumerate}
\item Singular isothermal sphere (SIS) model with $N_{\rm p}=1$:
\begin{equation}
\rho_{\rm SIS}(r)=\frac{\sigma_v^2}{2\pi Gr^2}
\end{equation}
with $\sigma_v$ the one-dimensional isothermal velocity dispersion.
\item  Isothermal-$\beta$ model with $N_{\rm p}=2$ \citep{1999PThPS.133....1H}:
\begin{equation}
\rho_{\rm iso}(r)=\frac{M_{\rm c}}{2\pi
 r_{\rm c}^3}\frac{3+(r/r_{\rm c})^2}{[1+(r/r_{\rm c})^2]^2}
\end{equation}
with $M_{\rm c}=M_{\rm 3D}(<r_{\rm c})$ the total mass enclosed within the core
      radius $r_{\rm c}$.
      Note $\rho_{\rm iso}(r)\propto r^{-2}$ at $r\gg r_{\rm c}$.
\item NFW model with $N_{\rm p}=2$: $\rho_{\rm NFW}(r)$ by Equation (\ref{eq:nfw}).
\item Baltz-Marshall-Oguri truncated-NFW model with $N_{\rm p}=2$ \citep[][BMO]{BMO}:
\begin{equation}
\rho_{\rm BMO}(r)=\frac{\rho_{\rm s}}{(r/r_{\rm s})(1+r/r_{\rm
 s})^2}\left[\frac{1}{1+(r/r_{\rm t})^2}\right]^2
\end{equation}
with $r_{\rm t}=2.6 r_{\rm vir} (\sim 3r_{\rm 200c})$ the truncation radius, 
where the ratio of the truncation to virial radius, 
$\tau_{\rm v}\equiv r_{\rm t}/r_{\rm vir}$, is fixed to the typical value 
for cluster-sized halos in the $\Lambda$CDM cosmology \citep{Oguri+Hamana2011}.
\item Einasto model with $N_{\rm p}=3$ \citep{Einasto1965}:
\begin{equation}
\rho_{\rm E}(r)=\rho_{\rm s}\exp\left[-\frac{2}{\alpha_{\rm E}}
			  \left(\frac{r}{r_{\rm s}}\right)^{\alpha_{\rm E}}\right]
\end{equation}
with $\alpha_{\rm E}$ the shape parameter describing the degree
      of curvature.  An Einasto profile with $\alpha_{\rm E}\approx 0.18$ closely resembles the
      NFW profile over roughly two decades in radius
      \citep{Ludlow+2013}. The logarithmic density gradient equals -2 at $r=r_{\rm s}$.
\end{enumerate}
The NFW, BMO, and Einasto density profiles 
represent a family of phenomenological models for DM halos
motivated by simulations and observations.
The SIS profile with $\rho(r)\propto r^{-2}$ is often adopted as a lens
model for its simplicity. 
The isothermal-$\beta$ model describes the total gravitating mass
profile for isothermal gas 
with a $\beta$ density profile
\citep{Cavaliere+Fusco-Femiano1978}.\footnote{For this model, 
$M_{\rm c}=(3\beta k_{\rm B} T r_{\rm c})/(2G\mu m_{\rm p})$ with  
$\beta$ the gas-density slope parameter, 
$T$ the gas temperature,
$k_{\rm B}$ the Boltzmann constant, 
$\mu$ the mean molecular weight, and $m_{\rm p}$ the proton mass.}

For the NFW, BMO, and Einasto models,
we define the halo concentration parameter
by $c_{200{\rm c}}=r_{200{\rm c}}/r_{\rm s}$.
We specify the NFW and BMO models with $(M_{200{\rm c}},c_{200{\rm c}})$, the Einasto
model with ($M_{200},c_{200},\alpha_{\rm E}$), 
the isothermal-$\beta$ model with ($M_{200{\rm c}},r_{\rm c}$)
and the SIS model with $M_{200{\rm c}}$.
For all cases, we can ignore the two-halo contribution to $\Delta\Sigma_+$,
which is estimated to be $\gamma_+\simlt 10^{-3}$ within our radial
range $R\simlt 2 r_{\rm vir}$ 
and is an order of magnitude smaller than the observed lensing signal
(see Figure \ref{fig:stackgt}). 

We constrain the model parameters with our 
$\llangle \widehat{\bDSigma_{+}}\rrangle$ profile and its full covariance matrix
${\cal C}_+$ (Section \ref{subsubsec:stackgt}).   We use Equation
(\ref{eq:stack_nlincor}) to make predictions for$\llangle
\widehat{\bDSigma_{+}}\rrangle$.
The $\chi^2$ minimization is performed using the {\sc minuit} minimization
package from the CERN program libraries. 
The best-fit parameters are reported in Table \ref{tab:mc200}, along with
the reduced $\chi^2$ and corresponding probability-to-exceed (PTE) values.
The NFW, BMO, and Einasto models provide excellent fits to the data with
PTE values of
0.66,
0.58, and
0.51,
respectively.  
The isothermal $\beta$ model yields a poorer but acceptable fit with a
PTE of  
0.33, 
while the SIS model gives an
unacceptable fit of
${\rm PTE}=4.7\times 10^{-3}$.
The SIS model is disfavored at $4.3\sigma$ significance from a likelihood-ratio
test based on the differenced $\chi^2$ values 
$\Delta\chi^2 \equiv \chi^2_{\rm SIS, min}-\chi^2_{\rm NFW,min} \simeq
18.6$ relative to the NFW model.
If the truncation radius of the BMO model is allowed to vary, 
we obtain $\tau_{\rm v}=(2\pm 106)\times 10^2$
(${\rm PTE}=0.56$),
which however makes the model essentially equivalent to the NFW profile.

In what follows, we will focus on our best models among those studied here,
namely the NFW, Einasto, and BMO density profiles. 
These models constrain $c_{200{\rm c}}$ in the
range ($1\sigma$),  
$3.2\simlt c_{200{\rm c}} \simlt 4.4$ 
($c_{200{\rm c}}=4.01^{+0.35}_{-0.32}$ for NFW), 
for our 16 X-ray-selected clusters
with $M_{200{\rm c}}=(1.3\pm 0.1)\times 10^{15}M_\odot\,h_{70}^{-1}$.
For the NFW model, we find $r_{\rm s}\simeq 360$\,kpc\,$h^{-1}$, so that
our data cover 
the range $0.6\simlt R/r_{\rm s}\simlt 10$. 

\subsubsection{Uncertainty in Halo Profile Shape}
\label{subsubsec:clash_masssys}


\begin{figure}[!htb] 
 \begin{center}
 \includegraphics[width=0.4\textwidth,angle=0,clip]{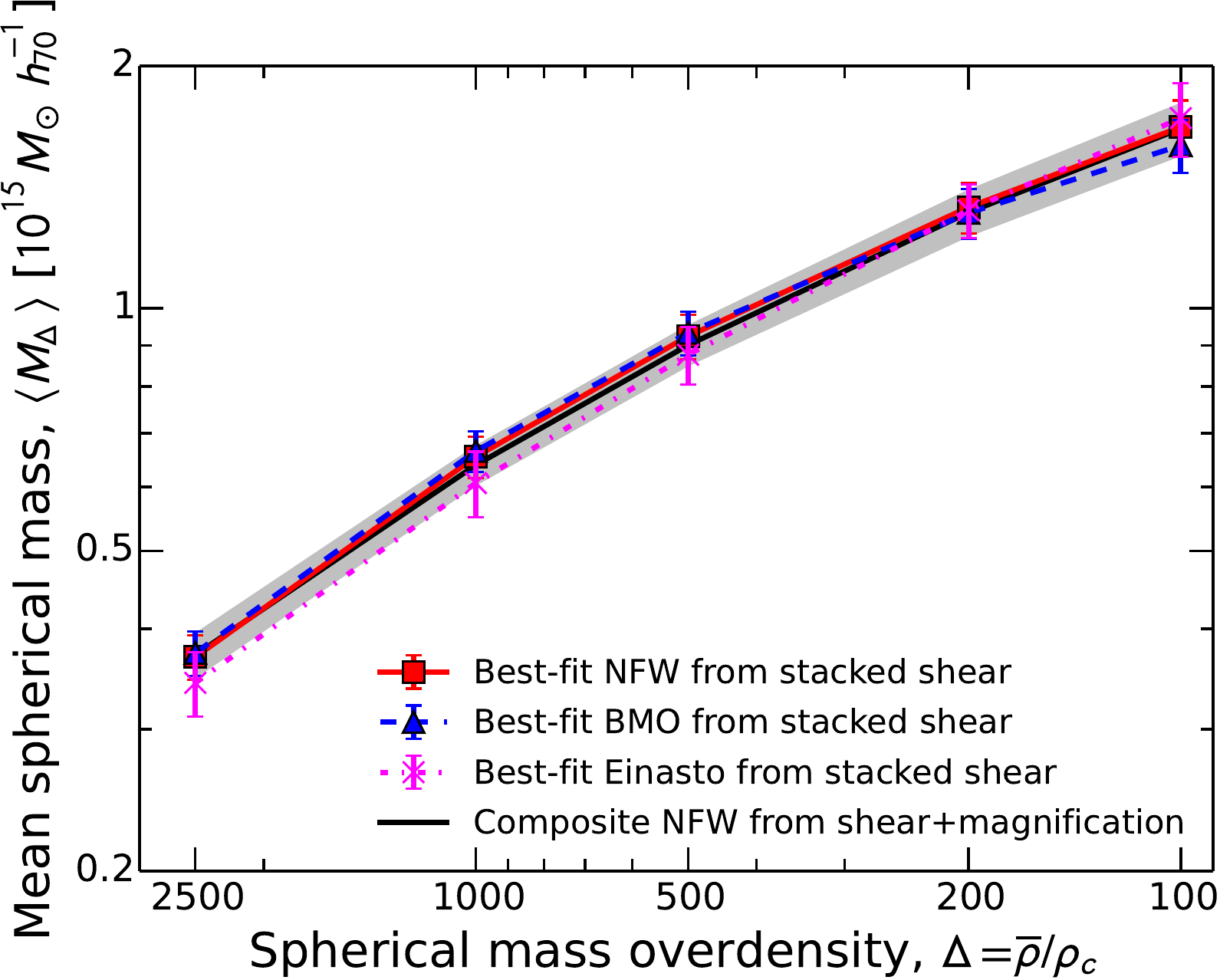} 
 \end{center}
\caption{
Mean cumulative mass profiles $M_\Delta$ of the ensemble of
 16 CLASH X-ray-selected clusters derived from different
 weak-lensing analysis methods,
shown at five characteristic values of the spherical mass overdensity,
$\Delta\equiv \rho(<r)/\rho_{\rm c}$.
The squares, triangles, and crosses with error bars, respectively, are
 the best-fit  NFW, truncated-NFW \citep[][BMO]{BMO}, and Einasto
 profiles with $1\sigma$ uncertainty derived from the stacked
 shear-only analysis (Figures \ref{fig:stackgt}), 
 demonstrating the effects of the uncertainty in halo profile shape.
The black-solid line and gray-shaded area show the mean and
 $1\sigma$ uncertainty of 
the composite profile $\llangle M_\Delta\rrangle$
from a weighted average of NFW fits to individual mass profiles
 $\Sigma(R)$ reconstructed from the shear+magnification constraints
(Figure \ref{fig:wldata}). 
\label{fig:mdelta}
}
\end{figure} 

To examine the impact of the uncertainty in profile shape,
we compare in Figure \ref{fig:mdelta} the spherical mass profiles
$M_{\Delta_{\rm c}}$ for the best-fit NFW, BMO, and Einasto models at
several overdensities.
These profiles are nearly identical at $\Delta_{\rm c}\ge 200$.
At $\Delta_{\rm c}=100$,
the BMO-to-NFW and Einasto-to-NFW mass ratios
are 
$0.95\pm 0.10$  
and
$1.03\pm 0.14$, respectively,
both consistent with unity.

Also shown in Figure \ref{fig:mdelta} is the composite profile 
$\llangle M_{\Delta}\rrangle$ for our sample
constructed from a weighted average of 
the individual mass estimates (Section \ref{subsec:mass}) constrained by
the combined shear+magnification measurements 
(Figures \ref{fig:wldata} and \ref{fig:mdata}).
Specifically, $\llangle M_{\Delta}\rrangle$ is defined by
\begin{equation}
\label{eq:Mstack}
\llangle M_{\Delta}\rrangle = 
\frac{\sum_{n=1}^{N_{\rm cl}} {\rm tr}({\cal W}_{+,n})\, M_{\Delta,n}}
{\sum_{n=1}^{N_{\rm cl}} {\rm tr}({\cal W}_{+,n})}
\end{equation}
($N_{\rm cl}=16$),
using the shear-sensitivity kernel ${\rm tr}({\cal W}_{+})$.
At $\Delta_{\rm c}=200$, 
$\llangle M_{200{\rm c}}\rrangle=(1.32\pm 0.08)\times 10^{15}M_\odot\,h_{70}^{-1}$,
in excellent agreement with the best-fit NFW halo mass,
$M_{200{\rm c}}=1.34^{+0.10}_{-0.09}\times 10^{15}M_\odot\,h_{70}^{-1}$, 
from the stacked shear-only analysis
(Table \ref{tab:mc200}).
These comparisons show that our results are robust against different
ensemble-averaging techniques and data combinations, and that
the uncertainty in profile shape does not significantly affect our
cluster mass measurement within the virial radius.
Similarly, the composite NFW prediction for 
$\llangle \bDSigma_+\rrangle$ is shown in Figure
\ref{fig:stackgt}, demonstrating consistency between 
the shear and magnification measurements.


On the other hand, the effects of baryonic physics can in principle
affect the mass density profile in the central high-density region of
the cluster. Here  
we turn to assess the possible impact of adiabatic contraction
on the total measured mass profile \citep{Gnedin+2004,Okabe+2013}
by introducing a central point mass into our modeling.
The results for the combined NFW and point-mass (NFW+pm) model are
listed in Table \ref{tab:mc200}.
For the central point-mass component,
we obtain $M_{\rm p}=(11\pm 33)\times 10^{12}M_\odot\,h_{70}^{-1}$, or 
a $1\sigma$ upper limit of $M_{\rm p} < 44\times 10^{12}M_\odot\,h_{70}^{-1}$
within $R_{\rm min}=200\,$kpc\,$h^{-1}$.
Including this additional degree of freedom to account for the central
unresolved mass component does not significantly change the best-fit mass and
concentration parameters for our data (Table \ref{tab:mc200}); however, it 
reduces the lower bound on $c_{200}$, allowing for somewhat lower
concentrations 
($c_{200{\rm c}}\simgt 3.2$ at $1\sigma$).

\subsection{CLASH Stacked Mass Profile Analysis}
\label{subsec:clash_stackwl}


\begin{figure}[!htb] 
 \begin{center}
 \includegraphics[width=0.45\textwidth,angle=0,clip]{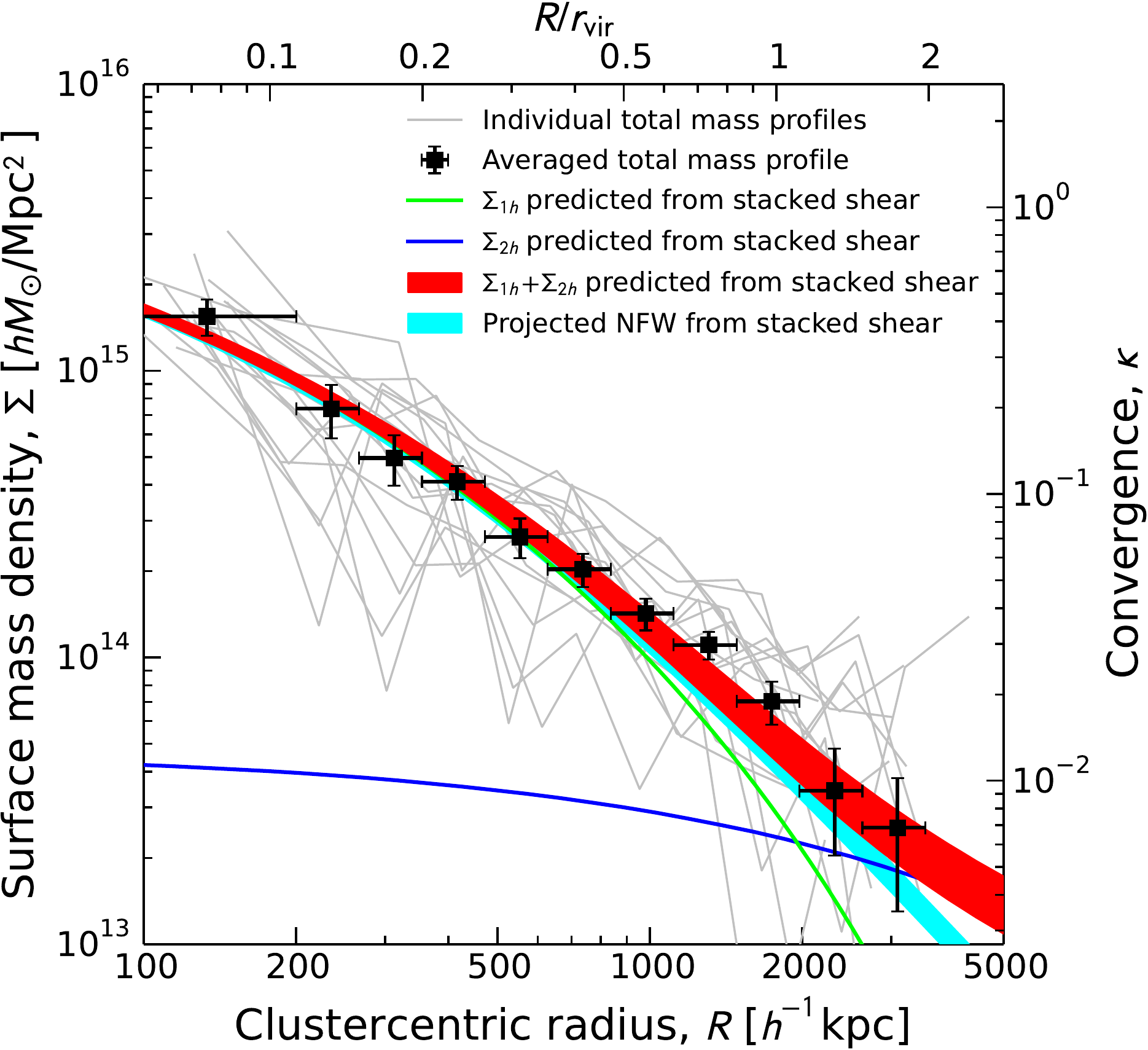} 
 \end{center} 
\caption{
The averaged total projected mass profile
(black squares) of the X-ray-selected subsample, 
which is obtained by
 stacking individual mass profiles  (gray
 lines; Figure \ref{fig:mdata}) derived from our 
 shear+magnification data (Figure \ref{fig:wldata}). 
The green- and blue-solid lines, respectively, are the 
one-halo ($\Sigma_{\rm 1h}$) and 
two-halo ($\Sigma_{\rm 2h}$) contributions
predicted by our best-fit model from the
 stacked shear-only analysis (Figure  \ref{fig:stackgt}).
The red-shaded area shows the sum of the predicted 
$\Sigma_{\rm 1h}$ and $\Sigma_{\rm 2h}$
components (68\% CL), in agreement
 with the observed total mass profile 
based on the combination of shear and magnification.
The projected NFW model (cyan-shaded area,  68\% CL) based on the stacked
 shear-only constraints slightly underpredicts the observed mass profile
 at $R\simgt r_{\rm vir}$. 
The scale on the right vertical axis indicates the corresponding lensing
 convergence,  
$\kappa = \llangle \Sigma_{\rm c}^{-1}\rrangle\llangle \Sigma\rrangle$,
scaled to the mean depth of weak-lensing observations. 
\label{fig:stackwl}
}
\end{figure}

We show in Figure \ref{fig:stackwl} the 
averaged total mass profile $\llangle\bSigma\rrangle$ 
which is obtained by stacking the individual cluster profiles $\bSigma$
derived from the shear+magnification constraints. 
We take $R_{\rm min}=200$\,kpc\,$h^{-1}$ and $R_{\rm max}=3500\,$kpc\,$h^{-1}$
to be consistent with the stacked shear-only analysis 
(Section \ref{subsec:clash_stackgt}).
The effective lensing sensitivity for this analysis is
$1/\llangle\Sigma_{\rm c}^{-1}\rrangle \simeq 3.98\times 10^{15}h M_\odot$\,Mpc$^{-2}$,
which is obtained using ${\rm tr}({\cal W}_n)$ as statistical weights
(Section \ref{subsubsec:stackwl}).
The averaged mass profile $\llangle\bSigma\rrangle$
is detected at a total S/N of $\simeq 21$,
which is $\simeq 16\%$ lower than that of the stacked distortion signal 
$\llangle\bDSigma_+\rrangle$ (Section \ref{subsubsec:clash_stackgt}).
This reduction in S/N is due  primarily to significant cluster-to-cluster
variations between the individual total $\bSigma$ profiles.  
We have also checked that our stacking results are insensitive to
 whether or not individual cluster profiles are scaled to their 
$r_{\rm  vir}$
prior to averaging albeit some variance in the outermost radial bin.


Here we examine the consistency of our ensemble-averaged 
{\it shear-only} and {\it shear+magnification} 
results in the context of the standard $\Lambda$CDM model.
To do this, we employ the halo-model prescriptions 
of \citet{Oguri+Takada2011} and \citet{Oguri+Hamana2011}
for computing the two-halo contribution $\Sigma_{\rm 2h}$
to the total projected matter distribution
$\Sigma=\Sigma_{\rm 1h}+\Sigma_{\rm 2h}$
expected from the stacked shear constraints on the one-halo component
$\Sigma_{\rm 1h}$.
For an ensemble of clusters with mass $M$ and redshift $z$,
the $\Sigma_{\rm 2h}$ component is computed by projecting
the two-halo term 
$\rho_{\rm 2h}(r) = \overline{\rho} b_{\rm h}(M)\xi_{\rm lin}(r)$ 
along the line of sight \citep[see Section 2.2 of][]{Oguri+Hamana2011}, 
with 
$\overline{\rho}$ the mean background density of the universe,
$b_{\rm h}(M)$ the linear halo bias,\footnote{The \citet{Tinker+2010} model is
given as a function of the interior overdensity $\Delta_m$ defined with
respect to the mean background density $\overline{\rho}(z)$ of the universe.  
In the present study we take $\Delta_{\rm c}=200$, corresponding to 
$\Delta_{\rm m}\sim 420$ at $z=0.35$ for our adopted cosmology.} 
and $\xi_{\rm lin}(r)$ the linear matter correlation function, 
all evaluated at $z=\llangle z_{\rm l}\rrangle$
in the WMAP seven-year cosmology (Section \ref{subsubsec:cmat}).
The two-halo term
is proportional to the product $b_{\rm h}(M)\sigma_8^2$, 
where $\sigma_8^2$ is the rms amplitude of linear mass fluctuations in a
sphere of comoving radius $r=8$\,Mpc\,$h^{-1}$.

To estimate the halo bias $b_{\rm h}(M)$ we adopt the model of
\citet{Tinker+2010} which is well calibrated using a large set of
$N$-body simulations. 
With this, we find 
$b_{\rm h} = 9.0 \pm 0.5 \pm 2$ for our best-fit mass model from the
stacked shear-only analysis,
where the former error reflects the uncertainty in the mass estimate and
the latter represents the model uncertainty of $\sim 20\%$ in the fitting
function of \citet{Tinker+2010} in the high-peak, high-bias regime.
We use our best-fit BMO model in Table
\ref{tab:mc200} to represent the $\Sigma_{\rm 1h}$ component.

In Figure \ref{fig:stackwl} we compare the observed $\llangle\bSigma\rrangle$
profile with the resulting shear-based predictions for the one-halo
($\Sigma_{\rm 1h}$), two-halo ($\Sigma_{\rm 2h}$), and total 
($\Sigma_{\rm 1h}+\Sigma_{\rm 2h}$) components. 
The two-halo component $\Sigma_{\rm 2h}$ slowly decreases from 
$\kappa_{\rm 2h} \equiv \llangle \Sigma^{-1}_{\rm c} \rrangle \Sigma_{\rm 2h} \sim 10^{-2}$ 
in the central region to 
$\kappa_{\rm 2h}\sim 4\times 10^{-3}$ at $R\sim 2r_{\rm vir}$,
and thus mimics a massive uniform sheet around the clusters.
Figure \ref{fig:stackwl} shows that the halo-model predictions for the total
signal agree well with our results, indicating good consistency
between our 
shear and magnification measurements in the context of the standard
$\Lambda$CDM model. 
On the other hand, the projected NFW model based on the stacked
shear-only constraints slightly underpredicts the observed mass profile
at $R\simgt r_{\rm vir}$ where the two-halo contribution is important.


\section{Discussion}
\label{sec:discussion}

\subsection{Mass Comparisons}
\label{subsec:mcomp}

In this subsection we compare our mass estimates for the
CLASH sample (Table \ref{tab:mass}) with those from previous studies.
More detailed statistical comparisons of cluster mass determinations for
the CLASH sample  
by different lensing, X-ray, SZE, and dynamical methods will be
presented in our forthcoming papers.

\subsubsection{Previous CLASH Lensing Results}

Now we compare our estimates of $M_{\rm vir}$ obtained by our uniform
analysis 
with those from our previous work on individual CLASH clusters,
namely
Abell\,2261 \citep{Coe+2012A2261}, 
MACS\,J1206.2-0847 \citep{Umetsu+2012},
and
MACS\,J0717.5+3745 \citep{Medezinski+2013}.
The major differences between this and previous weak-lensing analyses
are summarized as follows:
\begin{itemize}
\item We have applied a shear calibration
correction factor of $1/0.95\simeq 1.05$ (Section \ref{subsec:shape}), 
which was not included in our previous studies 
of Abell\,2261 and MACS\,J1206.2-0847.
\item The nonlinear correction given by Equation (\ref{eq:nlcor}) was
      not included in our previous work. 
\item All mean depth estimates have been uniformly performed here using a
      self-consistent photo-$z$ method as described in Section
      \ref{subsec:back}. 
\item A Bayesian inference approach has been used here
to measure cluster masses at several overdensities, by limiting
      the fitting range to $R\le 2$\,Mpc\,$h^{-1}$.  
The $\chi^2$ minimization was performed in our previous work to derive 
the best-fit virial mass and concentration parameters for the full
radial range. 
\end{itemize}
Besides, in earlier papers, we assumed a slightly different cosmology with
$(\Omega_{\rm m},\Omega_\Lambda)=(0.3,0.7)$. 
As a result, these changes lead to a typical increase of $\sim 10\%$ in
our $M_{\rm vir}$ estimates relative to our previous work.
This increase is primarily due to the inclusion of the shear calibration
correction, which translates into a typical increase of $\sim 8\%$ in
$M_{\rm vir}$.  On the other hand, the present analysis
takes into account systematic uncertainties $C^{\rm sys}$ (Equation
(\ref{eq:Csys}))
due primarily
to the residual mass-sheet degeneracy, providing more conservative error
estimates for clusters with poorer magnification constraints.

The following is a detailed comparison of each cluster:

{\it Abell\,2261} --- \citet{Coe+2012A2261} obtained 
$M_{\rm vir}=2.21^{+0.25}_{-0.23}\times 10^{15}M_\odot\,h_{70}^{-1}$
from an NFW fit to the mass profile from their
shear+magnification data
(Constraints (8) of their Table 4),
with an estimated correction of 
$\Delta M_{\rm vir}\simeq 0.15\times 10^{15}M_\odot\,h_{70}^{-1}$
for projection effects due to LSS as specifically observed
behind A2261. This is compared
to $M_{\rm vir}=(2.58\pm 0.54)\times 10^{15}M_\odot\,h_{70}^{-1}$ in
this work, corresponding to an increase of $17\%$ in the best-fit
$M_{\rm vir}$,
in which the revised estimates of mean depth result in a $\sim 5\%$
increase in mass.
The large increase in the uncertainty of $M_{\rm vir}$ is
primarily caused by the inclusion of $C^{\rm sys}$ (Equation
\ref{eq:Csys}) due to the residual mass-sheet uncertainty, which
dominates the error budget for the four outermost $\Sigma$ bins of
A2261. A large contribution from $C^{\rm sys}$ is generally expected for
low-$z$, high-mass clusters due to their large angular extent
on the sky.

{\it MACS\,J1206.2-0847} --- \citet{Umetsu+2012} found 
$M_{\rm vir}=1.64^{+0.49}_{-0.40}\times 10^{15}M_\odot\,h_{70}^{-1}$
based on their shear+magnification measurements (Method (2) of
their Table 5), compared to our estimate of 
$M_{\rm vir}=(1.87\pm 0.46)\times 10^{15}M_\odot\,h_{70}^{-1}$, which
represents an increase of $14\%$.
Their primary NFW model from full-lensing constraints (their Method (7)), 
including a correction for the surrounding LSS, yields 
$M_{\rm 500c}=(1.01\pm 0.15)\times 10^{15}M_\odot\,h_{70}^{-1}$
\citep[see also][]{Rozo+2014closing}, which agrees well with our estimate of $M_{\rm 500c}=(1.06\pm 0.21)\times 10^{15}M_\odot\,h_{70}^{-1}$.

{\it MACS\,J0717.5+3745} --- The cluster was recently studied by
\citet{Medezinski+2013}, who derived 
$M_{\rm vir}=3.19^{+0.63}_{-0.54}\times 10^{15}M_\odot\,h_{70}^{-1}$
based on their two-dimensional shear and azimuthally-averaged
magnification-bias measurements (their Table 5),\footnote{The 5\%
residual correction for shear calibration was included in the analysis
by \citet{Medezinski+2013}.} 
assuming a
composite model of an NFW halo and a constant component, the latter
accounts for the surrounding LSS.
Their estimated $M_{\rm vir}$ is $11\%$ lower than
our estimate of 
$M_{\rm vir}=(3.53\pm 0.60)\times 10^{15}M_\odot\,h_{70}^{-1}$,
as the correlated LSS contributions are partly absorbed into their
mass-sheet parameter. 

Finally, we compare our mass estimates 
of RXJ2248.7-4431 with the results from a weak-shear analysis of 
\citet{Gruen+2013RXJ2248} based on ESO/WFI data.
For this cluster we use the same co-added images
created by \citet{Gruen+2013RXJ2248}, but adopt 
substantially different approaches in our analysis 
(Section \ref{subsec:reduction}).
They derived mass estimates of this cluster at
several overdensities, 
$M_{\rm 200m}=33.1^{+9.6}_{-6.8}\times 10^{15}M_\odot\,h_{70}^{-1}$,
$M_{\rm 101c}=32.2^{+9.3}_{-6.6}\times 10^{15}M_\odot\,h_{70}^{-1} (\approx M_{\rm 100c})$,
$M_{\rm 200c}=22.8^{+6.6}_{-4.7}\times 10^{15}M_\odot\,h_{70}^{-1}$,
and
$M_{\rm 500c}=12.7^{+3.7}_{-2.6}\times 10^{15}M_\odot\,h_{70}^{-1}$,
which are all consistent within errors with ours
(Table \ref{tab:mass}).

\subsubsection{The Weighing the Giants Project}
\label{subsec:WtG}


\begin{figure}[!htb] 
 \begin{center}
 \includegraphics[width=0.4\textwidth,angle=0,clip]{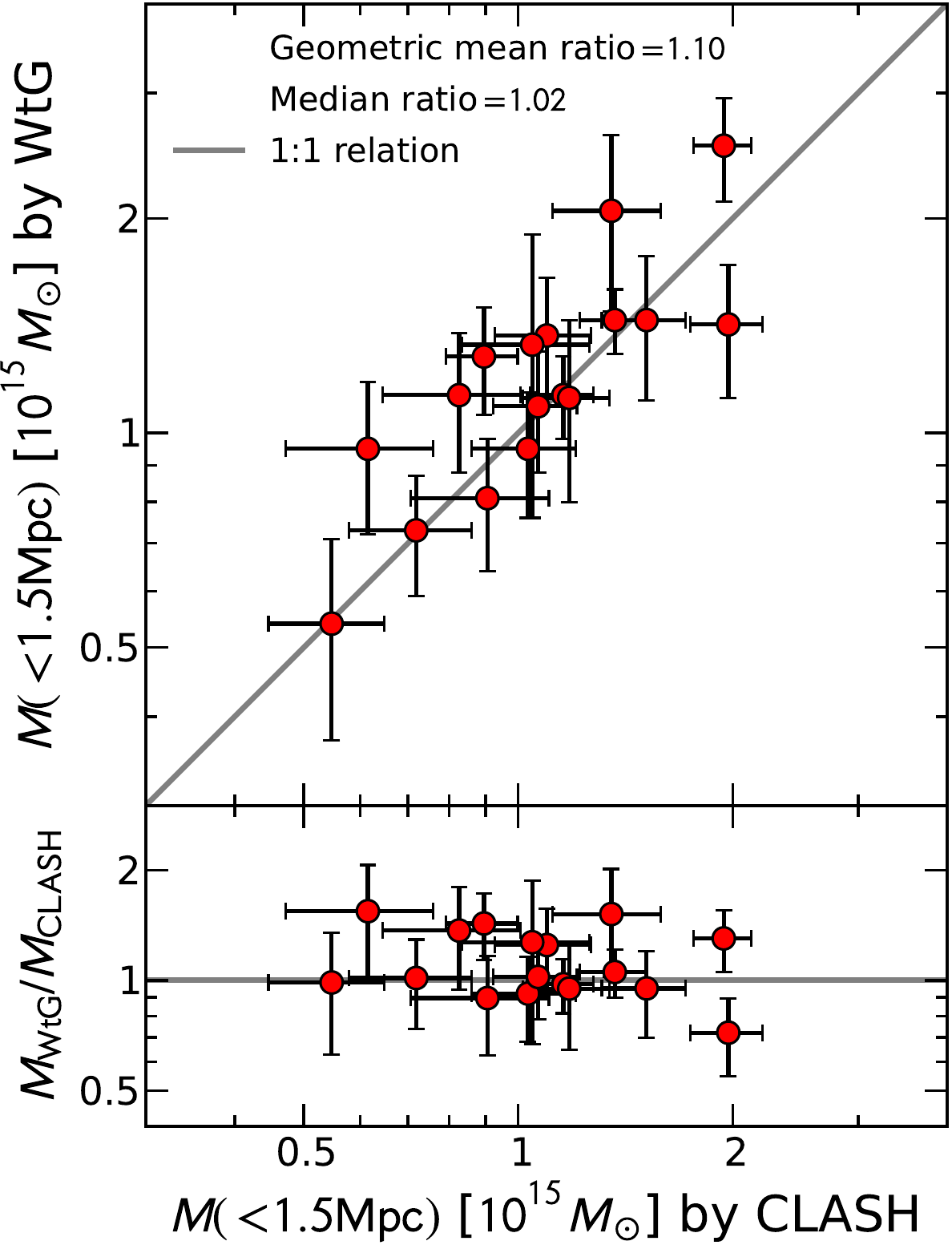}
 \end{center}
\caption{
\label{fig:WtG}
Comparison of Subaru shear-only mass estimates ($M_{\rm WtG}$) 
from \citet{WtG3}  to our Subaru shear+magnification results 
($M_{\rm  CLASH}$) for 17 clusters in common between the two studies
 (upper panel, circles with error bars).
For this comparison we measure the
 mass within a sphere of $r=1.5\,$Mpc$\,h_{70}^{-1}$ using the same
 cosmology 
$(h, \Omega_{\rm m}, \Omega_{\Lambda})=(0.7, 0.3, 0.7)$
adopted by \citet{WtG3}.
The solid line shows the one-to-one relation.
The lower panel shows $M_{\rm  WtG}/M_{\rm CLASH}$ of individual
 clusters against $M_{\rm CLASH}$.
}
\end{figure} 

The majority of the CLASH clusters were targeted as part of the WtG
project.
Recently, the WtG collaboration published results of their weak-lensing
shear mass measurements of 51 X-ray luminous clusters at 
$0.15\simlt z \simlt 0.7$ based on deep multi-color
Subaru/Suprime-Cam and CFHT/MegaPrime optical imaging \citep{WtG1,WtG2,WtG3}. 

Figure \ref{fig:WtG} shows a comparison of Subaru shear-based mass
estimates ($M_{\rm WtG}$) from \citet{WtG3} to 
our Subaru weak-lensing results.
There are 17 clusters in common between the two studies,
including 14 X-ray-selected and 3 high-magnification CLASH clusters.
To make this comparison,
we measure the mass ($M_{\rm CLASH}$) 
within a fixed physical radius of $r=1.5$\,Mpc\,$h_{70}^{-1}$
assuming the spherical NFW model and using the same cosmology 
$(h, \Omega_{\rm m}, \Omega_{\Lambda})=(0.7, 0.3, 0.7)$ adopted by \citet{WtG3}.
\citet{WtG3} derived cluster masses from NFW fits to their tangential
reduced-shear data  
over the radial range $R=0.75$--3\,Mpc\,$h_{70}^{-1}$, whereas our
mass measurements are based on the lensing convergence ($\kappa$) data 
over the range $R\le 2\,$Mpc\,$h^{-1}\simeq 2.9$\,Mpc\,$h_{70}^{-1}$,
constrained by the combined shear+magnification measurements. 
For this comparison, we use their results based on the photo-$z$
posterior probability distributions of individual galaxies where
available, otherwise those from their color-cut method.
For these 17 clusters,
we find a median $M_{\rm WtG}/M_{\rm CLASH}$ ratio of $1.02$ 
and an unweighted geometric mean (Section \ref{subsubsec:triaxiality})
of 
$\langle M_{\rm WtG}/M_{\rm CLASH}\rangle = 1.10\pm 0.09$.
For the error estimation,
we have neglected the correlation between the two axes
due to overlap of source galaxies used to measure the shear
because our analysis includes independent constraints from the
magnification effects on red galaxy counts.
Considering the systematic uncertainty of
$\pm 8\%$ in our overall mass calibration (Section
\ref{subsubsec:msys}), we find no significant offset between our
and WtG mass measurements.

\subsubsection{Rozo et al. (2014)}

\citet{Rozo+2014closing}
performed a self-consistent calibration of cluster scaling
relations between X-ray, SZE, and optical observables.
They used their calibrated mass--X-ray luminosity relation to predict
masses  ($M_{\rm R14}$)
within the overdensity radius $r_{\rm 500c}$
for the CLASH clusters at $z\le 0.4$ and
MACS\,J1206.2-0847 at $z=0.44$. 
For each cluster we calculate the mass ($M_{\rm CLASH}$) 
within the $r_{\rm 500c}$ of \citet{Rozo+2014closing}
assuming the spherical NFW model.

For 13 clusters in common with \citet{Rozo+2014closing}, we obtain 
a median mass ratio of $1.12$, and
an error-weighted geometric mean of 
$\langle M_{\rm R14}/M_{\rm CLASH}\rangle =1.13\pm 0.10$,
corresponding to a mass offset of 
$\Delta\ln M \equiv \langle \ln M_{\rm R14}\rangle - \langle \ln M_{\rm CLASH} \rangle =0.12\pm 0.09$. 
This offset is not significant compared to the overall 
calibration uncertainty of $\pm 8\%$ (Section \ref{subsubsec:msys}). 
Excluding two obvious outliers with $M_{\rm R14}/M_{\rm CLASH}\sim 2$
(RXJ\,2129.7+0005 and RXJ\,1532.9+3021), 
we find a median mass ratio of $1.11$ and a weighted geometric mean
ratio of $1.03\pm 0.10$, which corresponds to a mass offset of 
$\Delta\ln M = 0.03\pm 0.09$. 


\subsection{Cluster $c$--$M$ Relation}
\label{subsec:cm}


\begin{figure}[!htb] 
 \begin{center}
 \includegraphics[width=0.45\textwidth,clip]{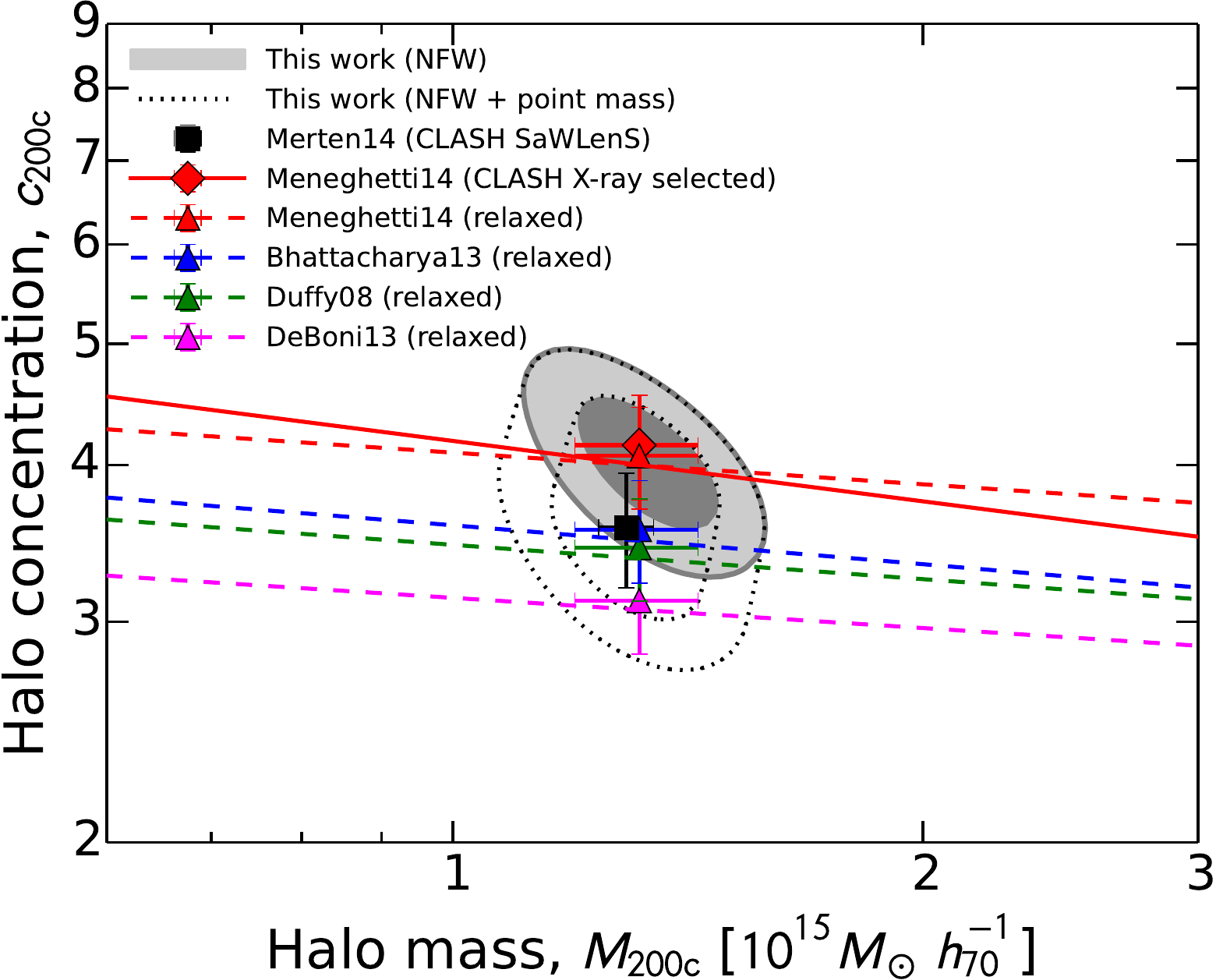} 
 \end{center}
\caption{
Stacked weak-shear constraints on
the mass and concentration 
of 16 CLASH X-ray-selected clusters
at $\llangle z_{\rm l}\rrangle\simeq 0.35$.
The gray contours show the 68.3\% and 95.4\% confidence levels in the
 $c$--$M$ plane for the spherical NFW fit.
Similarly, 
the black-dashed contours show the model fit including a central
point-mass component, accounting  for the possible impact of adiabatic
 contraction.
The triangles with error bars
show the predictions $\llangle c_{\rm 200c}\rrangle$ 
using the $c_{\rm 200c}(M_{\rm 200c},z)$ relations for relaxed halos in the literature.
The diamond with error bars represents the prediction for CLASH-like X-ray regular halos
\citep{Meneghetti2014clash}.
These predictions 
are obtained from the respective weighted averages of
 predicted $c$ values for the observed $(M_{200{\rm c}},z)$ distribution of
 our sample, where 
the $M_{200{\rm c}}$ masses are based on the 
 shear+magnification constraints.
For each model, the stacked model prediction is consistent with the corresponding
 $c$--$M$ relation 
evaluated at $z=\llangle z_{\rm l}\rrangle$ (solid: X-ray; dashed: relaxed).
The square represents the results from a strong- and weak-lensing 
({\sc SaWLenS}) analysis of 19 CLASH X-ray-selected clusters \citep{Merten2014clash}.
\label{fig:cm}
}
\end{figure}

Stacking the tangential distortion signal from a large sample of clusters
can provide useful integrated constraints on the halo concentration--mass ($c$--$M$)
relation. 
In Figure \ref{fig:cm} we summarize our stacked weak-shear constraints
in the $c$--$M$ plane
for our 16 CLASH X-ray-selected clusters. 

\subsubsection{Comparison with $\Lambda$CDM Predictions}
\label{subsubsec:lcdm}

\input{table8.tex}

We compare our results with $\Lambda$CDM predictions from numerical
simulations. 
Specifically, we examine the 
$c$--$M$ relations
obtained by \citet{Duffy+2008}, \citet{Bhatt+2013}, 
\citet{DeBoni+2013}, and M14 using the standard NFW model. 
Hence, we use our NFW results for a baseline comparison with
the numerical simulations.

To make a quantitative comparison between theory and observations, we
calculate, for a given $c(M,z)$ relation, the
sensitivity-weighted average $\llangle c_{200{\rm c}}\rrangle$
of concentrations for 
the observed mass and redshift distribution of our cluster sample, 
$\{M_{200{\rm c},n}\}_{n=1}^{N_{\rm cl}}$ (Table \ref{tab:mass})
and 
$\{z_{{\rm l},n}\}_{n=1}^{N_{\rm cl}}$ (Table \ref{tab:sample}),
by
\begin{equation}
\llangle c_{200{\rm c}}\rrangle = 
\frac{\sum_{n=1}^{N_{\rm cl}} {\rm tr}({\cal W}_{+,n})\, 
c_{200{\rm c}}(M_{200{\rm c},n},z_{{\rm l},n})}
{\sum_{n=1}^{N_{\rm cl}} {\rm tr}({\cal W}_{+,n})}
\end{equation}
which is marked at the average cluster mass $\llangle M_{200{\rm c}}\rrangle$
(Section \ref{subsubsec:clash_prof})
in Figure \ref{fig:cm}.  Note that the masses
$\{M_{200{\rm c},n}\}_{n=1}^{N_{\rm cl}}$ are constrained by the
combined shear+magnification data.
We use Monte-Carlo simulations to estimate the total uncertainty in
$\llangle c_{200{\rm c}}\rrangle$ taking into account both the observational
uncertainties in the cluster mass measurements (Section
\ref{subsec:mass})
 and the intrinsic scatter in the concentration. Here we assume a
 Gaussian intrinsic scatter of dispersion $\sigma_c=0.33c_{200{\rm c}}$ as
 found by \citet{Bhatt+2013}.

The resulting predictions are summarized in Table
\ref{tab:lcdm} and shown in Figure \ref{fig:cm}. 
Overall, the predicted concentrations $\llangle c_{\rm 200c}\rrangle$ 
for relaxed halos overlap well with
our range of allowed $c$ values,
$3.2\simlt c_{200{\rm c}}\simlt 4.4$ at $1\sigma$ (NFW and NFW+pm).
Our NFW results are in excellent agreement with the predicted
concentration 
from nonradiative hydrodynamical $N$-body simulations 
in a WMAP 7-year cosmology (M14, $\sigma_8=0.82$).
Our measurements are slightly higher than, but consistent with,
the DM-only predictions of 
\citet[][$\sigma_8=0.8$]{Bhatt+2013} and \citet[][$\sigma_8=0.796$]{Duffy+2008}.
We find a discrepancy of about $1.8\sigma$ 
between our NFW results and 
the DM-only prediction of \citet[][$\sigma_8=0.776$]{DeBoni+2013}. 
These differences can be partly explained by different
cosmological settings, such as the adopted values of $\sigma_8$,
as discussed in detail by M14.
Alternatively, this discrepancy can be reconciled if 
the NFW+pm model (Figure \ref{fig:cm}) is adopted to account for the possible
impact of unresolved adiabatic contraction (Section \ref{subsubsec:clash_prof}). 

Using the $c$--$M$ relations derived for the full population of halos
(including both relaxed and unrelaxed halos),
we find that the predicted concentrations 
are $4$--$9\%$ lower than those for the relaxed halos (Table \ref{tab:lcdm}).  
The full-sample predictions of the \citet{Bhatt+2013} and M14 models
are consistent with our results within $\sim 1\sigma$.

Finally, we examine the M14 predictions for a sample of CLASH-like
X-ray-regular halos, 
including the effects of the CLASH X-ray selection function and possible
biases due to projection effects. 
M14 characterized a sample of halos which
reproduces the observed distribution of X-ray regularity characteristics
in the CLASH X-ray-selected subsample.
They found that a large fraction
of this sample ($\sim 70\%$) is composed of relaxed halos, but it also
contains a non-negligible fraction of unrelaxed halos.
Matching their simulations to the individual CLASH clusters based on the
X-ray morphology,  M14 predict that the NFW concentrations recovered
from the lensing analysis of the CLASH X-ray-selected clusters are in
the range  
$3\simlt c_{\rm 200c}\simlt 6$.
We find this model provides an excellent match to the observed
concentration (Table \ref{tab:lcdm}). 

\subsubsection{Comparison with the CLASH {\sc SaWLenS} Analysis}
\label{subsubsec:sawlens}

For a massive cluster acting as a supercritical lens,
the strong- and weak-lensing regimes
contribute similar logarithmic radial coverage
\citep{Umetsu+2011stack}. 
Hence, adding strong-lensing information to weak-lensing
allows us to provide tighter constraints on the inner density profile. 
\citet{Merten2014clash} conducted a strong- and
weak-lensing \citep[{\sc SaWLenS} hereafter,][]{Merten+2009} 
analysis of 19 CLASH X-ray-selected clusters, 
including all clusters in our X-ray-selected subsample,
by combining data from 16-band {\it HST} imaging with the wide-field 
weak-shear data analyzed in the present study.
Their work is thus complementary to our wide-field weak-lensing
analysis which combines the shear and magnification effects.
\citet{Merten2014clash} derived a $c$--$M$ relation for their clusters,
finding a moderately-significant trend of  
decreasing $c_{\rm 200c}$ with increasing halo mass,
which is in good agreement with $\Lambda$CDM predictions.

The square in Figure \ref{fig:cm} represents the average concentration
predicted for our X-ray-selected subsample
using their $M_{\rm 200c}$ masses and best-fit $c(M,z)$ relation, demonstrating 
consistency between the results obtained with different
lensing methods.

\subsubsection{Comparison with Previous Stacked-lensing Studies}
\label{subsubsec:cm_comparison}


Our findings are in agreement with the results obtained by
\citet{Okabe+2013} 
who carried out a stacked weak-lensing analysis of
50 X-ray clusters 
($0.15<z_l<0.3$)
from $R=0.1$ to $2.8$\,Mpc\,$h^{-1}$,
using two-band imaging from Subaru/Suprime-Cam observations.
For their full sample,
\citet{Okabe+2013}  found a mean concentration of 
$c_{\rm 200c}=4.22^{+0.40}_{-0.36}$  
at their effective halo mass of
$M_{200{\rm c}} =(8.5\pm 0.6)\times 10^{14}M_\odot\,h_{70}^{-1}$,
which exceeds the predicted concentration from numerical simulations
\citep{Duffy+2008,Zhao+2009,Bhatt+2013,DeBoni+2013}. 

More recently, 
\citet{Covone+2014} used the CFHT Lensing Survey 
\citep[CFHTLenS,][]{Heymans+2012CFHTLenS}
shear catalog to measure the stacked shear signal of
1176 optically-selected clusters ($0.1 < z_{\rm l} < 0.6$) 
from $R=0.1$ to $20$\,Mpc\,$h^{-1}$,
in 6 bins of optical richness corresponding to the mass range 
$0.7 \simlt M_{\rm 200c}/(10^{14} M_\odot$\,$h_{70}^{-1}) \simlt 5$ at a
median redshift of $\overline{z}_l=0.36$.
The normalization of their $c(M)$ relation is 
higher than but marginally
consistent with the prediction  by \citet{Duffy+2008}, and is
in closer agreement with recent simulations by \citet{Bhatt+2013}, which
is qualitatively consistent with our results. 


\citet{Umetsu+2011stack} obtained an averaged total mass profile 
$\llangle \bSigma\rrangle$
of four similar-mass, strong-lensing clusters 
at $\llangle z_{\rm l}\rrangle \simeq 0.32$ 
using combined strong-lensing, weak-lensing shear and magnification measurements from
high-quality {\it HST} and Subaru observations.
These clusters display prominent strong-lensing features, characterized
by large Einstein radii of $\theta_{\rm Ein}\simgt 30\arcsec$ for a
fiducial source redshift of $z_s=2$.
\citet{Umetsu+2011stack} found that their stacked total mass profile is
well described by a single NFW profile over a wide radial range,
$R=40$--$2800$\,kpc\,$h^{-1}$.
The mean concentration is
$c_{\rm vir}=7.68^{+0.42}_{-0.40}$ at 
$M_{\rm vir}=2.20^{+0.16}_{-0.14}\times 10^{15} M_\odot\,h_{70}^{-1}$,
corresponding to the Einstein radius of 
$\theta_{\rm Ein}\simeq 36\arcsec$ ($z_s=2$),
which is compared to our CLASH X-ray-selected subsample with 
$c_{\rm vir}=5.0\pm 0.4$, 
$M_{\rm vir}=(1.58\pm 0.12)\times 10^{15}M_\odot\,h_{70}^{-1}$,
and 
$\theta_{\rm Ein} =  15\arcsec \pm 4\arcsec$ ($z_s=2$). 
Intriguingly, the results from these two stacking studies are in good
agreement with the $\Lambda$CDM prediction for 
the $c_{\rm vir}$--$\theta_{\rm Ein}$ relation based on semi-analytic
calculations of \citet[][see their Figure 10]{Oguri+2012SGAS}, in which
clusters with larger Einstein radii are statistically more concentrated
in projection.

\subsubsection{Impact of the Inclusion of Less Relaxed Clusters}
\label{subsubsec:cm_unrelax}

Here we examine the robustness of our results 
against the inclusion of less relaxed clusters.
To do this, we perform a stacked shear-only analysis by excluding those
clusters likely with a lesser degree of dynamical relaxation,
as indicated by their relatively higher degree of substructure 
\citep{Postman+2012CLASH,Lemze+2013CLASH}, namely
Abell\,209,
Abell\,2261, 
RXJ2248.7-4431
MACS\,J0329.7-0211, 
RXJ1347.5-1145,
and
MACS\,J0744.9+3927.
For the remaining subset of 10 X-ray regular clusters,
the best-fit NFW parameters are obtained as
$c_{\rm 200c}=3.9\pm 0.4$ and 
$M_{\rm 200c}=(1.30 \pm 0.12)\times 10^{15}M_\odot\,h_{70}^{-1}$, 
at $\llangle z_{\rm l}\rrangle = 0.334$.  
We thus find an only negligible change in the best-fit NFW parameters
compared to the total uncertainties.

We note that this subset exhibits a smaller level of cluster-to-cluster
variations in the tangential distortion signal, and
the total uncertainties in the stacked lensing signal are dominated by statistical noise.
Accordingly, we find the uncertainties in the derived parameters here are 
comparable to those found for our total X-ray-selected subsample of 16 clusters.


\subsubsection{Baryonic Effects}
\label{subsubsec:cm_baryon}

Our CLASH X-ray selection ($T_X>5$\,keV)
is designed to minimize the impact of baryonic effects 
on the cluster mass distribution.
The effects of baryonic physics can in principle impact the 
inner halo profile
\citep[$r\simlt 0.1r_{\rm vir}$;][]{Duffy+2010},
and thus modify the gravity-only  $c$--$M$ relation, especially for less
massive halos \citep{Duffy+2010,Bhatt+2013}.

In the present stacked shear-only analysis, we examined the possible
impact of adiabatic contraction  by introducing a central
point mass into our modeling.
From this we obtained a $1\sigma$ upper limit on the
unresolved central mass of 
$M_{\rm p}<44\times 10^{12}M_\odot$\,$h_{70}^{-1}$ 
within $R_{\rm min}=200$\,kpc\,$h^{-1}$
(NFW+pm in Table \ref{tab:mc200}).
We find this does not significantly impact the best-fit 
$M_{200{\rm c}}$ and $c_{200{\rm c}}$ values for our data, but allows
for somewhat lower concentrations, $c_{200{\rm c}}\simgt 3.2$ 
at $1\sigma$.
Our findings are consistent with the conclusions of \citet{Okabe+2013},
who obtained a tighter limit of 
$M_{\rm p}<17\times 10^{12}M_\odot\,h_{70}^{-1}$ 
within $R_{\rm min}=80\,$kpc\,$h^{-1}$ from their stacked weak-shear
analysis of 50 X-ray luminous clusters at $0.15<z<0.3$.


\subsection{Ensemble-averaged Cluster Mass Profile Shape}
\label{subsec:aveprof}

\subsubsection{Halo Mass Profile (1h term)} 
\label{subsubsec:1h}

Since the tangential shear is a measure of the radially-modulated
surface mass density, which is insensitive to sheet-like structures, the 
stacked shear-only analysis can provide powerful constraints on
the halo structure.
We find that the shape of the stacked shear profile
exhibits a steepening radial trend across the radial range 
200--3500\,kpc\,$h^{-1}$, which is
well described by 
the NFW, BMO (truncated-NFW), and Einasto models 
(Section \ref{subsubsec:clash_prof}), 
whereas the two-halo contribution
to $\Delta\Sigma_+$ is negligible across the radial range of our
observations. 

The Einasto shape parameter is constrained as 
$\alpha_{\rm E}=0.191^{+0.071}_{-0.068}$,
which agrees fairly well with
numerical simulations: 
$\alpha_{\rm E} = 0.175 \pm 0.046$ 
\cite[][the best-fit for the averaged profile of their 9 cluster-sized halos]{Gao+2012Phoenix},
$\alpha_{\rm E} = 0.172 \pm 0.032$ 
\cite[][the average and dispersion of their 19 dwarf- to cluster-sized halos]{Navarro+2004},
$\alpha_{\rm E} \sim 0.17$ 
\cite[][the median of their 6 cluster-sized halos]{Merritt+2006},
$\alpha_{\rm E} = 0.21\pm 0.07$ (M14; 
$\alpha_{\rm E}=0.24\pm 0.09$ when fitted to the projected total mass density
profiles).
The fitting formula given by \citet{Gao+2008} yields 
$\alpha_{\rm E}\sim 0.29$ for our high-mass clusters at 
$\llangle z_{\rm l}\rrangle \simeq 0.35$, which is consistent with our results
within $1.3\sigma$.
Our results are also consistent with the recent stacked weak-shear
analysis of 50 X-ray luminous clusters
by \citet{Okabe+2013}, who obtained $\alpha_{\rm E}=0.188^{+0.062}_{-0.058}$
for their sample with 
$M_{200{\rm c}} =(8.5\pm 0.6)\times 10^{14}M_\odot\,h_{70}^{-1}$.

%

Misidentification of cluster centers is a potential source
of systematic errors for stacked weak-lensing measurements on small
scales. 
\citet{George+2012COSMOS}
examined the impact of the
choice for the cluster center on the stacked weak-shear signal 
based on 129  X-ray-selected  galaxy groups at $0<z<1$ detected in the COSMOS field.
They show that the brightest or most massive galaxies near the X-ray centroids 
appear to best trace the center of mass of halos.
\citet{Zitrin+2012miscentering} analyzed the strong-lensing signature of 10,000 clusters from the
Gaussian Mixture Brightest Cluster Galaxy
\citep[GMBCG;][]{Hao+2010GMBCG} catalog,
finding a small mean offset of $\simeq 13$\,kpc\,$h^{-1}$ 
between the BCG and the smoothed optical light that is 
assumed to trace the DM in their analysis.

\citet{Johnston+2007b} demonstrated that the smoothing 
effects of miscentering on $\Delta\Sigma_+$
are much larger than on $\Sigma$, and 
produce a noticeable effect on $\Delta\Sigma_+$
out to 10 times the typical positional offset from the cluster mass
centroid
\citep{Johnston+2007b}. This is not surprising because 
$\Delta\Sigma_+$
is insensitive to flat sheet-like structures.
Here our CLASH X-ray-selection criteria ensure well-defined cluster centers, 
reducing the smoothing effects of cluster miscentering.
Assuming that the BCG--X-ray positional offset is a good proxy for the
offset from the mass centroid, the smoothing effects on
$\Delta\Sigma_+$
vanish at $R\simgt 10\sigma_{\rm  off}\sim 110$\,kpc\,$h^{-1}$ 
(Section \ref{subsec:center}), 
which is sufficiently smaller than the innermost measurement radius, 
$R_{\rm min}=200$\,kpc\,$h^{-1}$, for our stacked shear analysis.

\subsubsection{Total Mass Profile (1h+2h term)}
\label{subsubsec:1h+2h}


\begin{figure}[!htb] 
 \begin{center}
 \includegraphics[width=0.4\textwidth,angle=0,clip]{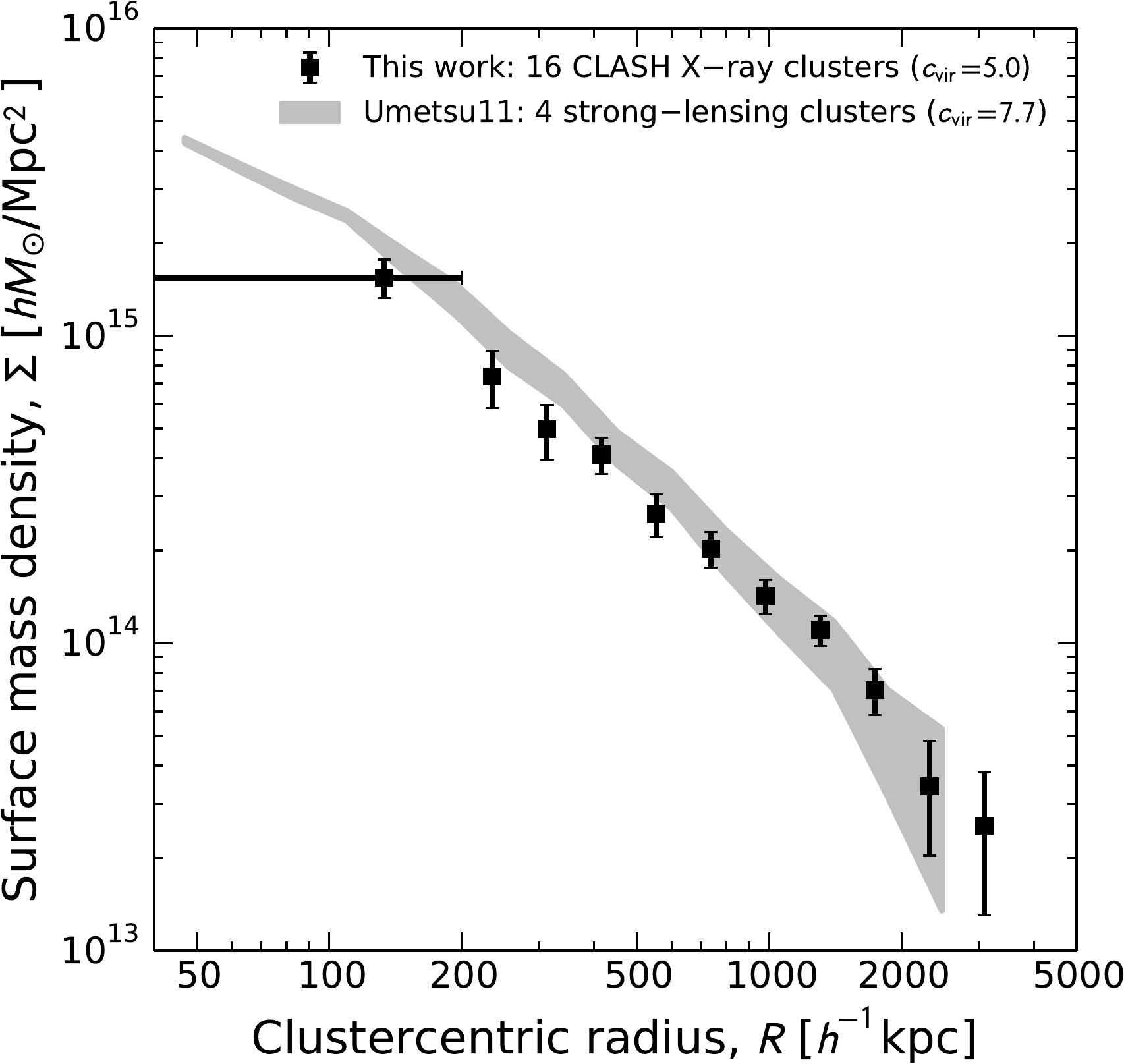}
 \end{center} 
\caption{
Comparison of stacked projected total mass profiles between different
 cluster samples.
The squares with error bars show the results 
(68\% CL, Figure \ref{fig:stackwl}) 
from our stacking analysis of 16 CLASH X-ray-selected clusters at
 $\llangle z_{\rm l}\rrangle\simeq 0.35$ 
based on the weak-lensing shear+magnification measurements.
The gray-shaded area  represents the averaged total mass profile
 (68\% CL)
of four similar-mass, strong-lensing-selected clusters at 
$\llangle z_{\rm l}\rrangle\simeq 0.32$ 
 obtained from a combined strong-lensing, weak-lensing shear and
 magnification analysis of \citet{Umetsu+2011stack}.
The mean concentration of the strong-lensing sample is 
$c_{\rm vir}= 7.7\pm  0.4$, compared to 
$c_{\rm vir}=5.0\pm 0.4$ for the CLASH X-ray-selected subsample.
\label{fig:kcomp}
}
\end{figure} 

We compare in Figure \ref{fig:kcomp} the ensemble-averaged projected
mass density profile $\llangle \bSigma\rrangle$ of our X-ray-selected
subsample with the results of \citet{Umetsu+2011stack} who analyzed 
combined strong-lensing, weak-lensing shear and magnification measurements 
of four strong-lensing-selected clusters
(Section \ref{subsubsec:cm_comparison}),
characterized by 
$c_{\rm vir}=7.68^{+0.42}_{-0.40}$
and
$M_{\rm vir}=2.20^{+0.16}_{-0.14}\times 10^{15} M_\odot\,h_{70}^{-1}$
at $\llangle z_{\rm l}\rrangle\simeq 0.32$.
This is translated into the halo bias factor,
$b_{\rm h}\simeq 10.9$,
which is only $\sim 20\%$ higher than that estimated
for our X-ray-selected subsample, $b_{\rm h}\simeq 9.0$ 
(Section \ref{subsec:clash_stackwl}).
In fact, the two cluster samples have similar
``peak heights'' in the linear (primordial) density field 
\citep{Tinker+2010}: 
$\simeq 3.8\sigma$ for our X-ray-selected subsample and 
$\simeq 4.1\sigma$ for the \citet{Umetsu+2011stack} sample.

Figure \ref{fig:kcomp} shows that the two samples have very
similar outer mass profiles at 
$R\simgt 1\,$Mpc\,$h^{-1} \sim 0.5r_{\rm vir}$, 
which are sensitive to the underlying mass
accretion rate or halo peak height \citep{Diemer+Kravtsov2014}.
At $R\simlt 400$\,kpc\,$h^{-1} (\sim r_{\rm s})$, on the other hand, 
we start to see systematic deviations between the two profiles,
reflecting different degrees of projected mass concentration.

As shown by high-resolution cluster simulations of \citet{Gao+2012Phoenix},
the asphericity of clusters can lead to large variations of up to a
factor of three in the projected density $\Sigma(R)$ at a given radius $R$,
depending on projection, especially within $R\sim 500$\,kpc\,$h^{-1}$
(see their Figure 9).
Such projection effects due to cluster asphericity could explain 
the high apparent concentration and
high central surface mass density of these four strong-lensing
clusters \citep[see also][]{Oguri+2012SGAS}.

\subsection{Systematic Errors}
\label{subsec:sys}

As described in Sections \ref{sec:method} and \ref{subsec:sysmass}, 
we have taken into account several major sources of uncertainties in our
error analysis. 
In this subsection we address 
other potential sources of systematic errors
and discuss their possible effects on our results.
In summary, we conclude that they are subdominant to the other sources
we have already addressed.

\subsubsection{Number Count Slopes}
\label{subsubsec:slope}

In the presence of magnification, one probes the number
counts at an effectively fainter limiting magnitude of
$m_{\rm lim}+2.5\log_{10}\mu$.   
The level of magnification is on average small in the weak
regime but for our innermost bin $\theta=[0.9\arcmin,1.2\arcmin]$ reaches a
median factor of $\mu\sim 1.7$,
corresponding to a magnitude shift of $\delta m\sim 0.6$.
Hence, we have implicitly assumed that the power-law behavior
in Equation (\ref{eq:magbias}) persists down to $\sim 0.6$\,mag fainter than
$m_{\rm lim}$ where the count slope may be shallower.  
For a given level
of count depletion, an overestimation of the count slope could lead to
an overestimation of the magnification, thus biasing the resulting mass
profile.  
However, the count slope $s_{\rm eff}$ for our data flattens 
slowly with depth varying from $s_{\rm eff}\sim 0.15$ to 
$s_{\rm eff}\sim 0.1$ from a typical magnitude limit of 
$m_{\rm lim}=25.4$ to $m_{\rm lim}+\delta m$ 
\citep[see also][]{Umetsu+2011}, 
so that
this introduces a small correction of only $7\%$ for the innermost bin,
much smaller compared to our noisy depletion measurements with a 
$\sim 54\%$ median uncertainty,
corresponding to $54\%/\sqrt{N_{\rm cl}}\sim 14\%$ when all clusters
are combined.
Therefore, we conclude that the effect of this correction is subdominant
with respect to the total uncertainty.

\subsubsection{Background-level Determination}
\label{subsubsec:backdens}


The background density parameter $\overline{n}_\mu$ for the
count-depletion analysis has been estimated from the red galaxy counts
in the outer region $\theta=[10\arcmin,\theta_{\rm max}]$ (Table
\ref{tab:wlsn}).
We find from the stacked mass profile  that the mean 
convergence at $\theta\ge 10\arcmin$,
where we have estimated $\overline{n}_\mu$,
is $\kappa = (8 \pm 4)\times 10^{-3}$ at $\llangle z_{\rm l}\rrangle \simeq 0.35$.
This corresponds to a depletion signal of
$\delta n_\mu/\overline{n}_\mu \approx (5\langle s_{\rm eff}\rangle-2)\kappa \sim -0.01$ 
with the mean count slope $\langle s_{\rm eff}\rangle\sim 0.15$ of
our sample, indicating that the estimated 
$\overline{n}_\mu$ is biased low by $1\%$.
This level of the signal offset, however, is smaller than the
calibration uncertainties of 
$\sigma(\overline{n}_\mu)/\overline{n}_\mu = (2-8)\%$ for an individual
cluster. 
Hence, for all clusters in our sample,
the offset signal lies within the prior range considered,
$[\overline{n}_\mu-\sigma(\overline{n}_\mu),\overline{n}_\mu+\sigma(\overline{n}_\mu)]$
 (Section \ref{subsubsec:likelihood}).
In fact, we find that the ML (best-fit) estimates of $\overline{n}_\mu$, 
as constrained by the combined shear+magnification data,
are on average $(1.0 \pm 0.6)\%$ larger than the
values estimated from the counts at $[10\arcmin,\theta_{\rm max}]$,
so that the lensing signal is consistently recovered.

This analysis demonstrates
that the background level determination is not critically sensitive to
our calibration prior on the background density parameter
$\overline{n}_\mu$, 
but driven by the combined shear+magnification data.\footnote{For
example,
the observable distortion in the nonlinear regime  is not invariant with adding a
mass-sheet component, so that the lensing constraints in the nonlinear
regime can help break the parameter degeneracy.}

\section{Summary and Conclusions}
\label{sec:summary}
 
We have presented a joint shear-and-magnification weak-lensing analysis of a sample of 
16 X-ray-regular and 4 high-magnification galaxy clusters 
at $0.19\simlt z\simlt 0.69$ 
targeted in the CLASH survey  \citep{Postman+2012CLASH}.
Our analysis uses
deep wide-field multi-color imaging
obtained primarily with Subaru/Suprime-Cam.

From a stacked shear-only analysis of the X-ray-selected subsample,
we have detected the ensemble-averaged
 lensing signal $\llangle\bDSigma_+\rrangle$
with a total S/N of $\simeq 25$  
in the radial range of $200$ to $3500$\,kpc\,$h^{-1}$
(Figure \ref{fig:stackgt}), 
providing integrated constraints on the halo profile
 shape and $c$--$M$ relation.
The shape of the stacked $\llangle \bDSigma_+\rrangle$ profile
exhibits a steepening radial
trend across the radial range,
which is well described by a family of
standard density profiles predicted for DM-dominated halos in
 gravitational equilibrium (Table \ref{tab:mc200}).
The best-fit Einasto shape
 parameter is $\alpha_{\rm E}=0.191^{+0.071}_{-0.068}$,  
which is consistent with
the NFW-equivalent Einasto parameter of $\sim 0.18$.


For the NFW model, we constrain the 
mean concentration of our X-ray-selected subsample
to lie in the range $c_{\rm 200c}=4.01^{+0.35}_{-0.32}$ 
at 
$M_{\rm 200c}=1.34^{+0.10}_{-0.09}\times 10^{15}M_\odot\,h_{70}^{-1}$
(Table \ref{tab:mc200}),
corresponding to the Einstein radius of $\theta_{\rm Ein}=15\arcsec\pm
4\arcsec$ ($z_s=2$).
Accounting for the CLASH selection function based on X-ray morphology 
and projection effects inherent in lensing observations (M14), we find an
excellent agreement between observations and theoretical
predictions  (Table \ref{tab:lcdm}). 
Our stacked constraints on the $c$--$M$ relation are 
slightly higher than but in agreement with
the results from the {\sc SaWLenS} analysis of 19
CLASH X-ray-selected clusters \citep{Merten2014clash}.
demonstrating consistency
between the results obtained with different lensing methods
 (Section \ref{subsubsec:sawlens}). 


We have reconstructed model-free projected mass profiles $\bSigma$ of 
all CLASH clusters (Figure \ref{fig:mdata}) from a joint likelihood
analysis of consistent shear-and-magnification measurements (Figure
\ref{fig:wldata}).
The cluster masses were estimated at several
characteristic radii by fitting the observed $\bSigma$ profiles
with a spherical NFW density profile
 (see Table \ref{tab:mass}). 
The results are subject to a systematic uncertainty of $\pm 8\%$
in the overall mass calibration (Section \ref{subsubsec:msys}). 

We have also derived an ensemble-averaged total projected mass profile
 $\llangle\bSigma\rrangle$ of our X-ray-selected subsample by
 stacking their individual mass profiles $\bSigma$ 
(Figure \ref{fig:stackwl}).
The averaged total mass profile is shown 
 to be consistent with our shear-based halo-model predictions for the total matter
 distribution $\Sigma_{\rm 1h}+\Sigma_{\rm 2h}$, 
including the effects of surrounding LSS as a two-halo
 term $\Sigma_{\rm 2h}$, thus establishing further consistency in the context
 of the standard $\Lambda$CDM model.  


An accurate determination of the cluster density profile over the full
radial range,
from a combination of strong- and weak-lensing information, 
 is crucial for testing DM and alternative gravity
paradigms.
The CLASH survey \citep{Postman+2012CLASH}
is designed to generate such multi-scale, multi-probe
lensing data using high-resolution 16-band {\it HST} imaging and wide-field ground-based
observations for a sizable sample of clusters \citep{Merten2014clash}.
A stacked cluster analysis, combining all lensing-related effects in the cluster
regime \citep{Umetsu2013}, 
will be a crucial next step toward a definitive determination of the
ensemble-averaged cluster mass profile from the inner core to beyond the virial radius,
providing a firm basis for a detailed comparison
with the $\Lambda$CDM paradigm and a wider examination of
alternative scenarios \citep{BEC2009,Narikawa+2013,Silva+2013}.


\acknowledgments
We thank the anonymous referee for the careful reading of
the manuscript and constructive suggestions.
We acknowledge fruitful discussions with
Nobuhiro Okabe,
Masamune Oguri, 
Mauro Sereno,
Jean Coupon,
and
Hitoshi Hanami.
We thank Ole Host for providing very helpful comments on the manuscript.
We are grateful to all members of the CLASH team who enabled us to carry
out the work.
We acknowledge the Subaru Support
Astronomers, plus
Justice  Bruursema,
Kai-Yang Lin,
and
Hiroaki Nishioka,
for assistance with our Subaru observations.
We thank Nick Kaiser and Masamune Oguri for making their 
{\sc imcat} and
{\sc glafic} packages publicly available.
This work is partially supported by the National Science Council of Taiwan
under the grant NSC100-2112-M-001-008-MY3
and by the Academia Sinica Career Development Award.
%
J.M. acknowledges  support from 
the Jet Propulsion Laboratory,
California Institute of Technology, under a contract with the National
Aeronautics and Space Administration.  
D.G. and S.S. were supported by SFB-Transregio 33 'The Dark Universe' by
the Deutsche Forschungsgemeinschaft (DFG) and the DFG cluster of
excellence 'Origin and Structure of the Universe'. 
Support for A.Z. was provided by NASA through Hubble Fellowship grant
\#HST-HF-51334.01-A awarded by STScI. 
The Dark Cosmology Centre is funded by the DNRF.

\input{ms.bbl}
\end{document}

%% file: table1.tex
\begin{deluxetable*}{lrrrrrcr}
\tablecolumns{8}
\tablecaption{
\label{tab:sample}
Cluster sample
}
\tablewidth{0pt}
\tablehead{
 \multicolumn{1}{c}{Cluster} &
 \multicolumn{1}{c}{$z_{\rm l}$} &
 \multicolumn{1}{c}{R.A. \tablenotemark{a}} &
 \multicolumn{1}{c}{Decl.\tablenotemark{a}} &
 \multicolumn{1}{c}{$k_{\rm B}T$\tablenotemark{b}} &
 \multicolumn{1}{c}{Offset\tablenotemark{c}}
 & \multicolumn{1}{c}{Filters\tablenotemark{d}}
 & \multicolumn{1}{c}{WL band\tablenotemark{e}}
 \\ 
\colhead{} &
\colhead{} &
\multicolumn{1}{c}{(J2000.0)} &
\multicolumn{1}{c}{(J2000.0)} &
\multicolumn{1}{c}{(keV)} &
\multicolumn{1}{c}{(kpc$/h$)} &
\colhead{} &
\colhead{(seeing [$\arcsec$])} 
}
\startdata
X-ray Selected:\\         ~~Abell 383 & $0.187$ & 02:48:03.40 & -03:31:44.9 & $6.5\pm0.24$ & $ 1.4$ & $ {\underline B_{\rm J}}V_{\rm J}{\underline R_{\rm C}}I_{\rm C}i'{\underline z'} $ &               $ i' $ ($ 0.57$)\\
         ~~Abell 209 & $0.206$ & 01:31:52.54 & -13:36:40.4 & $7.3\pm0.54$ & $ 3.9$ & $ {\underline B_{\rm J}}V_{\rm J}{\underline R_{\rm C}}i'{\underline z'} $ &        $ R_{\rm C} $ ($ 0.61$)\\
        ~~Abell 2261 & $0.224$ & 17:22:27.18 & +32:07:57.3 & $7.6\pm0.30$ & $ 4.0$ & $ {\underline B_{\rm J}} {\underline V_{\rm J}}{\underline R_{\rm C}}i^+z^+ $ &        $ R_{\rm C} $ ($ 0.56$)\\
    ~~RXJ2129.7+0005 & $0.234$ & 21:29:39.96 & +00:05:21.2 & $5.8\pm0.40$ & $ 6.3$ & $ {\underline B_{\rm J}}V_{\rm J}{\underline R_{\rm C}}i'{\underline z'} $ &        $ R_{\rm C} $ ($ 0.53$)\\
         ~~Abell 611 & $0.288$ & 08:00:56.82 & +36:03:23.6 & $7.9\pm0.35$ & $ 1.6$ & $ {\underline B_{\rm J}}V_{\rm J}{\underline R_{\rm C}}I_{\rm C}i'{\underline z'} $ &        $ R_{\rm C} $ ($ 0.65$)\\
       ~~MS2137-2353 & $0.313$ & 21:40:15.17 & -23:39:40.2 & $5.9\pm0.30$ & $ 1.7$ & $ {\underline B_{\rm J}}V_{\rm J}{\underline R_{\rm C}}I_{\rm C}{\underline z'} $ &        $ R_{\rm C} $ ($ 0.60$)\\
    ~~RXJ2248.7-4431 & $0.348$ & 22:48:43.96 & -44:31:51.3 & $12.4\pm0.60$ & $15.9$ & $ U_{877}{\underline B_{842}}V_{843}{\underline R_{844}}I_{879}z_{846}i^*{\underline z^*} $ &          $ R_{844} $ ($ 0.81$)\\
  ~~MACSJ1115.9+0129 & $0.352$ & 11:15:51.90 & +01:29:55.1 & $8.0\pm0.40$ & $ 9.5$ & $ {\underline B_{\rm J}}V_{\rm J}{\underline R_{\rm C}}I_{\rm C}{\underline z'} $ &        $ R_{\rm C} $ ($ 0.67$)\\
  ~~MACSJ1931.8-2635 & $0.352$ & 19:31:49.62 & -26:34:32.9 & $6.7\pm0.40$ & $ 4.3$ & $ {\underline B_{\rm J}}V_{\rm J}{\underline R_{\rm C}}I_{\rm C}{\underline z'} $ &        $ R_{\rm C} $ ($ 0.72$)\\
    ~~RXJ1532.9+3021 & $0.363$ & 15:32:53.78 & +30:20:59.4 & $5.5\pm0.40$ & $ 2.5$ & $ {\underline B_{\rm J}}V_{\rm J}{\underline R_{\rm C}}I_{\rm C}{\underline z'} $ &        $ R_{\rm C} $ ($ 0.57$)\\
  ~~MACSJ1720.3+3536 & $0.391$ & 17:20:16.78 & +35:36:26.5 & $6.6\pm0.40$ & $13.3$ & $ {\underline B_{\rm J}}V_{\rm J}{\underline R_{\rm C}}I_{\rm C}{\underline z'} $ &        $ R_{\rm C} $ ($ 0.79$)\\
  ~~MACSJ0429.6-0253 & $0.399$ & 04:29:36.05 & -02:53:06.1 & $6.0\pm0.44$ & $13.6$ & $ {\underline V_{\rm J}}{\underline R_{\rm C}}{\underline I_{\rm C}} $ &        $ R_{\rm C} $ ($ 0.75$)\\
  ~~MACSJ1206.2-0847 & $0.440$ & 12:06:12.15 & -08:48:03.4 & $10.8\pm0.60$ & $ 8.7$ & $ {\underline B_{\rm J}}V_{\rm J}{\underline R_{\rm C}}I_{\rm C}{\underline z'} $ &        $ I_{\rm C} $ ($ 0.71$)\\
  ~~MACSJ0329.7-0211 & $0.450$ & 03:29:41.56 & -02:11:46.1 & $8.0\pm0.50$ & $ 9.8$ & $ {\underline B_{\rm J}}V_{\rm J}{\underline R_{\rm C}}I_{\rm C}{\underline z'} $ &        $ R_{\rm C} $ ($ 0.47$)\\
    ~~RXJ1347.5-1145 & $0.451$ & 13:47:31.05 & -11:45:12.6 & $15.5\pm0.60$ & $29.4$ & $ g^\star{\underline B_{\rm J}}V_{\rm J}{\underline R_{\rm C}}I_{\rm C}{\underline z'} $ &        $ R_{\rm C} $ ($ 0.71$)\\
  ~~MACSJ0744.9+3927 & $0.686$ & 07:44:52.82 & +39:27:26.9 & $8.9\pm0.80$ & $12.6$ & $ {\underline B_{\rm J}}V_{\rm J}{\underline R_{\rm C}}I_{\rm C}i'{\underline z'} $ &        $ R_{\rm C} $ ($ 0.71$)\\
\hline High Magnification:\\  ~~MACSJ0416.1-2403 & $0.396$ & 04:16:08.38 & -24:04:20.8 & $7.5\pm0.80$ & $82.3$ & $ {\underline B_{\rm J}}{\underline R_{\rm C}}{\underline z'} $ &        $ R_{\rm C} $ ($ 0.55$)\\
  ~~MACSJ1149.5+2223 & $0.544$ & 11:49:35.69 & +22:23:54.6 & $8.7\pm0.90$ & $10.8$ & $ {\underline B_{\rm J}}V_{\rm J}{\underline R_{\rm C}}{\underline z'}K_{\rm S} $ &        $ R_{\rm C} $ ($ 0.80$)\\
  ~~MACSJ0717.5+3745 & $0.548$ & 07:17:32.63 & +37:44:59.7 & $12.5\pm0.70$ & $100.2$ & $ u^\star{\underline B_{\rm J}}V_{\rm J}{\underline R_{\rm C}}i'{\underline z'}JK_{\rm S} $ &        $ R_{\rm C} $ ($ 0.79$)\\
  ~~MACSJ0647.7+7015 & $0.584$ & 06:47:50.27 & +70:14:55.0 & $13.3\pm1.80$ & $25.4$ & $ {\underline B_{\rm J}}V_{\rm J}{\underline R_{\rm C}}I_{\rm C}i'{\underline z'} $ &        $ R_{\rm C} $ ($ 0.64$)
\enddata
\tablenotetext{a}{Optical cluster center.}
\tablenotetext{b}{X-ray temperature from \citet{Postman+2012CLASH}.}
\tablenotetext{c}{Projected offset between the X-ray and optical centers.}
\tablenotetext{d}{Multi-band filters used in our photometric, photo-$z$ and weak-lensing analysis. Bands used for CC selection are underlined. The majority of the data are from Subaru/Suprime-Cam. Bands from complementary facilities are described in Table \ref{tab:filters}.}\tablenotetext{e}{Band used for weak-lensing shape measurements and seeing FWHM in the full stack of images.}
\end{deluxetable*}

%% file: table2.tex
\begin{deluxetable}{lcl}
\tablecolumns{3}
\tablecaption{
\label{tab:filters}
Multi-band Filter Description
}
\tablewidth{0pt}
\tablehead{
 \multicolumn{1}{l}{Telescope/Instrument} &
 \multicolumn{1}{c}{Filter Name} &
 \multicolumn{1}{c}{Filter Description} 
}
\startdata
Subaru/Suprime-Cam 
 & $B_{\rm J}$ & Johnson $B$-band \\
 & $V_{\rm J}$ & Johnson $V$-band \\
 & $R_{\rm C}$ & Cousins $R$-band \\
 & $I_{\rm C}$ & Cousins $I$-band \\
 & $i$         &    SDSS $i$-band \\
 & $z'$        &    SDSS $z$-band \\
 \hline \\[-5pt]
 
KPNO Mayall/MOSAIC-I
& $i^+$        &    SDSS $i$-band \\
& $z^+$        &    SDSS $z$-band \\
\hline \\[-5pt]

ESO/WFI
 & $U_{877}$ & $U$/50-band \\
 & $B_{842}$ & Johnson $B$-band \\
 & $V_{843}$ & Johnson $V$-band \\
 & $R_{844}$ & Cousins $R$-band \\
 & $I_{879}$ & Cousins $I$-band \\
 & $z_{846}$ &  $z{+}$/61-band \\
 \hline \\[-5pt]
 
Magellan/IMACS 
& $i^*$        &    SDSS $i$-band \\
& $z^*$        &    SDSS $z$-band \\
 \hline \\[-5pt]
 
CFHT/MegaPrime
& $u^\star$    &    SDSS $u$-band\\
& $g^\star$    &    SDSS $g$-band\\
 \hline \\[-5pt]
 
CFHT/WIRCam
& $J$            &    NIR $J$-band\\
& $K_{\rm s}$    &    NIR $K_{\rm S}$-band
\enddata
\end{deluxetable}

%% file: table3.tex
\begin{deluxetable}{lrrrrr}
\tablecolumns{6}
\tablecaption{
\label{tab:wlsample}
Background galaxy samples for weak-lensing shape measurements
}
\tablewidth{0pt}
\tablehead{
 \multicolumn{1}{c}{Cluster} &
 \multicolumn{1}{c}{$N_g$} &
 \multicolumn{1}{c}{$\overline{n}_g$\tablenotemark{a}} &
 \multicolumn{1}{c}{$\overline{z}_{\rm eff}$\tablenotemark{b}} &
 \multicolumn{1}{c}{$\langle D_{\rm ls}/D_{\rm s}\rangle$} &
 \multicolumn{1}{c}{$f_{W}$}
}
\startdata
X-ray Selected:\\         ~~Abell 383 &   7062 & $ 9.3$ & $1.16$ & $0.79 \pm 0.04$ & $1.01$\\
         ~~Abell 209 &  14694 & $15.8$ & $0.94$ & $0.74 \pm 0.04$ & $1.04$\\
        ~~Abell 2261 &  15429 & $18.1$ & $0.88$ & $0.70 \pm 0.04$ & $1.05$\\
    ~~RXJ2129.7+0005 &  20104 & $21.1$ & $1.16$ & $0.75 \pm 0.04$ & $1.02$\\
         ~~Abell 611 &   7872 & $ 8.5$ & $1.13$ & $0.68 \pm 0.03$ & $1.07$\\
       ~~MS2137-2353 &   9864 & $10.3$ & $1.23$ & $0.68 \pm 0.03$ & $1.02$\\
    ~~RXJ2248.7-4431 &   4008 & $ 4.6$ & $1.05$ & $0.60 \pm 0.03$ & $1.15$\\
  ~~MACSJ1115.9+0129 &  13621 & $12.7$ & $1.15$ & $0.63 \pm 0.03$ & $1.03$\\
  ~~MACSJ1931.8-2635 &   4343 & $ 4.9$ & $0.93$ & $0.56 \pm 0.03$ & $1.06$\\
    ~~RXJ1532.9+3021 &  13270 & $16.2$ & $1.15$ & $0.61 \pm 0.03$ & $1.05$\\
  ~~MACSJ1720.3+3536 &   9855 & $12.0$ & $1.12$ & $0.58 \pm 0.03$ & $1.04$\\
  ~~MACSJ0429.6-0253 &   9990 & $11.7$ & $1.30$ & $0.62 \pm 0.06$ & $1.04$\\
  ~~MACSJ1206.2-0847 &  12719 & $13.3$ & $1.13$ & $0.54 \pm 0.03$ & $1.06$\\
  ~~MACSJ0329.7-0211 &  25427 & $29.3$ & $1.18$ & $0.54 \pm 0.03$ & $1.06$\\
    ~~RXJ1347.5-1145 &   9393 & $ 7.9$ & $1.17$ & $0.54 \pm 0.03$ & $1.06$\\
  ~~MACSJ0744.9+3927 &   7561 & $ 8.0$ & $1.41$ & $0.42 \pm 0.02$ & $1.15$\\
\hline High Magnification:\\  ~~MACSJ0416.1-2403 &  21241 & $24.9$ & $1.24$ & $0.61 \pm 0.03$ & $1.01$\\
  ~~MACSJ1149.5+2223 &  14016 & $14.2$ & $1.04$ & $0.41 \pm 0.02$ & $1.24$\\
  ~~MACSJ0717.5+3745 &   9724 & $11.1$ & $1.26$ & $0.48 \pm 0.02$ & $1.09$\\
  ~~MACSJ0647.7+7015 &   7339 & $10.2$ & $1.27$ & $0.45 \pm 0.02$ & $1.09$
\enddata
\tablenotetext{a}{Mean surface number density of background galaxies per arcmin$^2$.}
\tablenotetext{b}{Effective source redshift corresponding to the mean lensing depth $\langle\beta\rangle = \langle D_{\rm ls}/D_{\rm s} \rangle$ of the sample, defined as $\beta(\overline{z}_{\rm eff})=\langle\beta\rangle$.}
\end{deluxetable}

%% file: table4.tex
\begin{deluxetable*}{lrrrrrr}
\tablecolumns{7}
\tablecaption{
\label{tab:magsample}
Background galaxy samples for magnification bias measurements
}
\tablewidth{0pt}
\tablehead{
 \multicolumn{1}{c}{Cluster} &
 \multicolumn{1}{c}{$m_{\rm lim}$\tablenotemark{a}} &
 \multicolumn{1}{c}{$N_\mu$} &
 \multicolumn{1}{c}{$\overline{n}_{\mu}$\tablenotemark{b}} &
 \multicolumn{1}{c}{$s_{\rm eff}$\tablenotemark{c}} &
 \multicolumn{1}{c}{$\overline{z}_{\rm eff}$\tablenotemark{d}} &
 \multicolumn{1}{c}{$\langle D_{\rm ls}/D_{\rm s}\rangle$} 
 \\ 
\colhead{} &
\multicolumn{1}{c}{(AB mag)} &
\multicolumn{1}{c}{} &
\multicolumn{1}{c}{(arcmin$^{-2}$)} &
\multicolumn{1}{c}{} &
\multicolumn{1}{c}{} &
\multicolumn{1}{c}{} 
}
\startdata
X-ray Selected:\\         ~~Abell 383 & $25.2$ &  13763 & $13.3 \pm  0.4$ & $0.14 \pm 0.05$ & $1.23$ &  $0.80 \pm 0.04$\\
         ~~Abell 209 & $25.1$ &  12860 & $13.1 \pm  0.4$ & $0.14 \pm 0.05$ & $1.03$ &  $0.75 \pm 0.04$\\
        ~~Abell 2261 & $25.6$ &  17610 & $20.0 \pm  0.4$ & $0.14 \pm 0.04$ & $1.26$ &  $0.77 \pm 0.04$\\
    ~~RXJ2129.7+0005 & $25.6$ &  13467 & $13.9 \pm  0.5$ & $0.19 \pm 0.05$ & $1.05$ &  $0.73 \pm 0.04$\\
         ~~Abell 611 & $25.6$ &   7982 & $ 9.7 \pm  0.5$ & $0.20 \pm 0.06$ & $1.16$ &  $0.69 \pm 0.03$\\
       ~~MS2137-2353 & $25.6$ &  18095 & $17.2 \pm  1.5$ & $0.13 \pm 0.04$ & $1.05$ &  $0.64 \pm 0.03$\\
    ~~RXJ2248.7-4431 & $24.1$ &   2685 & $ 4.4 \pm  0.3$ & $0.11 \pm 0.11$ & $1.18$ &  $0.64 \pm 0.03$\\
  ~~MACSJ1115.9+0129 & $24.9$ &  13109 & $14.0 \pm  0.5$ & $0.14 \pm 0.05$ & $0.98$ &  $0.58 \pm 0.03$\\
  ~~MACSJ1931.8-2635 & $24.1$ &   5556 & $ 6.1 \pm  0.4$ & $0.20 \pm 0.07$ & $0.90$ &  $0.55 \pm 0.03$\\
    ~~RXJ1532.9+3021 & $25.4$ &  18653 & $14.3 \pm  0.5$ & $0.17 \pm 0.04$ & $0.99$ &  $0.57 \pm 0.03$\\
  ~~MACSJ1720.3+3536 & $25.2$ &  17804 & $16.3 \pm  0.6$ & $0.16 \pm 0.04$ & $1.05$ &  $0.56 \pm 0.03$\\
  ~~MACSJ0429.6-0253 & $25.4$ &  13521 & $12.3 \pm  0.4$ & $0.19 \pm 0.05$ & $1.05$ &  $0.55 \pm 0.06$\\
  ~~MACSJ1206.2-0847 & $24.6$ &  13252 & $11.4 \pm  0.4$ & $0.13 \pm 0.05$ & $1.04$ &  $0.51 \pm 0.03$\\
  ~~MACSJ0329.7-0211 & $25.4$ &  21192 & $22.1 \pm  0.5$ & $0.13 \pm 0.04$ & $1.01$ &  $0.49 \pm 0.02$\\
    ~~RXJ1347.5-1145 & $25.6$ &  15017 & $14.2 \pm  0.5$ & $0.14 \pm 0.04$ & $1.04$ &  $0.50 \pm 0.02$\\
  ~~MACSJ0744.9+3927 & $25.6$ &  17165 & $15.5 \pm  0.5$ & $0.15 \pm 0.04$ & $1.23$ &  $0.36 \pm 0.02$\\
\hline High Magnification:\\  ~~MACSJ0416.1-2403 & $25.6$ &  27068 & $27.2 \pm  0.6$ & $0.17 \pm 0.03$ & $1.07$ &  $0.56 \pm 0.03$\\
  ~~MACSJ1149.5+2223 & $25.6$ &  19271 & $18.3 \pm  0.6$ & $0.15 \pm 0.04$ & $1.06$ &  $0.41 \pm 0.02$\\
  ~~MACSJ0717.5+3745 & $25.6$ &  11641 & $11.9 \pm  0.4$ & $0.12 \pm 0.05$ & $1.15$ &  $0.45 \pm 0.02$\\
  ~~MACSJ0647.7+7015 & $25.6$ &  15043 & $14.3 \pm  0.5$ & $0.13 \pm 0.04$ & $1.17$ &  $0.42 \pm 0.02$
\enddata
\tablenotetext{a}{Fainter magnitude cut of the background sample. Apparent magnitude cuts are applied in the reddest CC-selection band available for each cluster to avoid incompleteness near the detection limit.}
\tablenotetext{b}{Coverage- and mask-corrected normalization of unlensed background source counts.}
\tablenotetext{c}{Slope of the unlensed background source counts $s_{\rm eff}=d\log_{10} \overline{N}_{\mu}(<m)/dm$.}
\tablenotetext{d}{Effective source redshift corresponding to the mean lensing depth of the sample, defined as $\beta(\overline{z}_{\rm eff})=\langle\beta\rangle$.}
\end{deluxetable*}

%% file: table5.tex
\begin{deluxetable}{lrrr}
\tablecolumns{4}
\tablecaption{
\label{tab:wlsn}
Cluster weak-lensing radial profiles
}
\tablewidth{0pt}
\tablehead{
 \multicolumn{1}{c}{Cluster} &
 \multicolumn{2}{c}{S/N} &
 \multicolumn{1}{c}{$\chi^2_{\rm min}$\tablenotemark{a}} 
 \\ 
\colhead{} &
\multicolumn{1}{c}{$g_+$} &
\multicolumn{1}{c}{$n_\mu$} &
\colhead{}
}
\startdata
X-ray Selected:\\         ~~Abell 383 & $  9.2$ & $  3.1$ & $ 5.8$\\
         ~~Abell 209 & $ 14.1$ & $  3.0$ & $ 5.7$\\
        ~~Abell 2261 & $ 16.4$ & $  8.1$ & $13.8$\\
    ~~RXJ2129.7+0005 & $ 10.3$ & $  2.5$ & $ 9.1$\\
         ~~Abell 611 & $  7.7$ & $  2.8$ & $ 2.2$\\
       ~~MS2137-2353 & $  8.5$ & $  4.0$ & $ 7.5$\\
    ~~RXJ2248.7-4431 & $  6.9$ & $  4.3$ & $ 5.1$\\
  ~~MACSJ1115.9+0129 & $  9.2$ & $  3.9$ & $ 4.5$\\
  ~~MACSJ1931.8-2635 & $  4.8$ & $  3.8$ & $ 8.0$\\
    ~~RXJ1532.9+3021 & $  6.6$ & $  5.0$ & $14.3$\\
  ~~MACSJ1720.3+3536 & $  7.7$ & $  4.4$ & $ 9.6$\\
  ~~MACSJ0429.6-0253 & $  7.7$ & $  4.0$ & $14.4$\\
  ~~MACSJ1206.2-0847 & $  9.5$ & $  4.2$ & $ 5.8$\\
  ~~MACSJ0329.7-0211 & $ 12.9$ & $  4.3$ & $ 9.6$\\
    ~~RXJ1347.5-1145 & $  9.7$ & $  5.8$ & $ 8.1$\\
  ~~MACSJ0744.9+3927 & $  8.7$ & $  3.6$ & $10.5$\\
\hline High Magnification:\\  ~~MACSJ0416.1-2403 & $ 10.7$ & $  3.8$ & $ 9.6$\\
  ~~MACSJ1149.5+2223 & $  9.6$ & $  2.4$ & $ 5.7$\\
  ~~MACSJ0717.5+3745 & $ 12.3$ & $  5.5$ & $ 6.1$\\
  ~~MACSJ0647.7+7015 & $  6.8$ & $  4.3$ & $11.1$
\enddata
\tablecomments{The lensing radial profiles are calculated in $N=10$ discrete radial bins over the radial range of $[\theta_{\rm min},\theta_{\rm max}]$, with a logarithmic radial spacing of $\Delta\ln\theta=\ln(\theta_{\rm max}/\theta_{\rm min})/N$. For all clusters, $\theta_{\rm min}=0.9\arcmin$ and $\theta_{\rm max}=16\arcmin$, except $\theta_{\rm max}=14\arcmin$ for RXJ2248.7-4431 observed with ESO/WFI.}
\tablenotetext{a}{Minimum $\chi^2$ obtained for the mass-profile solution.  For all clusters, the number of degrees of freedom is $2N-(N+1)=9$.}\end{deluxetable}

%% file: table6.tex
\begin{deluxetable*}{lrrrrrr}
\tablecolumns{7}
\tablecaption{
\label{tab:mass}
Cluster mass estimates based on joint weak-lensing shear+magnification measurements
}
\tablewidth{0pt}
\tablehead{
 \multicolumn{1}{c}{Cluster} &
 \multicolumn{1}{c}{$M_{500{\rm c}}$} &
 \multicolumn{1}{c}{$M_{200{\rm c}}$} &
 \multicolumn{1}{c}{$M_{\rm vir}$\tablenotemark{a}} &
 \multicolumn{1}{c}{$M_{100{\rm c}}$} &
 \multicolumn{1}{c}{$M_{200{\rm m}}$} &
 \multicolumn{1}{c}{$M(<1.5{\rm Mpc})$}
 \\ 
\colhead{} &
\multicolumn{1}{c}{($10^{14}M_{\odot}$)} &
\multicolumn{1}{c}{($10^{14}M_{\odot}$)} &
\multicolumn{1}{c}{($10^{14}M_{\odot}$)} &
\multicolumn{1}{c}{($10^{14}M_{\odot}$)} &
\multicolumn{1}{c}{($10^{14}M_{\odot}$)} &
\multicolumn{1}{c}{($10^{14}M_{\odot}$)}  
}
\startdata
X-ray Selected:\\         ~~Abell 383 & $ 6.1 \pm  1.5$ & $ 8.1 \pm  2.2$ & $ 9.4 \pm  2.8$ & $ 9.7 \pm  2.8$ & $10.3 \pm  3.1$ & $ 7.1 \pm  1.4$\\
         ~~Abell 209 & $11.6 \pm  1.8$ & $17.6 \pm  3.0$ & $21.9 \pm  4.0$ & $22.8 \pm  4.2$ & $24.7 \pm  4.7$ & $11.6 \pm  1.2$\\
        ~~Abell 2261 & $14.7 \pm  2.4$ & $21.3 \pm  4.1$ & $25.8 \pm  5.4$ & $26.8 \pm  5.8$ & $28.6 \pm  6.3$ & $13.7 \pm  1.5$\\
    ~~RXJ2129.7+0005 & $ 4.1 \pm  0.9$ & $ 5.5 \pm  1.4$ & $ 6.4 \pm  1.7$ & $ 6.6 \pm  1.8$ & $ 6.9 \pm  1.9$ & $ 5.3 \pm  1.0$\\
         ~~Abell 611 & $ 9.5 \pm  2.2$ & $14.1 \pm  3.9$ & $17.1 \pm  5.2$ & $18.1 \pm  5.7$ & $18.8 \pm  6.0$ & $10.3 \pm  1.7$\\
       ~~MS2137-2353 & $ 7.4 \pm  2.4$ & $12.4 \pm  4.8$ & $15.7 \pm  6.7$ & $17.0 \pm  7.5$ & $17.7 \pm  7.9$ & $ 9.0 \pm  2.0$\\
    ~~RXJ2248.7-4431 & $11.1 \pm  3.2$ & $20.3 \pm  6.7$ & $26.7 \pm  9.9$ & $29.5 \pm 11.5$ & $30.2 \pm 12.0$ & $12.0 \pm  2.0$\\
  ~~MACSJ1115.9+0129 & $ 9.3 \pm  2.0$ & $15.6 \pm  3.4$ & $19.6 \pm  4.5$ & $21.4 \pm  5.1$ & $21.8 \pm  5.2$ & $10.7 \pm  1.4$\\
  ~~MACSJ1931.8-2635 & $10.2 \pm  3.9$ & $14.8 \pm  6.4$ & $17.4 \pm  8.2$ & $18.5 \pm  9.0$ & $18.8 \pm  9.2$ & $11.0 \pm  2.9$\\
    ~~RXJ1532.9+3021 & $ 5.3 \pm  1.3$ & $ 7.1 \pm  1.9$ & $ 8.0 \pm  2.2$ & $ 8.4 \pm  2.4$ & $ 8.5 \pm  2.4$ & $ 6.6 \pm  1.3$\\
  ~~MACSJ1720.3+3536 & $10.1 \pm  2.0$ & $13.5 \pm  3.1$ & $15.3 \pm  3.7$ & $16.2 \pm  4.0$ & $16.2 \pm  4.0$ & $11.0 \pm  1.7$\\
  ~~MACSJ0429.6-0253 & $ 6.8 \pm  1.9$ & $ 9.4 \pm  3.0$ & $10.8 \pm  3.8$ & $11.5 \pm  4.2$ & $11.5 \pm  4.1$ & $ 8.3 \pm  1.8$\\
  ~~MACSJ1206.2-0847 & $10.6 \pm  2.1$ & $15.9 \pm  3.6$ & $18.7 \pm  4.6$ & $20.4 \pm  5.2$ & $20.1 \pm  5.1$ & $11.8 \pm  1.6$\\
  ~~MACSJ0329.7-0211 & $ 7.7 \pm  1.1$ & $10.0 \pm  1.5$ & $11.1 \pm  1.7$ & $11.7 \pm  1.8$ & $11.6 \pm  1.8$ & $ 9.0 \pm  1.0$\\
    ~~RXJ1347.5-1145 & $21.9 \pm  3.8$ & $29.5 \pm  6.1$ & $33.4 \pm  7.4$ & $35.6 \pm  8.2$ & $35.1 \pm  8.0$ & $19.7 \pm  2.3$\\
  ~~MACSJ0744.9+3927 & $11.2 \pm  2.9$ & $17.5 \pm  4.7$ & $20.2 \pm  5.8$ & $23.1 \pm  7.0$ & $21.0 \pm  6.1$ & $13.5 \pm  2.3$\\
\hline High Magnification:\\  ~~MACSJ0416.1-2403 & $ 7.0 \pm  1.3$ & $10.4 \pm  2.2$ & $12.4 \pm  2.8$ & $13.3 \pm  3.2$ & $13.3 \pm  3.1$ & $ 8.7 \pm  1.2$\\
  ~~MACSJ1149.5+2223 & $14.2 \pm  3.4$ & $25.4 \pm  5.2$ & $31.5 \pm  6.5$ & $36.4 \pm  7.7$ & $34.0 \pm  7.1$ & $15.1 \pm  2.1$\\
  ~~MACSJ0717.5+3745 & $20.9 \pm  2.9$ & $30.7 \pm  4.9$ & $35.3 \pm  6.0$ & $38.8 \pm  6.8$ & $37.1 \pm  6.4$ & $19.4 \pm  1.8$\\
  ~~MACSJ0647.7+7015 & $ 7.7 \pm  2.7$ & $13.2 \pm  4.2$ & $16.0 \pm  5.3$ & $18.5 \pm  6.4$ & $17.0 \pm  5.8$ & $10.5 \pm  2.1$
\enddata
\tablecomments{Cluster mass estimates $M_{\rm 3D}(<r)$ derived from single spherical NFW fits to individual projected mass density profiles (Figure \ref{fig:mdata}) reconstructed from combined shear+magnification measurements.  The fitting radial range is restricted to $R\le 2$\,Mpc\,$h^{-1}$. All quantities in the table are given in physical units assuming a concordance cosmology of $h=0.7$, $\Omega_{\rm m}=0.27$, and $\Omega_{\Lambda}=0.73$. The results are subject to a systematic uncertainty of $\pm 8\%$ in the overall mass calibration (Section \ref{subsubsec:msys}).}
\tablenotetext{a}{Virial overdensity $\Delta_{\rm vir}$ based on the spherical collapse model (see Appendix A of \citet{Kitayama+Suto1996}). For our redshift range $0.187\le z\le 0.686$, $\Delta_{\rm vir}$ ranges approximately from $\simeq 110$ to 140 with respect to the critical density of the universe at the cluster redshift.}\end{deluxetable*}

%% file: table7.tex
\begin{deluxetable*}{lccccc}
\tablecolumns{6}
\tablecaption{
\label{tab:mc200}
Best-fit models for the stacked distortion profile of the X-ray-selected subsample
}
\tablewidth{0pt}
\tablehead{
 \multicolumn{1}{c}{Model} &
 \multicolumn{1}{c}{$M_{200{\rm c}}$} &
 \multicolumn{1}{c}{$c_{200{\rm c}}$} &
 \multicolumn{1}{c}{Structural parameter} &
 \multicolumn{1}{c}{$\chi^2_{\rm min}/{\rm dof}$\tablenotemark{a}} &
 \multicolumn{1}{c}{PTE\tablenotemark{b}}
 \\ 
\multicolumn{1}{c}{} &
\multicolumn{1}{c}{($10^{14}M_{\odot}\,h_{70}^{-1}$)} &
\multicolumn{1}{c}{} &
\multicolumn{1}{c}{} &
\multicolumn{1}{c}{} &
\multicolumn{1}{c}{}
}
\startdata
                 SIS & $10.4^{+0.6}_{-0.6}$ & --- & --- & $25.4/10$ & $0.00$\\
  Isothermal $\beta$ & $14.2^{+1.2}_{-1.2}$ & --- & $r_{200{\rm c}}/r_c = 16.5^{+2.4}_{-1.6}$ & $10.3/9$ & $0.33$\\
                 NFW & $13.4^{+1.0}_{-0.9}$ & $4.01^{+0.35}_{-0.32}$ &                  --- & $6.8/9$ & $0.66$\\
              NFW+pm & $13.1^{+1.2}_{-1.1}$ & $3.66^{+0.68}_{-0.50}$ & $M_p= (  11\pm   33)\times 10^{12}M_\odot$ & $6.7/8$ & $0.57$\\
                 BMO & $13.1^{+0.9}_{-0.9}$ & $3.73^{+0.33}_{-0.31}$ &                  --- & $7.6/9$ & $0.58$\\
             Einasto & $13.2^{+1.0}_{-1.0}$ & $3.73^{+0.43}_{-0.52}$ & $\alpha_E= 0.191^{+0.071}_{-0.068}$ & $7.3/8$ & $0.51$
\enddata
\tablenotetext{a}{Minimum $\chi^2$ per degrees of freedom (dof).}
\tablenotetext{b}{Probability to exceed (PTE) the given $\chi^2_{\rm min}/{\rm dof}$ based on the standard $\chi^2$ probability distribution function.}
\end{deluxetable*}

%% file: table8.tex
\begin{deluxetable*}{lcccccc}
\tablecolumns{7}
\tablecaption{
\label{tab:lcdm}
Comparison with numerical simulations
}
\tablewidth{0pt}
\tablehead{
 \multicolumn{1}{c}{Model} &
 \multicolumn{2}{c}{Relaxed sample} &
 \multicolumn{2}{c}{Full sample} &
 \multicolumn{2}{c}{X-ray regular sample} 
 \\ 
\multicolumn{1}{c}{} &
\multicolumn{1}{c}{$\llangle c_{200{\rm c}}\rrangle$} &
\multicolumn{1}{c}{$c^{\rm (obs)}/c^{\rm (sim)}$} &
\multicolumn{1}{c}{$\llangle c_{200{\rm c}}\rrangle$} &
\multicolumn{1}{c}{$c^{\rm (obs)}/c^{\rm (sim)}$} &
\multicolumn{1}{c}{$\llangle c_{200{\rm c}}\rrangle$} &
\multicolumn{1}{c}{$c^{\rm (obs)}/c^{\rm (sim)}$}
}
\startdata
           Duffy et al. (2008) & $3.44\pm 0.32$ & $1.17\pm 0.15$ & $3.02\pm 0.28$ & $1.33\pm 0.17$ &        --- &        ---\\
    Bhattacharya et al. (2013) & $3.55\pm 0.33$ & $1.13\pm 0.14$ & $3.43\pm 0.32$ & $1.17\pm 0.15$ &        --- &        ---\\
         De Boni et al. (2013) & $3.12\pm 0.29$ & $1.29\pm 0.16$ & $2.75\pm 0.26$ & $1.46\pm 0.18$ &        --- &        ---\\
      Meneghetti et al. (2014) &$4.07\pm 0.38$ & $0.99\pm 0.12$ & $3.77\pm 0.35$ & $1.06\pm 0.13$ & $4.15\pm 0.40$ & $0.97\pm 0.12$
\enddata
\tablecomments{Predicted concentration $\llangle c_{\rm 200c}\rrangle$ and observed-to-predicted ratio $c^{(\rm obs)}/c^{(\rm sim)}$ for the 16 CLASH X-ray-selected clusters. The best-fit NFW model is used for a baseline comparison with the numerical simulations. Here the theoretical predictions from \citet{Duffy+2008}, \citet{Bhatt+2013}, and \citet{DeBoni+2013} are based on DM-only simulations, and those from \citet{Meneghetti2014clash} are based on nonradiative simualtions of DM and baryons.}
\end{deluxetable*}